\documentclass{ptephy}

\usepackage{graphics} 

\begin{document}

\title{Neutrino Spectroscopy with Atoms and Molecules}

\author{\name{Atsushi Fukumi}{1}, 
\name{Susumu Kuma}{2},
\name{Yuki Miyamoto}{3},
\name{Kyo Nakajima}{2},
\name{Itsuo Nakano}{3,4}, 
\name{Hajime Nanjo}{5},
\name{Chiaki Ohae}{3},
\name{Noboru Sasao}{2,3}\thanks{Corresponding author.},
\name{Minoru Tanaka}{6},    
\name{Takashi Taniguchi}{2}\thanks{Deceased.},
\name{Satoshi Uetake}{3,4}, 
\name{Tomonari Wakabayashi}{7},
\name{Takuya Yamaguchi}{3},
\name{Akihiro Yoshimi}{2,3}
and 
\name{Motohiko Yoshimura}{4}
}

\address{
\affil{1}{Kawasaki College of Allied Health Professions, Kurashiki, Okayama 701-0194, Japan}
\affil{2}{Research Core for Extreme Quantum World, Okayama University, Okayama 700-8530, Japan}
\affil{3}{Graduate School of Natural Science and Technology, Okayama University, Okayama 700-8530, Japan}
\affil{4}{Center of Quantum Universe, Okayama University, Okayama 700-8530, Japan}
\affil{5}{Department of Physics, Kyoto University, Kyoto 606-8502, Japan}
\affil{6}{Department of Physics, Osaka University, Toyonaka, Osaka 560-0043, Japan}
\affil{7}{Department of Chemistry, Kinki University, Higashi-Osaka, Osaka 577-8502, Japan}
\email{ sasao@fphy.hep.okayama-u.ac.jp}
}

\begin{abstract}%
We systematically investigate the new experimental method of using atoms or molecules
to measure important parameters of neutrinos still undetermined;
the absolute mass scale, the mass hierarchy pattern (normal or inverted),
the neutrino mass type (Majorana or Dirac), and
the CP violating phases including Majorana phases.
Most of these observables are difficult to measure in
neutrino oscillation experiments.

There are advantages of atomic targets, 
due to the closeness of available atomic energies  to
anticipated neutrino masses, over nuclear target experiments
such as the end point spectrum of $\beta$ decay and
two-electron line spectrum in the neutrinoless double $\beta$ decay,
both of which address some of the overlapping objectives
with atomic/molecular experiments.
Disadvantage of using atomic targets, the smallness of rates, 
is overcome by the macro-coherent amplification mechanism.

The atomic or molecular process we use is a cooperative
deexcitation  of a collective body of atoms
in a metastable level $|e\rangle$ emitting a neutrino pair and a photon;
$|e\rangle \rightarrow |g\rangle + \gamma + \nu_i \nu_j$
where $\nu_i$'s are neutrino mass eigenstates.
The macro-coherence is developed by
trigger laser irradiation of two colors, which frequently causes  
two-photon process $|e\rangle \leftrightarrow |g\rangle + \gamma +\gamma\,,
|e\rangle + \gamma \leftrightarrow |g \rangle + \gamma$
inside the target.
We discuss 
important aspects of the macro-coherence development in detail,
by setting up the master equation for the target
Bloch vector (whose components are population difference and
medium polarization) and propagating electric field.
Our master equation includes effects of phase decoherence
of medium polarization and decay of population difference.

The spectral rate (the number of events per unit time) of macro-coherent radiative emission of neutrino pair 
has three parts, and is given by
a factorized formula of the form,
(overall $\omega$ independent rate denoted by $\Gamma_0$) 
$\times$ (spectral shape function denoted by $I(\omega)$) 
$\times$ (time evolving dynamical factor),
where $\omega$ is the photon energy.
The constant factor $\Gamma_0$ determines the overall rate in the
unit of 1/time, and for Xe it is of order,
$1\,\mathrm{Hz}\,(n/10^{22}\mathrm{cm}^{-3})^3\,(V/10^2\mathrm{cm}^3)$.
The dynamical factor is time dependent and is given by
the space integrated quantity over the entire
target, of  the product of magnitude squared
of coherent polarization and field strength
(in the unit of the maximally extractable energy density)
stored inside the target.
The asymptotic value of time evolving dynamical factor is given by
contribution of field condensate accompanied by
macroscopic coherence, which is calculated
using the static limit of the master equation.
With an appropriate choice of heavy target atoms or molecules
such as Xe and I$_2$ 
that has a large M1$\times$E1 matrix element between $|e\rangle $ and $|g\rangle $,
we show that one can determine three neutrino masses along with
distinction of the mass hierarchy pattern (normal or inverted)
by measuring the spectral shape $I(\omega)$.
If one uses a target of available energy of
a fraction of 1 eV, the most 
experimentally challenging observable, the
Majorana CP phases, may be determined, comparing detected rate 
with differences of theoretical expectations which exist at the level of several \%.
The Majorana CP violating phase is expected crucial to the 
understanding of the matter-antimatter imbalance of our universe.
Our master equation, when  applied to E1$\times$E1 transition such
as pH$_2$ vibrational $Xv=1 \rightarrow 0$, can describe explosive PSR events
in which most of the energy stored in $|e\rangle$ is released in
duration of order a few nano seconds.

The present paper is intended to be self-contained explaining
some details of related theoretical works in the past, and further reports
new simulations and
our ongoing experimental efforts of the project to realize the neutrino mass
spectroscopy using atoms/molecules.

\end{abstract}
\subjectindex{neutrino,  Majorana particle, lepto-genesis, macro-coherence, paired super-radiance}
\maketitle

\vspace{-10mm} 
{\bf Contents} 
\begin{center}
\begin{tabular}{llr}
\hline
\hline
 1.   &  Introduction and overview & p.\pageref{Sec:Introduction} \\
      &  1.1\ \ Remaining important problems in neutrino physics and our objective & \\
      &  1.2\ \ Radiative emission of neutrino pair (RENP) & \\ 
      &  1.3\ \ Paired super-radiance (PSR) to be controlled  & \\
      &  1.4\ \ More on RENP and PSR & \\
      &  1.5\ \ Relation to cosmology and outlook & \\     
 2.   &  Theoretical aspects of paired super-radiance & p.\pageref{Sec:PSR-Theory} \\
      &  2.1\ \ Super-radiance and extension to two-photon emission process & \\
      &  2.2\ \ Master equation for paired super-radiance & \\
      &  2.3\ \ Dynamics of PSR & \\
 3.   &  Theory of macro-coherent radiative emission of neutrino pair (RENP) & p.\pageref{Sec:RENP-Theory} \\
      &  3.1\ \ Coherent neutrino pair emission from atoms/molecules & \\
      &  3.2\ \ Neutrino properties extractable from the photon spectrum & \\
      &  3.3\ \ Estimation of the dynamical RENP factor based on asymptotic solution & \\
 4.   &  Experimental aspects of PSR and RENP & p.\pageref{Sec:Experiments} \\
       &  4.1\ \ Overview and strategy towards PSR/RENP experiments & p.\pageref{Subsec:Experimet-Overview} \\
       &  4.2\ \ PSR experiment with para-hydrogen molecule &  p.\pageref{Subsec:Experimet-pH2} \\
       &  4.3\ \ Towards RENP experiment with Xe &  p.\pageref{Subsec:Experimet-Xe} \\
 5.   &   Summary and prospects & p.\pageref{Sec:Summary} \\  
 \hline
\multicolumn{2}{l}{Appendix A:\ \ Electroweak interaction under nuclear Coulomb potential}  
&   p.\pageref{App:EW-interaction}\\
\multicolumn{2}{l}{Appendix B:\ \ Mathematical structure of Maxwell-Bloch equation}  
&   p.\pageref{App:MB-equations}\\
\multicolumn{2}{l}{Appendix C:\ \ Molecules for RENP}  
&   p.\pageref{App:Molecules-RENP}\\
\multicolumn{2}{l}{Appendix D:\ \ Coherence time measurements of para-hydrogen vibrational levels}  
&   p.\pageref{App:Experimet-pH2-T2}\\
\multicolumn{2}{l}{Appendix E:\ \ Experimental studies on PSR/RENP targets in condensed phases}  
&   p.\pageref{App:Condensed-phases}\\
\multicolumn{2}{l}{\hspace{0.3cm} Appendix E.1:\ \ Bismuth in neon matrix}  
&   p.\pageref{App:Bi-Neon}\\
\multicolumn{2}{l}{\hspace{0.3cm} Appendix E.2:\ \ HF molecule trapped in solid pH$_{2}$}  
&  p.\pageref{App:HF-molecule}  \\
\multicolumn{2}{l}{\hspace{0.3cm} Appendix E.3:\ \ Nitrogen atom in fullerene C$_{60}$}  
&  p.\pageref{App:N-C60}  \\
\hline
\hline
\end{tabular}
\end{center}

\section{Introduction and overview}\label{Sec:Introduction}
\newcommand{\cal}{\mathcal}
\subsection{Remaining important problems in neutrino physics
and our objective}
The present status of neutrino mass matrix is summarized by
the following central values  
measured  by oscillation experiments \cite{nu-matrix}, \cite{neutrino 2012}:
\begin{eqnarray}
&&
s_{12}^2 = 0.31\,, \hspace{0.5cm}
s_{23}^2 = 0.42  \,, \hspace{0.5cm}
s_{13}^2 = 0.024 \,, 
\label{neutrino mixing angle parameters}
\\ &&
\Delta m_{21}^2 = 7.5 \times 10^{-5} {\rm eV}^2 \,, \hspace{0.5cm}
|\Delta m_{31}^2| = 2.47 \times 10^{-3} {\rm eV}^2\,.
\label{oscillation results}
\end{eqnarray}
The usual notation of angle factors is used;
$s_{ij} = \sin \theta_{ij}$ and $c_{ij} = \cos \theta_{ij}$.
The definition of the neutrino mixing (given by $U$) 
and mass (${\cal M}_{\nu} $) matrix is  given by \cite{nu-matrix}
\begin{eqnarray}
&&
U = 
\left(
\begin{array}{ccc}
1 & 0 & 0 \\
0 &  c_{23}& s_{23} \\
0 & -s_{23} & c_{23}
\end{array}
\right)
\left(
\begin{array}{ccc}
c_{13} & 0 & s_{13}e^{-i\delta} \\
0 & 1 & 0 \\
-s_{13}e^{-i\delta} & 0  & c_{13}
\end{array}
\right)
\left(
\begin{array}{ccc}
c_{12} & s_{12} & 0 \\
-s_{12} & c_{12} & 0 \\
0 & 0 & 1
\end{array}
\right) P
\,,
\\ &&
P = 
\left(
\begin{array}{ccc}
 1& 0 &0  \\
0 & e^{i\alpha} & 0 \\
0 & 0 &  e^{i\beta}
\end{array}
\right)  \hspace{0.3cm} {\rm for \; Majorana \; neutrinos}\,,
\hspace{0.3cm} 
= 1  \hspace{0.3cm} {\rm for \; Dirac \; neutrinos}\,,
\\ &&
{\cal M}_{\nu} = U{\cal M}_D U^{\dagger}
\,,
\end{eqnarray}
(where ${\cal M}_D$ is the diagonalized mass matrix).
Neutrino masses are ordered by $m_3 > m_2 > m_1$
for the normal hierarchical mass pattern (NH) and
$ m_2 > m_1 > m_3$ for the inverted hierarchy (IH).
For convenience we define the smallest mass by
$m_0$, which is $= m_1$ for NH and $= m_3$ for IH.

The ongoing and planned experiments to measure the neutrino masses using
nuclei as targets are  in two directions;
(1) measurement of the beta spectrum near the end point sensitive to 
both Dirac and Majorana masses,
(2) neutrinoless double beta decay near the end point of
two electron energy sum, sensitive to Majorana masses alone.
In the neutrinoless double beta decay one attempts to measure the
following parameter combination called the effective neutrino mass
\cite{0nu mass};
\begin{eqnarray}
&&
\left|\sum_i m_i U_{ei}^2 \right|^2 =
m_3^2 s_{13}^4 + m_2^2 s_{12}^4 c_{13}^4
 + m_1^2c_{12}^4 c_{13}^4 + 2m_1 m_2s_{12}^2 c_{12}^2 c_{13}^4 \cos (2\alpha)
\nonumber \\ &&
\hspace*{1cm}
 + 2m_1m_3 s_{13}^2 c_{12}^2 c_{13}^2\cos 2(\beta-\delta) + 2m_2m_3 s_{13}^2s_{12}^2 c_{13}^2
 \cos2(\alpha-\beta+\delta)
\,,
\end{eqnarray}
using our convention of Majorana phases.
The best upper limit of neutrino mass scale is derived
from cosmological arguments, and is $\sim 0.58$eV (95\% CL) 
\cite{cosmology-mass-bound}.

Despite of this remarkable success in
neutrino physics,
there are still many important questions to be answered.
(1) Whether the nature favors
either of the neutrino mass type, Dirac (described by 4 component spinor equation) or 
Majorana mass (described by 2 component spinor),
is unknown despite of its vital importance to lepto-genesis theory
\cite{fy-86}, \cite{davidson-ibarra}.
(2) Two important parameters of the neutrino mass matrix ${\cal M}_{\nu} $,
the smallest mass $m_0$ and the CP violating phase $\delta$,
are inaccessible experimentally in the near future.
(3) A definite
principle of measuring the additional CP phases $\alpha, \beta$ 
\cite{majorana cp phases},\cite{0nu mass} intrinsic to
the Majorana neutrino has  not been proposed as yet.

These are challenged by our method of neutrino mass spectroscopy
\cite{my-rnpe},\cite{my-rnpe1},\cite{rnpe-pc}
using atoms or molecules instead of nuclei
as targets, as closely explained in the present article.

In the rest of this Section
we overview theoretical aspects by
summarizing essence of theoretical
sections, Sec.2 and Sec.3.
Experimental status of our project
is summarized in Sec.4.

\subsection{Radiative emission of neutrino pair (RENP)}
\begin{figure}
\begin{center}
\includegraphics[width=0.4\textwidth]{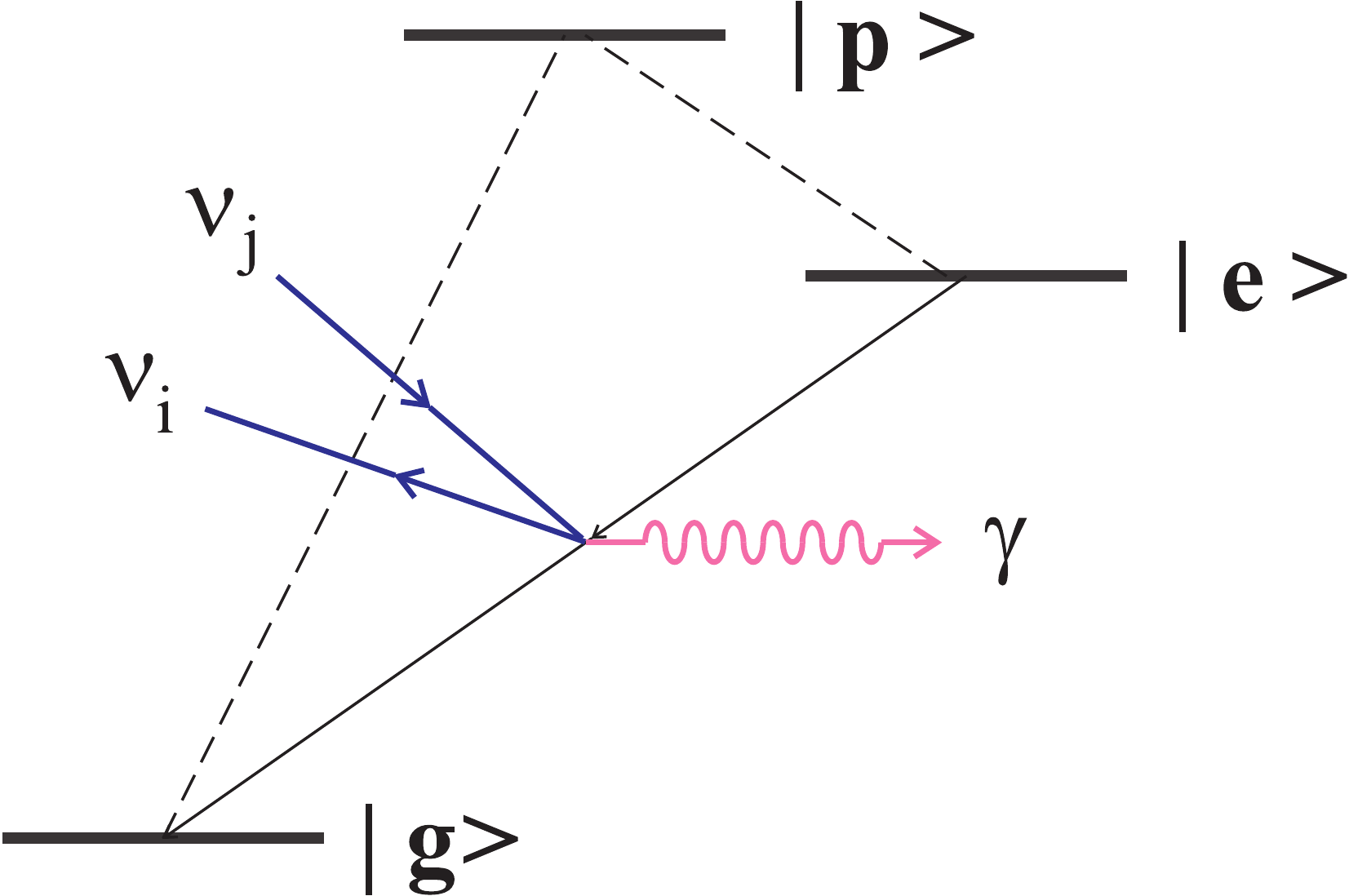}
\end{center}
   \caption{$\Lambda-$type  atomic level for RENP
$|e \rangle \rightarrow |g\rangle + \gamma + \nu_i\nu_j$
with $\nu_i$ a neutrino mass eigenstate.
   Dipole forbidden transition 
$|e \rangle \rightarrow |g\rangle + \gamma + \gamma$ 
may also occur via weak M1$\times$ E1  couplings to
virtual intermediate state $|p\rangle$.
}
   \label{lambda-type atom for renp}
\end{figure}

The atomic process we use for experiments
is $|e \rangle \rightarrow |g \rangle + \gamma + \nu_i\nu_j$,
as illustrated in Fig.\ref{lambda-type atom for renp}.
The process exists in the standard electroweak theory 
\cite{review of electroweak theory} without any doubt:
it occurs as a combined effect of weak interaction and QED,
a kind of second order perturbation process when
one regards four Fermi weak process as the first
order process. 
We denote this process by  RENP (radiative emission of neutrino pair).
The radiation-less pair emission $|e \rangle \rightarrow |g \rangle + \nu_i\nu_j$
 is faster as an
elementary process than RENP, but with the aid of macro-coherent
amplification by trigger laser irradiation RENP dominates over
the radiation-less process.
Moreover, the single emitted photon is a key to obtain needed information
on neutrinos, since emitted neutrinos are difficult to detect.
The atomic state $|e\rangle $ is assumed metastable, which
means to us that its lifetime roughly $> 1$msec
(its optimal value to be determined by repetition cycle of
excitation and trigger irradiation in actual experiments).
Besides a single photon $\gamma$ the final state has
two neutrino mass eigenstates of $\nu_i\,, i= 1, 2,3$.
The crucial key element of our experimental methods is
the ability of resolving neutrino mass eigenstates rather than
flavor eigenstates, as realized by the excellent frequency
resolution of used trigger laser.
This becomes possible by using the technique of
trigger laser irradiation in atomic processes.
There are step function like threshold rises in
the photon energy spectrum and six threshold locations  are at 
\begin{eqnarray}
&&
\omega_{ij} = \frac{\epsilon_{eg}}{2} -\frac{(m_i+m_j)^2}{2\epsilon_{eg}}
\,.
\end{eqnarray}
Here $\epsilon_{ab} = \epsilon_a - \epsilon_b$ is the atomic energy difference
between two states $|a\rangle \,, |b\rangle$.
Threshold locations sensitive to neutrino masses $m_i$
are separated by small photon energies.
For example, 
\(\:
(m_i+m_j)^2/(2\epsilon_{eg}) \sim 5
\:\)
meV for $m_i+m_j = 0.1$eV and $\epsilon_{eg} = 1$eV,
taking a typical atomic energy difference.
One can separate different mass eigenstates  
by fully exploiting the accuracy of frequency in the range of
$\omega \leq \omega_{11}$ (the largest threshold), used
as trigger laser in our proposed experiments.
In our approach one does not
need this order of precision of
detected photon energy.
From the continuous spectral shape of the single photon rate
one can determine both neutrino masses and mixing angles, along
with Majorana phases $\alpha, \beta$.

Atomic/molecular  targets have an advantage over conventionally used nuclei in their
closeness of released energy to neutrino masses (expected much smaller than 1 eV).
A demerit of these targets is the weakness of RENP rate $\propto G_F^2 \alpha \epsilon_{eg}^{n}$
with $\epsilon_{eg}$ the available energy of order eV and $n\sim 5$.
The smallness of rate is due to the small Fermi constant $G_F \sim 10^{-23}$eV$^{-2}$.
We use macro-coherence (giving the rate $n^2 V$ with
$n$ the target number density, $V$ the target volume) to overcome this problem.
When the number density $n$ is close to the Avogadro number per cm$^3$,
RENP rate may become measurable.

\begin{figure}
\begin{center}
\includegraphics[width=0.6\textwidth]{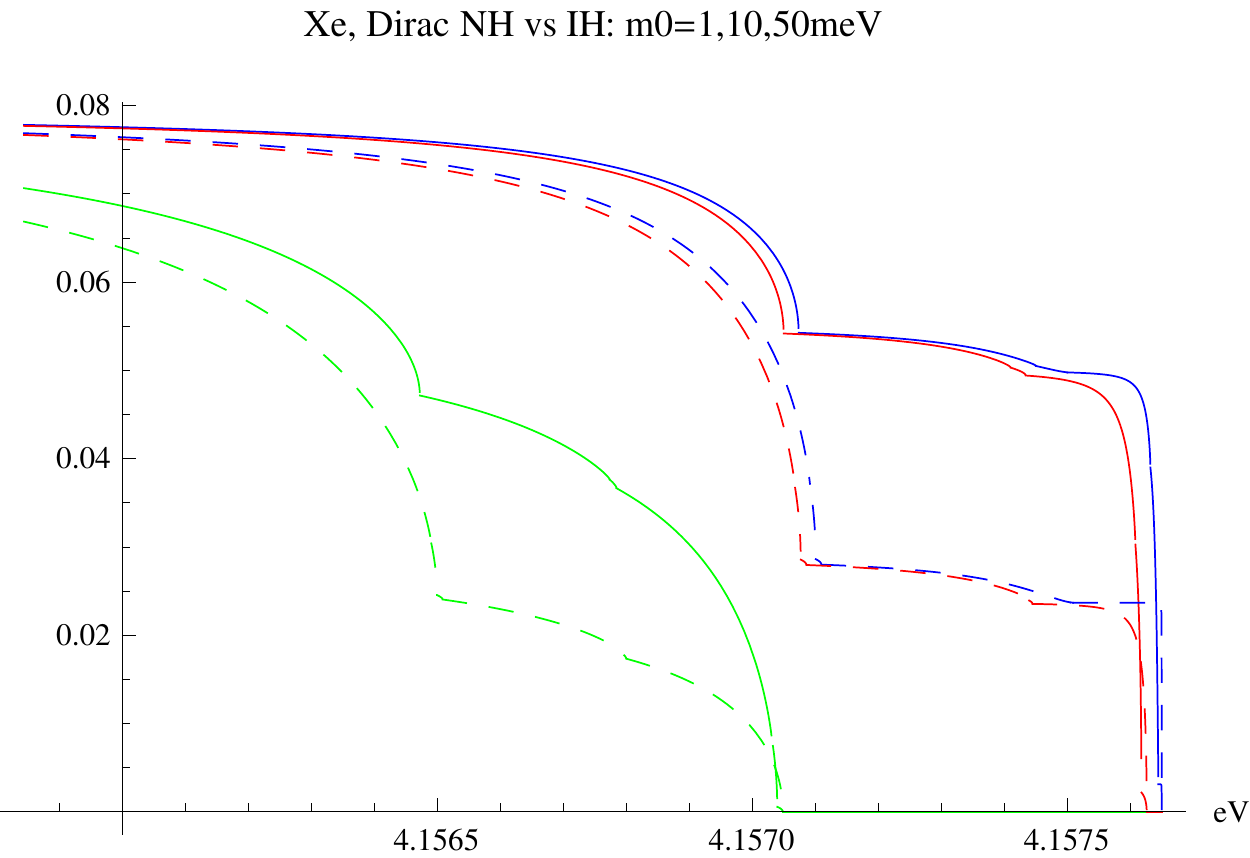}
\end{center}
\caption{RENP dimensionless spectrum function $I(\omega)$ 
near the neutrino pair emission thresholds from Xe level
$ 5p^5(^2P_{3/2})6s ^2[3/2]_2$.
 Neutrinos of the smallest mass of 1, 10 and 50 meV are taken
for the normal (solid curve) and the inverted (dashed curve) hierarchical mass pattern.
}
\label{xe renp ihnh}
\end{figure}

If the macro-coherent amplification works as expected,
the neutrino pair emission accompanied by the photon of energy $\omega$
occurs according to a time dependent rate formula of the factorized form,
\begin{eqnarray}
&&
\Gamma_{\gamma 2\nu}(\omega, t) = \Gamma_0 I(\omega) \eta_{\omega}(t)
\,.
\label{factorized rate}
\end{eqnarray}
The constant RENP rate $\Gamma_0$ may become of order 1 Hz
at the target number density $10^{22}$cm$^{-3}$
(the $\Gamma_0$ value scaling with the number density
$\propto n^3$) and the volume $10^{2}$cm$^{3}$
in the Xe example whose RENP spectrum $I(\omega) $ in
the threshold region is shown in Fig.\ref{xe renp ihnh}.
\footnote{ Neutrino parameters taken for this calculation are somewhat different from the most recent values given above, 	
Eq.~\ref{oscillation results} and are given in Sec.3. }
Relevant levels of Xe are
$| e \rangle = 5p^5(^2P_{3/2})6s ^2[3/2]_2$, 
metastable with lifetime of O[40]sec,
$|g \rangle = 5p^6\, ^1S_0$, and 
intermediate state $|p \rangle = 5p^5(^2P_{3/2})6s ^2[3/2]_1$.
The initial and the final states, $|e\rangle\,, |g\rangle$, have different parities and
the angular momentum difference $\Delta J = 2$.
Excited states here may be described by a pair state of
$6s$ electron and $5p$ hole: both of the initial metastable 
and intermediate states are given by the spin triplet pair with a large breaking of $LS $
coupling scheme.
The dynamical factor $\eta_{\omega}(t)$ shall be discussed later in Sec. 3
based on the solution of the master equation of Sec. 2.

From the spectrum feature of Fig.\ref{xe renp ihnh}
it should not be difficult to measure the absolute neutrino
mass scale and the distinction of normal (NH) and inverted (IH)
mass hierarchy of neutrino mass pattern.
The Majorana vs Dirac distinction is harder for Xe due to a large
energy level difference $\epsilon_{eg}\sim 8.3$ eV.
It is found in \cite{yb-x renp} that
the appropriate energy scale for measurements of Majorana
CP phases is a fraction of eV.

\begin{figure*}
 \begin{center}
\includegraphics[width=0.6\textwidth]{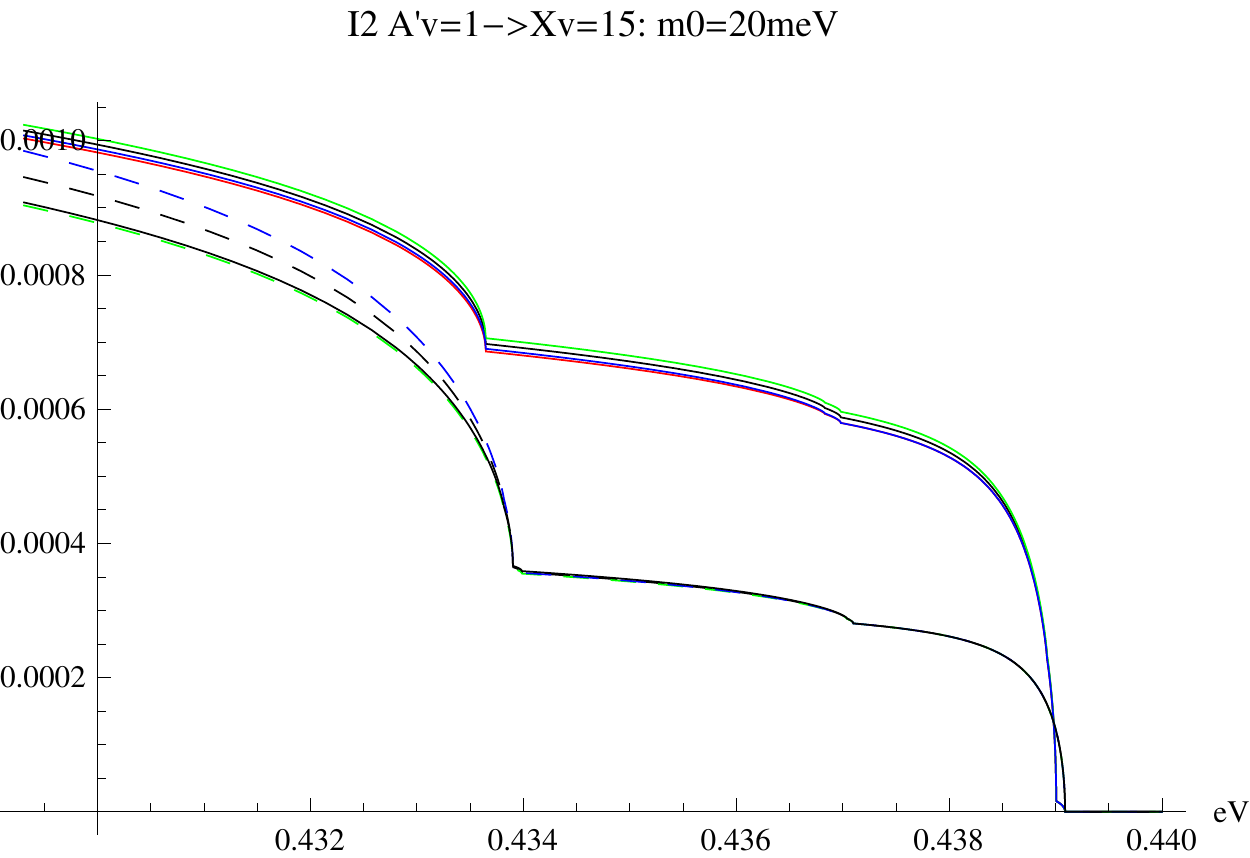}
 \end{center} 
   \caption{I$_2$ RENP spectrum between  A' v=1 and Xv=15.
   The Majorana  (in colored solid for NH and in colored dashed for IH) vs 
Dirac (in black solid for NH and in black dashed for IH) cases are compared.
Three Majorana CP phase
combinations $(\alpha, \beta-\delta) = (0,0)$ (in red),
$(\pi/2,0)$ (in green), and $(0,\pi/2)$ (in blue) are taken, with  the smallest neutrino mass 20 meV. 
The vertical scale is in arbitrary units.
}
   \label{i2 renp spectral md 20meV}
\end{figure*}

Attractive candidates of targets may be found in molecules.
Molecules are interesting due to a rich vibrational and rotational
band structure with much smaller level spacing than
the electronic ones.
We illustrate RENP spectrum calculation for I$_2$ 
electronic transition in Fig.\ref{i2 renp spectral md 20meV}.
Difference of spectrum rates for different CP Majorana phases is a several to 10 \% for IH.

%

\subsection{Paired super-radiance (PSR) to be controlled}
\begin{figure*}
 \begin{center}
\includegraphics[width=0.6\textwidth]{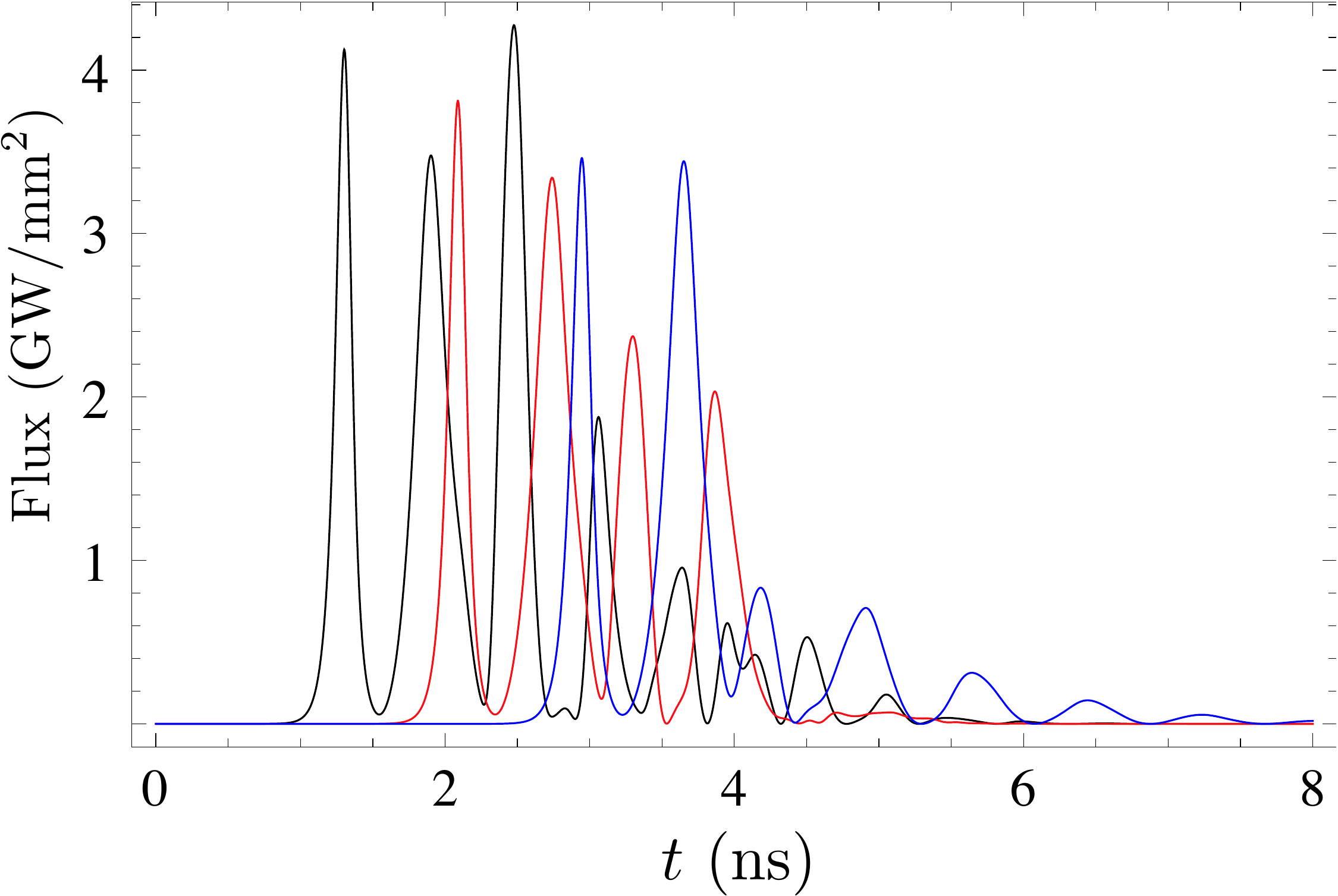}
\end{center}
\caption{
Time-evolving PSR output flux resulting from the
   symmetric trigger irradiation at two target ends, of the power range, 
$10^{-12}\sim 1$Wmm$^{-2}$, under the conditions of the target number density
$n=1\times 10^{21}$cm$^{-3}$,  target length $= 30 $cm, 
relaxation times $T_2=T_3=10, T_1=10^3$ ns's, and the initial polarization, $r_1^{(0)}=1$
and all other Bloch vector components = 0.
   Depicted outputs from 1 Wmm$^{-2}$ trigger power in black, 
from $10^{-6}$Wmm$^{-2}$ in  red,
   and from $10^{-12}$Wmm$^{-2}$ in  blue
are displaced almost equi-distantly in the first peak positions.
Vibrational transition $Xv=1 \rightarrow Xv=0$ of pH$_2$ is considered.
}
   \label{power dependence of PSR}  
\end{figure*}

It is crucial for the success of our method to
control a twin process, PSR (paired super-radiance),
$|e \rangle \rightarrow |g \rangle + \gamma + \gamma$. 
PSR event is interesting in itself.
With the macro-coherence (in which the coherent volume is not
wavelength limited unlike the single photon super-radiance (SR) \cite{sr review}),
typically exhibiting 
the back to back two photon emission with
equal energies at the half of atomic level difference $\epsilon_{eg}/2$
(under the trigger laser of the same frequency).
Emitted two photons are highly entangled especially for
$J=0 \rightarrow 0$ transition.
For long targets of the number density of metastable state over
$\sim 10^{20}$cm$^{-3}$, explosive PSR occurs if
the initial coherence between $|e\rangle$ and $|g\rangle$ is present \cite{psr dynamics}.
Explosive PSR event is characterized by instantaneous release
of energy stored in the upper level $|e\rangle$ into short
pulses of some time structure, taking place
with a time delay after weak trigger irradiation,
as illustrated in Fig.\ref{power dependence of PSR}
for vibrational pH$_2$ transition.
The largest instantaneous output/input power ratio
in this figure is $O[10^{21}]$.
The macro-coherent PSR is fundamentally different from
SR of a single photon emission process
in that the coherent region is extended beyond the
wavelength.

The macro-coherent amplification works for both PSR and RENP.
Its principle may be stated in simple terms as follows
\cite{macro-coherence}.
Coherent emission of particles (photons and neutrinos)
from a collective body of target atoms is characterized by 
quantum mechanical rate (probability per unit time, not necessarily time
independent)
which is a squared quantity of the  sum of amplitudes from
many atoms. When plane wave functions of emitted particles
are extracted, the rate from the collective body is proportional to
\begin{eqnarray}
&&
|\sum_L e^{i\sum_i \vec{k}_i\cdot(\vec{r}-\vec{r}_L)}{\cal A}_L (\vec{r}, t)|^2
\label{amplitude from collected atoms}
\,,
\end{eqnarray}
where $\vec{k}_i$ are momenta of emitted particles including those of
neutrinos and $\vec{r}_L$ is the atomic position.
The atomic amplitude part ${\cal A}_L(\vec{r}, t)$ is expected to be slowly varying
with $\vec{r}_L$ in the wavelength scale of $\vec{k}_i$.
Unlike the incoherent decay in which the phases of ${\cal A}_L (\vec{r}, t)$ from
different atoms are random
(resulting in the summed quantity 
Eq.~\ref{amplitude from collected atoms} of order $N$), the coherent process requires
a high level of phase coherence of this quantity to give order $N^2$
to Eq.~\ref{amplitude from collected atoms}.
When a single photon is involved as the only emitted particle of the process,
$\sum_i \vec{k}_i = \vec{k}$, 
the maximal coherence region is limited by the wavelength
$\sim 1/k$, the inverse of the wave number.
Nonetheless the coherent amplification $\propto N^2$ 
($N$ is the number of atoms within the wavelength limited coherent volume) leads to
an explosive collective decay as first discussed in
the celebrated paper of Dicke and later confirmed experimentally
(see the next chapter on more of this).
A typical effect of Dicke super-radiance is
the sudden de-excitation of all atoms in the wavelength limited coherent volume,
emitted photons confined in a narrow axial direction.
On the other hand,
the wavelength limitation is removed when more than two
particles are involved as in PSR and RENP and 
the momentum conservation $\sum_i \vec{k}_i=0$ holds: the
coherent volume may become truly macroscopic without
the wavelength limitation \cite{macro-coherence}.
In the case of PSR the momentum and the energy conservation
limits emitted two photons to back-to-back direction and
of equal energy $\epsilon_{eg}/2$, if no other phase
memory is present.
The termination of macro-coherent process may be much more
rapidly expedited than in the Dicke case.
Both super-radiance and macro-coherent amplification
is a highly dynamical process, and one cannot describe
its principal feature by a time constant rate.
One needs a time evolving dynamical equation, which we call
the master equation.

The master equation that describes PSR events including the
spatial grating effects (macroscopic polarization varying with the wavelength)
has been derived in \cite{psr dynamics}.
The equation is given in terms of the
 Bloch vector components $R_i(x,t)\,, i=1,2,3$ for medium polarization ($R_1 \pm i R_2$)
and the population difference ($R_3$),
and field mode envelopes  $\vec{E}_i(x,t)$ ($i=R, L$ denoting right- and
left-moving modes).
The single spatial direction, $x$-axis, is selected as the direction of trigger
irradiation.
Our master equation contains relaxation effects described by
three time constants, $T_3, T_2, T_1$ where the phase decoherence
times, $T_2$ for spatially homogeneous modes and $ T_3$ for 
spatial grating modes 
(both $\ll T_1$) are more important.
The population decay time $T_1$ may effectively
be taken infinitely large in our problem.
It is convenient to rescale these and length/time
variables $x,t$ using dimensionless variables.
The dimensionless Bloch vector is rescaled by dividing the
excited target number density $n$, the field envelope
strength by $\epsilon_{eg}n$, while the length/time by
$ t_* \equiv 2/(\epsilon_{eg}\alpha_{ge}n)$.
Here $\alpha_{ge} \propto$ the product of two transition dipole moments.
In the double limits of large relaxation times and
large target length,
explosive PSR events may occur with an effective
rate proportional to the stored energy in $|e\rangle$/ time duration.
Dependence of an effective PSR rate on the number density 
is then automatically $\propto n^2$ and
the macroscopic target volume is the relevant
coherent volume.
Thus, if the target length $\gg c t_*$
and relaxation time $\gg t_*$,
one may expect explosive PSR,
a new phenomenon which may find interesting
applications, for instance in quantum information.

Explosive events are not the only important
outcome of PSR phenomenon.
It turns out 
that static condensate remains after PSR emission and 
we expect that these condensates are described as 
steady state solutions of our master equation by
taking vanishing time derivatives, namely static
solutions.
If one neglects the spatial grating effect, 
these states are expected to become 
aggregate of many absolutely stable solitons 
as given in \cite{psr dynamics}.
The most important aspect of these condensate states is
their stability against two photon emission, and their
instability for RENP.
By its stability the condensate formation makes
the signal to the background ratio RENP/PSR large.
The condensate formation of large active region for RENP
is then the ideal target state for RENP experiments.
PSR (or rather two photon emission) has dual roles of
importance to realization of RENP;
first as a trigger to expedite RENP and second as a background of RENP
to be rejected.
We shall explain how these conflicting aspects are
reconciled by condensate formation having a very small
leakage flux at target ends.

\subsection{More on RENP and PSR}
Let us clarify further the mechanism of macro-coherent RENP
amplification.
The electroweak amplitude of RENP for a single atom is \cite{my-rnpe}
\begin{eqnarray}
&&
{\cal H}_{\gamma \nu} =
\frac{G_F (\vec{d})_{gp}\cdot\vec{E}(\vec{S})_{pe}\cdot\sum_{ij}a_{ij}\nu_j^{\dagger}\vec{\sigma}\nu_i}{\epsilon_{pg} -\omega}
 \,, \hspace{0.5cm}
a_{ij}=U_{ei}^*U_{ej} - \delta_{ij}/2
\,,
\label{rnpe amplitude}
\end{eqnarray}
where $\vec{S}$ and $\vec{d}$ are spin and dipole operators 
for atomic electron, and
$U$  is the unitary matrix relating the neutrino flavor, for instance
$\nu_e$, to the mass eigenstate $\nu_i$,
containing mixing angles and Majorana CP phases $\alpha, \beta$
like $U_{e2} \propto e^{i\alpha}\,, U_{e3} \propto e^{i(\beta-\delta)}$.
This effective hamiltonian is derived in the second order of
perturbation of the electroweak theory where
both $Z$ and $W$ exchange diagrams of weak interaction are involved \cite{my-rnpe}.
Three atomic levels are involved;
$|e\rangle$ for the initial,  $|p\rangle$ for the virtual intermediate,
and $|g\rangle$ for the final states.
Using the terminology of atomic physics, one would say that
RENP involves M1 $\times$ E1 transition.
Assume momentarily (proved by macro-coherence) both
the energy and the momentum conservation of 3-body RENP
process.
This determines six threshold energies
of neutrino pair emission at the photon energy
$\omega= \omega_{ij} =\epsilon_{eg}/2 - (m_i+m_j)^2/(2\epsilon_{eg}) $.
The threshold rises of rates are determined by the elements of
the neutrino mass matrix, including $\theta_{13}$ and Majorana phases $(\alpha, \beta)$,
as illustrated in Fig.\ref{xe renp ihnh}.

Let us explain how the Majorana/Dirac distinction comes out.
\footnote{
In what follows we explain how the interference term
arises for the pair emission of Majorana fermions using
the two-component formalism of \cite{my-rnpe1}.
In Sec. 3 a similar derivation of the interference
term is given using the four-component formalism
of the constrained, self-conjugate field $\psi^c = \psi$.
Two methods give identical results.
}

The Majorana field \cite{review of electroweak theory}, \cite{my-rnpe1}
can be decomposed in terms of plane wave modes as
\begin{eqnarray}
&&
\psi^M(\vec{x},t) = 
\sum_{i, \vec{p}} \left( u( \vec{p}) e^{-iE_i t + i\vec{p}\cdot\vec{x}} b_i( \vec{p}) 
+ u^c( \vec{p})e^{iE_i t - i\vec{p}\cdot\vec{x}}   b_i^{\dagger}( \vec{p})
\right)
\,,
\end{eqnarray}
where the annihilation $b_i( \vec{p})$ and creation $b^{\dagger}_i( \vec{p})$ operators of the same type appears in the expansion
(the index $i$ gives the $i-$th neutrino of mass $m_i$, and the helicity
summation is suppressed for simplicity).
The concrete form of the 2-component conjugate wave function 
$u^c \propto i\sigma_2 u^*$
is given in \cite{my-rnpe1}.
The Dirac case is different involving different type of operators $b_i(\vec{p})$ and $d^{\dagger}_i(\vec{p})$:
\begin{eqnarray}
&&
\psi^D(\vec{x},t) = \sum_{i, \vec{p}} \left( u(\vec{p}) e^{-iE_i t + i\vec{p}\cdot\vec{x}} b_i(\vec{p}) 
+ v(\vec{p}) e^{iE_i t - i\vec{p}\cdot\vec{x}}  d_i^{\dagger}(\vec{p})
\right)
\,.
\end{eqnarray}
Neutrino pair  emission amplitude of modes $i\vec{p}_1, j\vec{p}_2$ contains two terms
in the case of Majorana particle:
\begin{eqnarray}
&&
b_i^{\dagger}b_j^{\dagger}
\left( a_{ij} u^*(\vec{p}_1)u^c(\vec{p}_2) - a_{ji}u^*(\vec{p}_2)u^c(\vec{p}_1) \right)
\,, 
\end{eqnarray}
and its rate involves
\begin{eqnarray}
&&
 \frac{1}{2}|\left( a_{ij} u^*(\vec{p}_1)u^c(\vec{p}_2) - a_{ji}u^*(\vec{p}_2)u^c(\vec{p}_1) \right|^2
\nonumber \\ &&
=  \frac{1}{2}|a_{ij}|^2\left( |\psi(1,2)|^2 + |\psi(2,1)|^2 \right)
- \Re (a_{ij})^2\left( \psi(1,2) \psi(2,1)^*
\right)
\,, 
\end{eqnarray}
where the relation $a_{ji} = (a_{ij})^*$ is used and 
$\psi(1,2)  = u^*(\vec{p}_1)u^c(\vec{p}_2)$.
Result of the helicity sum $\sum \left( \psi(1,2) \psi(2,1)^*\right)$
is given in \cite{my-rnpe1}, which then gives
the interference term $\propto \Re (a_{ij})^2$.
The first term $\propto |a_{ij}|^2$ is common to
the Dirac and the Majorana neutrino.

The macro-coherence is developed by irradiation
of two trigger lasers of frequencies $\omega_i$ with the relation
$\omega_1+\omega_2 = \epsilon_{eg}$.
The development of macro-coherence works for any
combination $(\omega_1,\omega_2)$ and not restricted to $\omega_1=\omega_2$.
The frequency at the red side $\omega_1$ ($ < \epsilon_{eg}/2$)
is set for detected RENP photon energy.
The effective interaction of two fields $\vec{E}_i$ 
of frequency $\omega_i$ with atoms is
given by
\begin{eqnarray}
&&
{\cal H}_{2\gamma} =
\frac{(\vec{d})_{gp}\cdot\vec{E}_1(\vec{m})_{pe}\cdot\vec{E}_2}{\epsilon_{pg} -\omega_2}
\equiv (\vec{E}_1)_i (\alpha_{ge})_{ij}(\vec{E}_2)_j
\,,
\label{2g effective hamiltonian}
\end{eqnarray}
where $\vec{m}=g e\vec{S}/2m_e$ is the magnetic dipole operator,
$\vec{S}$ being the electron spin operator.
$(\alpha_{ge})_{ij}$ is a tensor giving the interaction strength of PSR
\cite{psr dynamics} \cite{2g propagation}.
The added hamiltonian ${\cal H}_{2\gamma }+{\cal H}_{\gamma \nu}$
($\vec{E} = \vec{E}_1$ and $\omega=\omega_1$ in Eq.~\ref{rnpe amplitude}\,)
describes the atom-field-neutrino interaction.
We shall treat effects of ${\cal H}_{2\gamma }$ non-perturbatively,
solving this part exactly by numerical means, 
and ${\cal H}_{\gamma \nu}$ is treated in the first non-trivial order of perturbation.

Two trigger laser irradiation is designed for efficient coherence development.
Depending on the magnitude of the product of dipoles that
appears in Eq.~\ref{2g effective hamiltonian},
there may or may not be significant PSR emission.
What is important for RENP is the later stage after
PSR related activities.
The asymptotic state of fields and target atoms
in the latest stage of trigger irradiation
is described by static solutions of the master equation
for time evolution. In many cases there is a remnant state
consisting of field condensates accompanied with a large
coherent medium polarization.
In the limit of small $T_3$ decoherence time
(relaxation time for grating modes)
this condensate is expected to be identical to a
soliton discovered in \cite{psr dynamics} or their aggregate.
In any event the asymptotic target state is
stable against two-photon emission, but
RENP occurs from any point in the target.

Laser irradiation is continued until $\sim$ several
times the relaxation time $T_2$ (non-grating spatially homogeneous modes) 
of phase coherence and terminated there.
This cycle is repeated to accumulate detectable level of RENP photons.
During a cycle RENP photon is emitted within the whole space region
within the target length because of the instability of condensates against RENP.
On the other hand, PSR photons are emitted at two target ends 
due to a leakage flux, and not from the inside of medium,
because condensates inside the target do not emit PSR photons.
This way the signal to the background ratio  RENP/PSR is largely
enhanced by the small leakage energy flux
at two ends due to QED two-photon process.

The important dynamical factor $\eta_{\omega}(t)$ in Eq.~\ref{factorized rate}
is given by the bulk integral over all target atoms/molecules of 
the quantity, the absolute magnitude squared of macroscopic
polarization $\times$ the total field strength, both in
the dimensionless units, as fully explained in Sec. 3.
The time dependence of $\eta_{\omega}(t)$ disappears
in the asymptotic time limit, and this  limit is
described by the static solution of our master equation.
Non-trivial static limit exists and the asymptotic state consists of
the field condensate supported by a macroscopic
medium polarization.
The state is identified as the soliton of \cite{psr dynamics}
in some parameter limit.

Besides the large bulk/edge rate enhancement there exists
a method to selectively detect RENP against two photon
emission (namely weak perturbative PSR).
The magnetic field may be used to
verify parity violating (PV) effects such as
correlated emission of photons to the field axis, effect intrinsic to
the weak process.
Evidently PV effects such as emergence of circular polarization
and angular correlation of emitted photons to
atomic spin are critical to prove that the process involves
weak interaction.

We mention an important point for the target choice
of RENP.
There is a technical, but important reason why one 
has to look for heavy atoms or heavy molecules
as candidate targets.
This is explained for $J= 2 \rightarrow 0$ (whose example is the Xe case)
and $J= 0 \rightarrow 0$ (Yb case) transitions
in the following.
The M1$\times$E1 atomic transitions for RENP
go through the angular momentum change of
$J=2 \rightarrow 1$ and $J=1 \rightarrow 0$, or 
$J=0 \rightarrow 1$ and $J=1 \rightarrow 0$.
The M1 transition is governed by atomic matrix elements of
$\langle J\pm 1 | \vec{S} | J \rangle$.
It can be shown that these matrix elements vanish in
the limit of exact $LS$ coupling scheme.
The $LS$ coupling scheme, however, breaks down for heavy atoms/molecules
 \cite{textbook on molecules}, and the overtaking $jj$ coupling scheme evades this constraint,
which explains a possible large M1 matrix elements.

\subsection{Relation to cosmology and outlook}
Our RENP measurements are sensitive to Majorana CP phases
in the following combinations in the photon energy thresholds \cite{rnpe-pc};

\vspace{0.5cm}
\hspace*{1cm}
\begin{tabular}{c|c|c} 
(12) &  (13) & (23) \\ \hline
$c_{12}^2s_{12}^2c_{13}^4\cos 2\alpha$ 
&  $c_{12}^2c_{13}^2s_{13}^2\cos 2(\beta - \delta)$ 
& 
$s_{12}^2c_{13}^2s_{13}^2 \cos 2(\alpha - \beta + \delta)$  
\\ 
\end{tabular}
\vspace{0.5cm}

It would be of considerable interest to compare this
with the following combination of the lepton asymmetry
that often appears in lepto-genesis theory \cite{davidson-ibarra}:
\begin{eqnarray}
&&
\frac{3y_1^2}{4\pi} \left(
- 2 (\frac{m_3}{m_2})^3 s_{13}^2 \sin 2( \alpha - \beta + \delta)
+ \frac{m_1}{m_2} \sin (2\alpha)
\right)
\,.
\end{eqnarray}
While the overall factor $y_1$,
the Yukawa coupling related to the heavy
Majorana fermion, is unknown,
the other combination of neutrino mass parameters
has strong correlation to experimentally measurable
quantities in RENP.

If one succeeds in measurements of RENP,
one may hope and proceed to detect relic neutrinos of 1.9 K
by using the spectrum distortion due to the Pauli blocking effect caused by
the relic sea of cosmic neutrinos \cite{my-taka}.\\

Our experimental strategy towards precision neutrino mass spectroscopy
is first to prove the macro-coherence principle by QED process
of stronger E1 $\times$ E1,
namely the discovery of explosive PSR, and then to
control PSR and create ideal form of  condensates for preparation of RENP.
A good target  of E1 $\times$ E1 type
is  pH$_2$ vibrational transition 
$Xv = 1 \rightarrow 0$ \cite{psr dynamics},
whose molecular properties shall be given in the following sections.

In the rest of this article we shall give
detailed account of theories behind the neutrino mass spectroscopy
and our current experimental status towards this project.
We shall present material in a self-contained way
even at the risk of overlap with
our several past publications on this subject.
Moreover, the material is presented in order to clarify
how we develop experimental steps towards the goal
of precision neutrino mass spectroscopy.
This way we give detailed account of PSR and
condensate formation so crucial to our neutrino mass spectroscopy.

The rest of the article is organized as follows.
In the first two sections we develop theories of
PSR and RENP based on macro-coherent amplification
mechanism.
Throughout these theoretical sections 
the natural unit of $\hbar = c= 1$ is used.
In Sec. 4 experimental aspects of PSR
and RENP and
the  status of our experimental project are described.
Several Appendices give detailed accounts
that supplement the main text.


$\;$ \\ 
\section{Theoretical aspects of paired superradiance}\label{Sec:PSR-Theory}
The twin process of two-photon emission
$|e\rangle \rightarrow |g\rangle + \gamma + \gamma$ 
is important to experimental realization of RENP 
$|e\rangle \rightarrow |g\rangle + \gamma + \nu_i \nu_j$  from two reasons;
(1) QED proof of the macro-coherent amplification mechanism
replacing the much weaker neutrino process, and (2) 
control of this QED process to reduce the major background against RENP.
In this Section we shall focus on theoretical aspects of
macro-coherent two-photon emission, termed
paired super-radiance, or PSR in short.
PSR is very interesting and has its own merits in quantum optics.
The first observation of PSR is our primary goal
prior to commencement of RENP experiments.
Aspects more directly related to real experiments 
shall be discussed in Sec. 4.

We deal with atomic processes, either in PSR or in RENP.
The fundamental interaction of atomic electrons with
the transverse electromagnetic field and the neutrino field
is briefly summarized in Appendix A.
The upshot of this approach is to use
bound or nearly free electron wave functions 
under the influence of nuclear Coulomb potential for
computation of atomic transition matrix elements,
which prevails in the following discussions.

\begin{figure*}
\begin{center}
 \begin{minipage}{7.5cm}
	\includegraphics[width=0.8\textwidth]{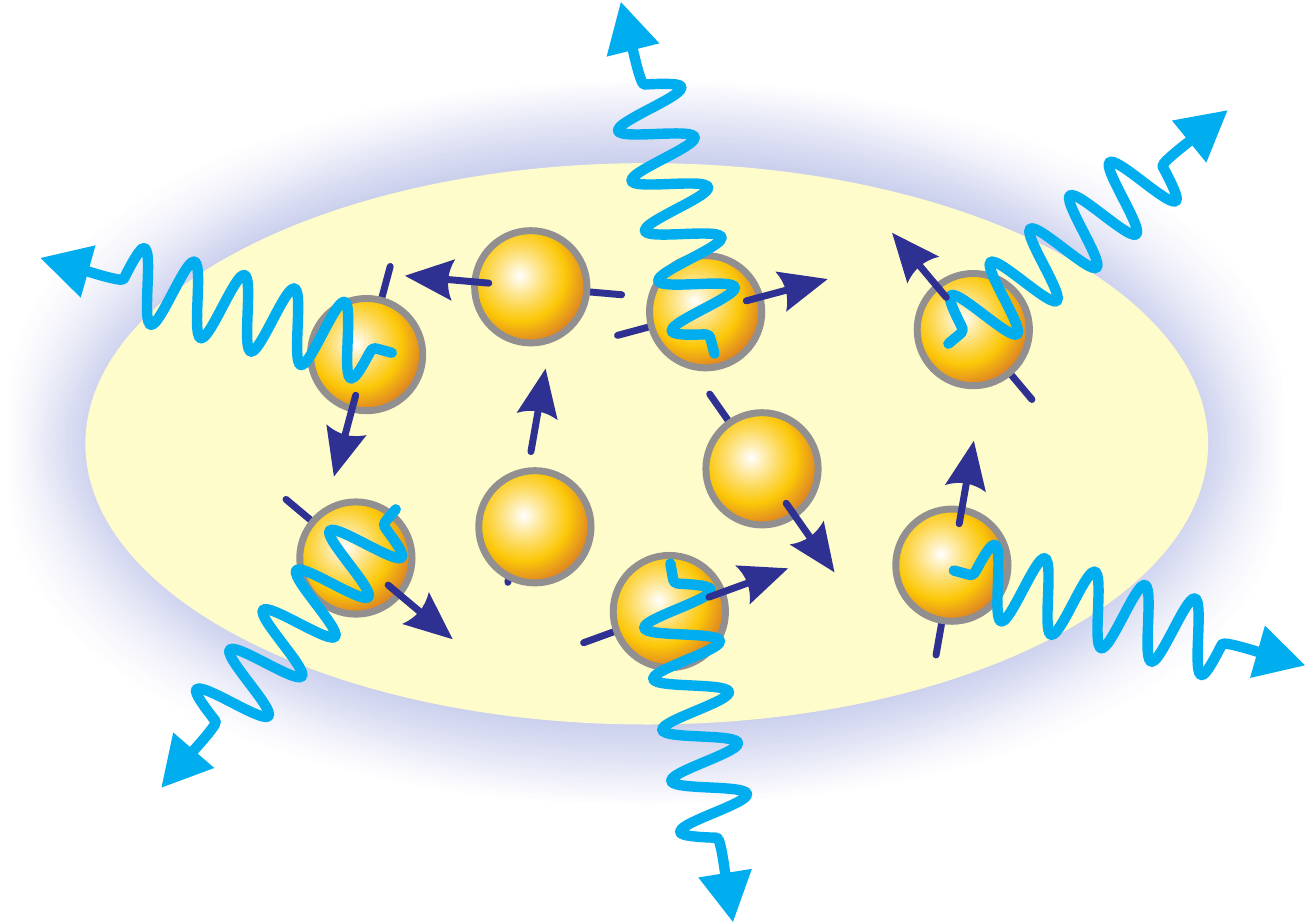}
   	\caption{Cartoon of spontaneous emission from a body of target atoms.}
   	\label{spontaneous emission cartoon}
 \end{minipage}
 \begin{minipage}{7.5cm}
	\includegraphics[width=\textwidth]{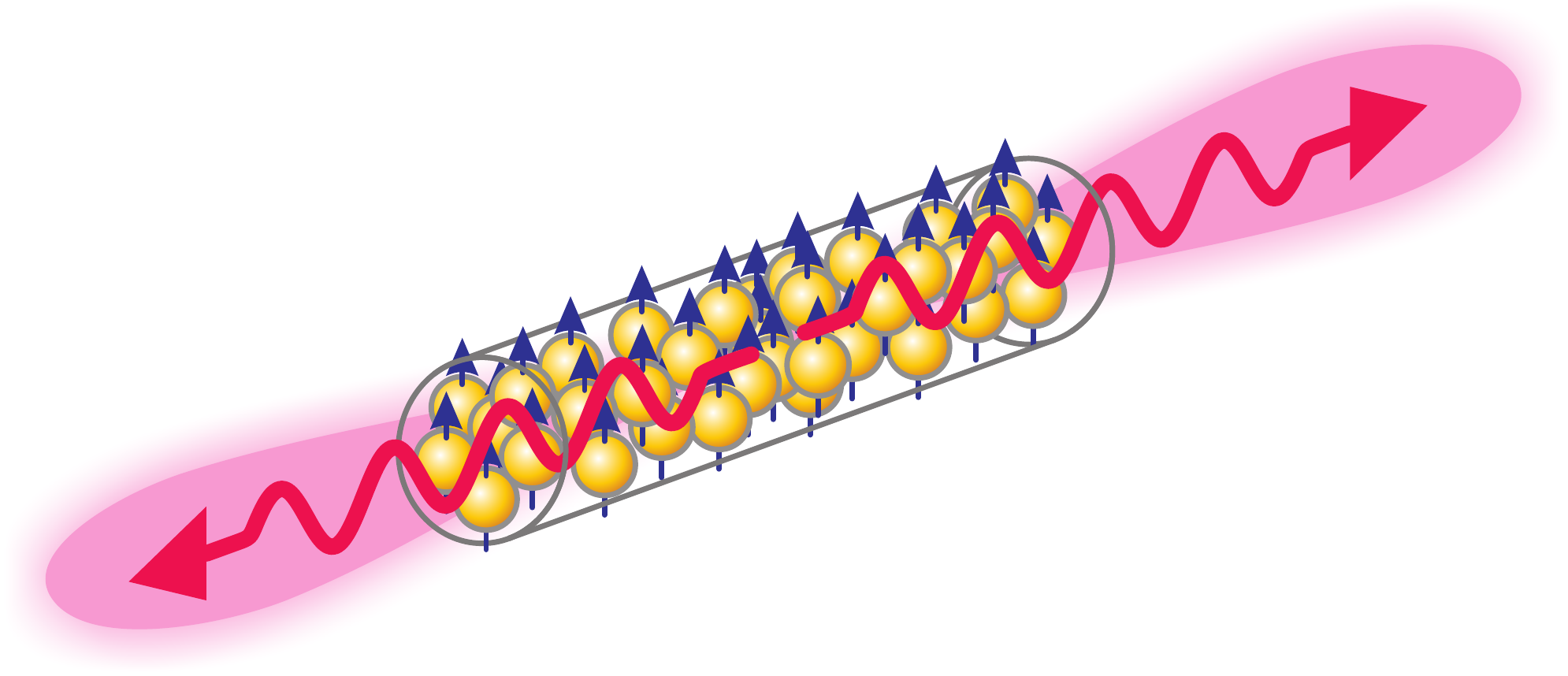}
   	\caption{Cartoon of explosive PSR.}
   	\label{psr cartoon}
 \end{minipage}
\end{center} 
\end{figure*}

It is useful first to compare PSR to the related process of single photon
super-radiance \cite{sr review}, or SR in short.
We shall thus begin with some rudimentary discussions on SR.
These two phenomena have similarities and
differences.
Cartoons showing spontaneous emission and paired super-radiance
are illustrated in Fig. \ref{spontaneous emission cartoon} 
and Fig. \ref{psr cartoon}.
At a superficial level this cartoon may give an impression
that PSR is similar to SR, but
the photon energy, the region of macro-coherence, and how they are realized
is entirely different, as explained below.

\subsection{Super-radiance and extension to two-photon emission process}
\subsubsection{Super-radiance}
The most common de-excitation process of excited atoms as occurs in a dilute
 gas is the spontaneous decay 
$|e\rangle \rightarrow |g\rangle + \gamma$, which arises
in the presence of a non-vanishing  electric dipole (E1) transition moment between
two relevant levels, $|e\rangle$ and $|g\rangle$.
The decay follows, as time $t$ increases, the exponential law $e^{-\Gamma t}$ with
a decay rate $\Gamma$ whose inverse is
a major portion of lifetime 
(the inverse lifetime is given by a sum of rates over other de-excitation processes).
The interaction hamiltonian that appears in the Fermi golden rule for the transition rate
contains a product of dipole matrix element 
and electric field $\vec{E}$ of emitted photon;
$H = \langle g|\vec{d}|e\rangle \cdot \vec{E}$.
The selection rule for this dipole-allowed transition is the parity
change and the angular momentum rule $\Delta J = \pm 1, 0$
(except the strictly forbidden $J= 0 \rightarrow 0$ transition) between two states.
A typical lifetime would be around 10 ns taking
atomic energy difference of 1 eV and atomic size of $10^{-8}$ cm.
Between the same parity states
the E1 transition is forbidden and
the dominant process of radiative decay may be a weaker magnetic dipole (M1) transition
which is caused for instance by an atomic operator $g e \vec{S}\cdot \vec{B}/(2m_e )$
where $\vec{S}$ is the electron spin operator and $ \vec{B}$ is the magnetic
field of emitted photon.
M1 decay rate is typically $\alpha \sim 1/100$ smaller than E1-allowed
decay rate.
Besides the exponential law
the spontaneous decay has an isotropic angular distribution of emitted photon
unless the initial state is spin polarized (even with spin polarization
the anisotropy of distribution is a minor effect).

These features of spontaneous decay are dramatically changed when
atoms decay cooperatively.
Dicke \cite{sr review} pointed out in 1954 the possibility
of what he termed super-radiance (SR).
The point here is that for $N$ excited atoms in a volume within the wavelength $\lambda$
of emitted photons there is no quantum mechanical way to distinguish a host atom of emitted photon.
Assuming no decoherence of phases due to interaction etc,
all $N$ atoms may decay with a definite phase relation preserved from the beginning
of the decay process.
Suppose that several photons emitted in an initial stage
have  nearly the same direction by a chance, for instance along the prolongation axis
in a cylindrical configuration of target atoms, 
then all these atoms may undergo cooperative decay.
Dicke mentioned a nice analogy of this coherent initial state
of $N (\gg 1)$ atoms to an eigenstate of total angular momentum.
In this picture two relevant states are regarded as spin up and spin down state.
Thus, $|e\rangle = |\uparrow \rangle\,, |g\rangle = |\downarrow \rangle$,
and the initial inverted state of $N$ atoms is given by
an eigenstate of the total angular momentum,
\(\:
|J, J \rangle = |\uparrow\rangle \times |\uparrow\rangle
\times |\uparrow\rangle \times \cdots \times |\uparrow\rangle
\,
\:\)
with $J=N/2$.
The radiative decay of the transition
$|e\rangle \rightarrow |g\rangle$ is governed by the lowering operator
$\tilde{J}_-$ of angular momenta.
If any phase decoherence does not occur during
subsequent radiative decays, then one would arrive at
a state again given by eigenstate of the total angular momentum,
\begin{eqnarray}
&&
|J\,, M \rangle = \tilde{J}_-^{J-M}|J\,, J \rangle
\,.
\end{eqnarray}
All these states satisfy the maximal symmetry under interchange of
atoms.
When $M $ becomes $ O[1] \ll N$, the factor
$ |\langle J, M-1|\tilde{J}_-|J,M \rangle|^2$,
hence the rate of the total de-excitation of the collective body,
 becomes proportional to $O[N^2]$, 
as verified by the well known relation,
\(\:
\tilde{J}_- |J\,, M \rangle = \sqrt{(J+M)(J-M+1)} |J\,, M-1 \rangle
\,.
\:\)
The enhancement of decay rate is hence due to $\sqrt{(J+M)(J-M+1)} = O[N]$ 
for $|M|\ll N$.
The underlying situation is that there are $O[N]$ ways to connect
initial and final eigenstates reflecting the indistinguishable nature
of host atoms.
This simple picture breaks down in the presence of de-phasing
such as van-der Waals interaction, because the maximal symmetry
is broken by interaction.

\begin{figure*}
 \begin{center}
\includegraphics[width=0.9\textwidth]{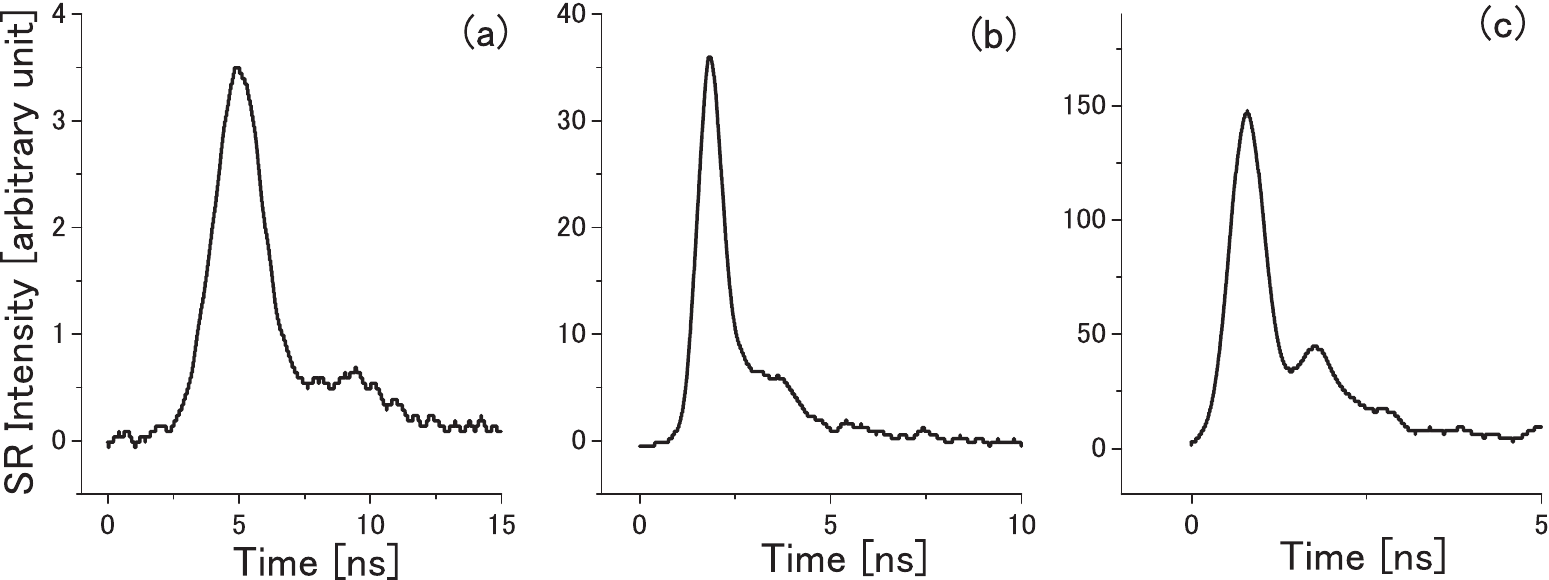}
\end{center} 
   \caption{
	Super-radiance signals \cite{ba ohae} observed in the  de-excitation process from 
        Ba $6s6p\; ^{1}\!P_{1}\,$ to $6s6d\; ^{1}\!D_{2}\,$ state. 
        Note the changes in delay times and peak heights as the density of the excited state increases 
        from (a) to (c).
   }
   \label{ba d-state sr}
\end{figure*}

Two essential features of SR lie in time profile and
directionality of emitted photons.
The temporal feature of the SR de-excitation as illustrated in
Fig.\ref{ba d-state sr}  (our own experimental result)
 is the early termination 
in the form of short pulse at time $\sim 1/(\Gamma N)$
and duration of a short time of the same order.
The initial time of cooperative decay, usually called
the delay time in the literature, may be defined as of this order $ 1/(\Gamma N)$.
Distribution of emitted direction is uni-directional
along the prolongation axis of the cylinder, taken as $x$ direction.
For simplicity we can ignore transverse field effect
parallel to $y, z$ directions
in the cylindrical configuration.

It would be appropriate now to explain fundamental dynamical
variables and their equations to further analyze the process.
We take the continuum limit of atom distribution within the target.
The fundamental dynamical variable of medium is then
the two-component complex wave function at each space point $\psi_a(x,t), a=e,g$,
and the field variable is electric field $E(x,t)$
(ignoring the vectorial nature).
Instead of the wave function it is more convenient to use
their bilinear forms, called the Bloch vector, 
$\vec{R}(x,t) \equiv \psi^*(x,t)\vec{\sigma}\psi(x,t)$
(due to the slow atomic motion the norm $\psi^*(x,t)\psi(x,t)$ 
expressing the atomic number density may be taken constant).
The Bloch vector is a linear combination of the density matrix element,
and it obeys in the pure quantum state
the von Neumann equation equivalent to the Schroedinger equation.
What is left to work out is then to write extended
von Neumann equation and the generalized Maxwell equation
incorporating atom-field interaction and
effects of relaxation in an open system,
generally described by two relaxation
time constants $T_i, i=1,2$ for the two-level system \cite{lindblad}.

Extensive works 
have been devoted to elucidate the origin of initiation for
SR. The commonly accepted view originating from
works of \cite{quantum initiation} is that the
inverted state of $N$ atoms generates quantum fluctuation of
medium polarization of $O[\sqrt{N}]$, inducing field emission due to
E1 coupling in the basic hamiltonian, and it finally develops into
a classical coherence evolution.
This gives rise to a natural time scale $1/\Gamma$ for initiation.
More concretely,
the origin of quantum fluctuation lies in the quantum algebraic relation
between the medium polarization $R_{\pm}$ and the population difference $R_3$
that prevails in the entire volume of target:
\begin{eqnarray}
&&
[R_+\,, R_-] = 2R_3 
\,,
\end{eqnarray}
as readily derived from the identity $| e\rangle \langle g| g \rangle \langle e |=
| e\rangle \langle e |$ etc at each atom.
In the completely inverted case $R_3$ is of order $N$, and one may
take the right hand side of this equation to be a c-number of this order.
This Heisenberg type of algebraic relation  then gives polarization 
a quantum fluctuation of RMS value of
$\langle R_+ R_- \rangle = O[N]$, even if its linear average value vanishes:
$\langle R_{\pm} \rangle = 0$.
This $O[\sqrt{N}]$ fluctuation of polarization necessarily couples to
field, and either zero point fluctuation of field or
spontaneously emitted single photon field may stimulate the growth
of medium polarization.

The classical development at later stages after quantum
initiation is well described by
a non-linear coupled set of partial differential equations \cite{sr review}
for medium polarization $R=R_-$, population difference $Z$, and field 
envelope $E$ (all functions of time $t$ and $x$),
\begin{eqnarray}
&&
\partial_t R = - id EZ \,,
\label{sr-mb 1}
\\ &&
\partial_t Z = i \frac{d}{2} (ER^* - E^*R) \,,
\label{sr-mb 2}
\\ &&
(\partial_t + \partial_x) E = i 2\pi \omega dn R \,,
\label{sr-mb 3}
\end{eqnarray}
where $n$ is the number density of atoms,
and $\omega$ is the frequency of emitted photon, while
$d$  is the dipole moment.
This set of equations is often called the Maxwell-Bloch equation.
The approximations made are (1) rotating wave approximation (RWA)
which omits rapidly oscillating terms in time,
(2) slowly varying envelope approximation (SVEA).
The initial condition for this set of differential equations
is given by random values of initial parameter
$R$ well described by a Gaussian distribution
of width $2\sqrt{N}$ with $N$ the total number of participating
atoms.
The mathematical structure of sine-Gordon equation
exists for this set of equations, which shall be discussed
in Appendix \ref{App:MB-equations} along with auto-modeling solutions.

Leaving aside the quantum origin, one can thus
clarify the spacetime evolution of SR signals,
using the classical Maxwell-Bloch equation and
taking a Gaussian ensemble of initial data set.
Solutions give a large fluctuation of the SR delay time,
pulse width, pulse height,
and it is
only meaningful to compare SR experiments with theory statistically, and not making
a shot by shot analysis.
Due to the quantum nature of initial stage of SR,
the phenomenon is sometimes called superfluorescence(SF) instead of SR.

The super-radiance was first observed in the infrared
region of molecular transition \cite{first sr}, and
subsequently in the optical region, including excited
atoms in solids \cite{sr review}.
It was found later \cite{triggered sr} that
the delay time may be shortened by irradiation of
trigger along the cylinder axis.
It is also expected that SR may be controlled to a certain extent.
We refer to two references in \cite{sr review}
for more comprehensive discussion on SR.

We show in Fig.\ref{ba d-state sr} our Ba experiment.
The global features of numerical solutions along with ringing of pulses 
may be understood by auto-modeling solutions of much simpler
ordinary differential equation, as
explained in Appendix \ref{App:MB-equations}.

An interesting and useful point to our RENP project is to
view SR as an cooperative enhancement of spontaneous decay 
in the presence of developed coherence.
The simple concept of constant decay rate,
namely time independent decay rate per unit time,
disappears in the phenomenon of super-radiance,
because all atoms in the excited state $|e \rangle$
nearly disappears after the pulse emission and
it is meaningless to discuss the latest stage of
evolution.
Another important point to subsequent discussion is
that the mechanism of our (triggered) PSR is more akin to the triggered SR
rather than SF.
This makes much easier to handle PSR without discussing the
stage of quantum initiation as needed for SF.
The trigger is essential to assist PSR, since
the elementary rate of two photon emission involves
an effective coupling in higher order of QED,
and the rate becomes $\sim 10^{16}$sec for para-H$_2 Xv=1$ two-photon
decay.
The quantum fluctuation of medium polarization
coupled to field by a much weaker effective constant than E1
is not sufficient for its macroscopic development
caused by fluctuating fields.
This is a fortunate aspect to theoretical analysis of PSR
dynamics:
the semi-classical approach becomes possible
in all stages of time evolution for polarization
development and amplification by trigger.

\subsection{Master equation for paired super-radiance}
\begin{figure*}
 \begin{center}
\includegraphics[width=0.4\textwidth]{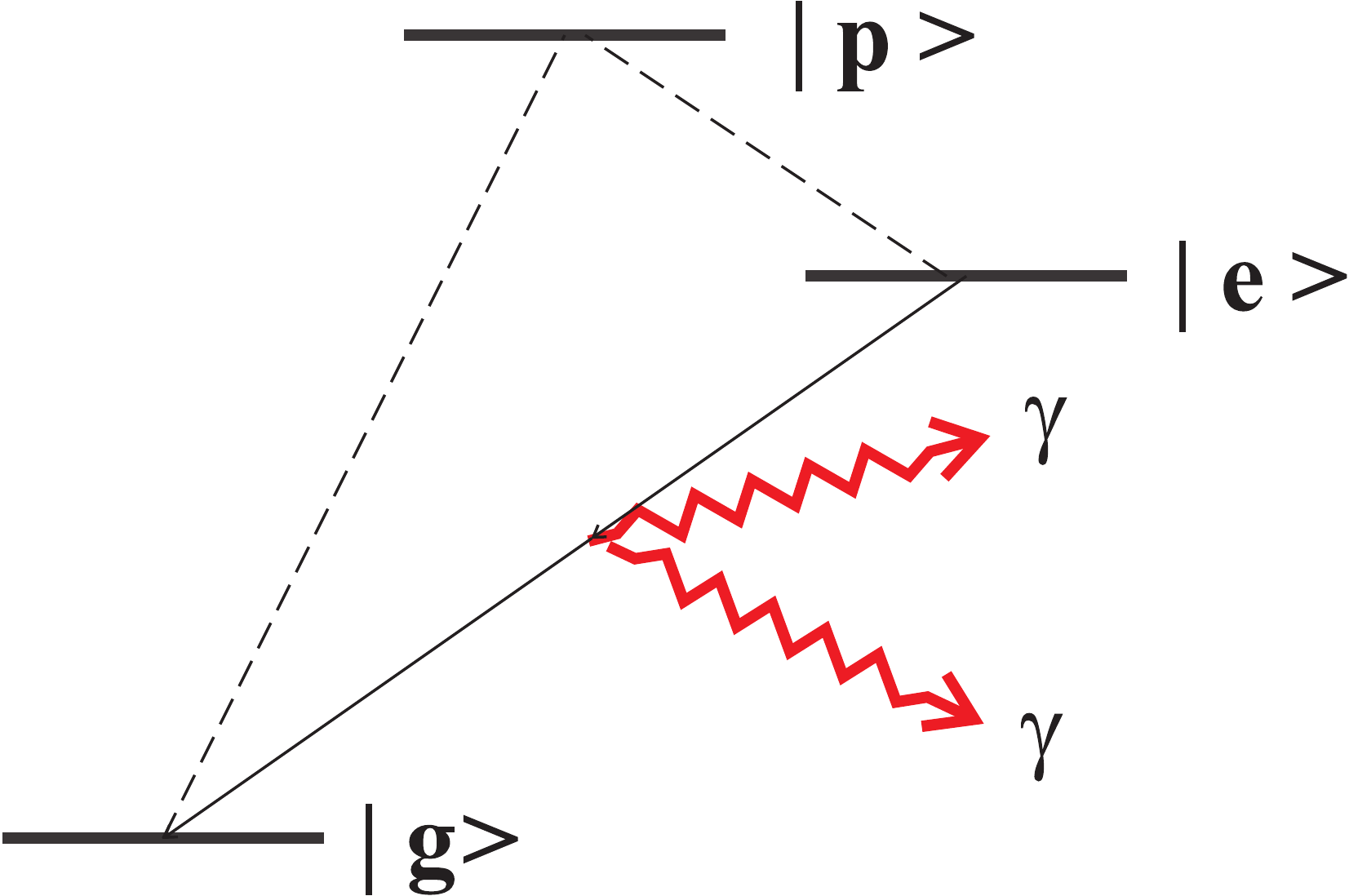}
\end{center} 
   \caption{Level structure for two-photon de-excitation.
   }
   \label{psr lambda-type atom}
\end{figure*}

We now turn to PSR.
Consider three level atom (or molecule) of level energies, 
$\epsilon_p > \epsilon_e >  \epsilon_g$, as shown in
Fig.\ref{psr lambda-type atom}.

When transition between two lower levels, $|e\rangle$ and $|g\rangle$,
is dipole forbidden, the dominant de-excitation process of $|e\rangle$ may be
two-photon decay $|e\rangle \rightarrow |g\rangle + \gamma \gamma$, 
It is described in terms of second order, $2\times 2$ matrix
hamiltonian \cite{psr dynamics},
\begin{eqnarray}
&&
 {\cal H}_I = -
\left(
\begin{array}{cc}
 \alpha_{ee} E^{+}E^{-}
& e^{i \epsilon_{eg}t} \alpha_{ge} (E^{+})^2  \\
e^{-i \epsilon_{eg} t} 
\alpha_{ge} (E^{-})^2
&   \alpha_{gg} E^{+}E^{-}
\end{array}
\right)
\,,
\label{stark hamiltonian}
\\ &&
\alpha_{ge} = \frac{2d_{pe}d_{pg}}
{\epsilon_{pg}+\epsilon_{pe}}
\,, \hspace{0.5cm}
\alpha_{aa}= \frac{2d_{pa}^2\epsilon_{pa}}{\epsilon_{pa}^2 - \omega^2}
\,, \hspace{0.5cm} (a = g\,, e)
\label{1 mode mu}
\,.
\end{eqnarray}
(A slightly different notation is used from \cite{psr dynamics}.)
Here 
\(\:
\epsilon_{ab} = \epsilon_a - \epsilon_b
\;,
\omega = \epsilon_{eg}/2
\,.
\:\)
The upper level $|p\rangle$ that has the largest coupling to
lower levels, $|e\rangle$ and $|g\rangle$, should be dominant,
but other competing intermediate states $|p\rangle$ may also contribute
to the hamiltonian $ {\cal H}_I$.
In the case of trigger for RENP M1 dipole coupling
is relevant to one of the two couplings, and
$d_{pa}$ is replaced by the magnetic dipole moment $\mu_{pa}$,
typically $O(1/100)$ smaller.
If we ignore $\omega$ dependence,
the quantities $\alpha_{ab}\,, a,b= e,g$ coincide with polarizability.
For simplicity we took isotropic medium and
linearly polarized fields, taking $\vec{E}^{\pm}$ as scalar
functions $E^{\pm}$.
The diagonal part $\propto \alpha_{aa}$ of this hamiltonian describes
AC Stark energy shifts, while off-diagonal parts $\propto \alpha_{ge}$ are
effective coupling for two photon emission and absorption.
This hamiltonian (\ref{stark hamiltonian}) has been derived 
adopting the Markov approximation
by eliminating the amplitude of level $|p \rangle$ in the three
level ($|e \rangle, |g\rangle, |p \rangle$) system.

The field $E$ is decomposed into positive and negative frequency
parts in Eq.~\ref{stark hamiltonian}, and it may contain multi-modes, in particular
two counter-propagating modes denoted by $E_R\,, E_L$ 
of the same incident frequency, thus $E=E_R+E_L$.
It is useful at this point
to recall the physical meaning of coupling constants $\alpha_{ab}$
in the interaction hamiltonian.
We first note that annihilation ($a_i$) and creation ($a_i^{\dagger}$) operators 
of photon modes are related to complex fields by
$E_i^+ = a_i \sqrt{\omega/2V}\,, E_i^- = a_i^{\dagger} \sqrt{\omega/2V}$
where $V$ is the quantization volume.
The important equations obtained after SVEA  
are written in terms of envelope functions:
\begin{eqnarray}
&& 
(\partial_t + \partial_x) E_R =
\frac{i\omega}{2}\left( (\frac{\alpha_{ee} + \alpha_{gg}}{2}
n + \frac{\alpha_{ee} - \alpha_{gg}}{2}R_3^{(0)})E_R 
+ \frac{\alpha_{ee} - \alpha_{gg}}{2}R_3^{(+)}E_L 
\right.
\nonumber \\ &&
\hspace*{1cm}
\left.
+ \alpha_{ge} 
\left( (R_1-iR_2)^{(0)} E_L^* + (R_1-iR_2)^{(+)}E_R^* \right)
\right)
\,,
\label{lr mode coupling 1}
\\ &&
(\partial_t - \partial_x) E_L =
\frac{i\omega}{2}\left( (\frac{\alpha_{ee} + \alpha_{gg}}{2}
n + \frac{\alpha_{ee} - \alpha_{gg}}{2}R_3^{(0)})E_L 
+ \frac{\alpha_{ee} - \alpha_{gg}}{2}R_3^{(-)}E_R 
\right.
\nonumber \\ &&
\hspace*{1cm}
\left.
+ \alpha_{ge} 
\left( (R_1-iR_2)^{(0)} E_R^* + (R_1-iR_2)^{(-)}E_L^* \right)
\right)
\,.
\label{lr mode coupling 2}
\end{eqnarray}
Quantities $R_{\pm}$ in the right hand side  are medium polarization and
$R_3$ is the population difference.
The right hand sides of  these equations give effects, all in bulk medium, 
of forward scattering  $\propto \frac{\alpha_{ee} + \alpha_{gg}}{2}
n + \frac{\alpha_{ee} - \alpha_{gg}}{2}R_3^{(0)}$,
backward scattering  $\propto \frac{\alpha_{ee} - \alpha_{gg}}{2}R_3^{(\pm)}$,  
RL(right left)- pair annihilation $\propto \alpha_{ge}  (R_1-iR_2)^{(0)} $,
and RR(right right)-, LL(left left)-pair annihilation  $\propto \alpha_{ge} (R_1-iR_2)^{(\pm)}$.
(The pair creation amplitudes appear in conjugate equations to those above.)
We have used quantities $R_i ^{(\pm)}e^{\pm 2i kx}$ as defined by 
decomposition of three major spatially varying components,
\begin{eqnarray}
&&
R_i = R_i^{(0)} + R_i^{(+)}e^{2i k x} + R_i^{(-)}e^{-2i k x}
\,, \hspace{0.5cm}
k = \omega > 0
\,.
\label{case of spatial grating}
\end{eqnarray}
Quantities $R_i ^{(\pm)}e^{\pm 2i kx}$
are what are called spatial grating in the literature
of non-linear optics.
The backward scattering terms, and RR,- LL-pair annihilation
and creation terms  are important only in the presence of
spatial grating of polarization.
Neglect of spatial grating is thus equivalent to retaining 
forward scattering  and RL-pair processes, and ignoring
all other terms whose effects represent propagation effects,
as noted in \cite{psr dynamics}.

When excitation to the metastable state $|e\rangle$
is done by high quality lasers, there may exist
a spatial grating in the initial Bloch vector components,
which can be incorporated as the initial condition
in our computation.
For instance, excitation by R-moving pump laser and
R-moving coupling laser may imprint phase factors
$e^{i\omega_p (t-x)}$ and $e^{-i\omega_c(t-x)}$
onto target atoms, 
and give the initial spatial grating 
$e^{-i(\omega_p-\omega_c)x}=e^{-i\epsilon_{eg}x}$ for
target atom at the position $x$.
This case corresponds to the initial grating of
$e^{-2i\omega x}$.
But even without initial spatial grating
two-photon absorption of trigger laser  by
non-grating components $R_i^{(0)}$ may generate spatial
grating terms, which thus requires the enlarged set
of dynamical variables including $R_i^{(\pm)}$.

The field equation is to be supplemented by the Bloch equation.
In two-photon process this reads as 
\begin{eqnarray}
\!\!\!\partial_t R_1 &\!=\!&
 (\alpha_{ee}-\alpha_{gg})E^+E^- R_2 - i\alpha_{ge} 
(e^{i\epsilon_{eg}t}E^+E^+ - e^{-i\epsilon_{eg}t}E^-E^-) R_3
-\frac{R_1}{T_2}
\,,
\label{bloch eq 1} 
\\
\!\!\!\partial_t R_2 &\!=\!& -(\alpha_{ee}-\alpha_{gg})E^+E^- R_1
+ \alpha_{ge} (e^{i\epsilon_{eg}t}E^+E^+ + e^{-i\epsilon_{eg}t}E^-E^-) R_3 
-\frac{R_2}{T_2}
\,,
\label{bloch eq 2}
\\
\!\!\!\partial_t R_3 &\!=\!&
 \alpha_{ge} \left(
i  (e^{i\epsilon_{eg}t}E^+E^+ - e^{-i\epsilon_{eg}t}E^-E^-  ) R_1
-  (e^{i\epsilon_{eg}t}E^+E^+ + e^{-i\epsilon_{eg}t}E^-E^-  )R_2
\right) 
-\frac{R_3 + n}{T_1},
\nonumber \\ &&
\label{bloch eq 3}
\\ &&
E^{\pm} = E_R^{\pm} + E_L^{\pm} \,.
\end{eqnarray}
Explicit decomposition of these equations into different grating
according to Eq.~\ref{case of spatial grating}
is given in \cite{psr dynamics}, using SVEA.
We have introduced two relaxation terms inversely proportional to
their time constants $T_i\,, i=1,2$. 
The origin of these relaxation terms is left untouched,
but they may be measured by experimental means.
They arise from interaction of subsystem, in this case
two states in $|e\rangle, |g\rangle$, with environment reservoir
such as a thermal bath.
This form is the most general relaxation terms
consistent with fundamental principles of quantum mechanics in the two-level problem
\cite{lindblad}.
\footnote{
See further below on the relaxation term $\propto 1/T_3$ when
spatial grating terms are present.
}
The phase decoherence time $T_2$
is much smaller and more important than the decay time $T_1$,
which may be taken infinitely large for our practical purpose.

It is both useful and convenient to introduce dimensionless quantities
and write the Maxwell-Bloch equation of our system in terms of
these dimensionless quantities.
The natural time and length scale determined by
the field equation is the inverse of $\alpha_{ab} \omega n$.
We shall define this as 
\begin{eqnarray}
&&
t_* = \frac{2}{\alpha_{ge}\epsilon_{eg}n}
\,.
\label{definition of t-star}
\end{eqnarray}
A somewhat different notation was used in \cite{psr dynamics}.
The natural field scale is then obtained from the Bloch equation
as
\begin{eqnarray}
&&
E_*^2 = \frac{1}{\alpha_{ge}t_*} = \epsilon_{eg}n
\,.
\end{eqnarray}
This $E_*^2$ is nothing but the energy density stored in the atomic system,
when the number density in $|e\rangle$ is equal to $n$.

It is useful to give numerical values for these quantities.
Take the example of para-H$_2$ molecule and its
vibrational two-photon transition $v=1 \rightarrow 0$ of the electronically ground state.
This example gives
\begin{eqnarray}
&&
t_* \sim 1 {\rm cm}\,\frac{10^{21}{\rm cm}^{-3}}{n}
\sim 40 {\rm ps} \,\frac{10^{21}{\rm cm}^{-3}}{n}
\,, \hspace{0.5cm}
E_*^2 \sim 60 {\rm GW mm}^{-2}\frac{n}{10^{21}{\rm cm}^{-3}}
\,.
\end{eqnarray}
Our dimensionless time/length and field units are then
\(\:
(\xi, \tau) =  ( x/t_*, t/t_*)
\,, \hspace{0.3cm}
e_i = E_i/E_*
\,.
\:\)
The Bloch vector components are rescaled by the target
number density, to give the dimensionless quantity $r_i = R_i/n$.
The $r_3$ value 1 means the complete inversion in the level
$|e\rangle$, and -1 the target in the ground state $|g\rangle$
completely.
The medium polarization $|r_1  \pm i r_2| =1$ implies the maximal
coherence.

We now give the master equation
when the spatial grating is present.
(the more general equation for two color trigger
is given in Appendix).
In the dimensionless unit they are
\begin{eqnarray}
&&
\partial_{\tau} r_1^{(0)} = 
 4\gamma_-(|e_R|^2  + |e_L|^2)r_2^{(0)}
+ 8 \Im (e_R e_L)r_3^{(0)}
+ 4\gamma_- e_R e_L^* r_2^{(-)} +  4\gamma_-e_L e_R^* r_2^{(+)}
\nonumber \\ && \hspace*{1.5cm}
- 2i (e_L^2- (e_R^*)^2) r_3^{(+)} - 2i (e_R^2- (e_L^*)^2) r_3^{(-)}
-\frac{r_1^{(0)}}{\tau_2} \,,
\label{rescaled bloch eq11} \\
&&
\partial_{\tau} r_1^{(+)} =4\gamma_- e_R e_L^* r_2^{(0)}
- 2i (e_R^2- (e_L^*)^2) r_3^{(0)}
+4\gamma_-(|e_R|^2  +|e_L|^2)r_2^{(+)} 
+  8 \Im (e_R e_L)r_3^{(+)}
-\frac{r_1^{(+)}}{\tau_3}\,, 
\nonumber \\
&&
\partial_{\tau} r_2^{(0)} = 
-4\gamma_-(|e_R|^2  + e_L|^2)r_1^{(0)} 
+ 8 \Re (e_R e_L)r_3^{(0)}
- 4\gamma_- e_R e_L^* r_1^{(-)}  - 4\gamma_- e_L e_R^* r_1^{(+)}
\nonumber \\ && \hspace*{1.5cm}
+ 2 (e_L^2 + (e_R^*)^2) r_3^{(+)} + 2 (e_R^2 + (e_L^*)^2) r_3^{(-)}
-\frac{r_2^{(0)}}{\tau_2}
\,,
\end{eqnarray}
\begin{equation}
\partial_{\tau} r_2^{(+)} =-4\gamma_-e_R e_L^* r_1^{(0)}
+ 2 (e_R^2 + (e_L^*)^2) r_3^{(0)}
-4\gamma_-(|e_R|^2  +|e_L|^2)r_1^{(+)} 
+  8 \Re (e_R e_L)r_3^{(+)}
-\frac{r_2^{(+)}}{\tau_3}\,, 
\end{equation}
\begin{eqnarray}
&&
\partial_{\tau} r_3^{(0)} = 
-8 \left( \Re (e_R e_L)r_2^{(0)}  + \Im (e_R e_L )r_1^{(0)}
\right)
+2i (e_R^2- (e_L^*)^2) r_1^{(-)} +2i (e_L^2- (e_R^*)^2) r_1^{(+)}
\nonumber \\ &&
\hspace*{1cm}
- 2 (e_L^2 + (e_R^*)^2) r_2^{(+)}  - 2 (e_R^2 + (e_L^*)^2) r_2^{(-)}
-\frac{r_3^{(0)}+1}{\tau_1}
\,,
\end{eqnarray}
\begin{equation}
\partial_{\tau} r_3^{(+)} =
2ir_1^{(0)} (e_R^2- (e_L^*)^2) - 2r_2^{(0)} (e_R^2 + (e_L^*)^2)
-8 \left( \Re (e_R e_L)r_2^{(+)}  + \Im (e_R e_L )r_1^{(+)}\right)
-\frac{r_3^{(+)}}{\tau_3}\,,
\\
\end{equation}
\begin{equation}
(\partial_{\tau} + \partial_{\xi})e_R = 
 \frac{i}{2} (\gamma_+ +  \gamma_- r_3^{(0)} ) e_R
  + \frac{i}{2}\gamma_- r_3^{(+)}e_L
+ \frac{i}{2}(r_1^{(0)} - ir_2^{(0)})e_L^*
+ \frac{i}{2}(r_1^{(+)} - ir_2^{(+)})e_R^*
\,, 
\end{equation}
\begin{equation}
(\partial_{\tau} - \partial_{\xi})e_L = 
 \frac{i}{2} (\gamma_+  +  \gamma_-r_3^{(0)} ) e_L
 + \frac{i}{2}\gamma_- r_3^{(-)}e_R
+ \frac{i}{2}(r_1^{(0)} - ir_2^{(0)})e_R^*
+ \frac{i}{2}(r_1^{(-)} - ir_2^{(-)})e_L^*
\,, 
\label{rescaled quantum field eq22}
\end{equation}
\begin{eqnarray}
\gamma_{\pm} = \frac{\alpha_{ee} \pm \alpha_{gg}}{2\alpha_{ge}}
\,.
\end{eqnarray}
Here $\tau_i =  T_i/t_*$ are relaxation times in
the dimensionless unit.
In our previous work two relaxation terms
are taken equal;  $T_3=T_2$, but relaxation
time $T_3$ for spatial grating terms may
in general differ from the phase de-coherence
time $T_2$ for non-grating terms,
as in the equations above.


In the limit of infinite relaxation times the conservation
law,
\begin{eqnarray}
&&
\partial_{\tau} \left(
(r_1^{(0)})^2 + (r_2^{(0)})^2 + (r_3^{(0)})^2
+2 r_1^{(-)}r_1^{(+)} + 2 r_2^{(-)}r_2^{(+)} + 2 r_3^{(-)}r_3^{(+)}
\right)= 0
\,,
\end{eqnarray}
holds at any point in the target.
Without spatial grating terms,
another important conservation law of energy exists in the
$T_1 \rightarrow \infty$ (but with finite $T_2, T_3$) limit.
In terms of quantities of physical dimensions,
it is
\begin{eqnarray}
&&
\frac{d}{dt}\int_0^{L}dx 
\left(\frac{\epsilon_{eg}}{2}R_3^{(0)} + 2(|E_R|^2 + |E_L|^2 )\right) =
- 2[|E_R|^2 - |E_L|^2]_{x=0}^L  
\,.
\end{eqnarray}
This expresses that exchange of energy between field and medium
is balanced by a net energy flux at two ends of target.
This conservation no longer holds
in the presence of spatial grating terms.
But correction due to this violation is expected small
because the spatially varying factor $e^{\pm 2i\omega x }$
averages out in the spatial integral above.

Finally, we note that the master equation for PSR
has the semi-classical nature:
PSR phenomena can occur without the quantum
initiation unlike SR if the trigger field is
irradiated.
The master equation may be solved numerically
under a variety of initial and boundary conditions.
The standard initial-boundary condition (IBC) we use
in the present work is the symmetric trigger
irradiation in which continuous wave (CW) trigger
lasers of counter-propagating directions and the equal frequency
are irradiated.
One can think of a few interesting cases of
IBC for the Bloch vector components such as
a complete inversion in $|e\rangle$, and
a large phase coherence between the two levels,
$|e\rangle$ and $|g\rangle$.
It turns out that these two cases give
quite different results.

\subsection{Dynamics of PSR}

For numerical simulations in this section, we have in mind 
two types of targets as exemplified by pH$_2$ and Xe,
which has E1 $\times $ E1 and M1 $\times $ E1.
These target atoms are candidates for PSR and RENP.
Their coupling matrices $(\alpha_{ab})$ and
other important parameters are as follows:
\begin{eqnarray}
&&
{\rm pH}_2 \,; \hspace{0.2cm} \epsilon_{eg}  = 0.52 {\rm eV}
\\
&&\hspace{1.0cm}
(\alpha_{ab} ) =
\left(
\begin{array}{cc}
0.87 & 0.055 \\
0.055 & 0.80
\end{array}
\right) 10^{-24} {\rm cm}^3
\,, \hspace{0.5cm}
\frac{1}{t_*} \sim \frac{1}{1 {\rm cm}}\frac{n}{10^{21}{\rm cm}^{-3}}
\,, \nonumber
\\ &&
{\rm Xe} ; 
\hspace{0.3cm}
|e \rangle =5p^5(^2P_{3/2})6s ^2[3/2]_2
\,, \hspace{0.3cm}
|g \rangle = 5p^6\, ^1S_0
\,, \hspace{0.3cm}
|p \rangle = 5p^5(^2P_{3/2})6s ^2[3/2]_1
\,,
\\ &&
\hspace{1.0cm}
\epsilon_{eg} =  8.3153 {\rm eV}
\,, \hspace{0.3cm}
\epsilon_{pe} = 0.1212 {\rm eV}
\,, \hspace{0.3cm}
\gamma_{pg} = 0.28 {\rm GHz}
\,,
\nonumber \\ &&
\hspace{1.0cm}
(\alpha_{ab} ) =
\left(
\begin{array}{cc}
-2.6 \times 10^{-9}&  0.47 \times 10^{-3}\\
0.47 \times 10^{-3}&  6.9
\end{array}
\right)
\; 10^{-24}{\rm cm}^3
\,, \hspace{0.5cm} 
\frac{1}{t_*} \sim \frac{1}{8{\rm cm}} \frac{n}{10^{21}{\rm cm}^{-3}}
\,.
\nonumber
\end{eqnarray}
Coupling parameters $\alpha_{ab}$ for pH$_2$
are given by the polarizability which
is justified when $\omega \ll \epsilon_{eg}$
\cite{psr dynamics}.

Results for fields and Bloch vector components 
have been numerically computed.
Previous computations \cite{psr dynamics} have been done
without taking into account effects of spatial grating.

\begin{figure*}
	\begin{center}
	\includegraphics[clip,width=0.5\textwidth]{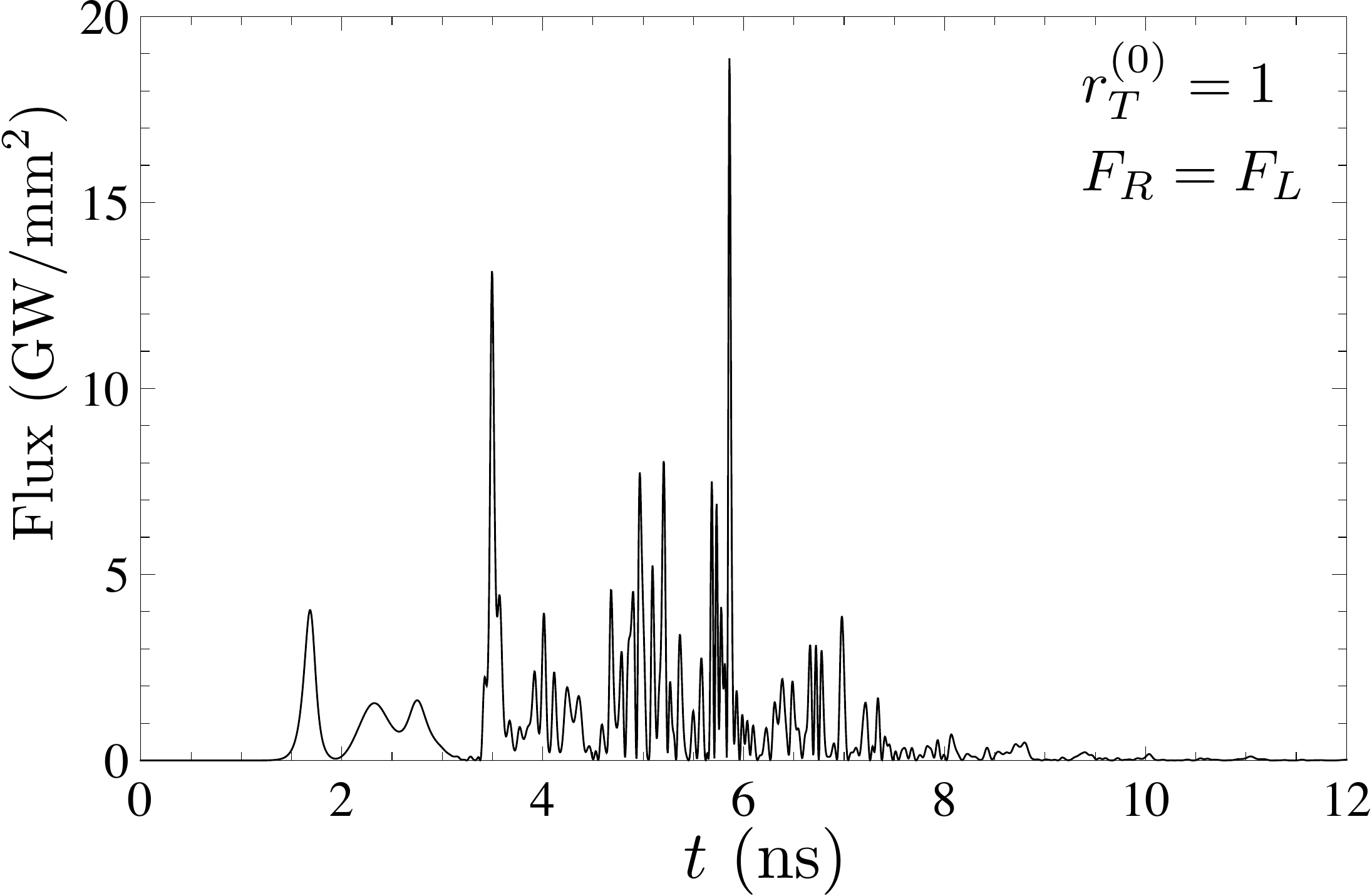}
	\end{center}
	\caption{
		Time-evolving PSR output fluxes at two target ends, resulting from the
   		symmetric trigger irradiation of the power 1 mWmm$^{-2}$ under the conditions of the target number density
		$n=1\times 10^{21}$cm$^{-3}$,  target length $= 100 $cm, 
		relaxation times $T_2=T_3=10, T_1=10^3$ ns's, and the initial polarization, 
		$r_T^{(0)}=1$, all other $r_i = 0$ (no initial grating assumed).
		About 40 \% of the stored energy in the upper level is released 
		from two ends in this example.
		Vibrational transition $Xv=1 \rightarrow Xv=0$ of pH$_2$ is considered.
                Right- and left-moving fluxes at two ends are identical.
	}
   	\label{1m-nograting-PSR}
\end{figure*}

\begin{figure*}
 	\begin{center}
	\includegraphics[clip,width=0.5\textwidth]{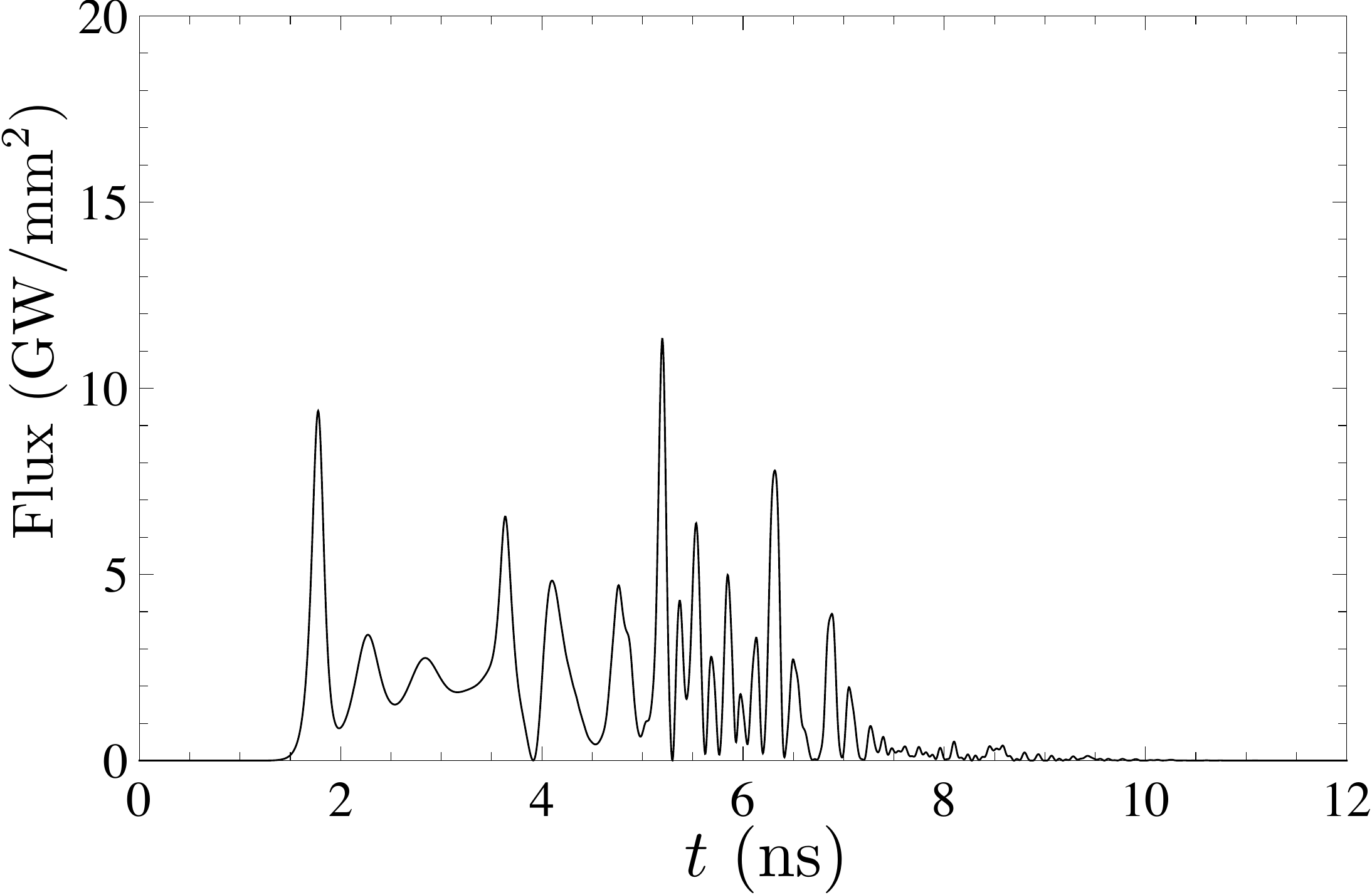}
	\end{center}
	\caption{Flux outputs when spatial grating terms
		are absent, using the same parameter set of Fig. \ref{1m-nograting-PSR} 
	}
   	\label{nograting effect}  
\end{figure*}

We first compare in Fig. \ref{1m-nograting-PSR}  
and Fig. \ref{nograting effect}  results of two calculations;
one with spatial grating modes included as dynamical
variables and another without them,
in both cases assuming no initial grating modes.
Results show a quantitative difference, but
qualitatively they show similar behaviors of
time evolution.
In any event it seems difficult to experimentally resolve this
level of temporal features.
All the rest of illustrated outputs are results including
grating modes as dynamical variables.


We present in Fig. \ref{1m-nograting-PSR} $\sim$ 
Fig. \ref{1m-fullgrating-PSR3D} (see also Fig. \ref{power dependence of PSR}   
in Sec.\ref{Sec:Introduction}),
numerical results for pH$_2$, including the spatial grating effect.
The quantity $r_T^{(i)} = r_1^{(i)} + i r_2^{(i)} $ is defined
for the total polarization of initial state in these computations.
Results of different initial conditions for the spatial grating
are shown; (1) case of full spatial grating without homogeneous
component of the coherence, (2) case of null spatial grating,
(3) mixture of grating and homogeneous components.
With full grating and without any homogeneous component
the output shows the greatest left-right (LR) asymmetry,
a one-sided flux despite of LR symmetric trigger irradiation.
The directionality of the output flux  satisfies the phase matching or
the momentum conservation assuming that the imprinted 
grating effectively gives medium a momentum.
The excitation laser is expected to imprint a certain
level of spatial grating, and it would be important
to control this grating for determining the asymmetry of
outputs.

\begin{figure*}
 \begin{center}
\includegraphics[clip,width=0.5\textwidth]{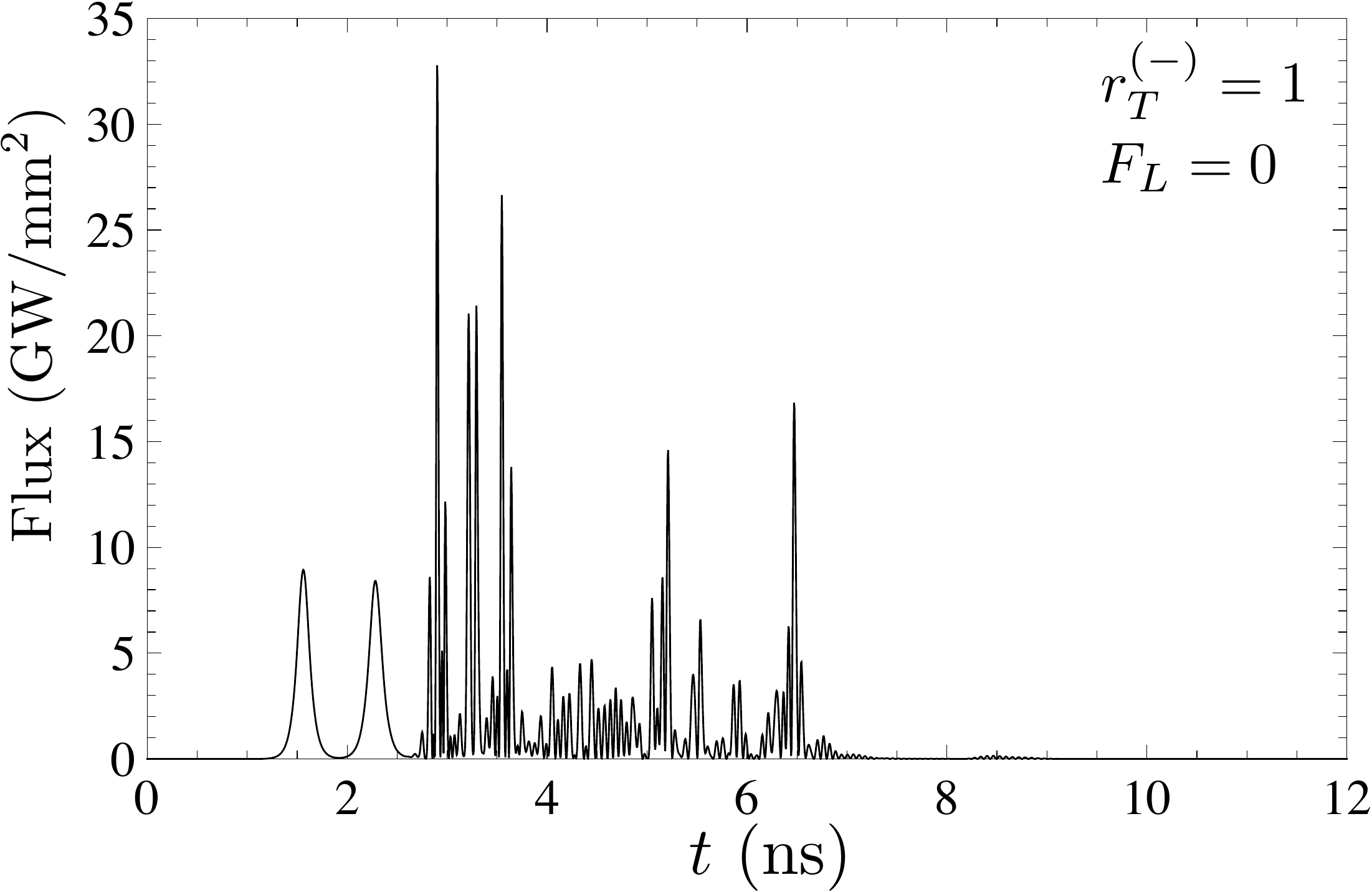}
\end{center}
\caption{
Time-evolving PSR right moving output
(left moving flux negligibly small), resulting from the
   symmetric trigger irradiation  of the power 1 mWmm$^{-2}$ 
under the conditions of the target number density
$n=1\times 10^{21}$cm$^{-3}$,  target length $= 100 $cm, 
relaxation times $T_2=T_3=10, T_1=10^3$ ns's, and the initial polarization
of fully grating mode, 
$r_T^{(-)}=1$, all other $r_i = 0$.
About 30 \% of the stored energy is released in this case.
Vibrational transition $Xv=1 \rightarrow Xv=0$ of pH$_2$ is considered.
}
   \label{1m-fullgrating-PSR}  
\end{figure*}


\begin{figure*}
 \begin{center}
\includegraphics[clip,width=0.9\textwidth]{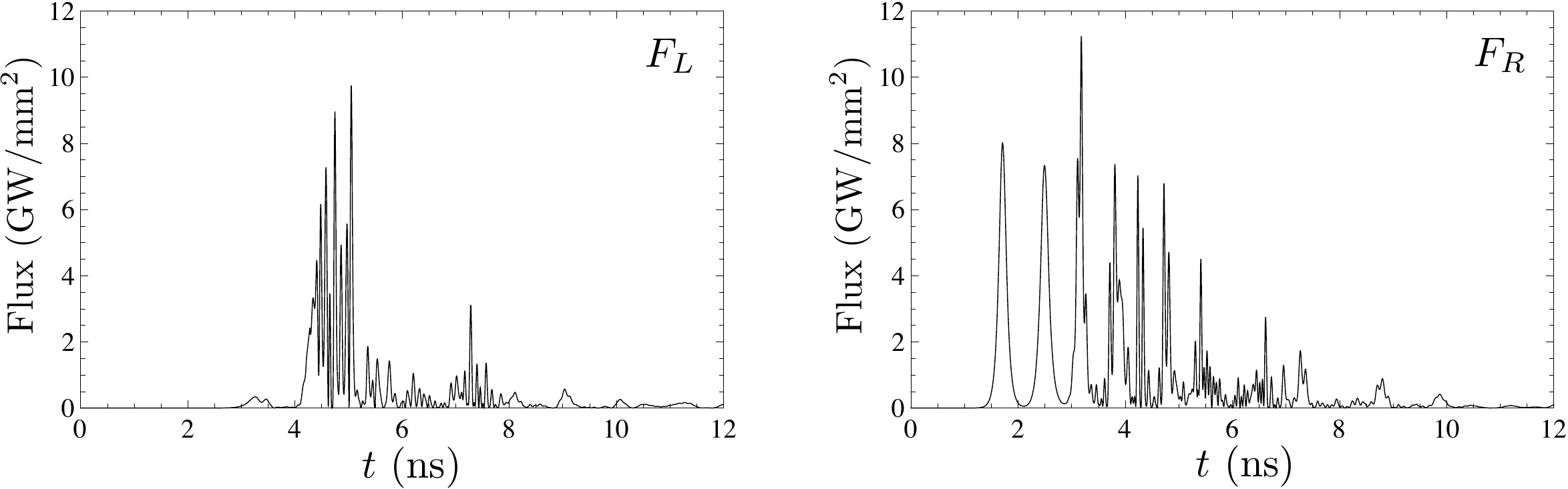}
\end{center}
\caption{
Time-evolving PSR output fluxes at the left target end
in the left panel and at the right end in the right panel, 
resulting from the
   symmetric trigger irradiation at two target ends of the power 1 mWmm$^{-2}$
under the conditions of the target number density
$n=1\times 10^{21}$cm$^{-3}$,  target length $= 100 $cm, 
relaxation times $T_2=T_3=10, T_1=10^3$ ns's, and the initial polarization, 
$r_T^{(-)}=0.9, r_T^{(0)}=0.1$, all other $r_i = 0$.
About 20\% from the right and 10\% from the left of
the stored energy are released.
Vibrational transition $Xv=1 \rightarrow Xv=0$ of pH$_2$ is considered.
}
   \label{1m-partialgrating-PSR}  
\end{figure*}

The space-time evolution of  
Bloch vector components are shown in Fig.\ref{1m-fullgrating-PSR3D}.
It is striking that even after dephasing times $T_2, T_3$
interesting spatial structures remain finite,
suggesting that explosive PSR event alone is
not the whole story of macro-coherent two-photon process.
Below we shall discuss static or steady state solutions
of the Maxwell-Bloch system.

Summarizing outputs,
we may say that
effects of grating modes are large when
initial grating modes are present,
but when they are absent,
differences are minor.

\begin{figure*}
 \begin{center}
\includegraphics[width=0.9\textwidth]{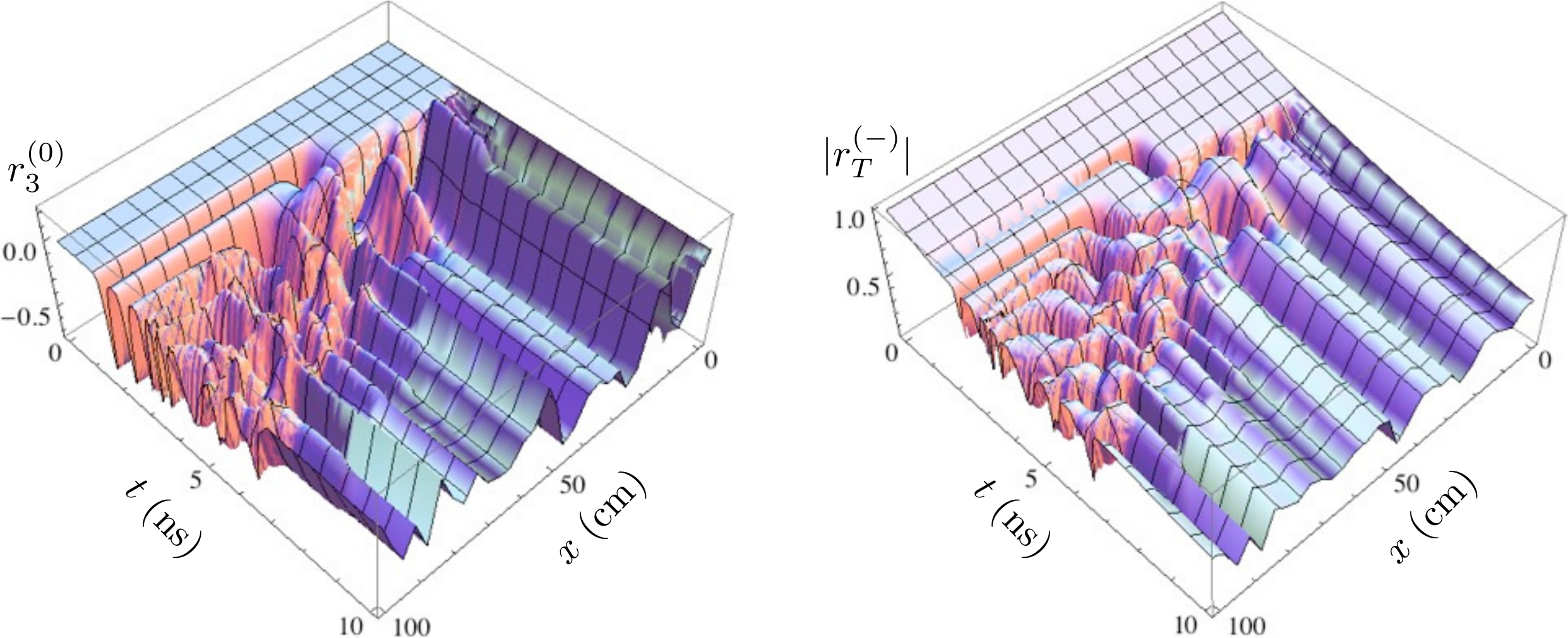}
\end{center}
\caption{
Space-time diagram of grating and non-grating
Bloch vector components 
corresponding to the parameter choice of
Fig. \ref{1m-fullgrating-PSR}.
}
   \label{1m-fullgrating-PSR3D}  
\end{figure*}

\begin{figure*}
 \begin{center}
\includegraphics[clip,width=0.5\textwidth]{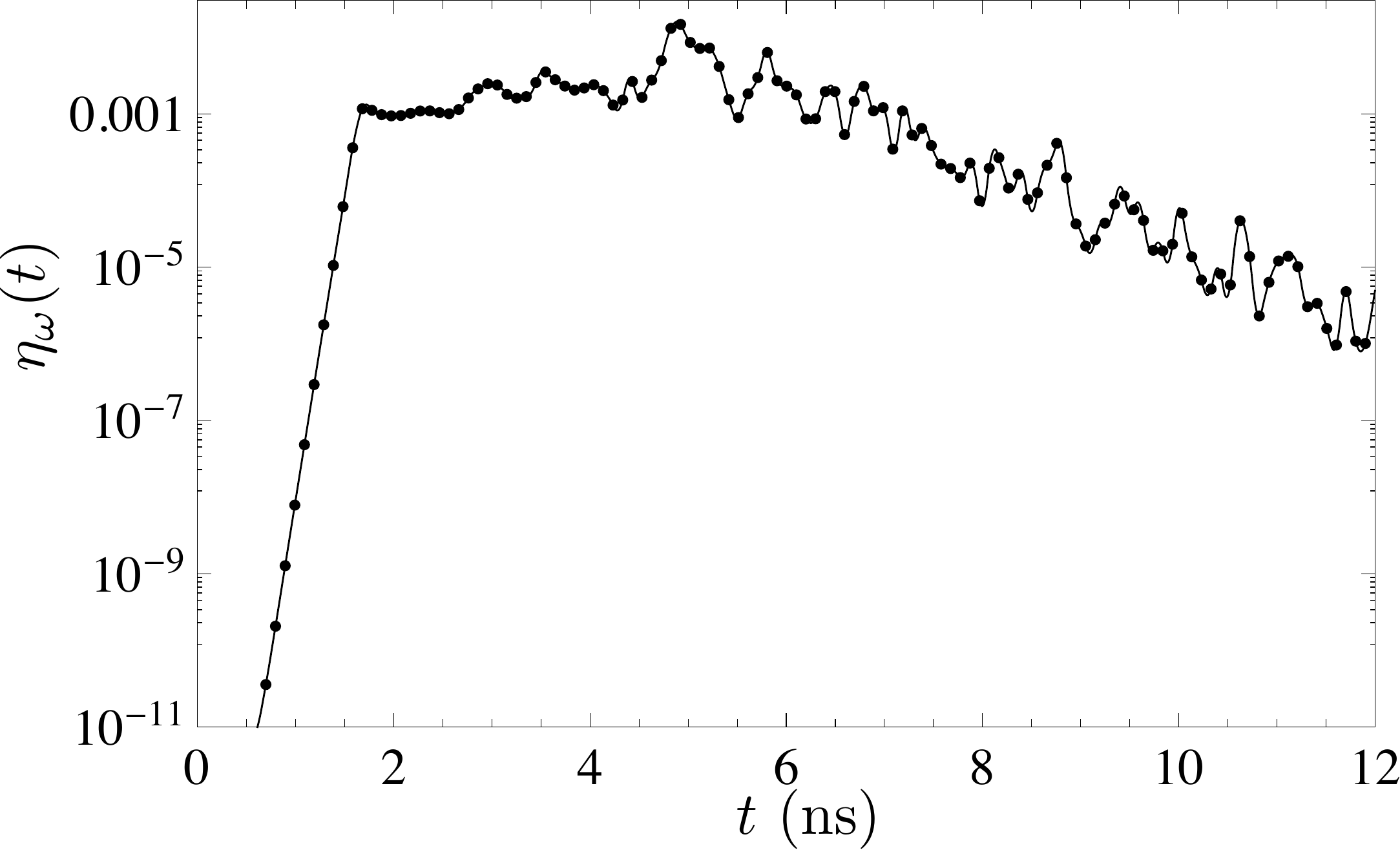}
\end{center}
\caption{
Potentiality of PSR emission given
by the spatial integral of $|\vec{E}^{-}(R_1-iR_2)|^2/(4\epsilon_{eg}n^3)$
corresponding to the parameter set of Fig.\ref{1m-nograting-PSR}.
}
   \label{eta for pH2}  
\end{figure*}

The local quantity  defined by
$|\vec{E}^{-}(R_1-iR_2)|^2/(4\epsilon_{eg}n^3)$ 
at each site
within the target
expresses a potentiality of PSR emission, 
which shall be reflected in fluxes at two ends in
later times.
The spatial integration of this local quantity is
denoted by $\eta_{\omega}(t)$, which gives an instantaneous
potentiality of PSR activity.
This global function is illustrated in
Fig.\ref{eta for pH2}
up to a time of order $T_2$.
The time duration of the plateau region roughly
gives the duration of explosive PSR emission.

In Sec. 4 we show the case of weak PSR outputs
in which the output flux is linearly proportional to
the trigger flux.
There is a set of threshold parameters that divides
the explosive and the linear PSR events.
But due to a high level of non-linearity it
is difficult to locate this threshold exactly.

The results so far are outputs for pH$_2$.
Xe results have different aspects, since their
two-photon coupling is caused by M1$\times$E1,
much weaker than E1$\times$E1 of pH$_2$
vibrational transition.

$\;$ \\ 
\section{Theory of macro-coherent radiative emission of neutrino pair (RENP)}
\label{Sec:RENP-Theory}
\begin{figure*}
\begin{center}
\includegraphics[clip,width=18em]{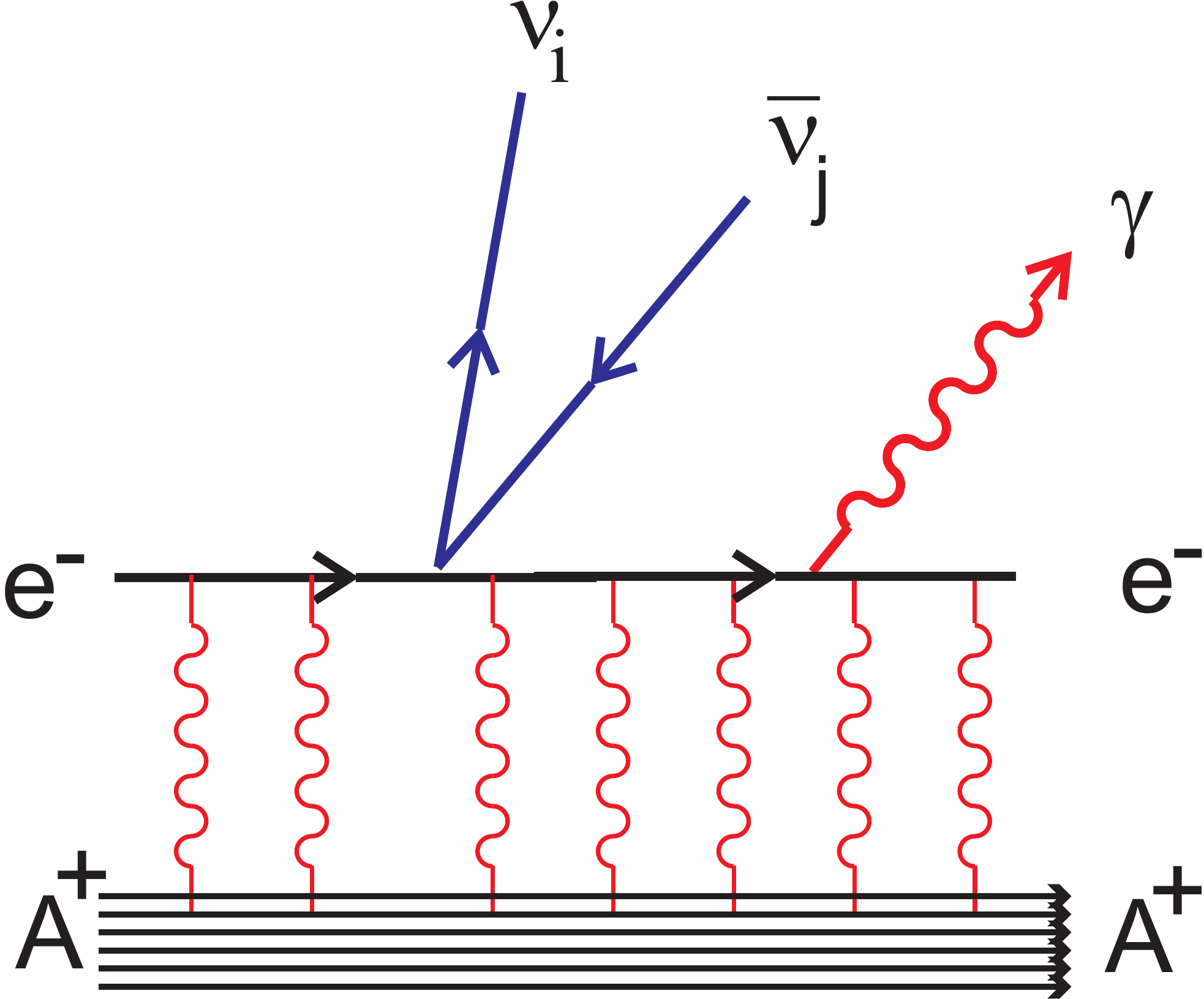}
\end{center}
\caption{RENP process. The many wavy lines between $e^{-}$ and $A^{+}$ represent frequent Coulomb interaction between atomic electron and nucleus.}
\label{FIG:Coulomb}
\end{figure*}

In this section, we present a theory of RENP \cite{my-rnpe}, \cite{my-rnpe1}
and numerical results
based on theoretical formulas thus derived. 
The RENP process is treated in the second-order perturbation
of QED and the weak four Fermi interaction, while the time evolution of 
the electromagnetic field in the target medium is dealt with non-perturbatively
in the semi-classical approximation.
A Feynman-like diagram of the elementary RENP process
is depicted in Fig.~\ref{FIG:Coulomb},
in which the nuclear Coulomb interaction is emphasized.
The non-perturbative part requires numerical solutions of
the master equation of Maxwell-Bloch type under
two color trigger irradiation.

In Appendix A a brief description of electroweak interaction
of atomic electron is given, which is the basis
of our following computations.

\subsection{Coherent neutrino pair emission from atoms / molecules}
We begin with a description of the elementary process 
$|e\rangle\to|g\rangle+\gamma(\vec{k})+\nu_i(\vec{p})+\bar\nu_j(\vec{p'})$,
in which an electron bounded in an atom of the macroscopic target
undergoes the QED and weak interactions. The amplitude is expressed as
\begin{equation}
\mathcal{A}_{ij}=\frac{i\mathcal{M}_d\mathcal{M}_W^{ij}}
                      {\epsilon_{pg}-\omega}\,
                 e^{-i(\vec{k}+\vec{p}+\vec{p'})\cdot\vec{x}_a}
                 (2\pi)\delta(\epsilon _{eg}-\omega-E_p-E_{p'})\,,
\label{RENPamp}
\end{equation}
where $\epsilon_{pg}=\epsilon_p-\epsilon_g$, $\omega=|\vec{k}|$ is 
the energy of the photon, $\mathcal{M}_{d(W)}$ denotes the E1 (weak) matrix 
element and the possible sum over the intermediate state $|p\rangle$ is to be
understood. We assume that the transition between 
$|e\rangle$ and $|p\rangle$ is the M1 type responsible to the neutrino pair 
emission and that between $|p\rangle$ and $|g\rangle$ is the E1 type 
responsible for the photon emission. We explicitly show the dependence on 
the position of the target atom $\vec{x}_a$ in Eq.~\ref{RENPamp}. 

For an ensemble of atoms, the total amplitude is the superposition of 
the contributions from all the atoms and proportional to 
$\sum_a \exp{[-i(\vec{k}+\vec{p}+\vec{p'})\cdot\vec{x}_a]}$.
For a macroscopic ensemble of $N$ atoms with a volume $V$, this summation 
approximates to $(N/V)(2\pi)^3\delta^3(\vec{k}+\vec{p}+\vec{p'})$, 
which means the momentum conservation among the emitted photon and 
the neutrino pair in addition to the energy conservation represented 
by the $\delta$ function in Eq.~\ref{RENPamp}. Thus, the kinematic 
configuration that satisfies the momentum conservation acquires an extra
enhancement in the rate. This is the essence of the macro-coherent 
amplification mechanism.

Due to the energy-momentum conservation, the RENP process with 
the macro-coherence resembles a relativistic three-body decay
of the parent particle of mass $\epsilon_{eg}$ into a photon and
a pair of neutrinos. However, the energy and the momentum of the photon
are dictated by the classical radiation field induced by the trigger laser.
Therefore, the emitted photon has no phase space
and we do not have to introduce the three-body phase space. Thus,
the differential rate of the process for a macroscopic ensemble of 
atoms is written as
\begin{equation}
d\Gamma_{ij}=n^2V\frac{|\mathcal{M}_d\mathcal{M}_W^{ij}|^2}
                      {(\epsilon_{pg}-\omega)^2}\,d\mathit{\Phi}_2\,,
\end{equation}
where $n$ is the number density of atoms, $d\mathit{\Phi}_2$ represents 
the Lorentz invariant two-body phase space of the neutrino pair, 
and the summations over the atomic spin states and the neutrino helicities 
are implicit. The explicit form of $d\mathit{\Phi}_2$ is given by
\begin{equation}
d\mathit{\Phi}_2=(2\pi)^4\delta^4(q-p-p')
                 \frac{d^3p}{(2\pi)^3 2E_p}\frac{d^3p'}{(2\pi)^3 2E_{p'}}\,,
\end{equation}
where $E_{p^{(\prime)}}=\sqrt{m_{i(j)}^2+\vec{p}^{(\prime)2}}$ is 
the neutrino energy of mass $m_{i(j)}$
and $q^\mu=(\epsilon_{eg}-\omega,-\vec{k})$ 
represents the four momentum of the neutrino pair.

The E1 matrix element is defined as 
$\mathcal{M}_d=-\langle g|\vec{d}|p\rangle\cdot\vec{E}$
with $\vec{E}$ being the radiation field developed in the target by
the trigger laser irradiation, while the weak matrix element is expressed by
\begin{equation}
\mathcal{M}_W^{ij}=
 \frac{G_F}{\sqrt{2}}
 \langle\nu_i(p,\lambda)\bar\nu_j(p',\lambda')|
  \sum_{a,b}\bar\nu_a\gamma_\mu(1-\gamma_5)\nu_b|0\rangle
   (v_{ab}J_V^\mu-a_{ab}J_A^\mu)\,,
\label{WeakME}
\end{equation}
where the atomic vector (axial-vector) current $J_V^\mu(J_A^\mu)$ is defined
by $J_V^\mu=\langle g|\bar e\gamma^\mu e|p\rangle$ 
($J_A^\mu=\langle g|\bar e\gamma^\mu\gamma_5 e|p\rangle$).
Coefficients $v_{ab}$ and $a_{ab}$ are given in terms of elements
of neutrino mass matrix, as given by Eq.~\ref{eq:v-a-coefficients}
of Appendix A.
In the non-relativistic approximation for the atomic wave functions, 
$J_V^\mu\simeq 0$ and $J_A^\mu\simeq(0,2\langle p|\vec{S}|e\rangle)$ 
in the rest frame of the atomic system. Hence, we neglect $J_V^\mu$ hereafter.
Evaluating the neutrino matrix element, we obtain
\begin{equation}
\mathcal{M}_W^{ij}=
 -\frac{G_F}{\sqrt{2}}
   \left(a_{ij}L^\mu_{ij}-\delta_M a_{ji}R^\mu_{ij}\right)J_{A\mu}\,,
\label{WeakAMP}
\end{equation}
where 
\begin{eqnarray}
&&
L^\mu_{ij}(R^\mu_{ij})=
       \bar u_i(p,\lambda)\gamma^\mu(1\mp\gamma_5)v_j(p',\lambda')
       \,,
       \label{chiral currents}
\end{eqnarray}
and
$\delta_M=0(1)$ for Dirac(Majorana) neutrinos.
The left-handed current $L_{ij}^\mu$, which comes from $(a,b)=(i,j)$
in Eq.~\ref{WeakME}, is common for both Dirac and Majorana cases;
while the right-handed current $R_{ij}^\mu$ from $(a,b)=(j,i)$ with
the use of the relation between the four-component spinor, 
$C\bar u^T=v$, is allowed only for the latter case. 
Technically, this chirality flip is understood as a property of 
the charge conjugation operation in the Dirac theory.

The square of the weak amplitude Eq.~\ref{WeakAMP} consists of four terms:
$L_{ij}^\mu L_{ij}^{\dagger\nu}$, $R_{ij}^\mu R_{ij}^{\dagger\nu}$,
$L_{ij}^\mu R_{ij}^{\dagger\nu}$ and $R_{ij}^\mu L_{ij}^{\dagger\nu}$.
The sum over the neutrino helicities and momenta leads to
\begin{eqnarray}
&&\int d\mathit{\Phi}_2 \sum_{\lambda,\lambda'}L_{ij}^\mu L_{ij}^{\dagger\nu}
=\int d\mathit{\Phi}_2 \sum_{\lambda,\lambda'}R_{ij}^\mu R_{ij}^{\dagger\nu}
\nonumber\\
&&
\hspace*{-1cm}
=\frac{\Delta_{ij}}{6\pi}
 \left[\left\{\Delta_{ij}^2
       -3\left(1-\frac{m_i^2+m_j^2}{q^2}\right)\right\}q^2g^{\mu\nu}
       +2\left\{1+\frac{m_i^2+m_j^2}{q^2}
                -2\frac{(m_i^2-m_j^2)^2}{q^4}\right\}q^\mu q^\nu\right]\,,
\end{eqnarray}
\begin{equation}
\int d\mathit{\Phi}_2 \sum_{\lambda,\lambda'}L_{ij}^\mu R_{ij}^{\dagger\nu}
=\int d\mathit{\Phi}_2 \sum_{\lambda,\lambda'}R_{ij}^\mu L_{ij}^{\dagger\nu}
=-\frac{\Delta_{ij}}{\pi}m_im_jg^{\mu\nu}\,,
\label{RLterm}
\end{equation}
where
\begin{equation}
 \Delta_{ij}^2=1-2\frac{m_i^2+m_j^2}{q^2}+\frac{(m_i^2-m_j^2)^2}{q^4}\,,
\end{equation}
and $q^2=\epsilon_{eg}(\epsilon_{eg}-\omega)$ is the invariant mass of
the neutrino pair. The cross terms Eq.~\ref{RLterm} 
exhibit the identical particle effect inherent in Majorana neutrinos.
They arise from the interference of two currents of
different chirality $L^\mu_{ij}$ and $R^\mu_{ij}$ 
defined in Eq.~\ref{chiral currents} and disappear in the massless 
limit as it should.

Thus, taking the statistical factor for the Majorana case into account,
the single photon spectral rate at frequency $\omega$ from the position
$x$ of unpolarized targets is given by
\begin{eqnarray}
&&
d\Gamma_{\gamma 2\nu}(\omega,x,t) = 
 \frac{G_F^2 V\,dx}{6\pi L}
 \left|\vec{E}^{-}(\omega,x,t)\frac{R_1(x,t)-iR_2(x,t)}{2}\right|^2
 C_{eg}(\omega) F(\omega)
\label{LocalRate}
\\ &&
C_{eg}(\omega) = 
\sum_p \frac{\langle g| \vec{d} |p\rangle \cdot \langle p| \vec{d} |g\rangle
\langle e| \vec{S} |p\rangle \cdot \langle p| \vec{S} |e\rangle}{(\epsilon_{pg} 
- \omega)^2}
\\ &&
\hspace*{-1cm}
F(\omega) = 
\sum_{ij}\Delta_{ij}(B_{ij} I_{ij}(\omega) - \delta_M B^M_{ij}m_im_j)
         \theta(\omega_{ij} - \omega)\,,\
B_{ij} = | a_{ij}|^2
\,, \ 
B^M_{ij} = \Re (a_{ij}^2) 
\,,
\label{Ffunc}\\ &&
\hspace*{-1cm}
I_{ij}(\omega)=
 \frac{q^2}{6}
  \left[2-\frac{m_i^2+m_j^2}{q^2}-\frac{(m_i^2-m_j^2)^2}{q^4}\right]
 +\frac{\omega^2}{9}
   \left[1+\frac{m_i^2+m_j^2}{q^2}-2\frac{(m_i^2-m_j^2)^2}{q^4}\right]\,,
\end{eqnarray}
where $\vec{E}^{-}(\omega,x,t)=\vec{E}_R^{-}+\vec{E}_L^{-}$ 
is the negative energy part of right- and left-moving fields within 
the target developed from trigger laser of frequency $\omega< \epsilon_{eg}/2$ 
and is limited in its magnitude by the maximum energy stored in the target, 
$|\vec{E}|^2\leq \epsilon_{eg}n$.

We note that the rate $d{\Gamma}_{\gamma 2\nu}$ is proportional to
the space-time dependent factor,
\begin{eqnarray}
&&
\left|\vec{E}^{-}\frac{R_1-iR_2}{2}\right|^2
\,,
\label{Tactivity}
\end{eqnarray}
which signifies how much the field-medium system
has been activated in favor of RENP.
It is thus important to obtain this factor as large
as possible in the target.
We define the following dimensionless quantity to express
the field-medium activity for RENP, 
\begin{equation}
\eta_\omega(\xi,\tau)=\frac{1}{\epsilon_{eg}n^3}
               \left|\vec{E}^{-}\frac{R_1-iR_2}{2}\right|^2
=\left|\left(e_R^*e^{-i\omega t_*\xi}+e_L^*e^{i\omega t_*\xi}\right)
       \frac{r_1-ir_2}{2}\right|^2,
\end{equation}
where $(\xi,\tau)=(x/t^*,t/t^*)$ with $t_*$ given in Eq.\ref{definition of t-star}, 
the slowly varying dimensionless envelope functions $e_{R,L}(\xi,\tau)$
are introduced as
\begin{equation}
E^-=\sqrt{\epsilon_{eg}n}
     \{e_R^*e^{i\omega t_*(\tau-\xi)}+e_L^*e^{i\omega t_*(\tau+\xi)}\},
\end{equation}
and the dimensionless Bloch vector components $r_i(\xi,\tau)$ are
defined by $R_i=nr_i$.
Further, using the grating modes 
$r_i=r_i^{(0)}+r_i^{(+)}e^{2ik t_*\xi}+r_i^{(-)}e^{-2ik t_*\xi}$,
we obtain
\begin{equation}
\eta_\omega(\xi,\tau)=\eta_\omega^R(\xi,\tau)+\eta_\omega^L(\xi,\tau),
\end{equation}
where 
\begin{equation}
\eta_\omega^{R(L)}(\xi,\tau)=
\frac{1}{4}\left[|e_{R(L)}|^2(|r_T^{(0)}|^2+|r_T^{(+)}|^2+|r_T^{(-)}|^2)
            +\Re\{e_R^*e_L(r_T^{(0)*}r_T^{(+)}+r_T^{(0)}r_T^{(-)*})\}\right]
\end{equation}
with $r_T^{(0,\pm)}=r_1^{(0,\pm)}+ir_2^{(0,\pm)}$.
We expect that the $\eta$ factor weakly depends on 
the frequency $\omega$ as is described below.

The formula  here gives rate of a kind of stimulated emission from
the developed trigger laser of field strength $|\vec{E}|^2$.
Although both the energy and the momentum
is conserved at the level of elementary process, 
the phase space factor is different
from a typical electron spectral shape of three body decay
into almost massless particles such as
the muon decay,
since the quantity $|\vec{E}|^2$ here is replaced by 
a kinematic factor $\propto \omega$ (electron energy) in the muon decay.  
Thus, the photon spectral rate as a function of $\omega$ is markedly 
different in the small energy region from that of muon decay.
Near the neutrino pair threshold region the spectrum resembles
that of muon decay, although the difference reflects effects of
finite neutrino masses.

As stated in Section 1,
RENP may occur at any point of target since
even condensate or soliton formation, capable of blocking PSR,
cannot stop RENP. RENP is a bulk process, while PSR occurs only at
target ends after soliton formation. 
The total RENP rate observed at both ends of the target 
is given by integrating Eq.~(\ref{LocalRate}) over $x$.
As a result, assuming the sum over possible intermediate states is 
dominated by a single state $|p\rangle$, we obtain
\begin{eqnarray}
&&\Gamma_{\gamma 2\nu}(\omega,t) = \Gamma_0 I(\omega)\eta_{\omega}(t)\,,
\label{rnpe spectrum rate gamma}\\
&&\Gamma_0=\frac{3n^2VG_F^2\gamma_{pg}\epsilon_{eg}n}{2\epsilon_{pg}^3}
            (2J_p+1)C_{ep}\,,
\end{eqnarray}
where $I(\omega)=F(\omega)/(\epsilon_{pg}-\omega)^2$, and $\gamma_{pg}$
denotes the E1 transition rate from $|p\rangle$ to $|g\rangle$.
The atomic spin factor $C_{ep}$ is defined by
\begin{equation}
\frac{1}{2J_e+1}\sum_{M_e}
\langle p M_p|\vec{S}|e M_e\rangle\cdot\langle e M_e|\vec{S}|p M_p'\rangle
=\delta_{M_pM_p'}C_{ep}\,,
\end{equation}
and $C_{ep}=2/3$ for Xe atom. The dynamical factor $\eta_{\omega}(t)$ 
is given by
\begin{equation}
\eta_{\omega}(t)=\eta_{\omega}^R(t)+\eta_{\omega}^L(t)\,,
\label{spatial integral}
\end{equation}
where
\begin{equation}
\eta_{\omega}^R(t)=
 \frac{t_*}{4L}\int^{L/t_*}_0\eta_\omega^R(\xi,t/t_*-L/t_*+\xi)d\xi\,,
\end{equation}
and
\begin{equation}
\eta_{\omega}^L(t)=
 \frac{t_*}{4L}\int^{L/t_*}_0\eta_\omega^L(\xi,t/t_*-\xi)d\xi\,,
\label{spatial integra3}
\end{equation}
correspond to the rates at the right and left ends respectively.
The overall rate $\Gamma_0$ has a dimension in 1/time, 
while $I(\omega)$ and $\eta_\omega(t)$ are dimension-less.

A nice feature of the formula Eq.~\ref{rnpe spectrum rate gamma} is that
one can separate the spectral feature given by $I(\omega)$ 
that serves to determine neutrino properties from
the absolute rate $\Gamma_0$ and the dynamical factor $\eta_{\omega}(t)$.  
The time dependent function $\eta_{\omega}(t)$
is sensitive to how the initial coherence between
$|e\rangle$ and $|g\rangle$ has been prepared and
to the quality of trigger laser.
Indeed, calculation of $\eta_{\omega}(t)$ requires numerical solutions
of the master equation.
We thus discuss two issues, the spectral shape and
the overall dynamical rate factor, separately in the following.

The limit of massless neutrinos gives the spectral form,
\begin{eqnarray}
&&
I(\omega; m_i = 0) =
\frac{\omega^2 - 6\epsilon_{eg}\omega +
3\epsilon_{eg}^2}{12(\epsilon_{pg}-\omega)^2}
\,,
\label{Iomegamassless}
\end{eqnarray}
%
where the prefactor of 
$\sum_{ij}B_{ij} =3/4$ 
is calculated using the unitarity of the neutrino mixing matrix.
On the other hand, near the threshold
these functions has the behavior of $\Delta_{ij}(\omega)$
$\propto \sqrt{\omega_{ij}-\omega}$.
\begin{table}
\centering \caption{\label{tab_aij} 
The threshold weight 
$B_{ij}=|a_{ij}|^2=|U^*_{ei}U_{ej}-\delta_{ij}/2|^2$.}
 \begin{tabular}{|c|c|c|c|c|c|}
  \hline
   $B_{11}$& $B_{22}$ & $B_{33}$ & 
   $B_{12}+B_{21}$ & $B_{23}+B_{32}$ & $B_{31}+B_{13}$ \\
  \hline
   $(c_{12}^2c_{13}^2-1/2)^2$ & 
   $(s_{12}^2c_{13}^2-1/2)^2$ & 
   $(s_{13}^2-1/2)^2$ &
   $2c_{12}^2s_{12}^2c_{13}^4$ &
   $2s_{12}^2c_{13}^2s_{13}^2$ &
   $2c_{12}^2c_{13}^2s_{13}^2$ \\
  \hline       
   0.0311 & 0.0401 & 0.227 & 0.405 & 0.0144 & 0.0325 \\
  \hline
   \end{tabular}
\end{table}

In what follows we first  calculate numerically (analytic results
are of little use to experimental design) the spectral shape 
$ I(\omega)$ for
two target candidates, Xe atom and I$_2$ molecule,
which greatly differ in their available energies, 8.32 eV
(Xe) and 0.88 eV (I$_2$ relevant vibrational transition).
These two targets have been selected for their
large M1$\times$E1 couplings and their large
difference in closeness to the minimum expected 
mass of neutrino heaviest pair $\sim 0.1$eV.
This makes it possible to determine
the relative easiness of undetermined
neutrino parameters in realistic experiments.

\subsection{Neutrino properties extractable from the photon spectrum}
Recent reactor and T2K experiments have
given compelling evidence for a sizable $\theta_{13}$,
which determines relative weights of neutrino pair
emission thresholds in our RENP process.
Here we summarize the neutrino parameters employed in the
following numerical calculation:
\footnote{These values are slightly different from those quoted in Eq.\ref{oscillation results}.
The main conclusions in this section should not be affected by these differences.}
\begin{equation}
\label{Delta2131}
\Delta m^2_{21} = 7.5 \times 10^{-5} \ \mathrm{eV}^2\,,\quad
|\Delta m^2_{31(32)}| = 2.32 \times 10^{-3} \ \mathrm{eV}^2\,,
\end{equation}
\begin{equation}
\label{sinsq1213}
\sin^2\theta_{12} = 0.31\,,\quad
\sin^2\theta_{13} = 0.025\,,\quad
\sin^2\theta_{23} = 0.42\,,
\end{equation}
for the NH(IH) case.
It is useful to recall weight factors Table(\ref{tab_aij}) given
in reference \cite{yb-x renp}.
These absolute weights are applied to the Dirac neutrino
pair production, while the rate of the Majorana neutrino
pair production at the crossed channel 
$\omega_{ij}, i\neq j$ contains the additional term
proportional to $B^M_{ij}$ in Eq.~\ref{Ffunc}, 
which depends on Majorana phase combinations,
\begin{equation}
\cos 2\alpha\,,\quad \cos 2(\beta-\delta)\,,\quad
\cos 2(\alpha - \beta+\delta)\,,
\label{cpv phase at crossing}
\end{equation}
at (1,2), (1,3) and (2,3) thresholds.

On the other hand, the threshold locations are separated by energy
differences $\epsilon_{eg}/2-\omega_{ij}=(m_i + m_j)^2/2\epsilon_{eg}$.
Hence the separation becomes larger as the atomic energy 
$\epsilon_{eg}$ is smaller for fixed neutrino masses.
A numerical illustration of the threshold locations is given below
for Xe atom in Fig.\ref{FIG:omXe}.

\subsubsection{Absolute masses and mass hierarchy}
\begin{figure*}
\begin{center}
\includegraphics[width=0.4\textwidth]{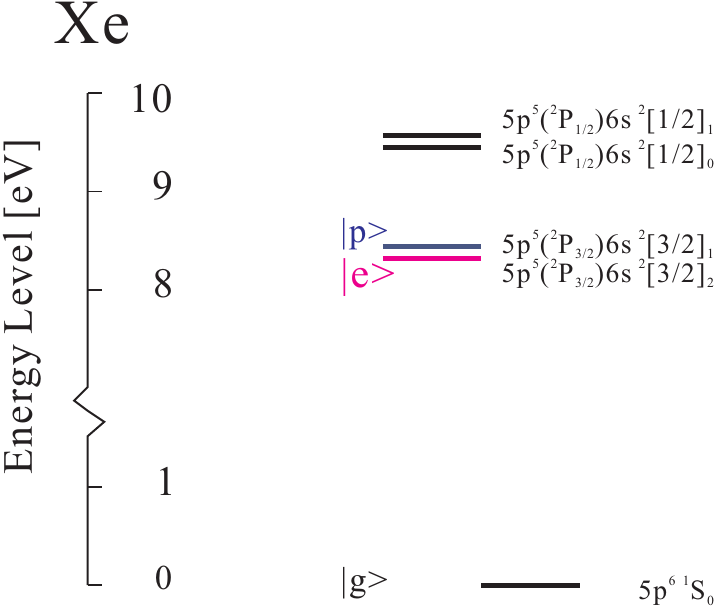}
\end{center} 
\caption{Xe energy level diagram for RENP experiment.}
\label{xe energy level for renp} 
\end{figure*}

Experiments are easier when detectable rates are larger.
The single photon spectral rates for macro-coherent RENP 
are larger for larger available energy if dipole and spin
matrix elements are the same.
Thus, we first calculate RENP rate for Xe which has a large energy 
difference between the first excited state, as seen in 
Fig.~\ref{xe energy level for renp}, which is metastable, and
the ground state. The overall rate for Xe is estimated as 
$\Gamma_0 \sim 1\,\mathrm{Hz}\,
 (n/10^{22}\mathrm{cm}^{-3})^3(V/10^2\,\mathrm{cm}^3)$.

\begin{figure*}
\includegraphics[clip,width=18em]{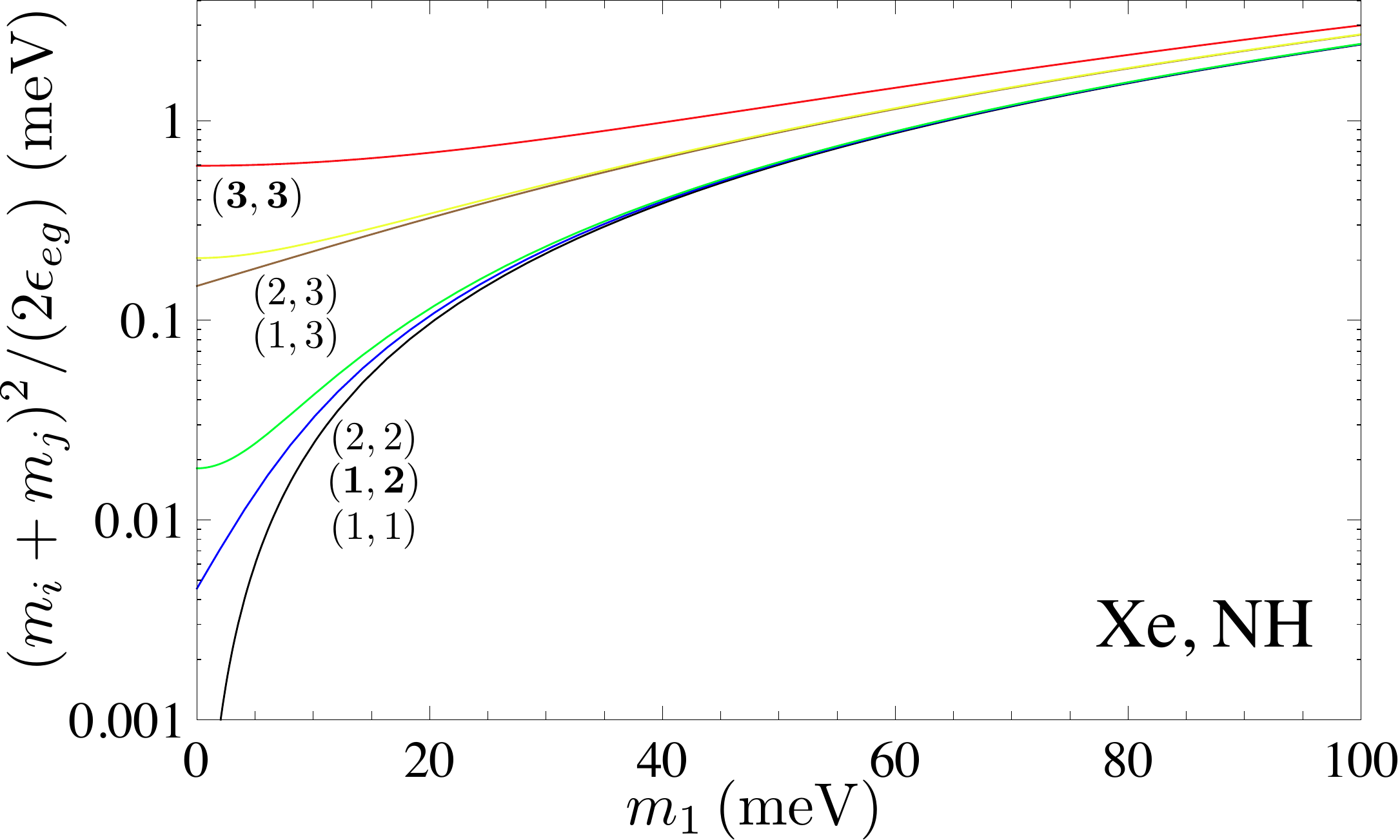}
\hspace*{2em}
\includegraphics[clip,width=18em]{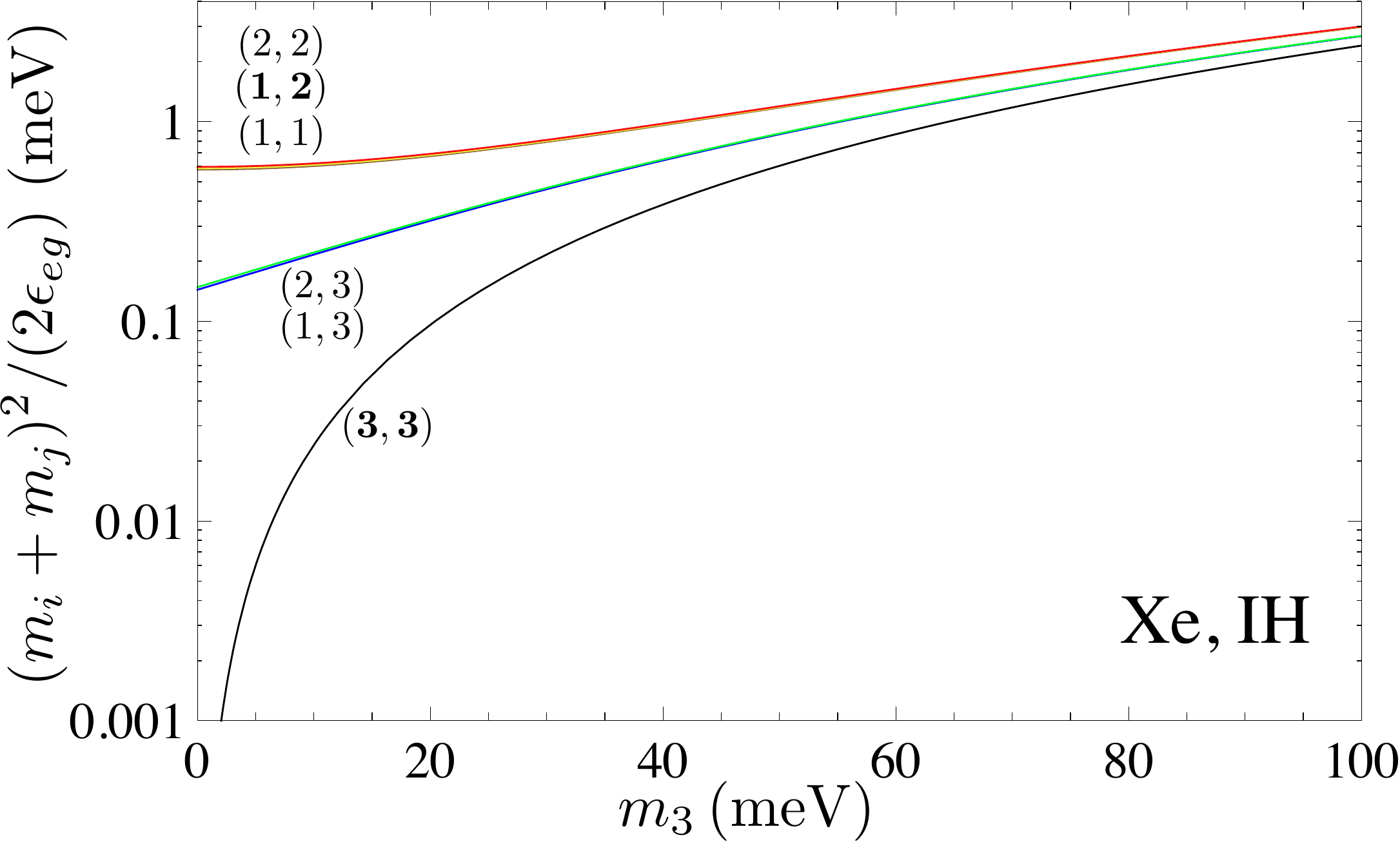} 
\caption{The locations of the six thresholds, the distances in energy
         from the half of level difference $\epsilon_{eg}/2$ given in
         meV unit, as functions of the smallest neutrino mass for 
         the case of Xe atom.
         The left(right) panel shows the NH(IH) case.         
         The pairs of numbers in parentheses indicate the thresholds and
         two prominent thresholds are printed in the bold font:
         (1,1) in black, (1,2) in blue, (2,2) in green, (1,3) in brown,
         (2,3) in yellow and (3,3) in red for the NH case;
         (3,3) in black, (1,3) in blue, (2,3) in green, (1,1) in brown,
         (1,2) in yellow and (2,2) in red for the IH case.}
\label{FIG:omXe}
\end{figure*}

In Fig.~\ref{FIG:omXe}, we show the threshold locations, 
that is, the threshold displacements from the maximal photon energy for 
the massless neutrinos, as functions of the lightest neutrino mass.
The multiplet structure is clear: a triplet, a doublet and a singlet 
for the NH case and the ordering of the multiplets is inverted
for the IH case.
Since the energy resolution of trigger lasers is not worse than
$O(10^{-6})\,\mathrm{eV}$, the identification of thresholds is 
possible provided that we have a sufficient statistics of data.
The expected number of events at each threshold is proportional to
the corresponding weight factor shown in Table \ref{tab_aij}.
The threshold of (1,2) has the largest weight and (3,3) is
the second one. Observing these thresholds, we will be able to determine
the corresponding neutrino masses and examine their consistency
with the known mass squared differences.

\begin{figure*}
\begin{center}
\includegraphics[width=0.6\textwidth]{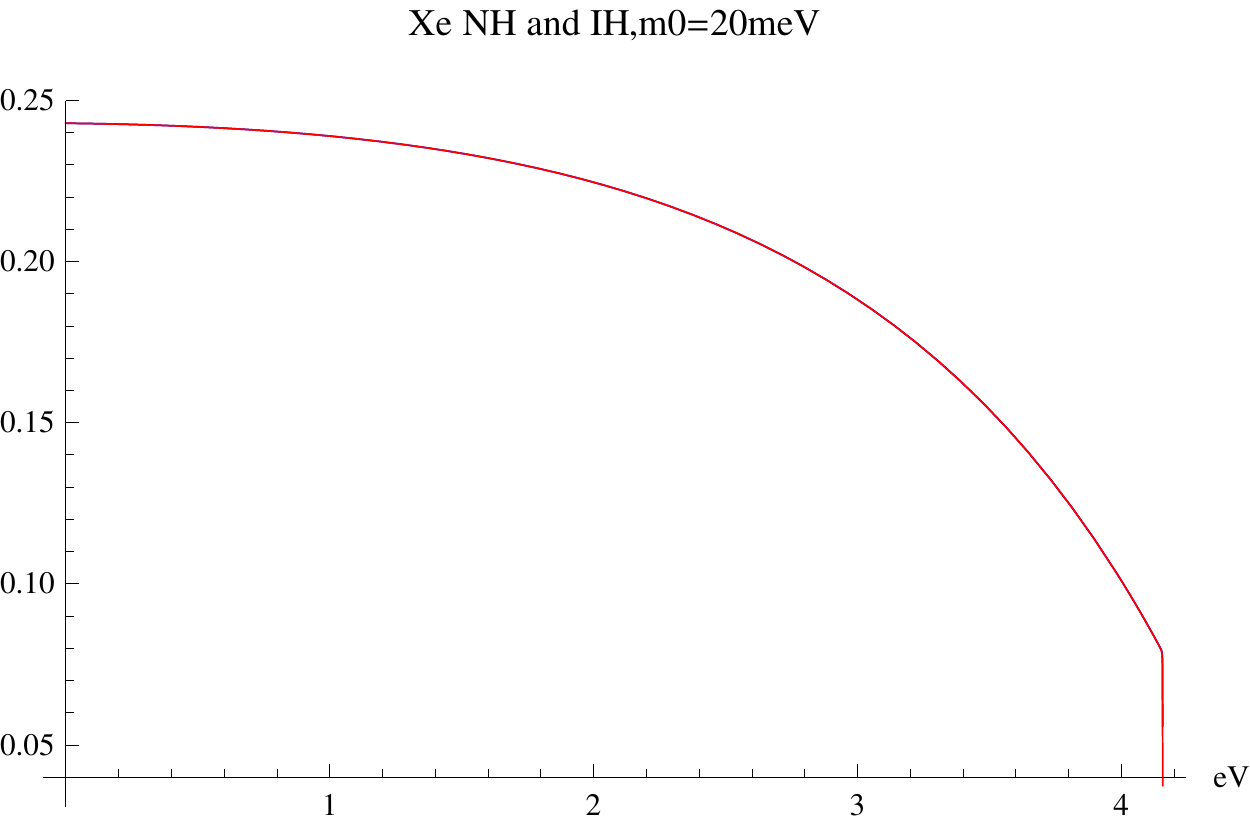}
\end{center} 
\caption{Xe RENP spectrum $I(\omega)$.
                In this global plot, the NH and IH spectra completely overlap each other, 
                and it is difficult to extract neutrino parameters.}
\label{xe renp spectral overall} 
\end{figure*}

\begin{figure*}
\begin{center}
\includegraphics[width=0.6\textwidth]{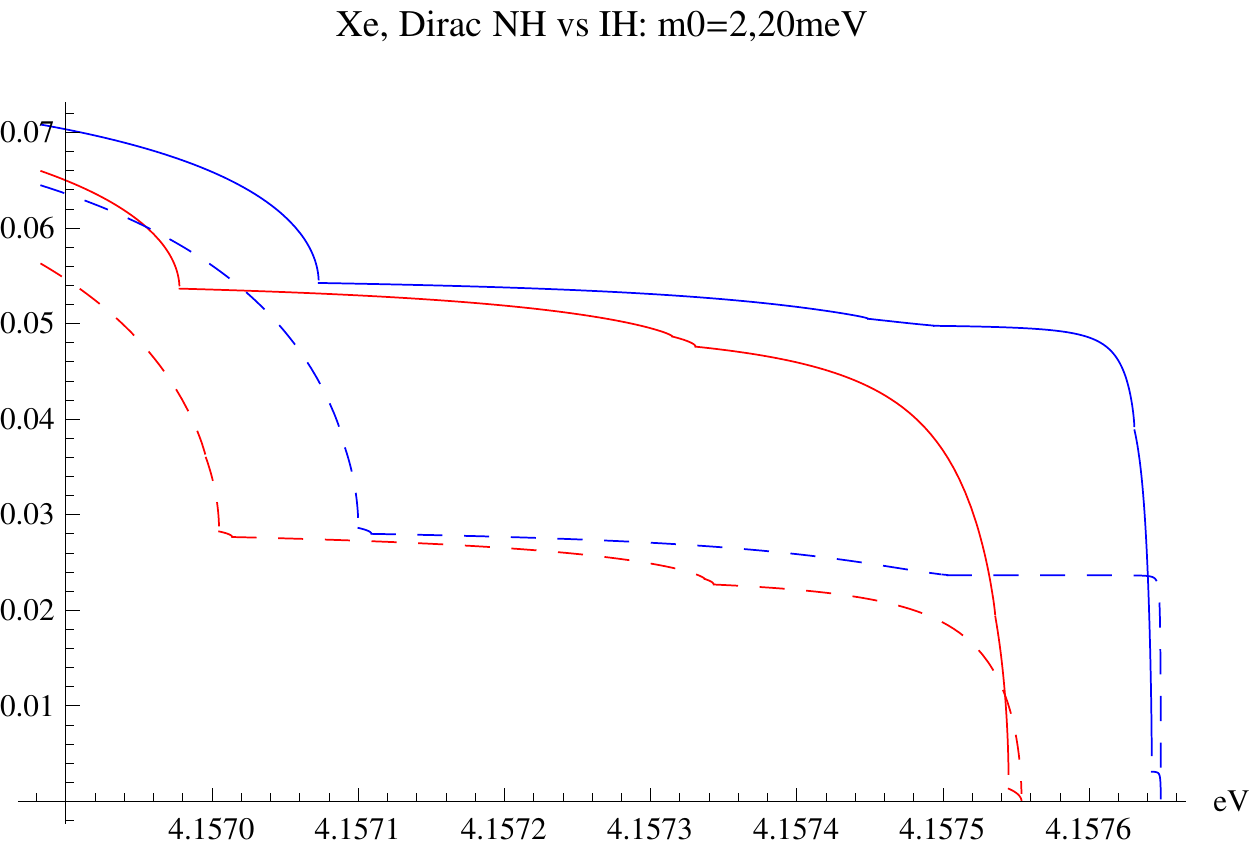}
\end{center} 
\caption{
Xe RENP spectrum $I(\omega)$ in the the threshold regions, 
using values of the smallest neutrino mass, 2 and 20 meV, for 
two cases of IH (in dashed curves) and NH (in solid). 
}
\label{xe renp spectral thresholds 2} 
\end{figure*}

The global spectral feature of RENP from Xe $J=2$ excited state
$5\mathrm{p}^5(^2\mathrm{P}_{3/2})6\mathrm{s}\,^2[3/2]_{J=2}$
is shown in Fig.~\ref{xe renp spectral overall},
while the enlarged threshold region is shown
in Fig.\ref{xe renp spectral thresholds 2}.
The locations of the thresholds 
corresponding to the three values of 
the smallest neutrino mass $m_0$ differ substantially.  
This feature can be used to determine the absolute neutrino mass.


There are two dominant thresholds and they are supposed to be
identified in a relatively earlier stage of experiment.
The most significant threshold comes from the emission of the neutrino
pair (1,2) and the second one is due to the (3,3) pair in both 
the NH and IH cases. The relative strength of these two thresholds
is a powerful tool to determine the hierarchy pattern: The threshold
at higher photon energy is stronger than the one at lower energy
in the NH case. While the former is weaker than the latter in the IH case.

\subsubsection{Dirac or Majorana}
Although Xe gives a good chance of
determining the overall neutrino mass scale or
the largest neutrino mass, along with distinction
of IH and NH, it would be too demanding to differentiate
Majorana neutrinos from Dirac neutrinos in Xe
experiment, since
their rate difference due to the interference term
is too small in this case.
We may characterize Xe RENP measurement  as
the first generation RENP experiment, since
it serves to identify the process itself along
with determination of an important neutrino parameter,
the largest neutrino mass,
although there are better targets for determination of
other neutrino parameters.
 
 In a recent paper \cite{yb-x renp}
 it was suggested that a smaller available energy of
 a fraction of eV, but still much larger 0.1 eV, 
might work for Dirac vs Majorana distinction.
 A candidate target is I$_2$ molecule, which has about $1/10$
 of Xe available energy.
The I$_2$ RENP experiment can be characterized by
the next generation experiment of precision neutrino mass
spectroscopy.
 
Before presenting RENP spectral shape from I$_2$ molecule,
it might be instructive to comment on molecular energy levels
and their associated wave functions \cite{textbook on molecules}.
As is well known, di-atomic molecules have vibrational and
rotational energies associated with relative atomic motion
besides electronic motions.
Electronic, vibrational, and rotational energies are
well separated due to three well separated velocities.
This makes it possible to compute electronic energy
levels assuming fixed positions of two nuclei,
leading to the concept of the potential energy curve of
each electronic configuration as a function of
nuclear distance.
Near the equilibrium position of nuclei the potential curve
may be approximated by a harmonic function whose
frequency $\omega_v$ gives energy levels of harmonic oscillator
$\omega_v (v+1/2)$.
We must deal with two potential curves corresponding to
two different electronic configurations, both for
M1 and E1 transitions.
This gives two overlapping integrals of vibrational
wave functions. Each of these overlaps is
called the Franck-Condon (FC) factor and calculated
assuming relevant vibrational quantum numbers
of respective electronic configuration.
Some rudimentary example of these is
illustrated in Appendix, using the Morse potential
for vibrational motion.
A work is in progress for accurate computation
of relevant potential curves for I$_2$, in collaboration with
theoretical chemists \cite{i2 renp rate}.

We have found that three relevant electronic states
for RENP are X (electronically ground state) for $|g\rangle$,
A' (metastable state) for $|e\rangle$, and the largest
overlap comes from A ($|p\rangle$).
Their experimental data are
taken from experimental values tabulated 
in Table \ref{tab_I2}. 
They are also used to construct the Morse potential in Appendix.

\begin{table}
\centering \caption{\label{tab_I2} 
$\mathrm{I}_2$ parameters.}
\begin{tabular}{|c|c|c|c|c|} \hline 
Electronic state of I$_2$& $D_e$(eV) & $T_e$(eV)  & $r_e$ ($10^{-8}$ cm) & 
$\omega_e$ (cm$^{-1}$) \\ \hline \hline
X $0_g^+ (^1\Sigma_g^+)$& 1.556 & 0 & 2.666 & 214.5 \\ \hline
A' $2u (^3\Pi_u)$& 0.311 & 1.245 & 3.079  &108.8  \\ \hline
A $1u (^3\Pi_u)$& 0.203 & 1.353  & 3.129 & 88.3 \\ \hline
\end{tabular}
\end{table}

\begin{figure*}
 \begin{center}
\includegraphics[width=0.6\textwidth]{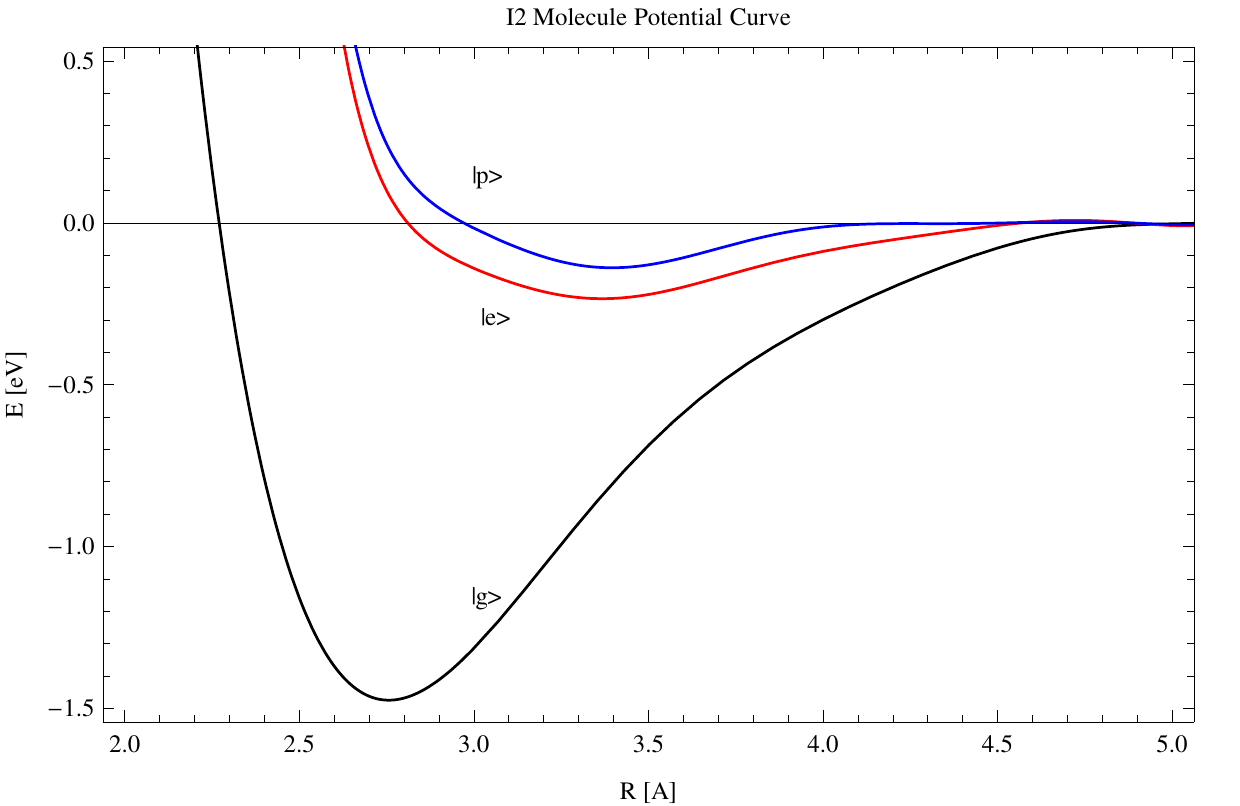}
 \end{center} 
   \caption{
   Conceptual diagram of potential curves of relevant electronic states of I$_2$  for RENP;
	$|e\rangle =A'\,, |g\rangle = X\,, |p\rangle = A$.
	Vibrational levels are suppressed for simplicity.
        The curves are drawn based on \cite{i2 spectral data}.}
   \label{i2 potential curves: conceptual} 
\end{figure*}

Here the energy unit $1 $cm$^{-1} = 1/8065.73$eV is used
for the curvature of the potential curve.
The dissociation energy $D_e$ is the energy difference between
the potential minimum at $T_e$ and level at dissociated atoms.
Conceptual diagram of relevant potential curves that
fit to these experimental data is shown 
in Fig. \ref{i2 potential curves: conceptual}.

We show a threshold feature of RENP photon spectra
for the transition between A'v=1 and Xv=15 of I$_2$ in 
Fig.~\ref{i2 renp spectral md 20meV}.
Although the difference in the spectral rates between Dirac and Majorana 
is not so significant even in I$_2$, it is possible to discriminate 
between them in principle.

\subsubsection{CP violating phases}
The rates of emission of 
$\nu_1\nu_3$ and of  $\nu_2 \nu_3$ 
are suppressed by the weights $B_{13}+B_{31}=0.033$ and  
$B_{23}+B_{32}=0.014$, respectively. Thus, 
it is most advantageous
to obtain information on the Majorana phase 
$\alpha$ at the (1,2) threshold, provided that the corresponding 
interference term is not suppressed by the smallness 
of the factor $m_1m_2$. 

We note that a linear combination of the 
CP violating phases given in Eq.~\ref{cpv phase at crossing} enter 
into the expression for the effective  Majorana mass measurable 
in the neutrinoless double beta decay:
\begin{eqnarray}
|\sum_i m_i U_{ei}^2|^2 &=&
5.8 \times 10^{-4}\,m_3^2 + 9.2\times10^{-2}\,m_2^2 + 4.5\times 10^{-1}\,m_1^2
 +4.1\times 10^{-1}\,m_1 m_2 \cos (2\alpha)
\nonumber \\ &&
 + 3.2\times 10^{-2}\,m_1m_3 \cos 2(\beta-\delta) 
+ 1.5\times 10^{-2}\,m_2m_3  \cos2(\alpha-\beta+\delta)
\,,
\end{eqnarray}
where the values given in Eq.~\ref{neutrino mixing angle parameters} are used for the mixing angle parameters.
The RENP experiment has an advantage
of independently determining all three CP factors Eq.~\ref{cpv phase at crossing}
instead of their single linear combination.


Sensitivity to the CP violating phases in I$_2$ RENP spectrum
is not as great as one might have expected, but
differences for different parameter sets $(\alpha, \beta)$
reaches several to 10\% for the inverted hierarchy case,
as seen in Fig.\ref{i2 renp spectral md 20meV} of Section 1.
For measurements of CP phases in the NH case
one should perhaps look for better candidate targets
of smaller available energy than I$_2$.

\begin{figure*}
 \begin{center}
\includegraphics[width=0.6\textwidth]{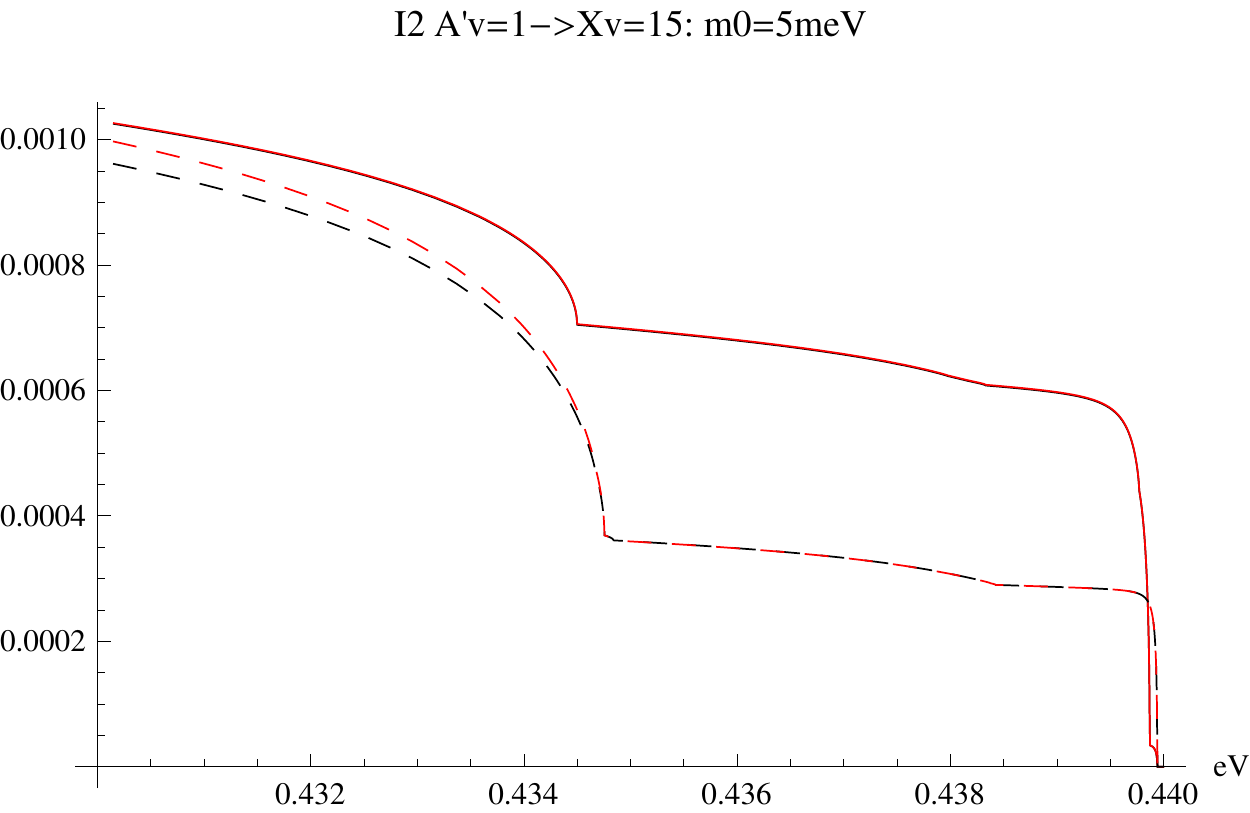}
 \end{center} 
   \caption{I$_2$ RENP spectrum between  A' v=1 and Xv=15.
   The Majorana  (in colored solid for NH and in colored dashed for IH) vs 
Dirac (in black solid for NH and in black dashed for IH) cases are compared.
A Majorana CP phase combination 
$(\pi/2,0)$ (in red) is taken, with  the smallest neutrino mass 5 meV. 
The vertical scale is in arbitrary units.
}
\label{i2 renp spectral md 5meV}
\end{figure*}

We show here I$_2$ RENP spectrum results 
for a smaller $m_0= 5$meV, which gives
smaller differences, as given
in Fig.\ref{i2 renp spectral md 5meV}.

%
%
\subsection{Estimation of the dynamical RENP factor based on asymptotic solution}
%

Our numerical outputs on RENP have so far been limited to the spectral shapes,
which are crucial to determine the neutrino parameters.
Some comments on the $\omega$ dependence of $\eta_{\omega}(t)$
in the fundamental formula 
$\Gamma_{\gamma 2\nu}(\omega, t) = \Gamma_0 I(\omega) \eta_{\omega}(t)$
are now in order.
Numerical simulations are necessary to compute
this quantity at each photon energy $\omega$, solving the master
equation Eq.~\ref{rescaled bloch eq1} $\sim$ \ref{rescaled quantum field eq2} in Appendix B
for two-color trigger irradiation.
The $\omega$ dependence of $\eta_{\omega}(t)$ thus originates from
$\omega$ dependence of basic couplings $\alpha_{ab}(\omega)$
that appears in the master equation,
in particular through $\gamma_-=(\alpha_{ee}-\alpha_{gg})/2\alpha_{ge}$ and $\alpha_{ge}$.
As seen in the formulas for these in Appendix B,
both of these are smooth functions of $\omega$.
Furthermore, if the upper level $|p\rangle$ lies much
above $\epsilon_{eg}$ (the energy difference between
$ |e\rangle$ and $|g\rangle$), these $\alpha_{ab}(\omega)$'s are nearly
$\omega$ independent.
In any event, one expects that variation of $\eta_{\omega}(t)$ 
in the vicinity of each threshold $\omega_{ij}$ can be neglected,
and the theoretical formula of $I(\omega)$ reflects the
spectral shape of experimental data in the threshold regions.
This way one can extract important information of
neutrino parameters directly from measured data.
It is however desirable to measure
the quantity $\eta_{\omega}(t)$ or their asymptotic value at times
$\gg T_2, T_3$ for the overall fitting
to the entire spectral shape.
This method is described in Sec. 4.

Whether one can actually perform RENP experiments with
meaningful results depends on
the absolute magnitude of rate $\Gamma_0$:
if it is too small, the spectral shape computation would be useless.
For instance, although the Xe RENP rate $\Gamma_0$
is of order
$1\,\mathrm{Hz}\,(n/10^{22}\mathrm{cm}^{-3})^3\,(V/10^2\mathrm{cm}^3)$,
the dynamical factor $\eta_{\omega}(t)$ might give
a further large suppression. 

Due to the complexity of our non-linear system
our simulations so far are limited to the explosive cases and
the linear regime, mainly for pH$_2$.
We feel that the late time evolution in
the intermediate output regime 
(case of large M1$\times$E1 coupling) is very important for RENP, which
has not been explored extensively due to
technical complexity.
Moreover, the RENP target choice is very limited
so far: Xe and 
I$_2$ (whose absolute rate is under study\cite{i2 renp rate}).
The case of Xe transition involves a smaller
$\alpha_{ge}$ (due to M1$\times$E1 nature of RENP), 
but a larger $\gamma_- \sim - 7300$
(the corresponding number for  pH$_2 \sim 0.6$),
which makes numerical simulations for $\eta_{\omega}(t)$ harder.
Under this circumstance our conclusion on the absolute
rate is premature and we shall further update the simulation.
Nonetheless, it would be useful to summarize what
we have done so far.

We expect that the final stage of field and
target state within the medium
is described by static solutions of
our master equation, which gives
a long time behavior and is stable
beyond the largest relaxation time $T_1$.
These static states are solutions obtained by
taking vanishing time derivatives of the master equation,
Eq.~\ref{rescaled bloch eq1} $\sim$ \ref{rescaled quantum field eq2} in Appendix B.
We first solve Bloch vector components, both grating
and spatially homogeneous modes, in terms fields.
These are inserted into the field equations to
search for static solutions by solving
the coupled system of ordinary differential equations.

\begin{figure*}
 \begin{center}
 \includegraphics[width=0.6\textwidth]{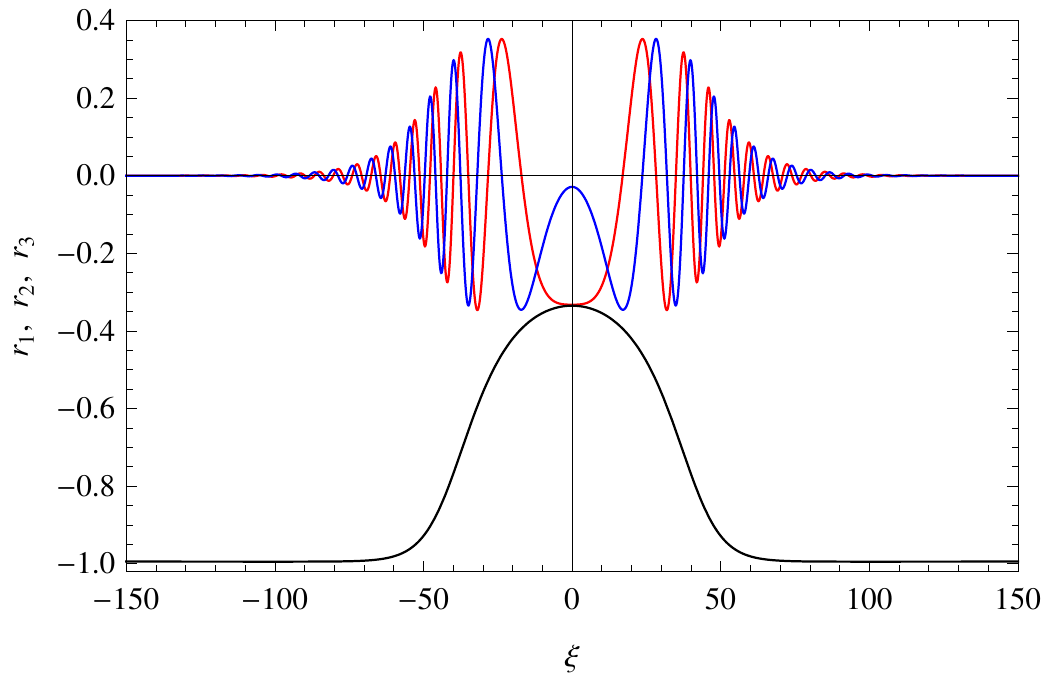}
 \end{center} 
   \caption{Example of an asymptotic solution for the input parameters
$\gamma_{\pm}=1, \tau_{1}=200, \tau_{2}=100, \tau_{3}=0.001$.
The red, blue and black curves show $r_{1}^{(0)}$, 
$r_{2}^{(0)}$ and  $r_{3}^{(0)}$, respectively.}
   \label{asymptotic 1}
\end{figure*}

\begin{figure*}
 \begin{center}
 \includegraphics[width=0.6\textwidth]{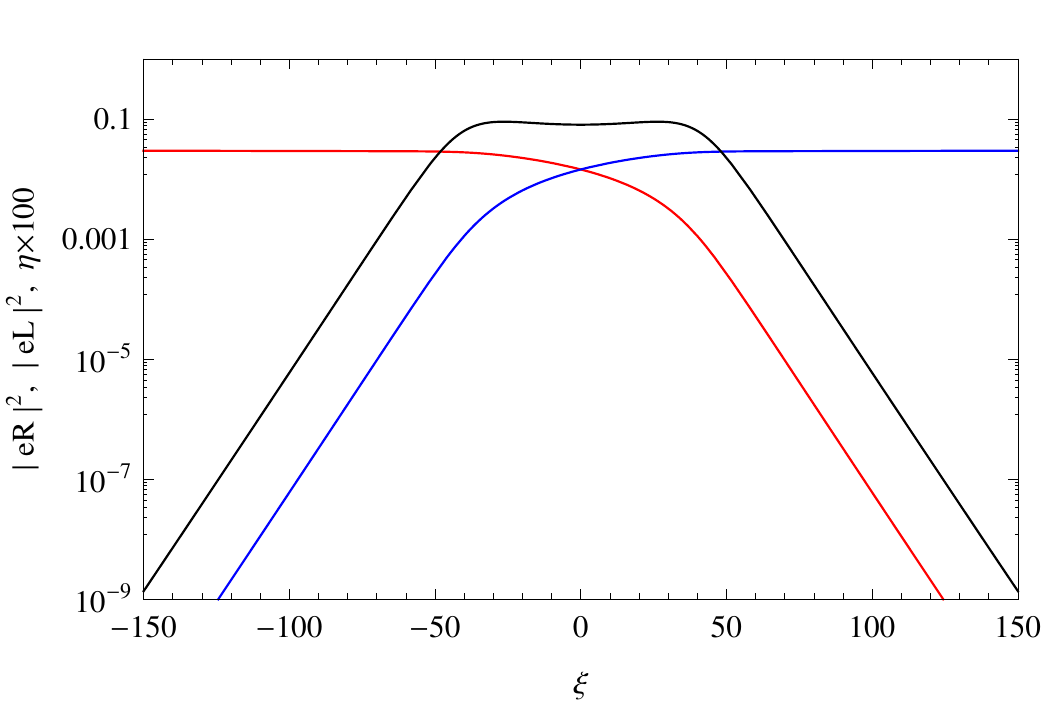}
 \end{center} 
   \caption{Field intensity and the dynamical factor
$\eta_{\omega}$ for the same parameter as in Fig.\ref{asymptotic 1}
The red (blue) curve shows $e_{R}^2$ ($e_{L}^2$), and the black line is
$\eta_{\omega} \times 100$.}
   \label{asymptotic 2}
\end{figure*}

Results of spatial profiles of the static
asymptotic solutions are illustrated in Fig.\ref{asymptotic 1}
and Fig.\ref{asymptotic 2} in dimensionless units.
The master equation for the single color mode,
Eq.~\ref{rescaled bloch eq11} $\sim$ \ref{rescaled quantum field eq22},
has been used for simplicity.
Due to the choice of a small $\tau_3$
this solution resembles the soliton solution
given in \cite{psr dynamics},
but it gives a relatively large $\eta_{\omega}$
(defined by Eq.~\ref{spatial integral} $\sim$ \ref{spatial integra3}), too.
This asymptotic case gives a time independent dynamical factor;
$ \eta_{\omega}(t) = \eta_{\omega}$(constant).
Whether a large $\eta_{\omega}$ factor is realized or not
in actual experimental situation needs further
study, both from theoretical and experimental
sides.
Moreover, it should be clarified how the time-evolving
dynamical factor $\eta_{\omega}(t)$ approaches the asymptotic constant $\eta_{\omega}$.
Also, relevance of the field condensate making up the asymptotic $\eta_{\omega}$
value to solitons in \cite{psr dynamics} should be
investigated.
(The soliton is distinguished from a mere condensate by
the presence of topological quantum number.)

$\;$ \\ 
\section{Experimental aspects of PSR and RENP}\label{Sec:Experiments}

\subsection{Overview and strategy towards PSR/RENP experiments}\label{Subsec:Experimet-Overview}
\subsubsection{Experimental principle of the neutrino spectroscopy with atoms}
We begin this Section by reviewing the experimental principle of our neutrino spectroscopy.
First of all, we prepare a collection of excited state of atoms or molecules.
We then observe photon spectrum in a radiative emission of neutrino pair (RENP) process denoted by
\begin{equation}
	| e \;\rangle \rightarrow | g \;\rangle  + \gamma + \nu_{i} + \nu_{j},
\end{equation}
where $\nu_{i}$'s are neutrino mass eigenstates.
Rich physics information exists in the photon spectrum, especially near the threshold regions given by 
\begin{equation}
	\omega_{ij}= \frac{\epsilon_{eg}}{2}-\frac{(m_{i}+m_{j})^2}{2 \epsilon_{eg}},
\end{equation}
where $\epsilon_{eg}$ is the energy difference between 	$|e \;\rangle$ and $|g \;\rangle$, and 
$m_{i(j)}$ denotes the neutrino masses.
The physics observables include
\begin{itemize}
\item neutrino absolute mass scale,
\item mass hierarchy pattern, normal or inverted,
\item mass type, Dirac or Majorana,
\item CP violating phases, $\alpha$ and $\beta - \delta$,
\end{itemize}
among others. Detailed discussion of the physics objectives can be found in Sec. 3.

A key notion in our experimental principle is the macro-coherent amplification mechanism.
In a word, the amplification is based on their collective de-excitation among all of the coherent atoms. 
It enhances not only the rate of the de-excitation process but also guarantees a ``momentum conservation law" among the emitted particles (photon and two neutrinos in the case of RENP). 
The physics information can be extracted from the photon spectrum with the help of this momentum conservation law.
Since we crucially rely upon this macro-coherent amplification mechanism, we plan to prove its validity by a separate experiment. 
We discuss this proof-of-principle experiment in more detail below.
The readers are referred to Sec. 2 for detailed discussion on the theory of the macro-coherent amplification.

Another important feature in our experiment is use of the trigger laser. 
It initiates coherent de-excitation process, and in effect selects the portion of the photon spectrum we like to investigate. 
In this sense, our neutrino spectroscopy uses a ``narrow-band spectrometer".  
We note that energy resolution of the spectrometer is determined by the laser line width, 
which can be much less than a few hundred MHz, or $\sim 1\,\mu$eV.
Detector energy resolution is not critically needed for spectroscopic information, but it helps to reject background signals.
The role of the trigger laser goes beyond the initiation of coherent de-excitation process; it helps developing condensate formation, another key notion in the RENP neutrino spectroscopy. 
The condensate, a system of coherent atoms and fields, is an ideal target state for RENP process. 

As discussed in  Sec. 3 in detail, the final RENP rate $\Gamma_{\gamma 2 \nu}$ is given by
\begin{equation}
	\Gamma_{\gamma 2 \nu}(\omega,t)=\Gamma_{0} I(\omega) \eta_{\omega}(t)
\end{equation}
where $\Gamma_{0}$ is the overall constant rate and $I(\omega)$ is 
	the spectrum function containing physics information.
The third factor $\eta_{\omega}$, termed as the dynamical RENP factor, is 
proportional to the squared product of the field $E^{-}$ and 
  the medium's coherence $(R_1-iR_2)/2$,
        averaged over along the target.
In order to extract $I(\omega)$, we need to divide experimentally observed rate $\Gamma_{\gamma 2 \nu}$
	by $\Gamma_{0}$ and $\eta_{\omega}$.
The constant $\Gamma_{0}$ can be calculated accurately while the RENP factor $\eta_{\omega}$
	may be obtained by detailed simulations.
It is, however, highly desirable to measure $\eta_{\omega}$ more directly.
We plan to determine it experimentally by measuring the coherence and 
	population difference ($r_{3}$) of the target medium as well as the field intensities 
        at both ends.
With aid of the master equations, $\eta_{\omega}$ can be determined from those quantities determined above.
One method of measuring coherence ($r_{1}$,$r_{2}$) for the case of Xe atoms is described in Sec. 4.3.

We detect single photon emission by an appropriate photon detector.
The detector may be UV to IR photo sensors, depending upon the value of initial energy gap between 
	metastable and ground states, but most likely we must use IR sensors 
        to investigate the detailed nature of neutrinos such as Majorana-Dirac distinction and CP phases.
We expect the photon emission direction to be along the trigger laser direction because macro-coherence mechanism preferentially develops the process with this mode.
Thus detector solid angle coverage need not be large.

\subsubsection{Paired Super-Radiance}
\begin{figure}
\begin{center}
\includegraphics[width=0.6\textwidth]{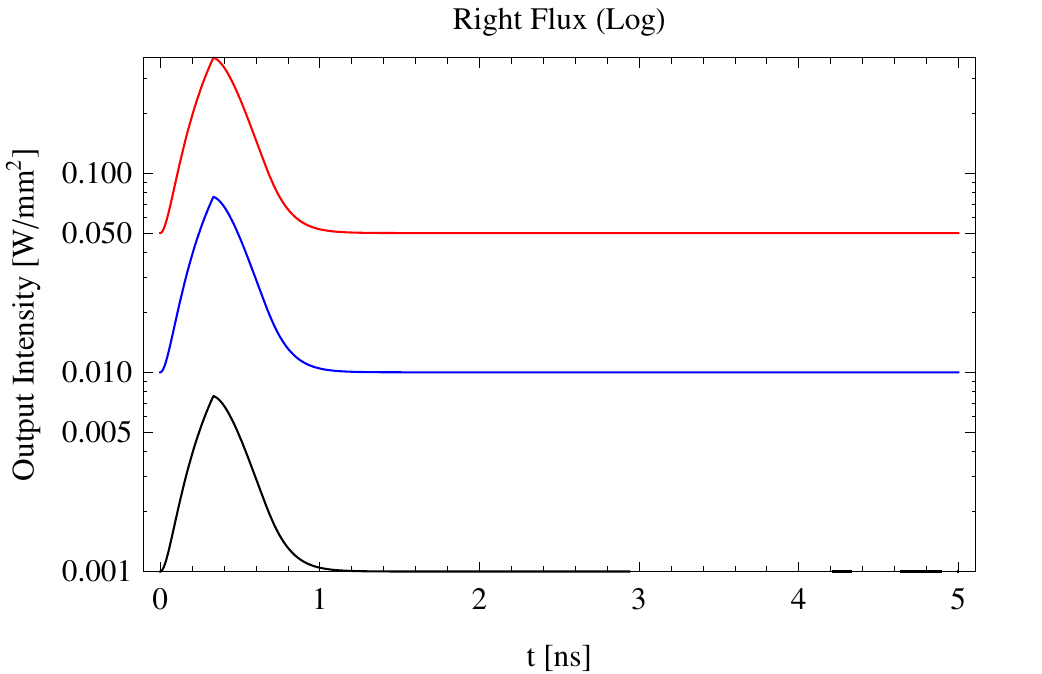}
\end{center}
\caption{Right-moving output flux from pH$_{2}$ vibrational excitation level ($Xv=1$). 
The initial coherence is assumed to be spatially homogeneous ($r_{1}^{(0)}=1$).
The black, blue and red lines correspond to trigger laser intensities of 1 mW/mm$^2$,
10 mW/mm$^2$, and 50 mW/mm$^2$.
Note logarithmic scale for the vertical axis. Left flux is identical to the right flux.}
\label{fig:pH2PSR30atmRightLog-Pure-Coherence}
\end{figure}

Exploitation of the macro-coherent amplification mechanism is an essential feature in our neutrino spectroscopy.
Since this is a new phenomenon not discussed in the past literature, we should first prove and study it in detail experimentally.
We like to do it with a QED process called paired super-radiance (PSR), a process which emits two photons 
in the decay from an excited state to the ground state ($|e \rangle \rightarrow |g \rangle + \gamma \gamma$). 
If the macro-coherent amplification is in operation, a collection of atoms emit coherently a pair of photons with the same energies. 
In particular, direction of the two photons is opposite (back-to-back) when the medium's initial macroscopic polarization (coherence) is spatially-homogeneous.
Thus observation of such radiations in a similar fashion to SR is an unambiguous proof of the principle.
There are several advantages of proving the new principle with PSR.
The most important is its rate: it is expected to be much larger than RENP since it is a QED interaction.
Characteristic features of the radiation, back-to-back with equal energies, make experimental distinction easy from other possible backgrounds.
The conditions and natures of PSR, mainly explosive PSR, are already shown in Sec. 2.
We summarize them here briefly, adding some more studies concerning to non-explosive events.
As shown in Sec. 2, explosive PSR events have the following characteristics:
\begin{itemize}
 \item Major part of energy stored initially in medium is released as two photons  
       in an extremely short period of time compared with medium's natural lifetime. 
 \item There exists certain threshold conditions above which explosive events take place.
 \item The threshold conditions are governed predominantly by target density and length, 
       initial coherence and de-phasing time, {\it etc.}. 
 \item Left-right symmetric (back-to-back) events occur when initial coherence is spatially-homogeneous 
       ({\it i.e.} $r_{1}^{(0)}$ or $r_{2}^{(0)} \simeq 1$ ) and trigger laser is injected from both ends.
 \item Asymmetry between left and right flux grows as  a larger initial value is assumed for the spatially-inhomogeneous coherence.
\end{itemize}
It is of great importance, from both theoretical and experimental view points, 
 to study what is expected if some of the threshold conditions are not fulfilled.
Below we show some results obtained by numerical studies performed with these motivations.
The process of interest is the same process treated in Sec. 2; de-excitation process 
	of para hydrogen molecules from a vibrationally-excited state ($Xv=1$)
        to the ground state ($Xv=0$).
Figure \ref{fig:pH2PSR30atmRightLog-Pure-Coherence} 
	shows right-moving output flux.
The three different curves in black, blue and red, correspond to 
three different input trigger laser intensities of 1 mW/mm$^2$,
10 mW/mm$^2$, and 50 mW/mm$^2$.
Other important parameters are
\begin{equation}
	n=8 \times 10^{20} \;[\mbox{cm}^{-3}], \quad
        L=10               \;[\mbox{cm}], \quad
        r^{(0)}_{1}=1, \quad
        T_{2}=T_{3}=0.2 \;[\mbox{ns}]
\end{equation}
where $r^{(0)}_{1}$ means initial coherence of spatial homogeneous mode, with all other
coherence set to zero.
The important fact to these results is the values of $T_{2}$ and $T_{3}$; here a factor 50 smaller value is assumed 
	than the typical values employed in Sec. 2.
As shown in Fig.\ref{fig:pH2PSR30atmRightLog-Pure-Coherence},
	events are found to have much simpler time structure
        compared with explosive events shown in Sec. 2.
Also their peak heights are much smaller and proportional to the input trigger intensity.
Although these events appear less dramatic, if observed, they bear great significance because 
	the two-photon process, which has the natural life time of $\sim 10^{16}$ sec, is seen 
        in a time period of nsec.
  
\begin{figure}
\begin{center}
\includegraphics[width=0.495\textwidth]{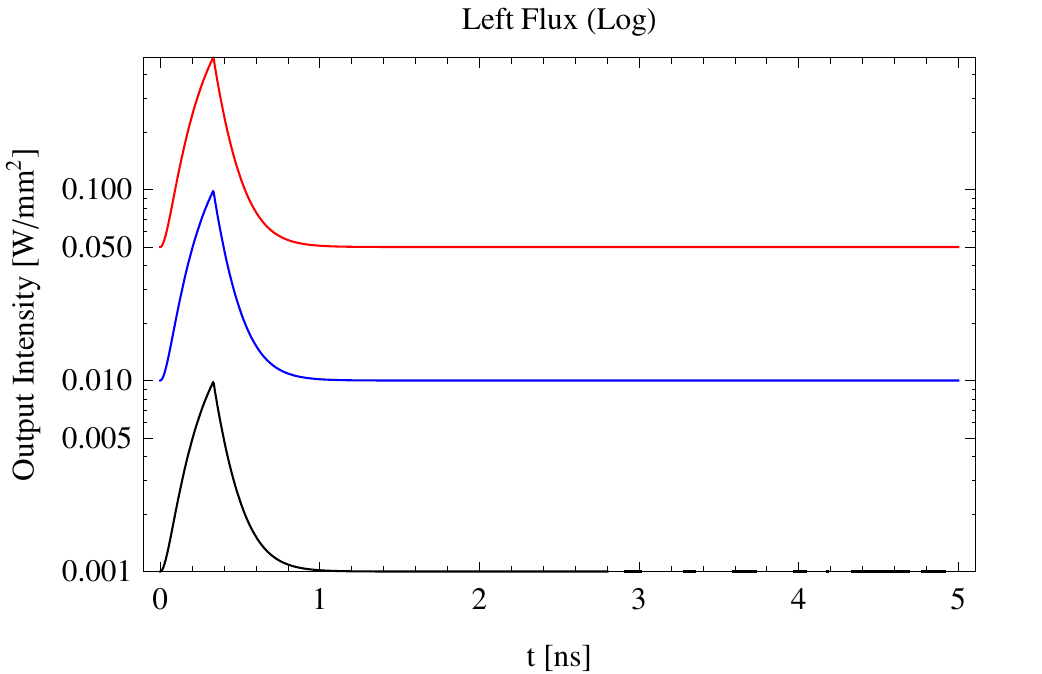}
\includegraphics[width=0.495\textwidth]{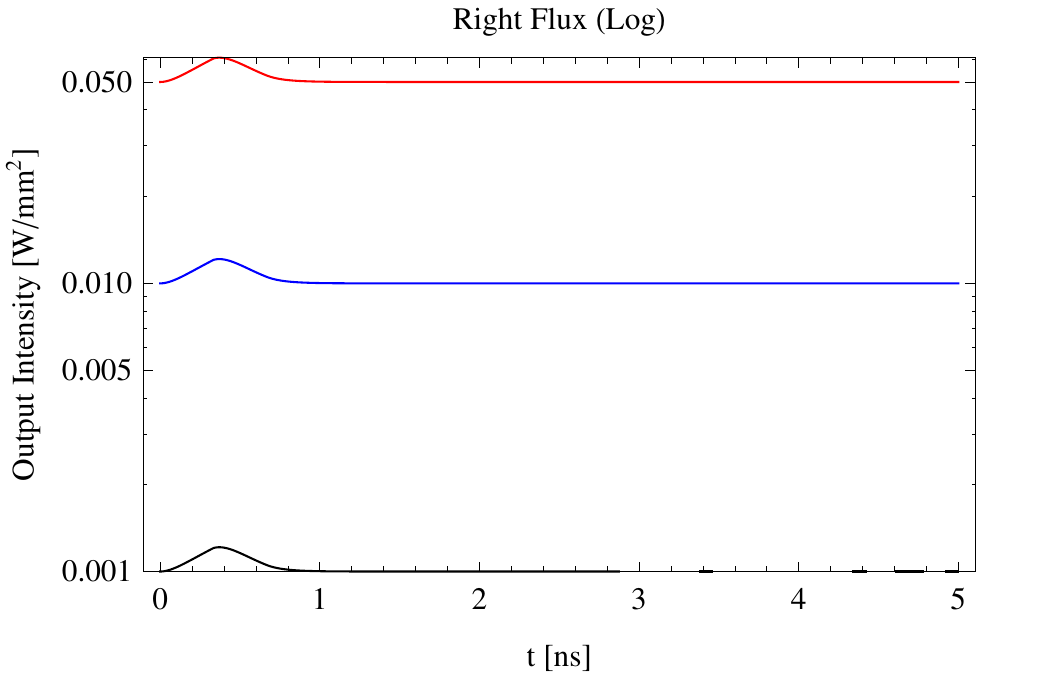}
\end{center}
\caption{Left and right moving output flux from pH$_{2}$ vibrational excitation level ($Xv=1$). 
The initial coherence assumed is $r_{1}^{(0)}=0.1$ and $r^{(+)}_{1}=0.9$.
The black, blue and red lines correspond to trigger laser intensities of 1 mW/mm$^2$,
10 mW/mm$^2$, and 50 mW/mm$^2$.
Note logarithmic scale for the vertical axis.}
\label{fig:pH2PSR30atmLeftLog-Mixed-Coherence}
\end{figure}

Initial coherence of spatial grating modes, $r_{1,2}^{\pm}$, affects substantially 
	partition between left and right output flux, as already pointed out in Sec. 2.
Figure \ref{fig:pH2PSR30atmLeftLog-Mixed-Coherence} 
	shows an output flux when the initial coherence is set to
\begin{equation}
	r^{(0)}_{1}=0.1, \quad r^{(+)}_{1}=0.9, \quad
\end{equation}
with all other parameters left same as in Fig.\ref{fig:pH2PSR30atmRightLog-Pure-Coherence}.
In this case, the right output flux is much smaller than the left. 
The fact remains true that the output flux is proportional to the input laser intensity.



The requirements for the PSR target, different from those for RENP in certain aspects, can be summarized as follows.
\begin{itemize}
\item The initial state must be metastable (E1 forbidden), and must have at least one intermediate state of higher energy 
connecting between initial and final (ground) states via E1 $\times$ E1 allowed transitions.
\item Both longitudinal ($T_{1}$) and transverse ($T_{2},\; T_{3}$) relaxation times must be longer than the PSR characteristic time scale, the time required to develop collective and coherent de-excitation process.  
\item  The initial atomic state must be prepared to have large coherence.
\item The atomic density in the initial state must exceed a certain threshold value.
\end{itemize}
At present we plan to realize a PSR experiment with para-hydrogen (pH$_2$) molecules, and details of the experiment will be described in the subsequent section.
We would like to end this topic by emphasizing two characteristic features of the PSR experiment.
First, it is a type of an experiment in which one tries to realize required experimental parameters or conditions.
This is because, due to its highly non-linear nature, PSR event has a clear threshold above which signal can be observed without fail.
This is in contrast to an experiment in which events occur proportional to elapse time.
Second, the role of trigger laser is important as in the case of RENP.
As stated already, the trigger laser initiates collective and coherent de-excitation process.
Without the trigger laser, PSR events would not be initiated; this is in sharp contrast to the Dicke type super-radiance, 
in which quantum fluctuation can actually start coherent de-excitation process.

\subsubsection{ Experimental strategy toward neutrino spectroscopy with atoms }
There are two crucial features in realizing the neutrino spectroscopy with atoms/molecules.
One is the experimental proof of the macro-coherent amplification mechanism, 
and the other is formation of the stable condensate; both are related to the dynamical RENP factor $\eta_{\omega} (t)$.
Thus our strategy is to establish macro-coherent amplification mechanism by PSR first.
As was stated already, we plan to realize PSR experiments with vibrational levels of pH$_2$ molecules.
The current status of  PSR experiment with pH$_2$ is described in detail in Sec. 4.2.

Studies on condensate formation are one of the crucial step after establishing the PSR process.
The points of condensate studies are
\begin{itemize}
 \item condensate formation method
 \item time development of stored energy and coherence
 \item experimental stability against PSR.
\end{itemize}
These studies will be performed with Xe atom and other heavy atoms/molecules.
Simultaneously with this condensate study, 
target selection studies for the neutrino spectroscopy will be pursued. 
The requirements for the target are qualitatively the same as in the case of PSR 
except that the intermediate state(s) must connect the initial and final states via E1 and M1 transitions.
In addition, the initial and final energy difference has profound impact on the neutrino spectroscopy 
	since important physics objectives such as Dirac-Majorana distinction and/or 
        CP violating phase determination can only be achieved with energy difference 
        comparable to the neutrino mass scale ($<0.5$ eV).
Experimental studies on those new challenging subjects are described in Sec. \ref{Subsec:Experimet-Xe}. 
Results of the some preliminary studies using other methods are presented in Appendix E.

\subsection{PSR experiment with para-hydrogen molecule}\label{Subsec:Experimet-pH2}

In this Section we describe PSR experiment with pH$_{2}$ in detail. 
\begin{figure}
\begin{center}
\includegraphics[clip,width=0.35\textwidth]{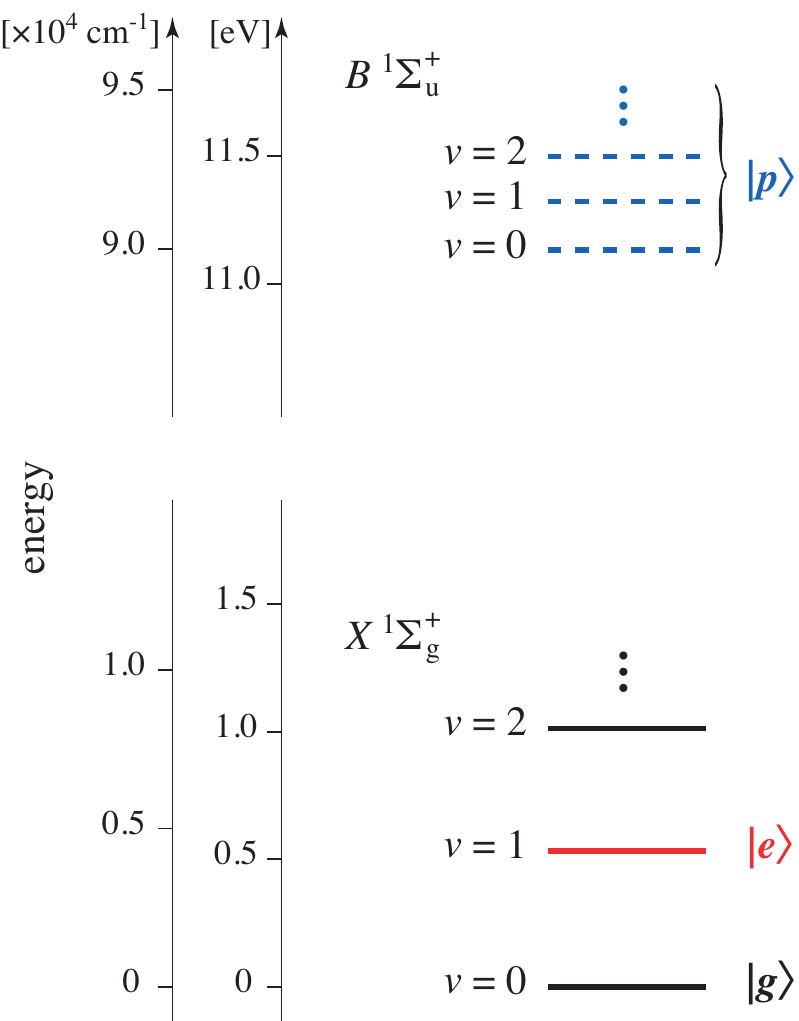}
\end{center}
\caption{Energy levels of pH$_{2}$ relevant to the PSR experiment.}
\label{fig:Energy levels of pH2}
\end{figure}
The diagram of the pH$_2$ energy levels relevant to this experiment is shown in Fig.~\ref{fig:Energy levels of pH2}.
The two-photon process from the vibrationally excited state ($Xv=1$) to
the ground state ($Xv=0$) is measured. 
We note vibrational transitions of the E1 type is strictly forbidden in homo-nuclear diatomic system.
Figure \ref{fig:pH2-setupX} shows a schematic diagram
of the experimental setup.
It mainly consists of three parts: 
(a) the target,
(b) the excitation laser system, and
(c) the trigger laser system.
The target is LN$_{2}$-temperature pH$_{2}$ gas contained in a cell.
The initial $v=1$ state is prepared by the excitation laser system.
The system consists of two color pulse lasers whose wavelengths are
532~nm and 683~nm. 
The trigger laser is a continuous wave (CW) laser and its wavelength
is $4.8$~$\mu$m.
The excitation laser beams are combined with the trigger laser
beam by a dichroic mirror (DCM) so that they propagate collinearly in
the target.
PSR outputs (left moving outputs) and excitation lasers
are dispersed by a prism and separately detected with a photo-diode
(PD). 
The backward components (i.e. right moving outputs) pass through the
Faraday rotator and are reflected by the polarizing beam splitter; then
they are detected by another PD.
In the following, details of the target and laser systems are described.


\begin{figure}
\begin{center}
\includegraphics[width=0.8\textwidth]{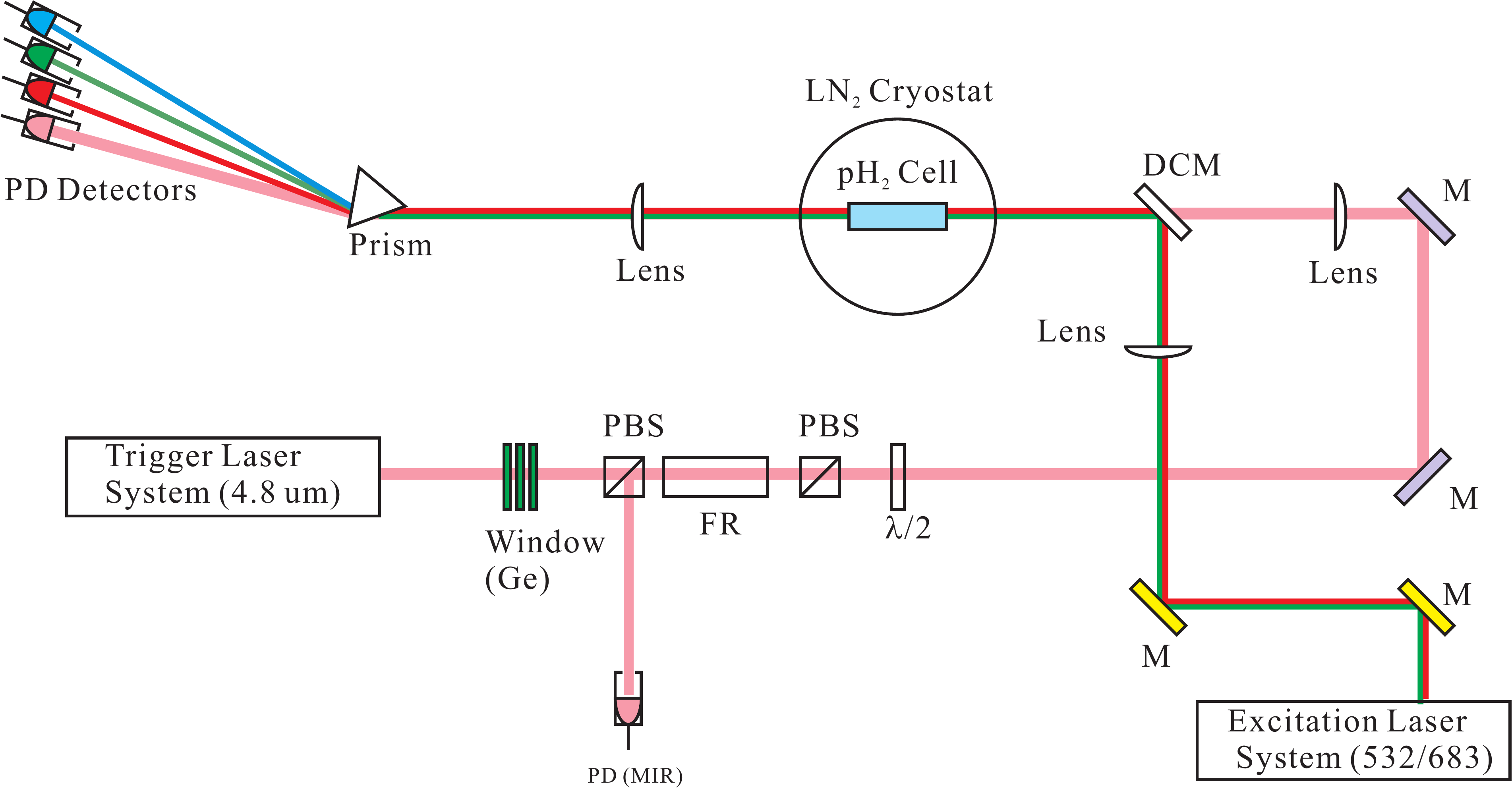}
\end{center}
\caption{Schematic diagram of the PSR experiment with pH$_{2}$.
FR: Faraday rotator, $\lambda/2$: half-wave plate, PD: photo-diode, M: mirror, DCM: dichroic mirrors, PBS: polarizing beam splitter.}
\label{fig:pH2-setupX}
\end{figure}

\subsubsection{Gas and solid para-hydrogen target}\label{Subsubsec:Experimet-pH2-Target}
Preparation of target, namely selection of atom/molecule species, choice of transition levels, 
 preparation of initial conditions {\it etc.}, is decisive factor for successful PSR experiments.
From the point of view of atom/molecule selection, important parameters to be considered are:
\begin{itemize}
\item Number density of atoms/molecules
\item Target length
\item Damage threshold (the maximum laser intensity which a medium can receive without damage)
\item Relaxation time ({\it T$_{1}$, T$_{2}$}).
\end{itemize}
Particularly important are the number of excited atoms/molecules and 
	coherence of ${\left|e\right>}$ and ${\left|g\right>}$. 
Both are determined by the number density of ground state atoms/molecules, 
target length and the excitation laser power. 
A higher density, a longer target and a higher damage threshold provide better conditions. 
However, the high-density and long-coherence are contradictory, 
	because frequent interaction in high-density samples makes the coherent time short;
        thus compromise is essential for this purpose.

For PSR, relevant transition between two levels should be E1 forbidden and 
	E1 $\times$ E1 allowed to have a large two-photon transition moment. 
Vibrational transitions of homo-nuclear diatomic molecules fulfill these conditions. 
Numerical simulation in Sec. 2 and Sec. 4.1 suggests that vibrational transition of para-hydrogen (pH$_{2}$) 
	with a large initial coherence is one of the ideal target for PSR experiments. 
Below properties of both gaseous and solid pH$_{2}$ are described and compared. 
Atoms or molecules trapped in solids (pH$_{2}$ for example) are also interesting targets for PSR/RENP; 
experimental studies in this direction are presented in Appendix E.

\begin{figure}
\begin{center}
\begin{minipage}{7.25cm}
	\includegraphics[width=\textwidth]{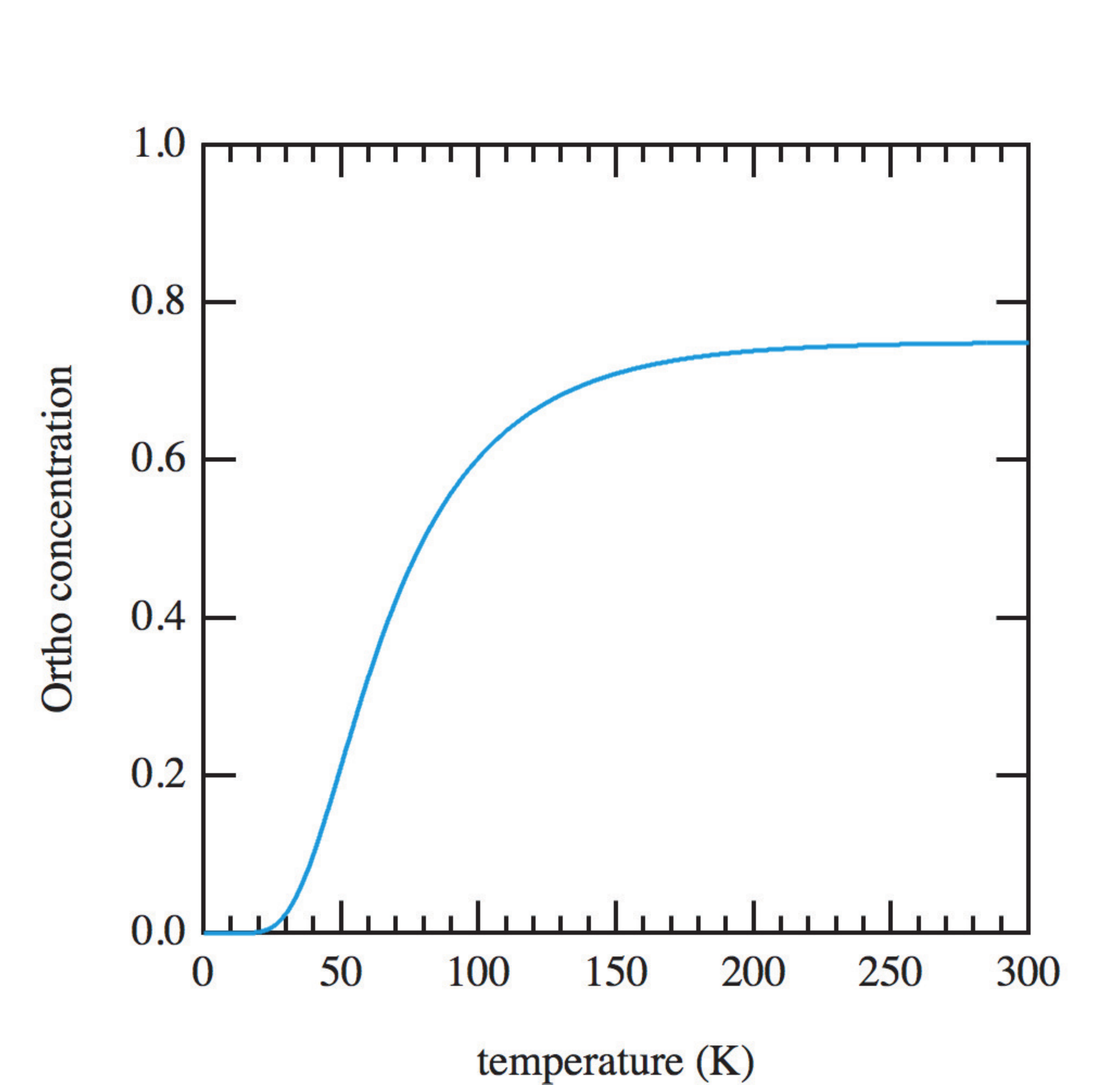}
	\caption{Ortho-para ratio (o/p) of thermally equilibrium hydrogen at 0 to 300 K. 
		The o/p can be controlled from almost zero to 0.75 by magnetic catalysts kept at corresponding temperature.}
	\label{hydrogen1}
\end{minipage}
\begin{minipage}{0.5cm}$\;$\end{minipage}
\begin{minipage}{7.25cm}
	\includegraphics[width=\textwidth]{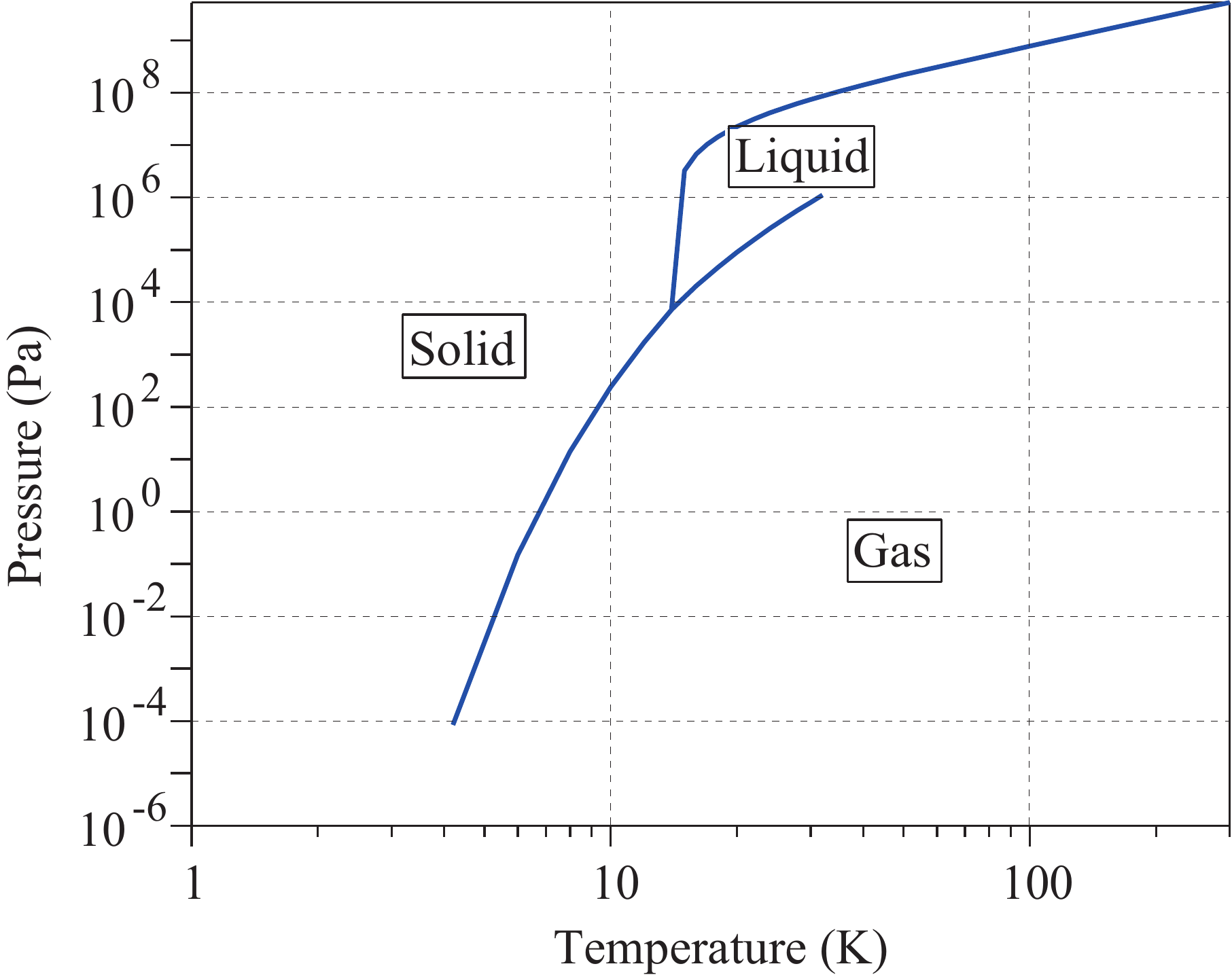}
	\caption{Phase diagram of eH$_{2}$.\cite{Hydrogen-Soures} Temperature and pressure at the triple point of eH$_{2}$ is about 13.8 K and 7000 Pa, respectively. Structure of solid pH$_{2}$ is h.c.p. at low temperature, low o/p.}
	\label{hydrogen2}
\end{minipage}
\end{center}
\end{figure}

\paragraph{Properties of molecular hydrogen}
The molecular hydrogen, the simplest diatomic molecules, has attracted great interest in various fields, both theoretical and experimental. 
Physical properties of the molecular hydrogen are well summarized in P. Clark Souers' book.\cite{Hydrogen-Soures} 
The nuclear spin isomers of hydrogen molecules show different nature. 
The nuclear spin of proton ({\it I} = 1/2) causes two spin states of molecular hydrogen, 
para ({\it I} = 0) and ortho ({\it I} =1). 
The nuclear spin wavefunction of pH$_{2}$ is anti-symmetric under the exchange of two identical protons 
	while that of ortho-hydrogen (oH$_{2}$) is symmetric. 
Therefore, the rotational quantum number {\it J} of pH$_{2}$ and oH$_{2}$ should be even and odd in the vibronic ground state, respectively, 
	because overall wavefunction should be anti-symmetric under the exchange of nuclei. 
Conversion between these two states is very slow without external magnetic fields so that it is possible to treat them as different molecules. 
On the other hand, ortho-para ratio (o/p) easily reaches thermal equilibrium with a magnetic catalyst. 
The o/p ratio of thermally equilibrium hydrogen (denoted by eH$_{2}$) at room temperature is 3:1 
	and it converges to zero at the limit of 0 K (Fig. \ref{hydrogen1}) and can be controlled from almost zero to 0.75 
        by passing gaseous or liquid hydrogen into magnetic catalysts kept at corresponding temperature. 
        It is worth noting that even at the triple point (13.8~K) o/p $\sim$ 10$^{-5}$, which is the lower limit by this method.

The phase diagram of eH$_{2}$ is shown in Fig.\ref{hydrogen2}.
The temperature and pressure at the triple point of eH$_{2}$ is about 13.8 K and 7000 Pa, respectively. 
The low melting point allows us to use gaseous sample at wide temperature ranges while solid sample should be kept around liquid Helium temperature. Structure of solid pH$_{2}$ is known to be hexagonal closed packing (h.c.p.) at low temperature and low o/p.
Spectroscopic data of molecular hydrogen are also well known. 
The light mass makes the rotational constants and vibrational frequency fairly large. Energy splitting between {\it J} = 0 and {\it J} = 2 (pH$_{2}$) and {\it J} = 1 and {\it J} =3 (oH$_{2}$) in the vibronic ground states is 354.38 cm$^{-1}$ (510 K) and 587.06 cm$^{-1}$ (845 K). 
This large energy separation strongly suppresses the population of rotationally excited states in thermal equilibrium at low temperatures. 
Even at 77 K, the Boltzman factor of the {\it J} = 2 state is less than 10$^{-2}$. The vibrational frequency is 4401.2 cm$^{-1}$. The first electronic excited state lies at  91700 cm$^{-1}$ above the ground state and can be treated as a far-resonant state in our experimental scheme. Rovibronic energy states are not largely changed in solid pH$_{2}$ because of weak inter-molecular interaction.
Hydrogen is homo-nuclear diatomic molecule with no permanent electric dipole moment so that both pure rotational transition and rovibrational transition are dipole-forbidden although weak electric quadrupole transition between {\it v} = 1 and {\it v} = 0 was observed with long optical path. \cite{Hydrogen-quad} On the other hand, Raman transition is allowed and has been studied extensively, especially vibrational Raman transition. 
Because of the large vibrational energy, molecular hydrogen is often used as a medium of a Raman shifter. 
Both solid and gaseous hydrogen have been investigated as coherent Raman media by adiabatic manipulations. \cite{Hydrogen-Harris1} \cite{Hydrogen-Harris2} \cite{Hydrogen-Hakuta} \cite{Hydrogen-Katsuragawa} 
These approaches are suggestive because coherence of system is constitutive for PSR.

\paragraph{Gaseous hydrogen}
The coherence time of vibrational states of gaseous hydrogen can be estimated from the linewidth of vibrational Raman transitions. 
The linewidth depends on the number density (pressure) and temperature. 
Hereafter, the relation of number density and coherent time is discussed.

In general, linewidth of gaseous sample is determined by
\begin{itemize}
\item Natural lifetime
\item Doppler broadening
\item Collisions (Pressure broadening)
\item Power broadening
\item Transit-time broadening.
\end{itemize}
The power broadening and transit-time broadening is not intrinsic linewidth of sample. 
It is to be noted that the type and nature of decoherence (homogeneous or inhomogeneous) are not distinguished here 
	and only total coherence time is discussed.
The parameter $T_{1}$ describes the lifetime due to the spontaneous emission and represented by the Einstein's A coefficient of samples. In the case of molecular hydrogen gas, its rovibrational transition is dipole-forbidden. It means that the lifetime of rovibrational excited states is substantially long. Actually, lifetime due to the quadrupole transition of the {\it v} = 1 state is in the order of 10$^{6}$ sec. \cite{Hydrogen-quad} Natural lifetime broadening can be neglected in what follows.

Doppler broadening, which is attributed to inhomogeneous Doppler effect caused by Maxwell-Boltzman distribution of molecular velocity, 
is dominant in low pressure gases. 
Neglecting a small relativistic correction, the full width at half maximum (FWHM) of Doppler broadening is given by
\begin{eqnarray}
\Delta\nu_{D}=\nu_{0}\;\;\sqrt[]{\mathstrut \frac{2kT \log(2)}{mc^{2}}}
\end{eqnarray}
where $\nu_{0}$, $T$, $m$ and $c$ is the transition frequency, temperature, mass of molecules and the velocity of light, respectively.

Collisional broadening, also called as pressure broadening, is dominant in high pressure gas. 
Detailed analysis of their lineshape is difficult because complicated inter-molecular interactions should be considered. \cite{Hydrogen-collision} However, its lineshape is approximately Lorentzian and its FWHM is proportional to the pressure. Pressure-broadening coefficients are species-dependent values and typically 10 MHz per Torr.

\begin{figure}
\begin{center}
\includegraphics[width=0.6\textwidth]{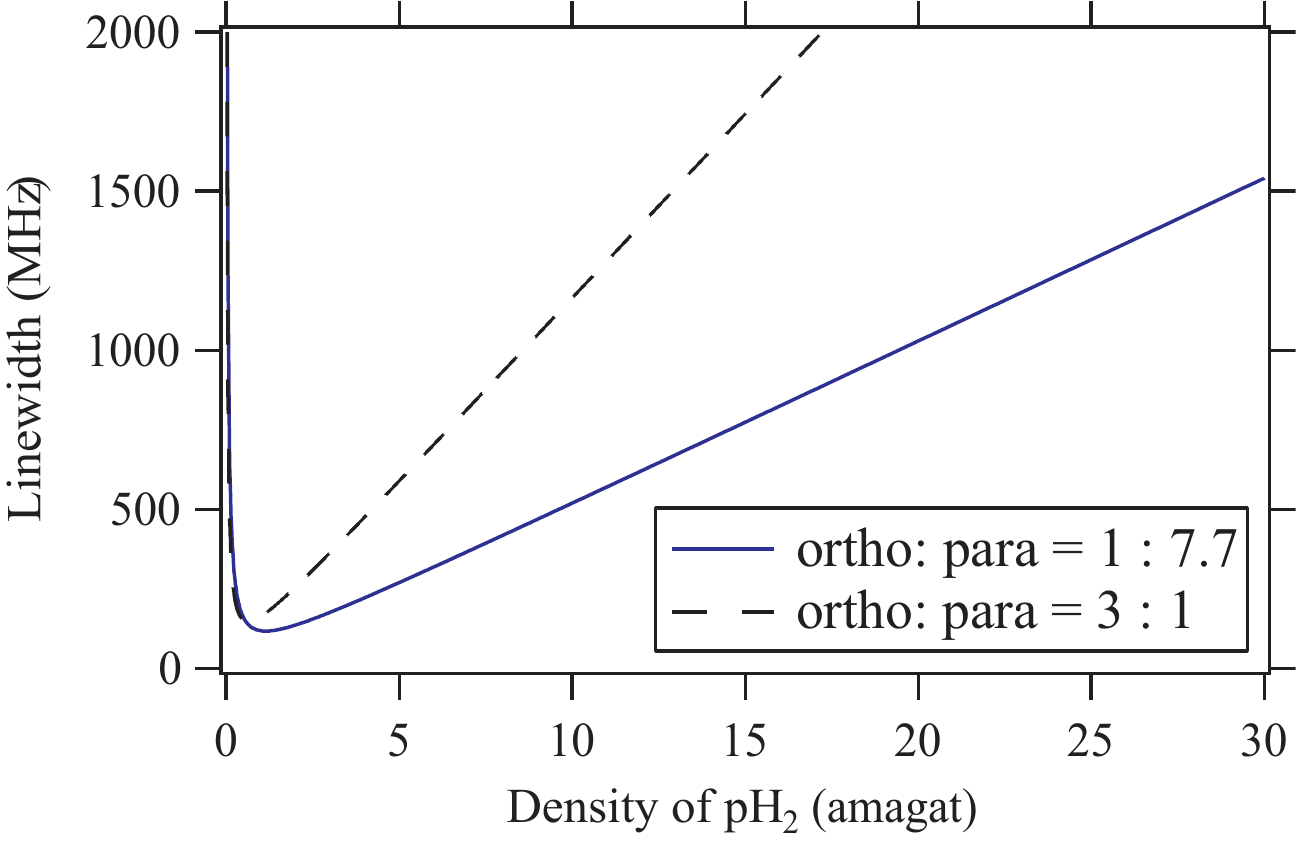}
\end{center}
\caption{Linewidth of Q$_{1}$(0) Raman transition of gaseous pH$_{2}$ at 81 K 
with ortho-para ratio of 1:7.7 and 3:1 as a function of density of pH$_{2}$.}
\label{hydrogen3}
\end{figure}

In the case of rovibrational transitions of molecular hydrogen gas, the pressure and temperature dependence of Raman linewidth is well studied. Bischel and Dyer \cite{Hydrogen-width} reported experimental linewidth and fitting coefficients using phenomenological equation as
\begin{eqnarray}
	\Delta\nu_{hydrogen}=\frac{A}{\rho}+B \times \rho
\end{eqnarray}
where $\rho$ is the total number density of pH$_{2}$ and oH$_{2}$ in amagat 
(1 amagat = 2.69 $\times$ 10$^{19}$ cm$^{-3}$) and $A$, $B$ is fitting coefficients.
(Table \ref{Hydrogen_width}). The second term corresponds to pressure broadening. 
The first term represent effect of Doppler broadening and "Dicke-narrowing". The Dicke-narrowing appears between the Doppler dominant region and the collision dominant region. This narrowing is due to a smaller Doppler shift caused by frequent velocity-changing collisions. 
Fig. \ref{hydrogen3} shows linewidth of the Q$_{1}$(0) Raman transition at 81 K and 
ortho:para = 1:7.7 and 3:1 as a function of density of pH$_{2}$.
%
The minimum linewidth of the o/p = 1:7.7 sample at 81 K is 117.4 MHz (corresponds to 2.7 ns) at 1.15 amagat (= 3.09 $\times$ 10$^{19}$ cm$^{-3}$). It is easy to prepare $\sim$ 10$^{21}$ cm$^{-3}$ pH$_{2}$ gas. However, $T_{2}$ is in the order of 0.1 ns at such high pressure.

Damage threshold by laser field of gaseous hydrogen depends on frequency of the applied field and is probably determined by non-resonant multi-photon ionization in most of our frequency region. Recent ultrafast experiments reported multi-photon ionization of hydrogen molecules by visible light is dominant at $>$ $\sim$ TW cm$^{-2}$.\cite{Hydrogen-multiphoton} 
It is easy to elongate the sample cell of gaseous hydrogen to 1 m or more and possible to achieve longer pass length by using a multi-pass cell or a cavity.

\begin{table}
\caption{Linewidth coefficients of Raman Q$_{1}$(0) transition of gaseous pH$_{2}$.\cite{Hydrogen-width}}
\label{Hydrogen_width}
\begin{center}
\begin{tabular}{llll}
\hline
T(K) & ortho:para & A (MHz amagat) & B (MHz /amagat)\\
\hline
81 & 3:1 & 189 $\pm$ 40 & 29.0 $\pm$ 1.0 \\
81 & 1:7.7 & 76 $\pm$ 6 & 45.4 $\pm$ 0.8 \\
298 & 3:1 & 257 $\pm$ 12 & 76.6 $\pm$ 0.8\\
\hline
\end{tabular}
\end{center}
\end{table}

\paragraph{Solid hydrogen}
Solid pH$_{2}$ is an attractive target from the point of coherent experiments because it fulfills high density and long coherence simultaneously. 
The number density of saturated solid pH$_{2}$ is about 2.6 $\times$ 10$^{22}$ cm$^{-3}$ at 4 K, which corresponds to that of gaseous sample at 1000 atm, 300 K. Due to weak interaction, not only vibrational motion but also rotational motion of hydrogen are quantized and coherence time is much longer than classical solids. The long coherence time of the excited vibrational state is estimated to be on the order of 10 ns from linewidth of stimulated Raman spectroscopy. \cite{Hydrogen-momose} The time-resolved coherent anti-Stokes Raman spectroscopy (TRCARS) also supported this value. 
Introduction of the TRCARS experiment and discussion of the long {\it T$_{2}$} of solid pH$_{2}$ are in Appendix D. 
Long {\it T$_{1}$} (not radiative) of the {\it v} = 1 state in solid pH$_{2}$ was also reported to be $\sim$ 40 $\mu$s at 4.8 K.\cite{Hydrogen-Li}

However, damage threshold of solid pH$_{2}$ was reported to be 180 MW cm$^{-2}$, which is fourth order smaller than that of gaseous sample. \cite{Hydrogen-Hakuta} Furthermore, it may be troublesome to prepare longer solid pH$_{2}$ than 10 cm with optically transparent quality. Application of multi-pass or cavity is also difficult because of scattering in the solid.

\paragraph{Comparison between pH$_{2}$ gas and solid targets}
 Table \ref{Hydrogen_comp} lists a set of parameters and their typical values relevant to the PSR experiment, comparing 
 between gaseous and solid pH$_{2}$.
 As seen, the solid phase is better from the view points of the number density and de-phasing time $T_2$.
 Demerit of using solid phase is its low damage threshold. 
 It limits attainable number density and
 initial coherence low; actually too low to observe PSR events with our current technique.
 On the other hand, the numerical simulation of Sec. 4.1 shows that the linear regime PSR may well be observed 
 with gas phase pH$_{2}$.
 We have thus chosen gas phase pH$_{2}$ aiming at the first observation of PSR events.
 It should be noted however that solid pH$_{2}$ is much more attractive once the damage threshold limitation is overcome.
 Some development efforts along this line are described below and Appendix D.
\begin{table}
\caption{Typical properties of gaseous and solid pH$_{2}$ Parameters of gas is at the minimum linewidth condition at 81 K. Parameters of the solid is at around 4 K.} 
\label{Hydrogen_comp}
\begin{center}
\begin{tabular}{lll}
\hline
 & gas & solid \\
\hline
Density (cm$^{-3}$) & 10$^{19}$ $\sim$ 10$^{21}$ & 2.6 $\times$ 10$^{22}$\\
{\it T$_{1}$} (s) & 10$^{6}$ & 4 $\times$ 10$^{-5}$ \\
{\it T$_{2}$} (ns) & 3 $\sim$ 0.1 & $\sim$10  \\
Damage threshold (W cm$^{-2}$) & $\sim$ 10$^{12}$ & $\sim$ 10$^{8}$\\
Target length (m) & 0.1 $\sim$ 1 & $\leq$ 0.1 \\
\hline
\end{tabular}
\end{center}
\end{table}

\paragraph{Experimental techniques}
In solid pH$_{2}$, the coherence time depends largely on the oH$_{2}$ concentration. 
The linewidth of vibrational Raman transition to the $v$ =1 state is about 10 MHz at o/p of 2000 ppm while this becomes 60 MHz at 20000 ppm.\cite{Hydrogen-momose} 
Then, a highly purified pH$_{2}$ is desired. 
Purity of pH$_{2}$ is also important in gaseous pH$_{2}$ because FWHM of gaseous hydrogen 
	at the same pH$_{2}$ density is probably smaller for the pure pH$_{2}$ sample as seen in Fig. \ref{hydrogen3}, 
        although there is no precise data at various ortho-para ratios. 
Almost pure pH$_{2}$ can be obtained by passing hydrogen gas into magnetic catalysts such as FeO(OH) kept at about 14~K. 
The o/p ratio of prepared samples can be estimated in several ways. 
By measuring IR absorption spectra of solid hydrogen, the o/p ratio is determined by intensity of oH$_{2}$ induced pH$_{2}$ vibrational transition, which can be observed in solid phase. 
A typical value of ortho-para ratio in our laboratory is about 500 ppm.

\begin{figure}
\begin{center}
\includegraphics[width=0.4\textwidth]{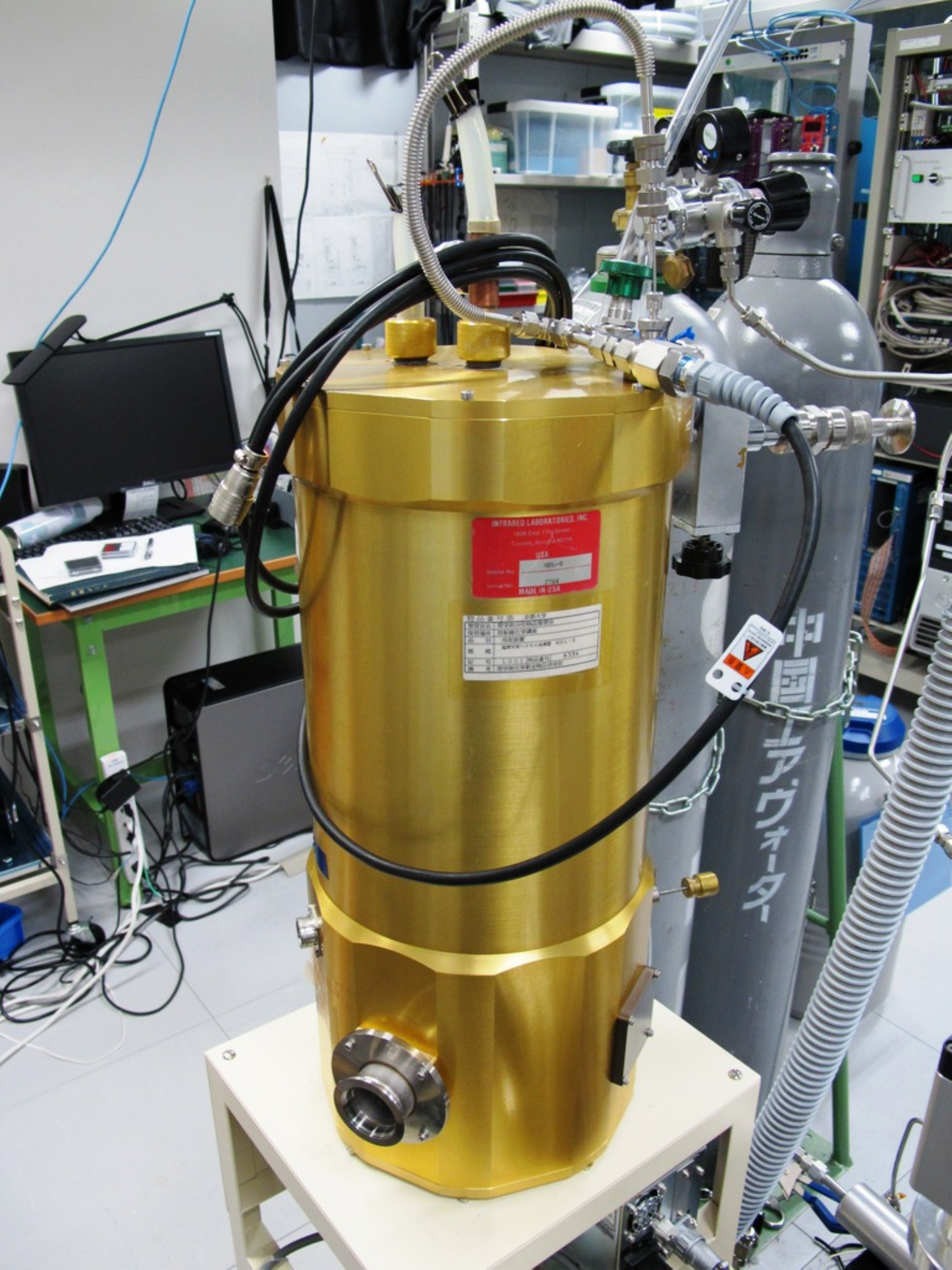}
\end{center}
\caption{The cryostat for gas experiments (Infrared Laboratories, ltd. HDL-8). Both liquid N$_{2}$ (77 K) and He (4.2 K) can be used as cryogenic liquids. Temperature of a cold stage can be controlled by a heater with a PID circuit. There are two view ports with IR windows for optical experiments.}
\label{hydrogen4}
\end{figure}

\begin{figure}
\begin{center}
\includegraphics[width=0.4\textwidth]{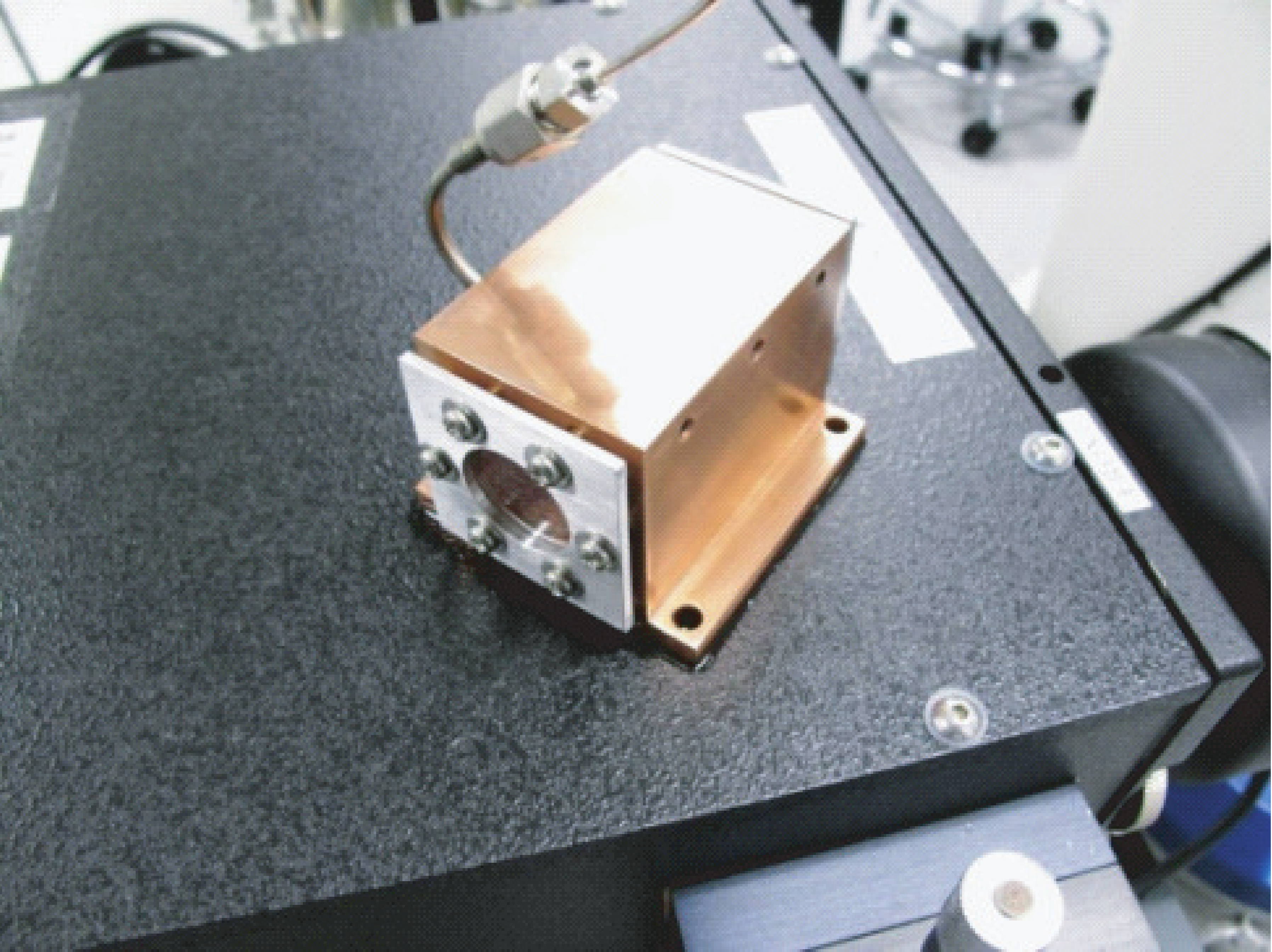}
\end{center}
\caption{The home-made gas cell of 7.5 cm length and 2 cm diameter applicable to $\sim$ 1 K and $\sim$ 500 kPa. The cell was made of oxygen-free copper and both sides were sealed by IR windows with indium gaskets for an optical path. Sample gas was installed via stainless tubing connected to cell with silver soldering.}
\label{hydrogen5}
\end{figure}

For gaseous experiments, a sample cell was attached to a cold stage of a cryostat  (Fig. \ref{hydrogen4}).
 The temperature of cell was measured by Si diodes and controlled by a heater with a PID circuit. 
 The cell was made of oxygen-free copper and both sides were sealed by IR windows with indium gaskets for an optical path. We usually use BaF$_{2}$ windows for both cell and cryostat because of its excellent IR transmittance. The optical path is 7.5 cm in length and 2 cm in diameter (Fig. \ref{hydrogen5}).
Sample gas was installed via stainless tubing connected to cell with silver soldering. 
Pressure capacity of the cell is around 500 kPa. For higher pressure experiments, 
	it is better to use small diameter cell and CaF$_{2}$ or Sapphire windows. 
For example, pressure capacity of a cell with 4 mm diameter and windows of 2 mm thickness is 8 MPa (CaF$_{2}$ windows) or 40 MPa (Sapphire windows).

For solid experiments, a vacuum chamber with a liquid Helium bath or a closed cycle refrigerator (cooling power should be larger than 1.0 W at 4 K) are used. 
The optical grade crystal is prepared typically by the closed cell method. 
This method was originally developed by Oka et al..\cite{Hydrogen-closed} 
The prepared almost pure pH$_{2}$ gas is introduced through a thin stainless steel tube into a copper cell cooled by a cryostat. 
The cell is similar to that shown in Fig. \ref{hydrogen5}. The inlet gas flow rate is controlled to keep the cell temperature at about 8 K. The solid pH$_{2}$ grows radially inward from the cell wall and forms a polycrystalline aggregate of h.c.p. crystals having their c axes normal to the wall of the cell. The sample thus grown is transparent. 
The crystals prepared by this method sometimes suffer from the cracking especially during the cooling cycle after the crystal growth and the inhomogeneity due to the polycrystalline structure.

To prepare higher quality crystals, pressurized liquid method was developed by Hakuta et al.\cite{Hydrogen-press} 
Crystals in the closed cell method are obtained by direct condensation from gaseous hydrogen. 
In contrast, this method produces crystals from the pressurized liquid phase by which crack free single crystal can be obtained. 
Solid pH$_{2}$ on the melting curve about 15 K has a similar molar volume as the solid at $\sim$ 4 K under the saturated vapor pressure. 
Thus, a homogeneous crystal is grown at about 15 K by pressurizing the liquid-phase first, and then the transparent crystal is cooled down to the experimental temperature. The obtained pH$_{2}$ crystal is transparent and uncracked and have a higher damage threshold for laser irradiation.

\subsubsection{Laser system for coherence preparation and trigger}
\label{Subsubsec:Experimet-pH2-Laser}
\newcommand{\ket}[1]{\big\vert\, #1\, \big\rangle}
%
In this section, we explain laser systems for the PSR experiment
using para-H$_2$ vibrational states.
We explain the excitation laser system first, and
the trigger laser system is described later.
Since we employ 
a nonlinear optical frequency-conversion technique 
for the laser systems,
we also explain this technique briefly.
For detail, the reader is referred to a standard textbook
such as 
Ref. \cite{RWBoyd:NonlinearOptics2c2} or 
Ref. \cite{Demtroder:LaserSpectroscopy41c58}.

\begin{figure}
\begin{center}
\includegraphics{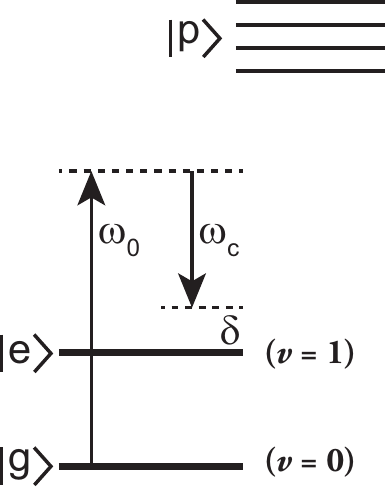}
\caption{Energy level diagram to prepare coherence $\rho_{eg}$ in an
 atomic or molecular system. Two color driving laser fields at
 frequencies of $\omega_0$ and $\omega_{c}$ are applied. A small 
 detuning of $\delta$ is negative in this diagram.
 When the detuning $\delta$ is appropriate and the intensity of driving
 lasers is sufficiently large, the magnitude of the coherence of this
 transition becomes its maximum value.
}
\label{fig:ramantransition}
\end{center}
\end{figure}

As described in Sec. \ref{Subsec:Experimet-Overview},
it is important to prepare a large atomic (or molecular) coherence in
the ground state $\ket{g}$ and the excited state $\ket{e}$ for the
experiment. 
To prepare a large coherence, we manipulate the stimulated Raman
process adiabatically as depicted in Fig. \ref{fig:ramantransition}.
For adiabatic manipulation, we apply two-color strong driving lasers
whose frequencies are slightly detuned from the two-photon resonance
of a selected Raman transition. When the detuning $\delta$ is
appropriate and the intensity of driving lasers is sufficiently
large, the magnitude of the coherence of this transition becomes its
maximum value $|\rho_{eg}| = 0.5$.

\begin{figure}
\begin{center}
\includegraphics{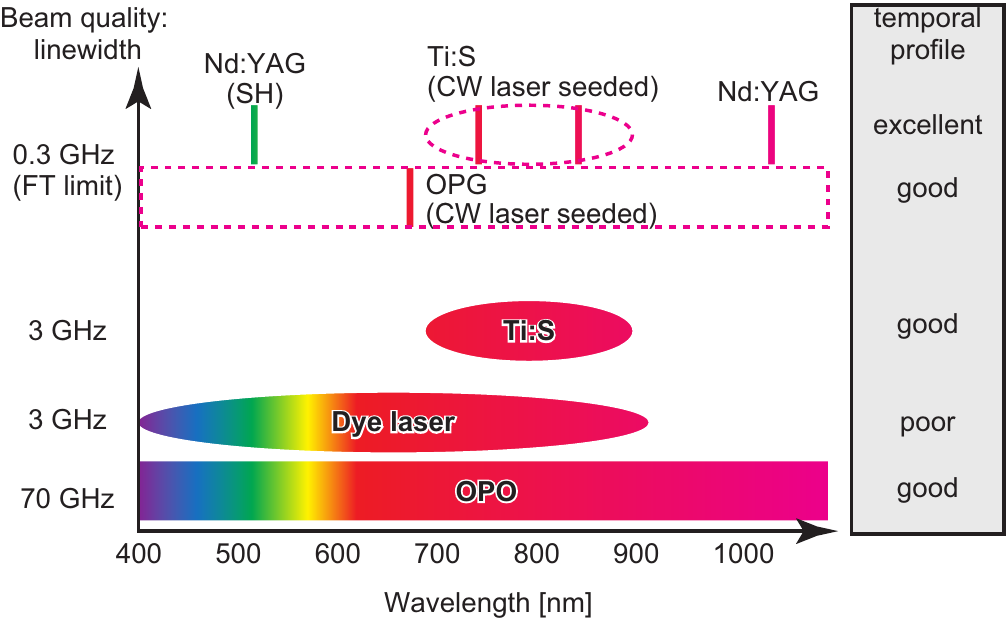}
\caption{Comparison of the beam quality (linewidth and temporal
 profile) and the spectral coverage range of available laser 
 technique. OPO: optical parametric oscillator; OPG: optical
 parametric generator; Ti:S: titanium sapphire laser;
 Nd:YAG: neodymium-doped yttrium aluminum garnet laser. The
 fundamental wavelength is 1064~nm and its second harmonic (SH) is
 532~nm;
 FT limit: Fourier-transform limit.
 The wavelength tuning range of dye lasers and OPOs are very wide,
 however, its linewidth is very broad: typically 20--500 times broader
 than the FT-limit. Injection seeding technique is useful to reduce
 linewidths of those light source.
}
\label{fig:availablelasers}
\end{center}
\end{figure}

In the experiment, high power and narrow-linewidth lasers are
required. The time duration of incident laser 
on the order of the target relaxation time (i.e. $T_2$)
is sufficient. Thus a nanosecond pulse laser is preferred for 
two reasons. First, its power is sufficiently high: a peak intensity of 
$>$1~GW$/$cm$^2$ with a modest beam radius of $0.1$~mm can easily be
obtained, much higher than a continuous-wave (CW) laser.
Second, the Fourier transform limited linewidth is on the order of
100~MHz, which is sufficiently narrow for the experiment.
Also a good temporal overlap of these two lasers is important.
Figure \ref{fig:availablelasers} depicts a comparison of 
the beam quality and wavelength coverage range 
of the available technologies.
The choice of the most relevant technology for our experiment
is a crucial issue.
For one of the driving lasers, an injection-seeded Nd:YAG laser is a
good choice 
because of its
availability of high power (up to several J, peak power of
hundreds~MW) and narrow linewidth.  
The wavelength of the other driving laser is
determined by the target atom or molecule. 
Here, we consider hydrogen molecule as a target system for the PSR
experiment. 
When we choose the first vibrational level as the excited state
$\ket{e}$ (energy of 4160~cm$^{-1}$) and a second harmonic of
Nd:YAG laser (wavelength of 532~nm) as one of the driving laser,
the wavelength of the other laser is 683~nm.

Two types of laser techniques are commercially available at this 
wavelength range: a dye laser and optical parametric oscillator (OPO). 
A dye laser may be often used, however,
such liquid laser is in general unstable and its linewidth is very broad.
Usually the linewidth of a dye laser is on the order of GHz:
20--30 times broader than the transform limit.
Another disadvantage of the dye laser is that frequent
maintenance (replacing the dye solution and re-optimization 
of laser cavity) is required.
%
An OPO is solid-state system and it offers very wide tunability
(205~nm to 2550~nm). 
However the linewidth of such OPO is much broader than that of dye laser.
Typically it is about 75~GHz: 300--500 times broader than the
transform limit.
%
Alternatively, a titanium sapphire (Ti:S) laser may be another choice.
It is a widely tunable all-solid-state laser system.
In particular, injection seeded Ti:S laser offers Fourier transform
limit linewidth and good beam quality.
Dual-wavelength injection seeded Ti:S laser was used to produce near
maximum molecular coherence \cite{TSuzuki:PRL2008}. In this case,
however, rotational Raman transition was used. Lasing
wavelength range of Ti:S laser is limited from 700~nm to 900~nm (about
3000~cm$^{-1}$). Thus this laser may be difficult to use for
vibrational Raman process of pH$_2$ because the Raman transition 
frequency is 4160~cm$^{-1}$.

Whereas those light sources mentioned above are unsuitable for the
experiment, injection seeded OPO or optical parametric 
generator (OPG) is highly 
suitable 
technique. 
Continuous-wave (CW) laser seeding technique is often used to generate
narrow linewidth coherent light 
\cite{JEBjorkholm:ApplPhysLett1969,%
MJTMilton:OptLett1994,AVSmith:JOSAB1995,JMBoon-Engering:OptLett1995,%
PBourdon:OptLett1995,NSrinivasan:JJAP1996,%
OVotava:JChemPhys1997,ABorsutzky:QuantumSemiclassOpt1997,%
SWu:OptCommun1999,WDKulatilaka:ApplPhysB2005%
}.
Injection seeding requires an external light source of modest power
(from a few $\mu$W to a few mW) and good spectral properties; also 
the tuning bandwidth is limited by seed laser. However it offers 
the advantage of transform limited or nearly transform limited linewidth.

\begin{figure}
\begin{center}
\includegraphics{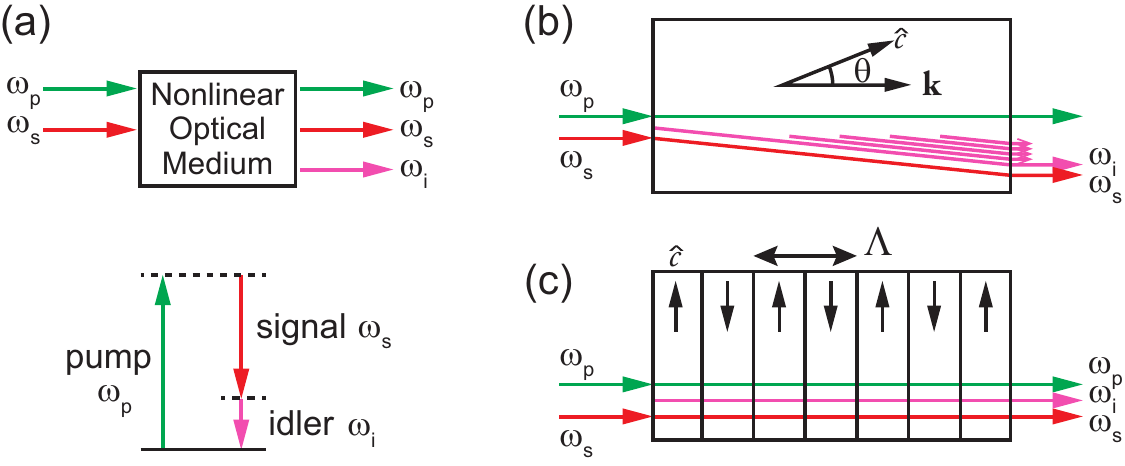}
\caption{(a) Schematic diagram of the injection-seeded optical parametric
 generation. Corresponding energy level diagram is also shown.
 Schematic representations of a parametric frequency conversion
 with (b) an angle-phase-matching homogeneous bulk crystal
 and (c) a periodically poled material. 
 In (b), the polarization of the signal and the idler waves are taken as 
 extraordinary. The generated signal and idler waves at each point in
 the crystal does not propagate collinearly with each other.
 In (c), the optical axis of the crystal
 alternates in orientation with period $\Lambda$ (on the order of
 5~$\mu$m to 10~$\mu$m). This case all waves collinearly propagate
 the crystal.
}
\label{fig:nonlinear}
\end{center}
\end{figure}

An injection seeding OPG uses the process of optical parametric
amplification, which is shown in Fig. \ref{fig:nonlinear}~(a). 
A strong pump wave at the frequency of $\omega_p$ and a weak
signal wave at the frequency of $\omega_s$ interact in a nonlinear
optical medium. As a result, an output wave at the difference
frequency $\omega_i = \omega_p - \omega_s$ is generated, 
and the signal wave is amplified. 
When the nonlinear medium is placed in an optical cavity, the presence
of feedback produces oscillation. In this case, the system is called OPO. 
Experimentally, OPGs are especially preferable for injection seeding 
because an OPG requires no optical cavity. 
Thus there is no need to actively control the cavity resonant
frequency to match the frequency of the seeding laser. This makes the
whole system simple.

To obtain a high conversion efficiency, a phase matching condition should be
fulfilled: 
$\vec{k_p} = \vec{k_s} + \vec{k_i}$, where $\vec{k_j}$ are the wave
vectors corresponding to the waves with frequencies $\omega_j$
($j=p,s,i$). The phase matching condition is interpreted as
momentum conservation for the three photons participating in the
nonlinear mixing process.
Conventionally the condition is fulfilled in birefringent crystals
that have two different refractive indices $n_o$ and $n_e$ for the
ordinary and the extraordinary waves. The ordinary wave is polarized
perpendicular to the plane containing the propagation vector $\vec{k}$
and the optical axis. On the other hand, the extraordinary wave is
polarized in the plane containing $\vec{k}$ and the optical axis.
While the ordinary refractive index $n_o$ does not depend on the
propagation direction, the extraordinary index $n_e$ depends on the
angle $\theta$ between the optical axis and $\vec{k}$.
Therefore careful tuning of the angle (i.e. the refractive index)
is required to establish the phase matching condition.
A serious drawback of using angle tuning is walkoff effect, which is
often observed in a birefringent crystal\cite{Hect:Opticsc84}.
Whenever the angle $\theta$ is different from 0 or 90 degrees, the
Poynting vector $\vec{S}$ and the propagation vector $\vec{k}$ are not
parallel for extraordinary waves. As a result, ordinary and
extraordinary waves with parallel propagation vectors quickly diverge
from one another as they propagate through the crystal, as depicted in
Fig. \ref{fig:nonlinear} (b). This walkoff
effect limits the spatial overlap of the two waves and decreases the
efficiency. Thus in general one cannot achieve a large single-pass
parametric gain with an angle-phase-matching bulk crystal even in the perfect
phase-matching condition.
To overcome low conversion efficiency, many systems 
that have been reported in literature 
used optical cavity to increase the pump power and the effective
interaction length. 
Also a pair of crystals is often used in order to compensate walkoff.

There is a relatively new technique known as quasi-phase-matching.
In quasi-phase-matching, a periodically-poled material whose
optical axis is inverted periodically as a function of position 
is used. The idea of quasi-phase-matching is schematically illustrated
in Fig. \ref{fig:nonlinear}~(c).  
The advantage of quasi-phase-matching is that there is no
walkoff effect. Thus one can use a longer crystal than an angle
phase-matching bulk crystal and therefore can get higher single-pass
conversion efficiency. 
Since the single-pass conversion efficiency is large, 
an optical cavity is not necessarily required. 
Therefore an injection seeding OPG with a periodically poled crystal
offers great simplicity of the system. 

A drawback of periodically poled crystals, as compared to
conventional bulk crystals, is 
that it is difficult to make a thick crystal with present technology.
This is because a high electric field (above 20~kV$/$mm) must be
applied to the crystal during the fabrication process of 
periodical inversion of crystal polarization.
There is no such restriction for bulk crystals, thus fabrication 
of thick crystals is feasible.
Since the optical parametric process uses a nonlinear effect,
high conversion efficiency can be obtained with high input pump
intensity.  
However, there is a certain damage threshold that limits maximum
input intensity for all nonlinear materials. 
In order to keep input intensity less than the damage threshold,
a thicker crystal (i.e. larger aperture crystal) is much better
for high power operation. 

There are various types of periodically poled crystals such as
lithium niobate (LiNbO$_3$, LN), lithium tantalate (LiTaO$_3$, LT),
and potassium titanyl phosphate (KTiOPO$_4$, KTP).
Among these crystals, LN has the largest nonlinear coefficient.
However it also has a lower photo-induced damage resistivity. 
Doping of MgO improves resistivity, but its damage threshold is
still not very high.
The nonlinear coefficient of KTP is lower than that of LN, and the 
damage threshold is almost the same as that of LN.
On the other hand, stoichiometric lithium tantalate (SLT) has
a higher damage resistivity than LN and KTP. Although its nonlinear
coefficient is about $2/3$ of LN, the damage threshold is twice as
large as LN \cite{NEYu:JJAP2004}. Thus one can expect a higher output
power by using SLT.

\begin{figure}
\begin{center}
\includegraphics{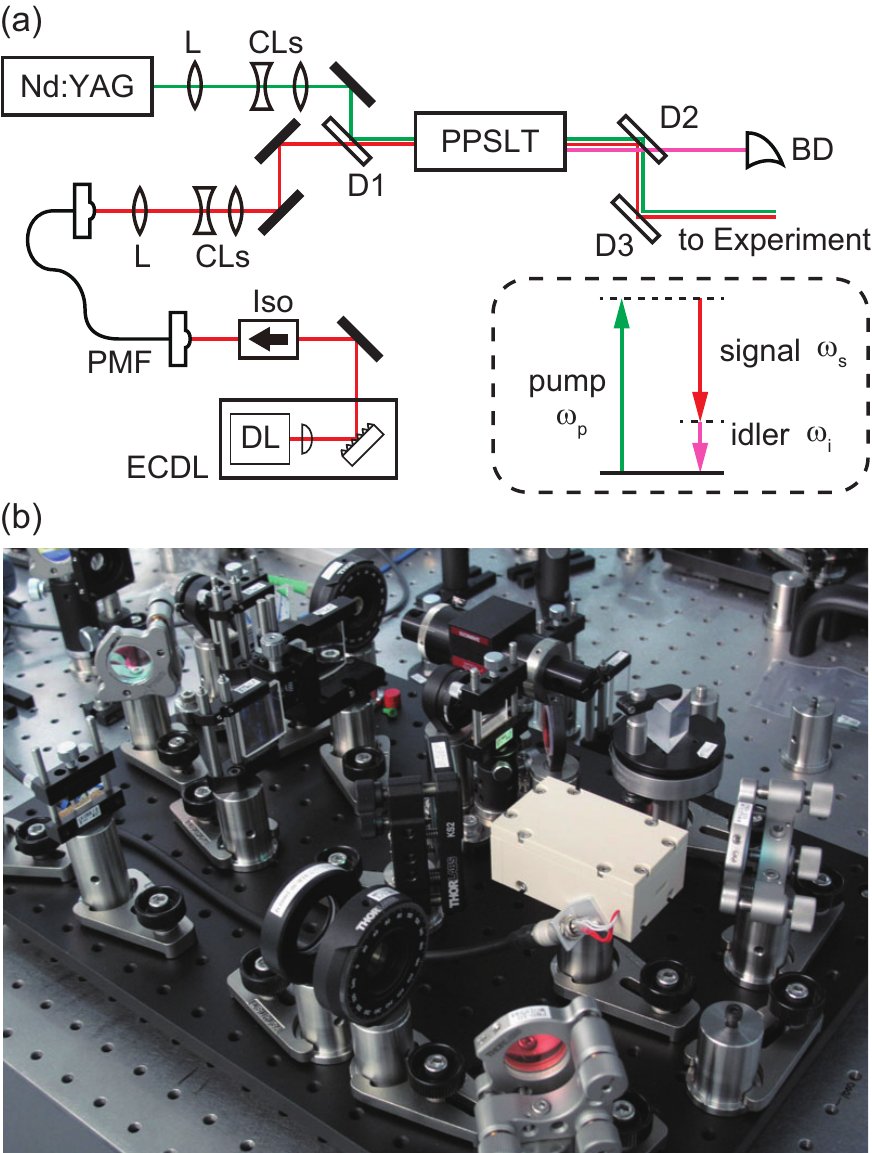}
\caption{(a) Schematic diagram and (b) Photograph of the 
 excitation laser system. 
 ECDL: extended cavity laser diode; 
 DL: diode laser chip;
 Nd:YAG: pulsed Nd:YAG laser;
 L: spherical lens;
 CLs: cylindrical lenses;
 D1: dichroic mirror used to combine pump laser (Nd:YAG) and
 continuous-wave seed laser (ECDL); 
 D2--D3: dichroic mirrors used to separate the pump plus the signal beam
 and the idler ($2.4$~$\mu$m); 
 BD: beam dumper;
 Iso: isolator;
 PMF: polarization maintaining optical fiber;
 PPSLT: periodically-poled stoichiometric lithium tantalate crystal.
 The PPSLT crystal is pumped by the second harmonic of a Q-switched
 injection-seeded Nd:YAG laser, and generates signal (683~nm) and
 idler waves. 
 }
\label{fig:lasersetup}
\end{center}
\end{figure}

\begin{figure}
\begin{center}
\includegraphics{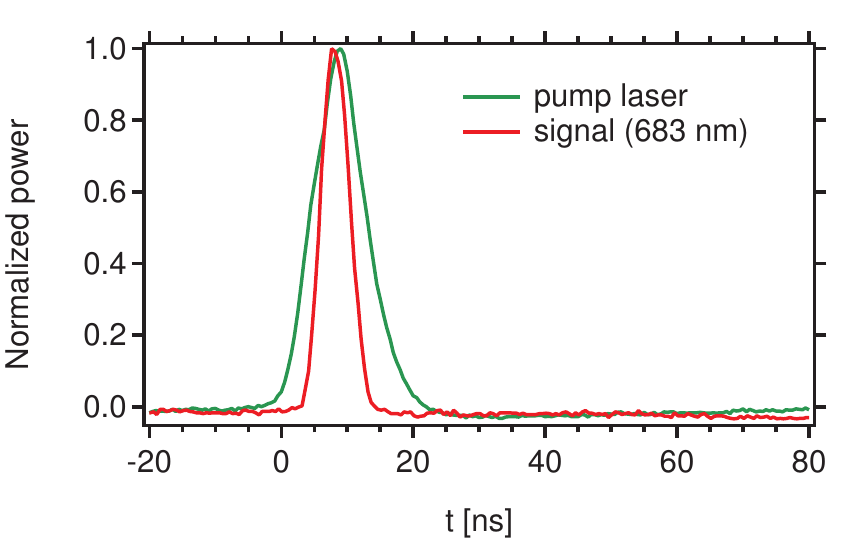}
\caption{Temporal profile of OPG output. Green line shows the
 temporal profile of the pump laser transmitted the crystal. Red 
 line shows that of the generated signal (683~nm) light.
}
\label{fig:temporal}
\end{center}
\end{figure}

We employed MgO doped PPSLT crystal (Oxide Corp., Q1532-O001) for 
our OPG system. The thickness of PPSLT is limited to 1~mm, so 
we used a 8~mm wide crystal in order to reduce the input intensity. 
A schematic diagram and photograph of the OPG system is shown in
Fig. \ref{fig:lasersetup}~(a) and (b).
The PPSLT crystal was pumped by the second harmonic of a Q-switched
injection-seeded Nd:YAG laser (Litron LPY642T). The Nd:YAG laser is
flash lamp pumped at repetition rate of 10~Hz with a true TEM$_{00}$
single transverse mode (M$^2 < 1.3$) and single longitudinal mode output. 
Pulse energies up to 130~mJ with a pulse duration of 8~ns is available
at 532~nm.  
The pump beam is expanded to elliptical shape ($0.68\times 3.7$~mm$^2$)
by means of cylindrical lenses.
For injection seeding, an extended cavity diode laser
(ECDL) in the Littrow configuration was used. We used a commercially available laser diode chip (TOPTICA
LD-0685-0050-3, no anti-reflection (AR) coating) for the ECDL. 
The output power of the ECDL is more than 10~mW with 
a typical mode-hop-free scanning range of 3~GHz.
A polarization maintaining optical fiber was used for transverse mode
cleaning. 
Typically, a fiber output of 2~mW was used for injection seeding.
The injection seeding beam was also expanded to elliptical shape 
($1\times 8$~mm$^2$).
The pump and injection seeding lasers are combined with a dichroic
mirror.
The PPSLT crystal is 40~mm long $\times$ 1~mm thick $\times$ 8~mm wide
(periodic domain width of 7~mm) and periodicity of $10.3$~$\mu$m.
Phase matching temperature at 683~nm was measured to be around
110$^\circ$C. The crystal is AR coated for three wavelengths: 
pump (532~nm), signal (683~nm), and idler (2.4~$\mu$m).
Figure \ref{fig:temporal} shows a typical temporal profile of
the transmitted pump beam and generated signal (683~nm) light. 
Good temporal overlap is obtained.


We next describe the trigger laser system.
According to the PSR simulation 
(Sec. \ref{Subsec:Experimet-Overview}), 
the power of trigger
laser is less important, however the narrow linewidth ($<$MHz) may be
crucial.  
The wavelength of the trigger laser is $4.8$~$\mu$m. 
While there are many lasers available at visible or near-infrared (NIR)
region, coherent light sources emitting mid-infrared (MIR) radiation is 
less common. 
Recently, quantum cascade lasers, which directly radiate mid-infrared
wavelength light, are commercially available.
However the linewidth is broad (about 45~MHz) and 
they are still expensive.
Moreover, peripheral optical devices in the MIR spectral region such as 
optical isolators, gratings, high-reflectivity low-loss mirrors for
high-finesse optical cavity, or a wavelength meter are also less
common.  
Thus there are practical drawbacks when one develops laser sources with
direct laser radiation devices in the MIR region.
Instead, a light source based on nonlinear optical process (i.e. 
difference-frequency generation) with a quasi-phase-matching material
pumped by NIR lasers is highly competitive. It allows to transfer the
high performance characteristics of the pump laser at NIR region to
the MIR: precise wavelength resolution and narrow linewidth. 

\begin{figure}
\begin{center}
\includegraphics{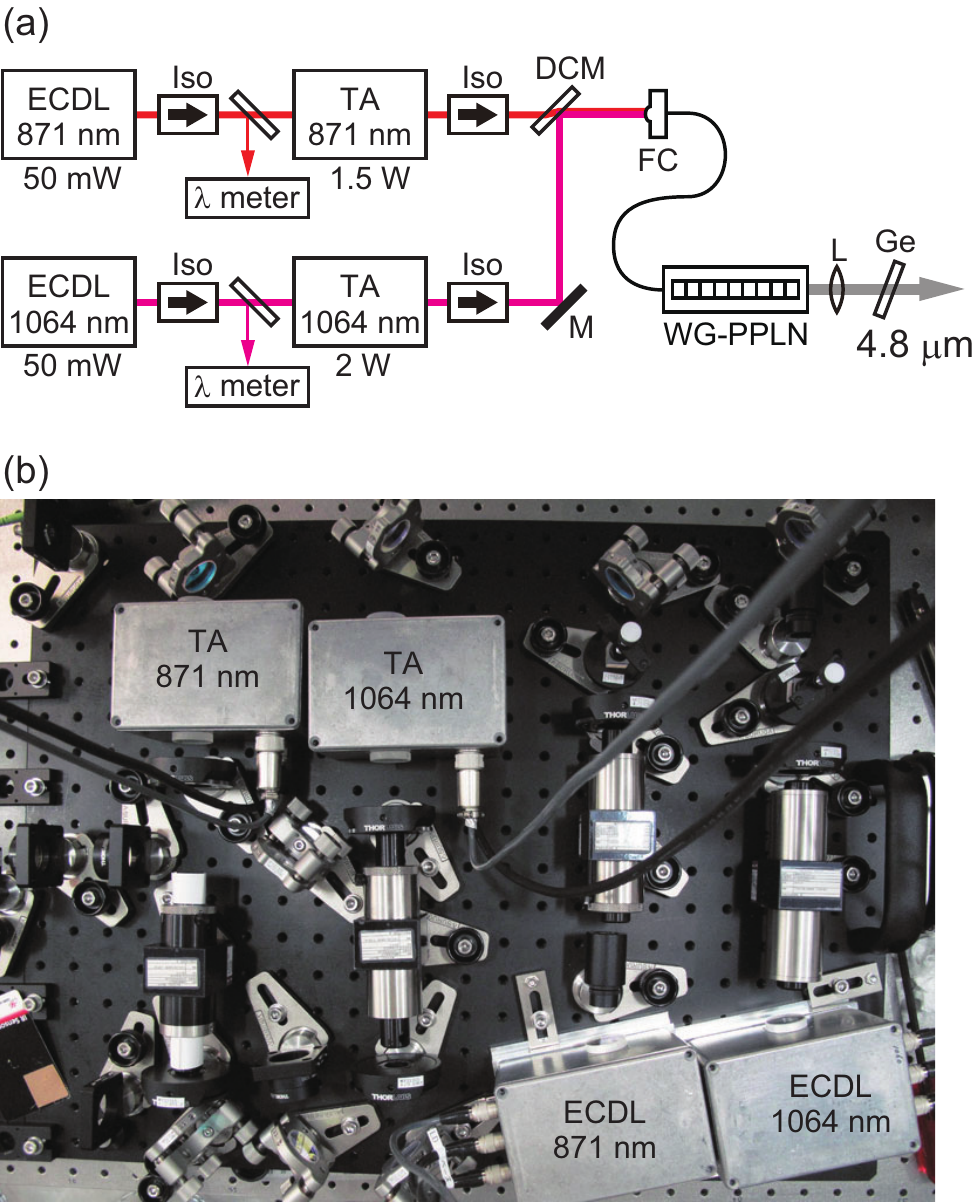}
\caption{Trigger laser system: (a) Block diagram of the trigger laser
 system. The wavelength of the trigger laser is $4.8$~$\mu$m. We
 employ difference-frequency generation (DFG) to obtain coherent light
 at this mid-infrared wavelength range. For pump lasers of DFG, we
 employ two master oscillator power 
 amplifier (MOPA) systems at wavelength of 871~nm and 1064~nm,
 respectively. Both lasers are coupled into optical fiber and then
 coupled to the waveguide PPLN crystal. Typical
 output power from each device is indicated in the figure. 
 (b) Photograph of the system. Note that MOPA systems are only shown
 in the photograph.
 ECDL: extended cavity laser diode; 
 TA: tapered amplifier;
 Iso: isolator;
 DCM: dichroic mirror used to combine two pump lasers;
 FC: fiber coupler;
 WG-PPLN: waveguide PP-MgO:LN crystal;
 L: collimator lens;
 Ge: Germanium filter.
 }
\label{fig:triggerlaser}
\end{center}
\end{figure}

Figure \ref{fig:triggerlaser} shows the block diagram and the
photograph of the trigger
laser system. We employ two of master oscillator power amplifier
(MOPA) systems as pump sources of DFG.
The wavelength of the MIR light is determined by measuring 
wavelengths of both pump lasers by a wavelength meter (HighFinesse,
WS6-200). One of the pump laser is combined with the other by DCM and
coupled into optical fiber. We employ waveguide PP-MgO:LN crystal
(NTT electronics, WD4800-000-A-B-C). A great advantage of waveguide PPLN
device is that one can obtain longer interaction length with higher
beam intensity, which yields two or three orders of
magnitude larger single-pass frequency-conversion efficiencies
than no-waveguide crystals. 
This is especially effective for CW laser because its beam intensity is
relatively low.


\subsection{Towards RENP experiment with Xe}
\label{Subsec:Experimet-Xe}

\subsubsection{RENP with Xe atoms}
\begin{figure}[htb]
\begin{center}
\includegraphics[width=0.6\textwidth]{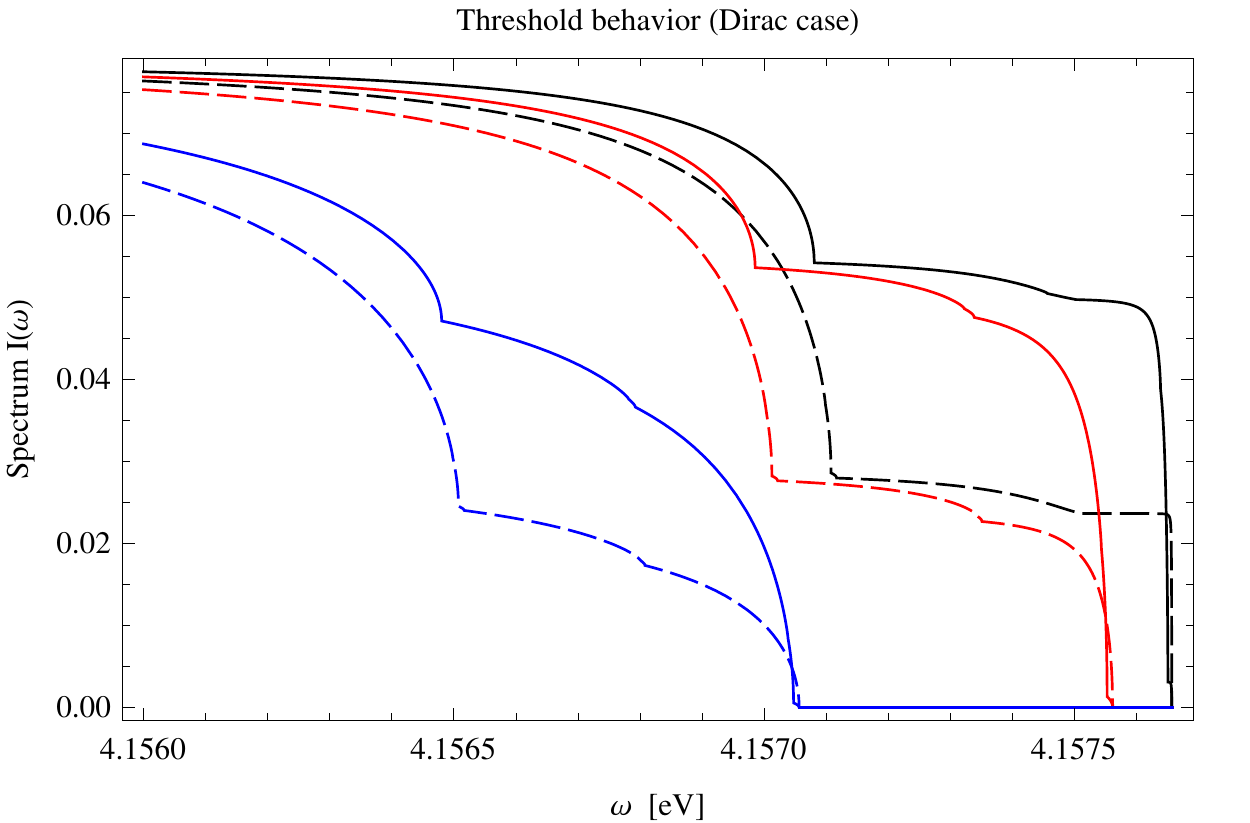}
\caption{Examples of Xe RENP spectra near threshold.
Assumed are Dirac neutrinos with the minimum masses $m_{0}$ of 2 (black),
20 (red) and 50 (blue) in the normal (solid) or inverted (dashed)
hierarchy.}
\label{fig-RENP}
\end{center}
\end{figure}

As described in Sec. 1,  Xe atom is one of  the important candidates for studying the RENP. 
The lowest excited state $5p^{5}(^2P_{3/2})6s\ ^{2}[3/2]_{2}$ with total electron angular momentum $J=2$ in Xe atom
can decay only through M2 transition to the ground state $5p^{6} (^{1}S_{0})$, and thus this
$J=2$ state is metastable with lifetime of $O(40)$ s~\cite{Walhout1994}.  The RENP process in Xe atom 
would be recognized
by observing a characteristic decay spectrum of emitted photon from the metastable state, which has 
distinctive threshold at the photon energy of 4.16 eV, half of the energy difference between the metastable
and ground states. Figure~\ref{fig-RENP} shows such typical RENP spectrum, where different threshold 
locations $\omega_{ij}$ appear
depending on mass scale of participating neutrino pair.   

In order to successfully observe the RENP process, understanding and precise control of a coherent 
state of atoms interacting with static field condensate is inevitably important~\cite{my-rnpe}.
Therefore the investigation on feasibility and further research/development is needed for the preparation and confirmation of the 
atomic coherence between the metastable and the ground state in Xe atom. The fundamental characteristics 
of metastable state in Xe, numerical calculation of achievable coherence, and the prospect for 
experiments are described below.

\subsubsection{Properties of Xe atom in gas phase}
The atomic coherence between the metastable and the ground state in Xe can be produced by applying 
two light-fields ("pumping" and "coupling or sometime called stokes") on so-called $\Lambda$-type three-level atomic system.
However, the achievable atomic coherence tends to be limited from several origins, which broaden 
in effect the line widths in the relevant atomic transitions. Because these line widths are affected by 
physical and spectroscopic properties of Xe atom, these parameters are summarized here.
  
The metastable state of Xe atom has long lifetime over 40 s for even isotopes in low-pressure gas phase. 
The radiative lifetimes of 42.9(9) s and 42.4(13) s for $^{132}$Xe and $^{136}$Xe respectively, are reported
where the observation was performed in MOT (Magneto-optical trap) experiment~\cite{Walhout1994}. 
The density of Xe atom can be increased up to $2.5\times 10^{19}\ /{\rm cm}^{3}$ with a pressure of 1 atm
at room temperature. If the gas is cooled to 190 K (boiling point at pressure of 3 atom) and the 
pressure is increased to 3 atom, the number density reaches $1.1\times 10^{20}\ /{\rm cm}^{3}$. 
The atomic collision should be considered for relaxation of metastable atoms in relatively high density 
gas phase we here consider for our purpose ($>10^{19}\ {\rm /cm^{3}}$).
The collisional relaxations (atom-atom and atom-wall) of the metastable Xe in a
gas cell are also reported, where the observed relaxation rate takes minimum ($\sim 10^{2}\ {\rm s}^{-1}$; 
depending on cell size) at a pressure of 0.2 - 1 Torr and reaches to $1\times 10^{4}\ {\rm s}^{-1}$ at 10 
Torr~\cite{Barbet1975}. From this reported result, 
the collisional relaxation time of the metastable state of $O(0.1\sim 0.01)\ {\rm \mu s}$ is expected at a pressure of 1 atm
that is enough long for experimental RENP studies. 

\begin{figure}[htb]
\begin{center}
\includegraphics[width=0.5\textwidth]{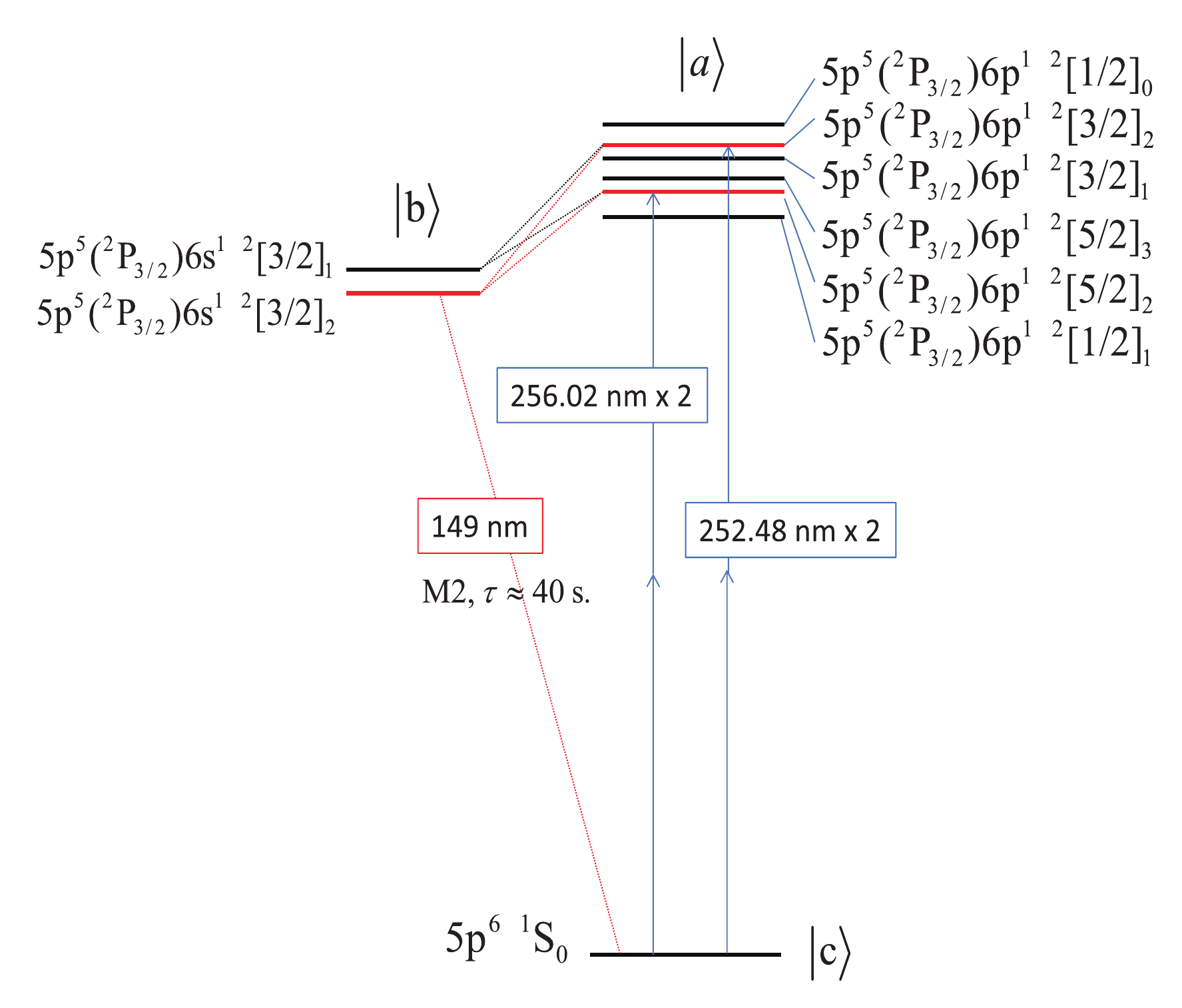}
\caption{Schematic energy level for populating Xe atom into metastable state.}
\label{xe-level}
\end{center}
\end{figure}

Populating Xe atoms into the metastable state is performed through two-photon excitation from
the ground state to $5p^{5}(^{2}P_{3/2})6p$ states as shown in Fig.~\ref{xe-level}. Among these $6p$-blanches the two states, 
$5p^{5}(^{2}P_{3/2})6p\ ^{2}[3/2]_{2}$ and $5p^{5}(^2P_{3/2})6p\ ^{2}[5/2]_{2}$, are allowed both for the two-photon excitation 
from the ground state and single-photon E1 transition to the metastable state in the $\Lambda$-type three-state system.  
Both of these 6p-states have E1-transitions to the $6s\ [3/2]_{2}$ (metastable) and $6s\ [3/2]_{1}$ which promptly
decays into the ground state with E1 transition. The consecutive E1 decays through $6s\ [3/2]_{1}$ state is
one of sources for relaxations and decoherences in the three-level system.
The spectroscopic parameters for these two states including the wavelength $\lambda_{tp}$ for two-photon
and $\lambda_{6sJ-6p}$ for the 6p-6s transitions,  
transition probabilities $A_{6sJ\leftarrow 6p}$, are summarized in Table~\ref{tab1}.
\begin{table}[htb]
\caption{The parameters of relevant excited 6p states for populating Xe atoms into the metastable state  
$5p^{5}(2P_{3/2})6s\ ^{2}[3/2]_{2}$. The values of transition probabilities $A_{6s_{J}\leftarrow 6p}$ are taken from 
\cite{Aymar1978}.}
\label{tab1}
\begin{center}
\begin{tabular}{lcccccc}
\hline
\multicolumn{1}{c}{-} & \multicolumn{1}{c}{$E_{6p}$} & \multicolumn{1}{c}{$\lambda_{tp}$}
& \multicolumn{1}{c}{$\lambda_{6s2-6p}$} & \multicolumn{1}{c}{$A_{6s2\leftarrow 6p}$} 
& \multicolumn{1}{c}{$\lambda_{6s1-6p}$} & \multicolumn{1}{c}{$A_{6s1\leftarrow 6p}$}  \\
\multicolumn{1}{c}{-} & \multicolumn{1}{c}{(eV)} & \multicolumn{1}{c}{(nm)}
& \multicolumn{1}{c}{(nm)} & \multicolumn{1}{c}{$({\rm s^{-1}})$} 
& \multicolumn{1}{c}{(nm)} & \multicolumn{1}{c}{$({\rm s^{-1}})$}  \\
\hline
 $5p^{5}(2P_{3/2})6p\ ^{2}[3/2]_{2}$ & 9.821  & 252.5 & 823.4 & $2.24\times 10^{7}$ 
 & 895.5 & $ 9.42\times 10^{6}$ \\
 $5p^{5}(2P_{3/2})6p\ ^{2}[5/2]_{2}$ & 9.686 & 256.0 & 904.8 & $9.72\times 10^{6}$ 
 & 992.6 & $ 1.74\times 10^{7}$ \\
\hline
\end{tabular}
\end{center}
\end{table}
Among these two states, $5p^{5}(2P_{3/2})6p\ ^{2}[3/2]_{2}$ is preferable in terms of usability of wavelength
in IR region. Therefore this state is used in the following estimation for atomic coherences. 

Another mechanism of the relaxation is the ionization of Xe atoms in excited 6p or 6s states due to absorption of
third photon from the pump field. The first ionization energy of 12.1 eV for Xe is only 2.3 eV higher than the 6p state.
The ionization rates for these states are proportional to the light field strength.
There are some experimental and theoretical studies for ionization of Xe atoms in 6p states.
For the state $5p^{5}\ 6p [3/2]_{2}$, the observed ionization rate of  
$\Gamma_{\rm I}=3.0\cdot I_{252}\ [{\rm s^{-1}}]$ is reported, where $I_{252}$ is field strength of 252.5 nm in unit of 
${\rm W/cm^{2}}$~\cite{Chen1980}.  Although the metastable Xe atoms can also absorb the third photon to 
be ionized, the ionization rate of $\Gamma_{\rm I}=0.19\cdot I_{252}\ [{\rm s^{-1}}]$ roughly estimated by
the reported photoionization cross section~\cite{Kau1997}
is smaller than that of 6p state.

It must be noted also on the Doppler broadenings in each transition which affect the 
phase coherence in the relevant two states. At room temperature, the Doppler widths (FWHM) of
the 5p-6p and 6p-6s transitions are $2.6\ {\rm GHz}$ and $0.40\ {\rm GHz}$, respectively.

\subsubsection{Coherence preparation and its measurement}
Atomic coherence between the metastable and ground state in Xe atom is produced by two laser
fields, one for two-photon excitation from ground $\left| c \right\rangle$
to the 6p state $\left| a \right\rangle$ (pump field with frequency $\omega_{\rm p}$), another
for coupling between 6p and the metastable state $\left| b \right\rangle$ 
(coupling field with frequency $\omega_{\rm c}$). 
As described previously, the preparation process of coherence includes the two-photon excitation and
some relaxations and is then complex. Therefore, it is important to quantitatively estimate how large atomic
coherence can be achieved. We then perform numerical simulations for the coherence of metastable Xe. 

The time evolution of populations and coherences for $\Lambda$-type three-level system 
in the presence of the two laser fields are estimated by the optical-Bloch equations.
The optical Bloch equation is the homogeneous (spatially constant) version of the Maxwell-Bloch
equation in which the light-field is taken as given. 
The equations for the populations $\rho_{\rm aa}$, $\rho_{\rm bb}$ and $\rho_{\rm cc}$ in each state
are described as 
\begin{eqnarray}
\dot{\rho}_{\rm aa} &=& \frac{i\Omega_{\rm ab}}{2}(\tilde{\rho}_{\rm ab}-\tilde{\rho}_{\rm ba})+
\frac{i\Omega_{\rm ac}}{2}(\tilde{\rho}_{\rm ac}-\tilde{\rho}_{\rm ca})-(\Gamma^{\rm (rad-all)}_{a}+
\Gamma^{\rm (ion)}_{a}) \rho_{\rm aa} \\
\dot{\rho}_{\rm bb} &=& \frac{i\Omega_{\rm ab}}{2}(\tilde{\rho}_{\rm ba}-\tilde{\rho}_{\rm ab})-
(\Gamma^{\rm (ion)}_{\rm b}+\Gamma^{\rm col}_{\rm b})\rho_{\rm bb}
+\frac{\Gamma^{\rm (rad-meta)}_{\rm a}}{2}\rho_{\rm aa} \\
\dot{\rho}_{\rm cc} &=& \frac{i\Omega_{\rm ac}}{2}(\tilde{\rho}_{\rm ca}-\tilde{\rho}_{\rm ac})
+\frac{\Gamma^{\rm (rad)}_{\rm a-c}}{2}\rho_{\rm aa},
\end{eqnarray}
and similarly for the coherences, the following equations describe the time evolutions
\begin{eqnarray}
\dot{\tilde{\rho}}_{\rm ab} &=& -(\gamma_{\rm ab}-i\delta_{\rm ab})\tilde{\rho}_{\rm ab}+
\frac{i\Omega_{\rm ab}}{2}(\rho_{\rm aa}-\rho_{\rm bb})-\frac{i\Omega_{\rm ac}}{2} \tilde{\rho}_{\rm cb}\\
\dot{\tilde{\rho}}_{\rm ac} &=& -(\gamma_{\rm ac}-i\delta_{\rm ac})\tilde{\rho}_{\rm ac}+
\frac{i\Omega_{\rm ac}}{2}(\rho_{\rm aa}-\rho_{\rm cc})-\frac{i\Omega_{\rm ab}}{2} \tilde{\rho}_{\rm bc}\\
\dot{\tilde{\rho}}_{\rm cb} &=&  -(\gamma_{\rm cb}-i(\delta_{\rm ab}-\delta_{\rm ac}))\tilde{\rho}_{\rm cb}+
\frac{i\Omega_{\rm ab}}{2} \rho_{\rm ca}-\frac{i\Omega_{\rm ac}}{2} \rho_{\rm ab}.
\end{eqnarray}
Here the "slow" variables for coherences are introduced as
\begin{equation}
\tilde{\rho}_{\rm ab}=\rho_{\rm ab}e^{-i(\omega_{\rm ab}-\omega_{\rm c})t},\ 
\tilde{\rho}_{\rm ac}=\rho_{\rm ac}e^{-i(\omega_{\rm ac}-2\omega_{\rm p})t},\ 
\tilde{\rho}_{\rm cb}=\rho_{\rm cb }e^{-i \omega_{\rm cb} t},
\end{equation}
in order to move into appropriate rotating frames.
$\Omega_{\rm ac}$ and $\Omega_{\rm ab}=\mu_{\rm ab}E_{\rm c}/\hbar$ represent the Rabi 
frequencies for the pump and coupling field, respectively. 
$\Omega_{\rm ac}$ is defined as an effective two-photon Rabi frequency for the field at frequency of $\omega_{\rm p}$;
 $\Omega_{\rm ac}=\alpha_{\rm ac}E_{\rm p}^{2}/\hbar$, where 
$\alpha_{\rm ac}=\sum_{i}\mu_{1i}\mu_{i2}/2\hbar(\omega_{i}-\omega_{\rm p})$ and the $\mu_{ij}$ indicates
the corresponding dipole matrix elements. $E_{\rm p}$ and $E_{\rm c}$ are the amplitudes of the applied 
pump and coupling fields respectively, and these field amplitudes are described with corresponding
intensities $I$ of the incident lights as $E=\sqrt{2Z_{0}I}=27.5\sqrt{I({\rm W/cm^{2}})}\ {\rm V/cm}$. 
In these equations we include the spontaneous decay term
$\Gamma_{\rm a}^{\rm (rad-all)}$ and $\Gamma_{\rm a}^{\rm (rad-meta)}$ corresponding the decay 
rates from $\left| a \right\rangle$ to two 6s states and only to the metastable state 
$\left| b \right\rangle$. These decay rates take the values 
$\Gamma_{\rm a}^{\rm (rad-all)}=2.63\times10^{7}\ {\rm s^{-1}}$~\cite{Horiguchi1981} and 
$\Gamma_{\rm a}^{\rm (rad-meta)}=2.24\times10^{7}\ {\rm s^{-1}}$~\cite{Aymar1978}.
The state $\left| a \right\rangle$ tends to decay into the ground state $\left| c \right\rangle$ with
two E1 decays through $5p6s\ ^{2}[3/2]_{1}$ whose decay rate is estimated as
$\Gamma^{\rm (rad)}_{\rm a-c}=(1/\Gamma_{a\rightarrow 6s1}+1/\Gamma_{6s1\rightarrow c})^{-1}=9.1
\times10^{6}\ {\rm s^{-1}}$~\cite{Aymar1978,Morton2000}.
As described in the previous subsection, the states $\left| a \right\rangle$ and $\left| b \right\rangle$ also have decay
rates $\Gamma_{\rm a}^{\rm (ion)}$ and $\Gamma_{\rm b}^{\rm (ion)}$
through ionization by the pump field as relaxation processes.
The collisional loss of population in $\left| b \right\rangle$ is assumed 
$\Gamma_{\rm b}^{\rm (col)}=5\times10^{7}\ {\rm s^{-1}}$ (we here assume high density;  $> 1 \,{\rm atm}$).
In this calculation, the widths of Doppler broadening in each transition are included in  
the relaxation terms $\gamma_{ij}$ in coherence. These values are 
$\gamma_{\rm ac}=1.6\times 10^{10}\ {\rm s^{-1}}$,  $\gamma_{\rm ab}=2.5\times 10^{9}\ {\rm s^{-1}}$, 
and $\gamma_{\rm bc}=1.4 \times 10^{9}$ s$^{-1}$. The pressure broadening assumed to be $1\times 10^{9}\ {\rm s^{-1}}$ at a
pressure of 1 atm (room temperature) is also considered in these decoherence terms.
We assume that the detunings $\delta_{\rm ab}$ and $\delta_{\rm ac}$ in each field
are larger than the linewidths in the corresponding transitions, so that both of these transitions by the 
applied fields are adiabatic.

\begin{figure}[h]
\begin{center}
\includegraphics[width=0.45\textwidth]{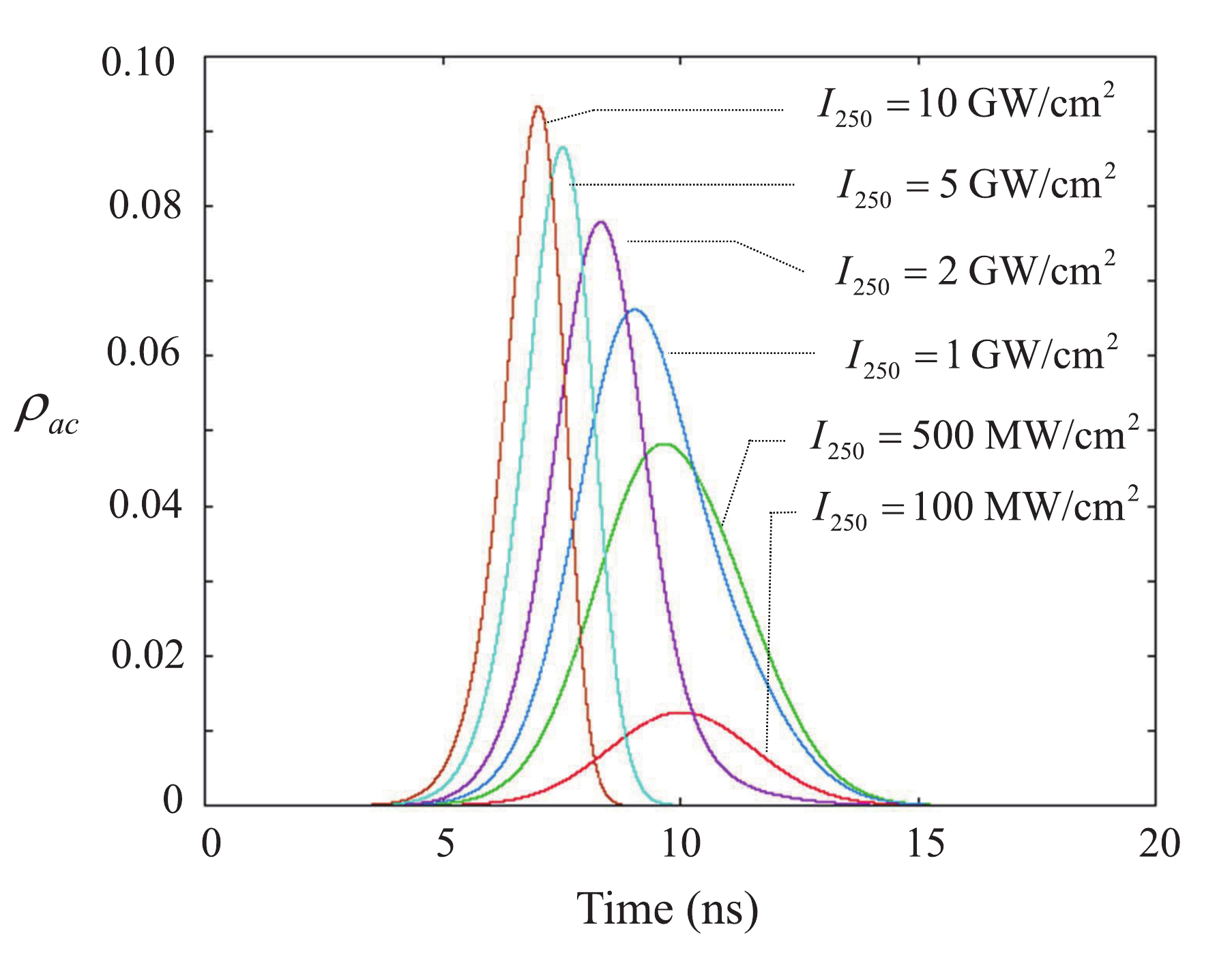}
\includegraphics[width=0.45\textwidth]{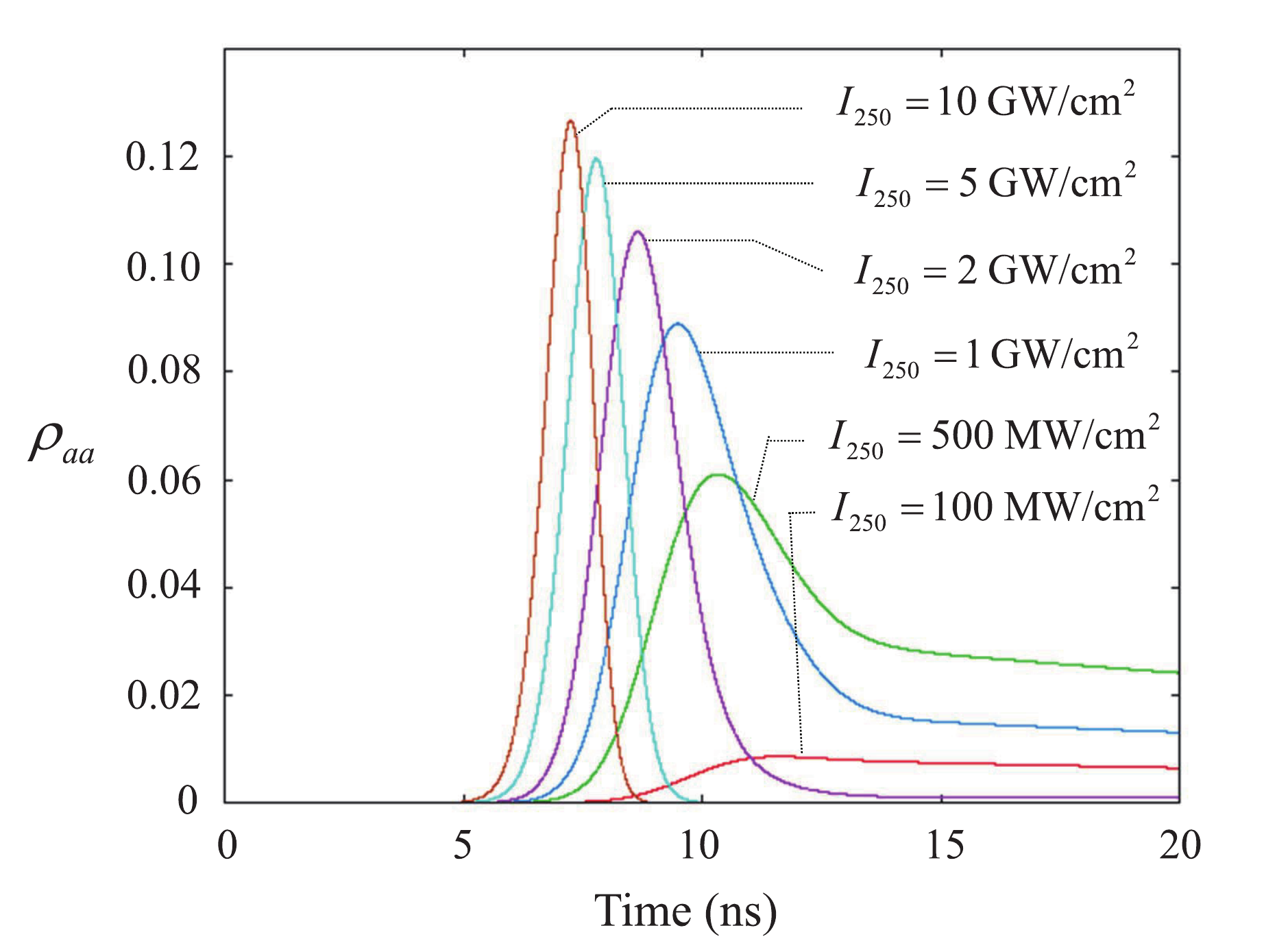}
\caption{Numerically calculated atomic coherence $\rho_{ac}$ between  
ground state and 6p state (left panel) and the population $\rho_{aa}$ (right panel) when
only the two-photon field (252 nm) are applied.}
\label{sim1}
\end{center}
\end{figure}

The dependence of population $\rho_{\rm aa}$ in the 6p state and the coherence $\rho_{\rm ac}$ between 
6p and 5p states on the intensity of the incident pump light is firstly estimated by the above equations. The
incident pump light is assumed to have a Gaussian envelope with a width of 5 ns and to have a peak intensity
at $t=10\ {\rm ns}$. 
The detuning $\delta_{\rm ac}$ of pump frequency is assumed to be $1.2\times10^{10}\ {\rm s^{-1}}$ 
(2.0 GHz) so as to be larger than the Doppler broadening 1.3 GHz (HWHM). 
Figure~\ref{sim1} shows time evolutions of $\rho_{\rm ac}$ and $\rho_{\rm aa}$ when the pump field is applied. 
The effective two-photon Rabi frequency $\Omega_{\rm ac}$ is calculated with 
the effective dipole moment 
$\alpha_{12}$. Here only $6s^{2}[3/2]_{1}$ state
is taken as the intermediate state $\left| i \right\rangle$ for calculating the $\alpha_{12}$.
Thus $\Omega_{\rm ac}$ is estimated as 
\begin{equation}
\Omega_{\rm ac} = 5.44\cdot I_{252}\  {\rm s^{-1}},
\end{equation}
where $I_{252}$ denotes the pump-field strength in unit of ${\rm W/cm^{2}}$.
These results shows that the effect of ionization loss starts to appear with the field strength
of $I_{252} \approx 1\ {\rm GW/cm^{2}}$, and that the coherence achieves 
$\rho_{\rm ac} \approx 0.06$ at that field. 
The achievable coherence tends to saturate around $\rho_{\rm ac} \approx 0.1$ due to the
ionization effect.

\begin{figure}[htb]
\begin{center}
\includegraphics[width=0.46\textwidth]{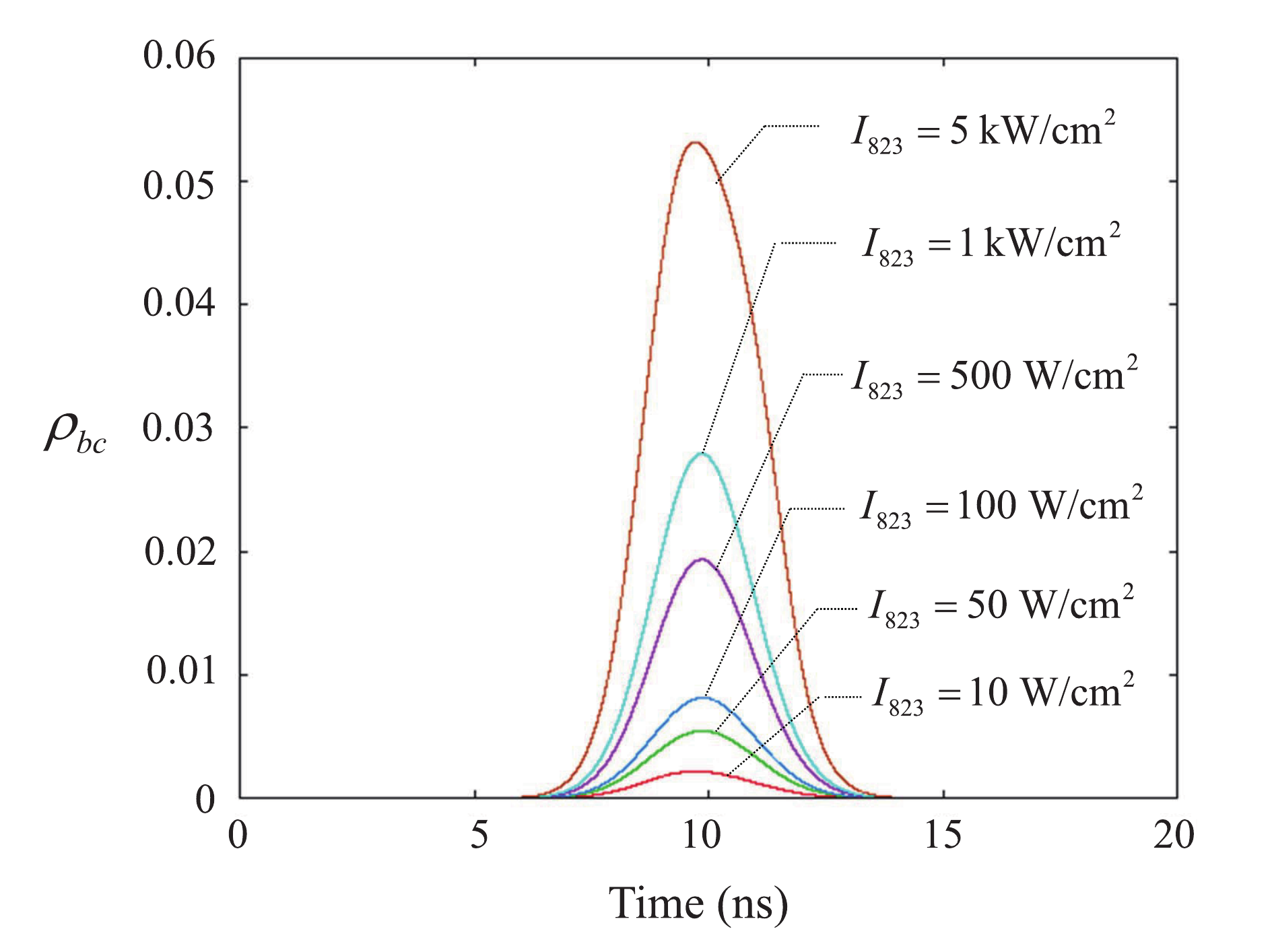}
\includegraphics[width=0.45\textwidth]{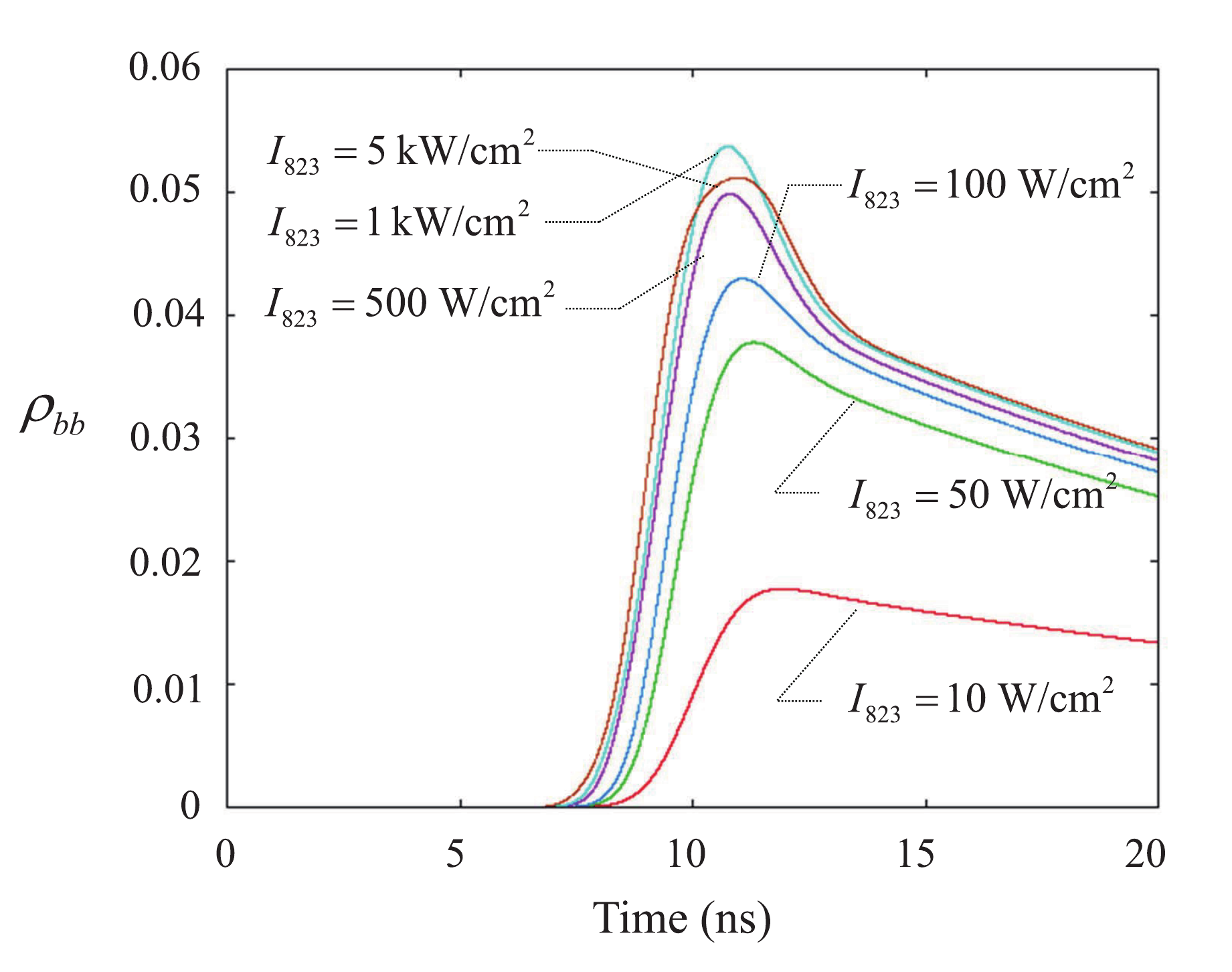}
\caption{Numerically calculated atomic coherence $\rho_{bc}$ between the 
the metastable and ground states.}
\label{sim2}
\end{center}
\end{figure}

The coherence $\rho_{\rm bc}$ between the metastable and the ground state is calculated 
by adding the coupling field of 823 nm to the above situation. 
The Rabi frequency for this transition can be calculated with the $A$-coefficient 
$A=2.24\times10^{7}\ {\rm s^{-1}}$~\cite{Aymar1978} in the transition as
\begin{equation}
\Omega_{\rm ab}=5.43\times 10^{8}\cdot \sqrt{I_{823}}\ {\rm s^{-1}} 
\end{equation}
where $I_{823}$ is the coupling-field strength in unit of ${\rm W/cm^{2}}$.
The estimated atomic coherences $\rho_{\rm bc}$ with 
$I_{823}=10\sim 5000\ {\rm W/cm^{2}}$ and $I_{252}=500\ {\rm MW/cm^{2}}$ are summarized 
in Fig.~\ref{sim2}. The populations $\rho_{\rm bb}$ in the metastable state are also shown in this
Figure. 
The detuning $\delta_{\rm bc}$ in coupling field is assumed to be $2.5\times 10^{9}\ {\rm s^{-1}}$
(0.4 GHz) which is larger than Doppler broadening 0.2 GHz (HWHM). 
The expected coherence increases to $\rho_{\rm bc}= 0.01\sim 0.1$ when the coupling Rabi frequency
$\Omega_{\rm ab}$ exceeds the two-photon Rabi frequency $\Omega_{\rm ac}$.
As shown in the results, the coherence $\rho_{\rm bc}= 0.05$ is available if the pump and coupling fields 
are increased to $I_{250}=500\ {\rm MW/cm^{2}}$ and $I_{823}=5\ {\rm kW/cm^{2}}$. In this condition each 
Rabi frequency takes $\Omega_{\rm ab}=3.8\times10^{10}\ {\rm s^{-1}}$ and 
$\Omega_{\rm ac}=2.7\times10^{9}\ {\rm s^{-1}}$. The population $\rho_{\rm bb}$
in the metastable state reaches $\sim 0.05$ under this condition.

\begin{figure}[htb]
\begin{center}
\includegraphics[width=0.7\textwidth]{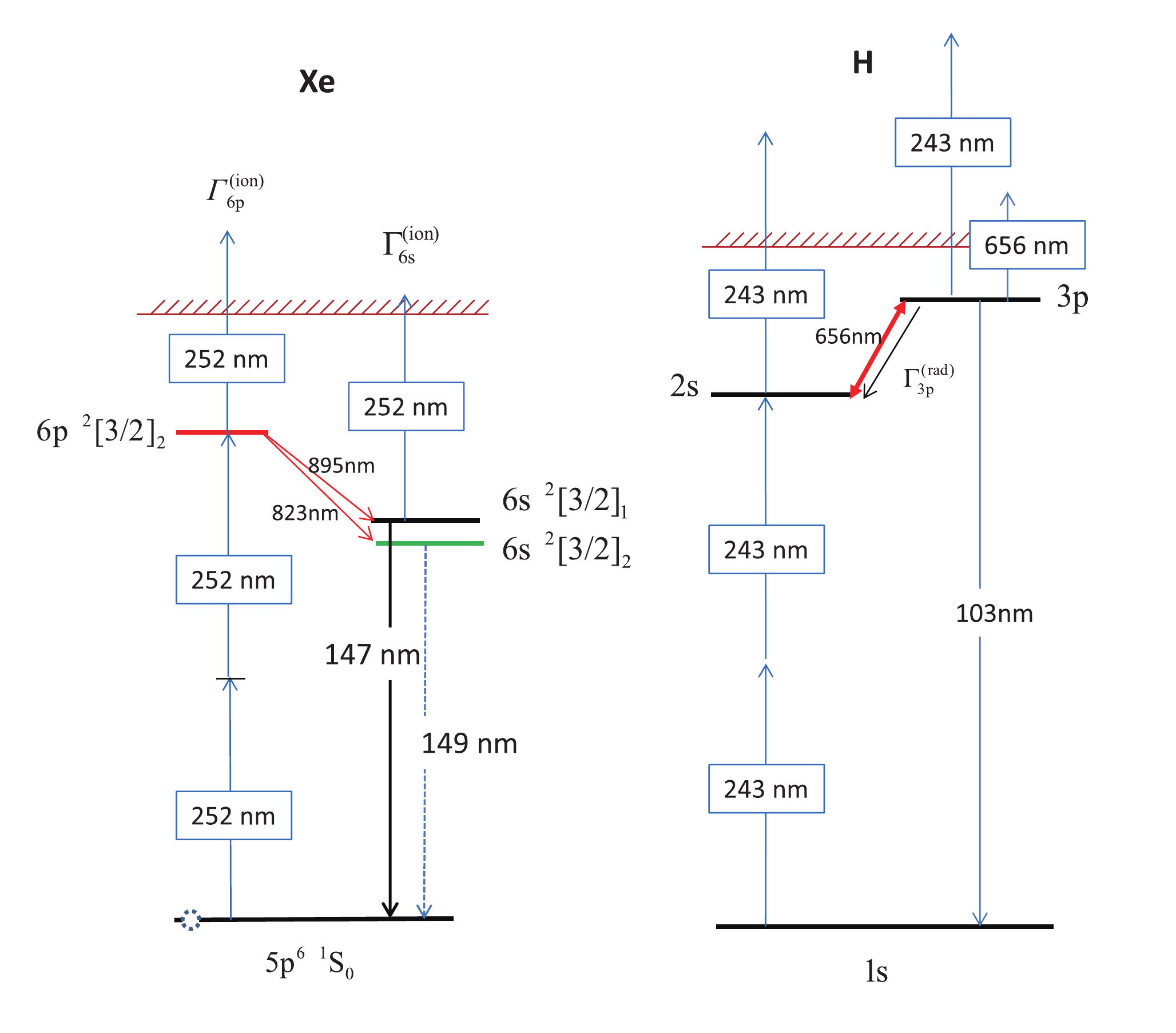}
\caption{Comparison of the $\Lambda$-type three-level systems between Xe and Hydrogen atom.
The two-photon excitation scheme and the ionization are common in these species.}
\label{xe-h}
\end{center}
\end{figure}

There would not be straightforward way to confirm presence of coherence $\rho_{\rm bc}$, because
there is no adequate transition path between $\left| b \right\rangle$ and $\left| c \right\rangle$
for coherence measurements such as coherent anti-stokes Raman scattering (CARS). This is a same
situation for both of the coherence between $6s _{2}$ ($\left| b \right\rangle$) and 
$5p^{6}$ ($\left| c \right\rangle$), and between $6s _{1}$ and $5p^{6}$.
However, it is known that the three level system behaves as the dressed state system when the strong coupling field is applied (Rabi frequency exceeds the line broadening), and this strong coupling between 
these states can be confirmed by observing that absorption of the radiation from the state  $\left| b \right\rangle$ to $\left| c \right\rangle$ is reduced, called
as Electromagnetically Induced Transparency (EIT)~\cite{Boiler1991}. 
A relevant case for this phenomenon was studied
with atomic hydrogen~\cite{Zhang1995} which has a similar $\Lambda$-type three-level structure as 
shown in Fig.~\ref{xe-h}. In both of these systems, the atoms are pumped to the excited state by two-photon transition and the atoms tend to be ionized by the applied fields.
In the EIT experiments with hydrogen atom investigated in~\cite{Zhang1995}, the transparency of the VUV radiation (103 nm) from the $3p$ to the ground state and the reduction of the photoion signal from the atomic 
system were observed when the coupling Rabi frequency $\Omega_{\rm 2s-3p}$ exceeds the line broadening 
$\delta_{\rm Doppler}=2.3\times 10^{11}\ {\rm s^{-1}}$. 
In Xe atom, this method can be applied to the three-level system of $\left| a \right\rangle$, $\left| c \right\rangle$ and $6s\ ^{2}[3/2]_{1}$ (instead of $\left| b \right\rangle$), where the transparency of the 
VUV radiation (147 nm) from the $6s\ ^{2}[3/2]_{1}$ is expected to be observed.
The required condition for the applied coupling field in this experiment is 
$\Omega_{\rm coupling} > \delta_{\rm Doppler}\sim 2\times 10^{10}\ {\rm s^{-1}}$ corresponding the coupling
field intensity of $> 2\ {\rm kW/cm^{2}}$. 
The coupling between the three states $\left| a \right\rangle$, $\left| b \right\rangle$ and 
$\left| c \right\rangle$, which is the target system in our studies, can be estimated from the observation in the above system and the numerical simulation. The reduction of the photoion signal with the EIT condition
would also be useful for evaluation of the atomic coherence as the experiments with hydrogen.

\subsubsection{Prospect for experiment}
As described above, the atomic coherence of the metastable state in Xe atom is expected to reach 
$\rho_{\rm bc}\sim 0.05$ which enables to study the RENP process
by the numerical simulations where the pump and the coupling fields lead adiabatic transitions. 
In order to perform the experiments for the coherence preparation and measurement,
the following experimental conditions are required:
\begin{itemize}
\item Linewidth of the applied fields is narrower than the line broadening
\item Smooth temporal profile for the applied fields
\item Smooth spacial profile for the applied fields.
\end{itemize}
The main broadening in Xe atom in gas phase arises from the Doppler effect, and estimated
$1.6\times 10^{10}\ {\rm s^{-1}}\ ({\rm 2.6 GHz (FWHM)})$ for two-photon transition and
$2.5\times 10^{9}\ {\rm s^{-1}}\ ({\rm 0.4 GHz (FWHM)})$ for coupling transition at room temperature.
The pressure broadening would not be negligible in the high-pressure atomic gas, which becomes 
$\sim 10^{9}\ {\rm s^{-1}}$ at the pressure of 1 atm. 
The required linewidth is then $< 100\ {\rm MHz}$. 
Therefore the laser system with enough narrow linewidth such as an injection seeded Ti:Sapphire
is required (Sec.4.2.2). As listed above the applied fields should have smooth spacial profiles for both 
of the pump and coupling, and these fields should be overlapped in space. 
The available field strength should be estimated with the realistic beam diameter for this overlapping.
If the beam diameters are assumed  that $\phi_{\rm pump}=0.2\ {\rm mm}$,  
$\phi_{\rm coupling}=0.5\ {\rm mm}$ and 
the pulse energy is assumed 10 mJ/pulse with a pulse width of 10 ns for both fields, the field strength 
can then reach 
$I_{\rm pump}=600\ {\rm MW/cm^{2}}$ and $I_{\rm coupling}=100\ {\rm MW/cm^{2}}$
with which the coherence $\rho_{\rm bc}$ becomes $> 0.05$.

\begin{figure}[htb]

\begin{center}
\includegraphics[width=0.5\textwidth]{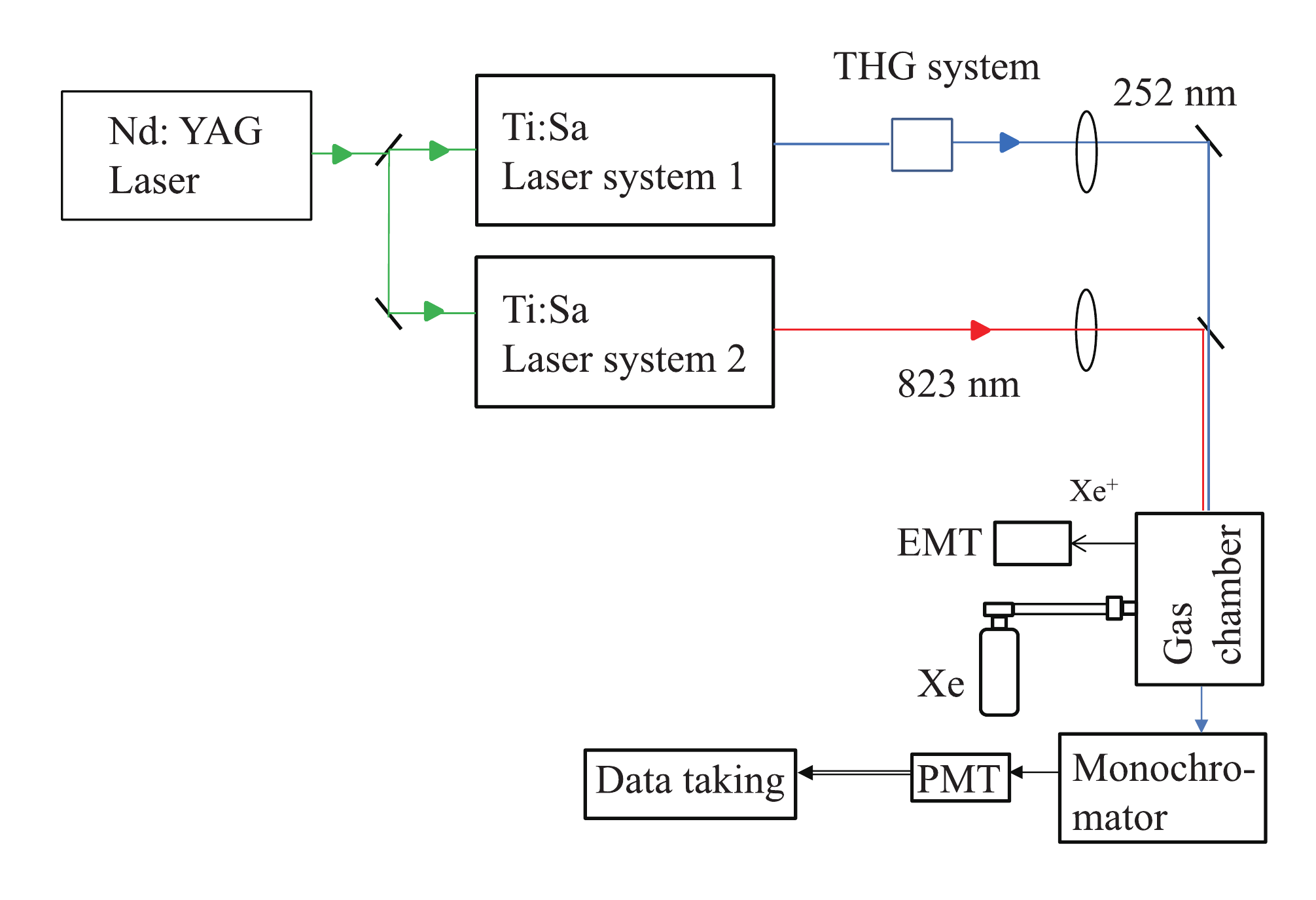}
\caption{The conceptual setup for the coherence preparation and measurement of Xe.}
\label{xe-setup}
\end{center}
\end{figure}

The conceptual setup for the experiment is shown in Fig.~\ref{xe-setup}. The high-quality pulsed beams for 
pump and coupling can be produced with Ti:S lasers injection seeded by Nd:YAG laser.
The pumping field of 252 nm is obtained through third harmonic generation (THG) by using BBO crystals 
from the output beam of a Ti:S Laser. The detectors includes a monochromator and a photomultiplier 
(PMT) for the light, and an electron multiplier (EMT) for the ionized Xe.
The fundamental experiment as R\&D for spectroscopy of the metastable Xe is now being 
performed with relatively broad linewidth lasers such as a pulsed Dye. It contains spectroscopy 
of transitions from the two-photon excited 6p states, investigation of populations in the metastable 
state, and lifetime of the metastable states. 


\begin{thebibliography}{99}
%
\bibitem{nu-matrix}
J. Beringer et al. (Particle Data Group), {\it Phys. Rev.}{\bf D86}, 010001(2012).
Our definition of Majorana phases are related to their
$\alpha_i$ by
\(\:
\alpha = (\alpha_2 - \alpha_1)/2\,,
\beta = - \alpha_1/2
\,.
\:\)

\bibitem{neutrino 2012}
G.L. Fogli et al,, 
arXiv: 1205,5254v3(2012).
Reports from Daya Bay, RENO, Double Chooz, and
T2K at International Conference on Neutrino Physics
and Astrophysics, "Neutrino 2012"
at Kyoto, available at web-site
neu2012.kek.jp.

\bibitem{0nu mass}
M.~Doi, T.~Kotani, H.~Nishiura, K.~Okuda and E.~Takasugi,
{\it Phsy. Lett.}{\bf 102B}, 323(1981).

\bibitem{cosmology-mass-bound}
E. Komatsu et al.,
{\it ApJS}{\bf 192}, 18(2011).

\bibitem{fy-86}
M. Fukugita and T. Yanagida,
{\it Phys. Lett. } {\bf B 174} 45 (1986). 


\bibitem{davidson-ibarra}
S. Davidson and A. Ibarra,
{\it Nucl. Phys. }{\bf B648}, 345 (2003), and references therein.


\bibitem{majorana cp phases}
S.M. Bilenky, J. Hosek, and S.T. Petcov,
{\it Phys. Lett.}{\bf B94},495(1980).
J. Schechter and J.W.F. Valle,
{\it Phys.Rev.}{\bf D22}, 2227(1980).

\bibitem{my-rnpe}
M. Yoshimura,
{\it  Phys. Lett.}{\bf B699},123(2011).

\bibitem{my-rnpe1}
M. Yoshimura,
{\it Phys. Rev.}{\bf D75}, 113007(2007).

\bibitem{rnpe-pc}
M. Yoshimura, A. Fukumi, N. Sasao,
T. Yamaguchi,
{\it Progr. Theo. Phys.}{\bf 123},523(2010).

\bibitem{review of electroweak theory}
I.J.R. Aitchison and A.J.G. Hey,
{\it Gauge Theories in Particle Physics},
Vol.1 and 2, 3rd edition (informa, 2003).


\bibitem{yb-x renp}
D.N. Dinh, S. Petcov, N. Sasao, M. Tanaka,
and M. Yoshimura,
{\it Observables in neutrino mass spectroscopy using atoms}.

\bibitem{sr review}
For a review of both the theory and
experiments of superradiance,
M. Benedict, A.M. Ermolaev, V.A. Malyshev, I.V. Sokolov, and
E.D. Trifonov,
{\it Super-radiance Multiatomic coherent emission},
Informa (1996).
For a formal aspect of the theory,
M. Gross and S. Haroche, {\it Phys.Rep.}{\bf 93}, 301(1982).
The original suggestion of superradiance is due to
R.H. Dicke, {\it Phys. Rev.}{\bf 93}, 99(1954).

\bibitem{psr dynamics}
M. Yoshimura, N. Sasao, and M. Tanaka,
{\it Phys.Rev. }{\bf A86},013812(2012),
and
{\it Dynamics of paired superradiance},
arXiv:1203.5394[quan-ph] (2012).

\bibitem{macro-coherence}
M. Yoshimura, C. Ohae, A. Fukumi, K. Nakajima, I. Nakano,
H. Nanjo, and N. Sasao,
{\it Macro-coherent two photon and radiative
neutrino pair emission}, arXiv 805.1970[hep-ph](2008).

M. Yoshimura, {\it Neutrino Spectroscopy using Atoms (SPAN)},
in Proceedings of 4th NO-VE International Workshop,
edited by M. Baldo Ceolin(2008).

\bibitem{2g propagation}
L.M. Narducci, W,W. Eidson, P. Furcinitti, and D.C. Eteson,
{\it Phys. Rev.}{\bf A 16}, 1665 (1977);
M. Yoshimura,
{\it  Progr.Theor.Phys.}{\bf 125}, 149 (2011).

\bibitem{textbook on molecules}
For instance,
B.H. Bransden and C.J. Joachain,
{\it Physics of Atoms and Molecules},
2nd edition, Prentice Hall (2003).

\bibitem{my-taka}
T. Takahashi and M. Yoshimura,
{\it Effect of Relic Neutrino on Neutrino Pair Emission
from Metastable Atoms},
hep-ph/0703019.

\bibitem{lindblad}
G.~Lindblad, 
Comm.\ Math.\ Phys.\ \textbf{48}, 119 (1976).


\bibitem{quantum initiation}
F. Haake, H. King, G.S. Schoeder, J. Haus, and R. Glauber,
{\it Phys. Rev.}{\bf A 20}, 2047(1979).
D. Polder, M.F.H. Schuurmans, and Q.H.F. Vrehen,
{\it Phys. Rev.}{\bf A 19}, 1192(1979).

\bibitem{first sr}
N. Skribanowitz, I.P. Herman, J.C. McGillivray, and M.S. Feld,
{\it Phys. Rev. Lett.}{\bf 30}, 309(1973).
For a review, see Chapter 2 of {\it Super-radiance Multiatomic coherent emission}
\cite{sr review}.

\bibitem{triggered sr}
Q.H.F. Vrehen and M.F.H. Schuurmans,
{\it Phys. Rev. Lett.}{\bf 42}, 224(1979).
N.W. Carlson, D.J. Jackson, A.L. Schaswlow,
M. Gross, and S. Haroche, 
{\it Opt. Commun.}{\bf 32}, 350(1980).

\bibitem{ba ohae}
C.~Ohae {\it et al.}, work in progress

\bibitem{i2 renp rate}
M. Tashiro, M. Ehara, S. Kuma, Y. Miyamoto, 
N. Sasao, S. Uetake, and M. Yoshimura,
work in progress.

\bibitem{i2 spectral data}
W.A. de Jong, L. Visscher, and W.C. Nieuwpoort,
{\it J. Chem. Phys.}{\bf 107}, 21(1997) and references therein.

\bibitem{Hydrogen-Soures}
	P. Clark Souers:
			{\it Hydrogen properties for Fusion Energy} (University of California Press, 1986).

\bibitem{Hydrogen-quad}
	U. Fink, T. A. Wiggins and D. H. Rank:
		J. Mol. Spectrosc. {\bf 18} (1965) 384

\bibitem{Hydrogen-Harris1}
	S. E. Harris, and A. V. Sokolov:
		Phys Rev. A {\bf 55} (1997) R4019

\bibitem{Hydrogen-Harris2}
	D. D. Yavuz, D. R.Walker, M.Y. Shverdin, G.Y. Yin, and S. E. Harris:
		Phys Rev. Lett. {\bf 91} (2003) 233602

\bibitem{Hydrogen-Hakuta}
	J. Q. Liang, M. Katsuragawa, Fam Le Kien, and K. Hakuta:
		Phys Rev. Lett. {\bf 85} (2000) 2474

\bibitem{Hydrogen-Katsuragawa}
	M. Katsuragawa, K. Yokoyama, and T. Onose:
		Opt. Express {\bf 13} (2005) 5628

\bibitem{Hydrogen-collision}
	J. -M. Hartmann, C. Boulet, D. Robert:
			{\it Collisional effects on molecular spectra} (Elsevier, 2008).

\bibitem{Hydrogen-width}
	W. B. Bischel and M. J. Dyer:
		Phys Rev. A {\bf 33} (1986) 3113

\bibitem{Hydrogen-multiphoton}
	Y. V. Vanne and A. Saenz:
		Phys Rev. A {\bf 80} (2009) 053422

\bibitem{Hydrogen-momose}
	T. Momose and T. Oka:
		J. Low. Temp. Phys. {\bf 139} (2005) 515

\bibitem{Hydrogen-Li}
	J. Z. Li, M. Suzuki, M. Katsuragawa, and K. Hakuta:
		J. Chem. Phys. {\bf 115} (2001) 930

\bibitem{Hydrogen-closed}
	T. Oka:
		Annu. Rev. Phys. Chem. {\bf 44} (1993) 299

\bibitem{Hydrogen-press}
	M. Suzuki, M. Katsuragawa, R. S. D. Sihombing, J. Z. Li, K. Hakuta:
		J. Low. Temp. Phys. {\bf 111} (1998) 463

\bibitem{RWBoyd:NonlinearOptics2c2}
Robert~W. Boyd,
\newblock {\em Nonlinear Optics}, Chapter 2
\newblock  (Academic Press, second edition, 2003).

\bibitem{Demtroder:LaserSpectroscopy41c58}
W.~Demtr{\"{o}}der,
\newblock {\em Laser Spectroscopy, volume 1}, Chapter 5.8
\newblock  (Springer, Berlin, fourth edition, 2008).

\bibitem{TSuzuki:PRL2008}
Takayuki Suzuki, Masataka Hirai, and Masayuki Katsuragawa, Phys. Rev. Lett.,
  {\bf 101}, 243602 (Dec 2008).

\bibitem{JEBjorkholm:ApplPhysLett1969}
J.~E. Bjorkholm and H.~G. Danielmeyer, Appl. Phys. Lett., {\bf 15}(6), 171--173
  (1969).

\bibitem{JMBoon-Engering:OptLett1995}
J.~M. Boon-Engering, W.~E. van~der Veer, J.~W. Gerritsen, and W.~Hogervorst,
  Opt. Lett., {\bf 20}(4), 380--382 (Feb 1995).

\bibitem{ABorsutzky:QuantumSemiclassOpt1997}
A~Borsutzky, Quantum Semiclass. Opt., {\bf 9}(2), 191 (1997).

\bibitem{PBourdon:OptLett1995}
P.~Bourdon, M.~P\'{e}alat, and V.~I. Fabelinsky, Opt. Lett., {\bf 20}(5),
  474--476 (Mar 1995).

\bibitem{WDKulatilaka:ApplPhysB2005}
W.~D. Kulatilaka, T.~N. Anderson, T.~L. Bougher, and R.~P. Lucht, Applied
  Physics B: Lasers and Optics, {\bf 80}, 669--680 (2005).

\bibitem{MJTMilton:OptLett1994}
M.~J.~T. Milton, T.~D. Gardiner, G.~Chourdakis, and P.~T. Woods, Opt. Lett.,
  {\bf 19}(4), 281--283 (Feb 1994).

\bibitem{AVSmith:JOSAB1995}
A.~V. Smith, W.~J. Alford, T.~D. Raymond, and Mark~S. Bowers, J. Opt. Soc. Am.
  B, {\bf 12}(11), 2253--2267 (Nov 1995).

\bibitem{NSrinivasan:JJAP1996}
Narayanan Srinivasan, Takashi Kimura, Hiromitsu Kiriyama, Masanobu Yamanaka,
  Yasukazu Izawa, Sadao Nakai, and Chiyoe Yamanaka, Jpn. J. Appl. Phys., {\bf
  35}(Part 1, No. 6A), 3457--3458 (1996).

\bibitem{OVotava:JChemPhys1997}
Ondrej Votava, Joanna~R. Fair, David~F. Plusquellic, Eberhard Riedle, and
  David~J. Nesbitt, J. Chem. Phys., {\bf 107}(21), 8854--8865 (1997).

\bibitem{SWu:OptCommun1999}
Sheng Wu, Vadym~A Kapinus, and Geoffrey~A Blake, Optics Communications, {\bf
  159}, 74 -- 79 (1999).

\bibitem{Hect:Opticsc84}
Eugene Hecht,
\newblock {\em Optics}, Chapter 8.4
\newblock  (Addison Wesley, San Francisco, fourth edition, 2002).

\bibitem{NEYu:JJAP2004}
Nan~Ei Yu, Sunao Kurimura, Yoshiyuki Nomura, and Kenji Kitamura, Jpn. J. Appl.
  Phys., {\bf 43}(10A), L1265--L1267 (2004).
  
\bibitem{Walhout1994} M. Walhout, A. Witte, and S.L. Rolston, Phys. Rev. Lett. 72 (1994) 2843.

\bibitem{Barbet1975} A. Barbet, N. Sadeghi and J.C. Pebay-Peyroula, J. Phys. B 8 (1975) 1776.

\bibitem{Aymar1978} M. Aymar and M. Coulombe, At. DATA and Nucl. DATA Tab. 21 (1978) 537.

\bibitem{Chen1980} C.H. Chen, G.S Hurst and M.G. Payne, Chem. Phys. Lett. 75 (1980) 473.

\bibitem{Kau1997} R. Kau, I.D. Petrov, V.L. Sukhorukov, H. Hotop, Z. Phys. D 39 (1997) 267.

\bibitem{Horiguchi1981} H. Horiguchi, R.S.F. Chang, and D.W. Setser, J. Chem. Phys. 75 (1981) 1207.

\bibitem{Morton2000} D.C. Morton, Astrophys. J. Suppl. Ser. 130 (2000) 403.

\bibitem{Boiler1991} K.-J. Boiler, A. Imamoglu, and S. E. Harris, Phys. Rev. Lett. 66 (1991) 2593. 

\bibitem{Zhang1995} G. Z. Zhang, M. Katsuragawa, K. Hakuta, R.I. Thompson and 
B.P. StoicheffPhys. Rev. A 52 (1995) 1584.

\bibitem{sakurai}
J.J. Sakurai,
{\it Advanced Quantum Mechanics},
Addison-Wesley (1967).


\bibitem{Burnham-Chiao}
D.C. Burnham and R.Y. Chiao,
{\it Phys. Rev.}{\bf 188}, 660(1969).


\bibitem{Hydrogen-Iodine}
	D. R. Stull:
		Ind. Eng. Chem. {\bf 39} (1947) 540

\bibitem{Hydrogen-Iodine2}
	J. B. Koffend, A. M. Sibai, and R. Bacis:
		J. Physique {\bf 43} (1982) 1639

\bibitem{Hydrogen-Iodine3}
	R. B\"{o}hling, J. Langen, and U. Schurath:
		Chem. Phys. {\bf 130} (1989) 419

\bibitem{Momose92}
	T.\ Momose, D.\ P.\ Weliky, and T.\ Oka,
	J.\ Mol.\ Spectrosc. \textbf{153}, 760 (1992). 

\bibitem{Kuroda03}
	K.\ Kuroda, A.\ Koreeda, S.\ Takayanagi, M.\ Suzuki, and K.\ Hakuta,
	Phys.\ Rev.\ B \textbf{67}, 184303 (2003). 

\bibitem{Abram80a}
	I.\ I.\ Abram, R.\ M.\ Hochstrasser, J.\ E.\ Kohl, and M. G. Semack,
	Chem.\ Phys.\ Lett. \textbf{71}, 405 (1980). 

\bibitem{Li98}
	J.\ Z.\ Li, M.\ Katsuragawa, M.\ Suzuki, and K.\ Hakuta,
	Phys.\ Rev.\ A \textbf{58}, R58 (1998). 

\bibitem{Laubereau78}
	A.\ Laubereau and W.\ Kaiser,
	Rev.\ Mod.\ Phys. \textbf{50}, 607 (1978) and references therein.

\bibitem{Perera11}
	M.\ Perera, B.\ A.\ Tom, Y.\ Miyamoto. M.\ W.\ Porambo, L.\ E.\ Moore, W.\ R.\ Evans, T.\ Momose, and B.\ J.\ McCall,
	Opt.\ Lett. \textbf{36}, 840 (2011). 

\bibitem{Prior72}
	W.\ R.\ C.\ Prior and E.\ J.\ Allin,
	Can.\ J.\ Phys. \textbf{50}, 1471 (1972). 

\bibitem{Kuma12}
	S.\ Kuma, Y.\ Miyamoto, K.\ Nakajima, A.\ Fukumi, K.\ Kawaguchi, I.\ Nakano, N.\ Sasao, M.\ Tanaka, J.\ Tang, T.\ Taniguchi, S.\ Uetake, T.\ Wakabayashi, A.\ Yoshimi, and Y.\ Yoshimura, 
	submitted to J. Chem. Phys.

\bibitem{Kien03}
	F.\ L.\ Kien, A.\ Koreeda, K.\ Kuroda, M.\ Suzuki, and K.\ Hakuta,
	Jpn.\ J.\ Appl.\ Phys. \textbf{42}, 3483 (2003). 

\bibitem{McCumber63}
	D.\ E.\ McCumber and M.\ D.\ Sturge,
	J.\ Appl.\ Phys. \textbf{34}, 1682 (1963). 

\bibitem{Carman70}
	R.\ L.\ Carman, F.\ Shimizu, C.\ S.\ Wang, and N.\ Bloembergen,
	Phys.\ Rev.\ A \textbf{2}, 60 (1970). 

\bibitem{Wallace74}
	B.\ A.\ Wallace and H.\ Meyer,
	J.\ Low Temp.\ Phys. \textbf{15}, 297 (1974).
        
\bibitem{Katsuragawa02}
	M.\ Katsuragawa, J.\ Q.\ Liang, Fam Le Kien, and K.\ Hakuta,
	Phys.\ Rev.\ A \textbf{65}, 025801 (2002). 

\bibitem{BondybeyCPL1980}
	V. E. Bondybey, G. P. Schwartz, J. E. Griffiths, and J. H. English, 
		Chem. Phys. Lett., {\bf 76}, 30 (1980).

\bibitem{CremerApplPhysB1984}
	C. Cremer and G. Gerber, 
		Appl. Phys. B, {\bf 35}, 7 (1984).

\bibitem{XuPRL2011} 
	C.-Y. Xu, S.-M. Hu, J. Singh, K. Bailey, Z.-T. Lu, P. Mueller, T. P. O'Conner, and U. Welp, 
		Phys. Rev. Lett., {\bf 107}, 093001 (2011).
        
\bibitem{Hydrogen-CH4}
	S. Tam, M. E. Fajardo, H. Katsuki, H. Hoshina, T. Wakabayashi, and T. Momose, 
		J. Chem. Phys. {\bf 111}, 4191 (1999).

\bibitem{Hydrogen-deposition}
	M. E. Fajardo and S. Tam:
		J. Chem. Phys. {\bf 108} (1998) 4237

\bibitem{Hydrogen-HF}
	Y. Miyamoto, H. Ooe, S. Kuma, K. Kawaguchi, K. Nakajima, I. Nakano, N. Sasao, J. Tang, T. Taniguchi, and M. Yoshimura, 
		J. Phys. Chem. A, {\bf 115},14254 (2011).

\bibitem{CampbellAdvMater1999} 
	E. E. B. Campbell, S. Couris, M. Fanti, E. Koudoumas, N. Krawez, and F. Zerbetto, 
		Adv. Mater., {\bf 11}, 405 (1999).

\bibitem{ChaiJPC1991} 
	Y. Chai, T. Guo, C. M. Jin, R. E. Haufler, L. P. F. Chibante, J. Fure, L. Wang, J. M. Alford, and R. E. Smalley, 
		J. Phys. Chem., {\bf 95}, 7564 (1991).

\bibitem{AlmeidaMurphyPRL1996} 
	T. Almeida-Murphy, Th. Pawlik, A. Weidinger, M. H\"one, R. Alcala, and J.-M. Spaeth, 
		Phys. Rev. Lett., {\bf 77}, 1075 (1996).

\bibitem{MortonJCP2006} 
	J. J. L. Morton, A. M. Tyryshkin, A. Ardavan, K. Porfyrakis, S. A. Lyon, and G. A. Briggs, 
		J. Chem. Phys., {\bf 124}, 014508 (2006).

\bibitem{feld}
J.C. McGillivray and M.S. Feld,
{\it Phys. Rev.}{\bf A 14}, 1169(1976). 

\bibitem{breather soliton}
M. Yoshimura, N. Sasao, and M. Tanaka,
paper in preparation.



                
  















\end{thebibliography}

$\;$ \\ $\;$ \\
\section{Summary and prospects}\label{Sec:Summary}
In the present work we first explained in detail
theoretical principles of the neutrino mass
spectroscopy with atoms and molecules,
which aims at determination of
undetermined important neutrino parameters.
The key idea for measurements is to
amplify otherwise small rates of
radiative emission of neutrino pair (RENP),
$|e\rangle \rightarrow |g \rangle + \gamma + \nu_i \nu_j$
(with $\nu_i,\nu_j$  mass eigenstates), by
developing the macro-coherent medium
polarization among target
atoms/molecules strongly coupled  to fields inside the medium,
in order to stimulate
the process cooperatively.
The amplification is realized by trigger
irradiation of two colors into a well prepared target state
of good phase coherence between atoms in two
relevant levels, $|e\rangle$ and $|g\rangle$. 
The amplification also works
for two-photon emission called paired super-radiance
(PSR), $|e\rangle \rightarrow |g \rangle + \gamma +\gamma$.
Detailed account of the master equation for PSR
and coherence development for RENP is presented
and results of numerical simulations on the pH$_2$
$Xv=1$ vibrational transition have
been presented both for explosive PSR events and 
events in the weaker linear regime.
Our master equation includes effects of phase decoherence
of medium polarization and decay of population difference.

By selecting a metastable state $|e\rangle$ forbidden to
decay to lower levels via E1 transition, one can obtain a large E1$\times$E1 
two-photon PSR rate for $|e\rangle \rightarrow |g\rangle$.
A good example of this feature is pH$_2$ vibrational
transition $Xv=1 \rightarrow 0$.
The PSR event may occur explosively, as shown in
Figure 4 of Section 1, if the target relaxation time $T_2$
is larger than some number, for the target number
density $n=10^{21}$cm$^{-3}$ this number being
of order a few to several nano seconds.
Even for smaller $T_2$'s, the enhanced output over
the input trigger is expected as discussed in Section 4.
Discovery of these PSR events serves to prove
our principle of macro-coherence, a necessary prerequisite
for RENP experiments.

The RENP spectral rate is given by
the factorized formula 
$\Gamma_{\gamma 2\nu}(\omega) = \Gamma_0 I(\omega)\eta_{\omega}(t)$.
The factor $\Gamma_0$ determines the overall rate in the
unit of 1/time, and for Xe it is of order,
$1\,\mathrm{Hz}\,(n/10^{22}\mathrm{cm}^{-3})^3\,(V/10^2\mathrm{cm}^3)$.
The spectral shape as a function of photon energy
$\omega$ is given by $I(\omega)$, which serves to
determine the neutrino parameters.
Detailed photon spectrum calculations
of Xe and I$_2$ transitions, having
available energy 8.3 eV and 0.88 eV, respectively,
have been given to explain how the parameter
determination is made possible.
The dynamical factor $\eta_{\omega}(t)$ is
the space integrated quantity over the entire
target of $|\vec{E}^{-}(R_1-iR_2)|^2/(4\epsilon_{eg}n^3)$, the dimensionless
  squared product of the coherent polarization $(R_1-iR_2)/2$ and 
  the field $E^{(-)}$ inside the target.

The dynamical factor $\eta_{\omega}(t)$ is
highly time dependent in the phase of large
PSR related activities.
It asymptotically approaches a constant value
of the stable state made of field condensate supported by
coherent medium polarization.
For RENP it is important to utilize this
factor as large as possible.
How the condensate state is described is explained
and a sample simulation has been presented,
leaving more detailed investigation to future
works.
If the amplification works as expected,
the absolute neutrino mass determination
along with distinction of mass hierarchy 
pattern (normal or inverted) 
becomes feasible. Other
important neutrino parameters such
as Majorana/Dirac distinction along with
CP violating phases may also
be determined if the initial metastable
atomic state lies close to the ground level,
preferably within a fraction of eV. 

Heavy atoms/molecules which
have a large breaking of the $LS$ 
coupling scheme are favored for
RENP target candidates, since they may
give a large enough amplitude of M1$\times$E1 
type for RENP between $|e\rangle$ and $|g\rangle$ .

In the later half of the paper
experimental status of our project
has been reported, along with an overview of
the entire experimental strategy.

Our basic strategy towards precision neutrino mass spectroscopy is first to prove the macro-coherent amplification principle by the PSR process.
To this end, we have chosen gas-phase para-hydrogen (pH$_2$) as a target, and focus on the E1 forbidden transition from the vibrational excited state Xv=1 to the ground state Xv=0.
As is discussed in Sec.2 and 4, PSR is highly non-linear phenomenon, and can be observed only when a certain set of initial conditions are realized.
In the case of pH$_2$, they are the number density ($n$), the initial coherence ($r_{1}$ or $ r_{2}$), and dephasing time ($T_2$ and $T_3$).
We expect all necessary condition can be fulfilled in the current technology available to us at least for observing PSR events in the linear regime.
As described in Sec.4, all components are prepared, and ready for the experiments.

It is crucial to achieve a large value of the RENP dynamical factor $\eta_{\omega}(t)$ to realize high precision mass spectroscopy. 
We plan to study evolution of macroscopic coherence and formation of field condensates, both of which directly affects 
$\eta_{\omega}(t)$, using Xe atom as an experimental platform.
Details of the current status on the Xe experiment can be found in Sec.4.3.
Condensed-phase targets are highly desirable not only for RENP experiments but also for observation of explosive PSR events. 
We are seeking and trying to develop high density solid targets with long dephasing time. 
Some preliminary results of our efforts along this line are presented in Appendix D and E.

It is evident that
we need much more detailed simulations
and experimental studies on
how best we should design
this challenging experiment.
Some entirely new idea is welcome and
may further enlighten
prospects towards this important objective
of neutrino physics.

\vspace{10mm}
\paragraph{Note added in proof}
After the original submission of the manuscript
some of us succeeded in deriving  breathing solitons
which are useful as an ideal target state for radiative neutrino pair emission (RENP).
This shall be discussed elsewhere in a separate work \cite{breather soliton}.

\vfill\pagebreak

\section*{Acknowledgment}
We should like to thank M. Katsuragawa, K. Kawaguchi, Y. Kubozono, T. Momose,
S.T. Petcov, and S. Yabushita for
many helpful discussions and comments on this project.
This research was partially supported by Grant-in-Aid for Scientific Research A (21244032),
Grant-in-Aid for Challenging Exploratory Research (24654132),
and Grant-in-Aid for Scientific Research on Innovative Areas "Extreme quantum world opened up by atoms"
(21104002)
from the Ministry of Education, Culture, Sports, Science, and Technology of Japan.\\[10mm]

\appendix

\section{Electroweak interaction under nuclear Coulomb potential}
\label{App:EW-interaction}

\subsection{Electrons in atom and their electromagnetic interaction}

Electrons in an atom are described by solutions of
the Dirac equation under the nuclear Coulomb potential 
\cite{sakurai}
augmented by interaction
with other electrons in the atom.
Bound and nearly free (modified by the nuclear Coulomb
field) electron states thus obtained
form a complete set of states that may be used
as an expansion basis of electron field 
denoted by a four component operator $\psi_e(\vec{x},t)$.
The interaction of atomic electrons with the
electromagnetic field $\vec{A}(\vec{x},t)$ is best described in physical terms by
using the radiation gauge, as given for instance
in standard textbooks of QED such as \cite{sakurai}.
The radiation gauge field satisfies 
$\vec{\nabla}\cdot \vec{A} = 0$ and possesses
two transverse degrees of freedom representing
two transverse waves of photon.
In this gauge the static Coulomb interaction is
separated from the transverse field interaction,
hence is convenient for our purpose.

In the present work we deal with interaction of
atomic electrons with the transverse electromagnetic
field $\vec{A}$ and three massive neutrino field $\nu_i$.
This way we can derive an effective hamiltonian 
of atomic electrons in the electroweak theory
\cite{review of electroweak theory}
we need for our purpose.

In most cases of atomic electrons relativistic
effects are minor and atomic electrons have
bound energies much less than the electron 
rest mass. Or equivalently, the average electron
velocity within an atom is much less than the light velocity.
This makes non-relativistic treatment sufficient,
although in some cases relativistic corrections
such as $LS$ coupling are important.
Under this circumstance the electron field 
operator can be decomposed into two parts;
a large and a small two-component fields.
The major large two-component field is denoted by
a new operator $\phi_e(\vec{x},t)$.

As is well known \cite{sakurai},
the electromagnetic interaction of atomic electrons
is governed by interaction hamiltonian (or rather the hamiltonian
density to be spatially integrated), 
\begin{eqnarray}
&&
e \phi_e^{\dagger}(\vec{x},t)(-i \vec{\nabla})\cdot \vec{A}(\vec{x},t)\phi_e(\vec{x},t)
\,.
\end{eqnarray}
When this interaction hamiltonian is applied to
atomic process, it gives a vertex amplitude for
the atomic transition $|i\rangle \rightarrow |f\rangle$,
\begin{eqnarray}
&&
e \varphi_f^{\dagger}(\vec{x},t)(-i \vec{\nabla})\cdot \vec{A}(\vec{x},t)\varphi_i(\vec{x},t)
\,,
\end{eqnarray}
where $\varphi_a$ is a wave function solving the nuclear Coulomb potential
problem \cite{textbook on molecules}.

A further simplification arises by taking the long wavelength limit
of emitted photons
(the wavelength of photon much larger than the atomic size), 
and this leads to what is called the E1 transition
in the literature \cite{sakurai}, \cite{textbook on molecules},
where one can replace the vertex amplitude above  by 
\begin{eqnarray}
&&
\varphi_f^{\dagger}(\vec{x},t)\vec{d}\cdot \vec{E}(\vec{x},t)\varphi_i(\vec{x},t)
\,, \hspace{0.5cm}
\vec{d} = e\vec{x}
\,.
\end{eqnarray}
In the many-electron system in atom
the dipole operator $\vec{d}$ should be replaced by  a summed
quantity over all contributing electrons at position $\vec{x}_n$;
$\vec{d} = e\sum_n\vec{x}_n$.
The wave functions $\varphi$ should also be replaced by
those of the multi-electron states \cite{textbook on molecules}.

\subsection{Quantized massive neutrino fields}
We have two possibilities to describe massive neutrinos: 
one is the Dirac type and another is the Majorana one. 
The neutrino and the anti-neutrino are distinct particles
in the Dirac case, while they are identical in the Majorana case.
RENP final states discriminate these cases due to the identical particle
effect as explained in Secs.~\ref{Sec:Introduction} and \ref{Sec:RENP-Theory}.
The neutrino expansion base may
be taken, effectively and conveniently,
as plane wave functions, since the neutrino interaction with
atomic nuclei is very weak.

For the Dirac type, a quantized free neutrino field is
represented as
\begin{equation}
\psi^D(x)=\sum_{p,\lambda}\left[b(p,\lambda) u(p,\lambda) e^{-ipx}+
                                d^\dagger(p,\lambda) v(p,\lambda) e^{ipx}
                          \right]\,,
\end{equation}
where $b(p,\lambda)$ is the annihilation operator of the neutrino with
a momentum $p$ and a helicity $\lambda$, and $d^\dagger(p,\lambda)$
denotes the creation operator of the anti-neutrino. The four-component
spinors $u(p,\lambda) e^{-ipx}$ and $v(p,\lambda) e^{ipx}$ are solutions of the Dirac equation. 
We employ the following convention for the neutrino momentum sum,
\begin{equation}
\sum_p=\int\frac{d^3p}{(2\pi)^3 2E_p}\,,
\end{equation}
where $E_p=\sqrt{m^2+\vec{p}^2}$ is the energy of neutrino of mass $m$.
The Dirac field $\psi^D$ either annihilates a neutrino or creates 
an anti-neutrino. A free Majorana neutrino field 
is quantized as
\begin{equation}
\psi^M(x)=\sum_{p,\lambda}\left[b(p,\lambda) u(p,\lambda) e^{-ipx}+
                                  b^\dagger(p,\lambda) v(p,\lambda) e^{ipx}
                          \right]\,,
\end{equation}
where $b$ ($b^\dagger$) represents the annihilation (creation) operator 
of the Majorana neutrino and 
$\psi^M$ satisfies $\psi^M=C(\overline{\psi^M})^T$
with $C$ being the charge conjugation matrix.
\footnote{
This $\psi^M$ effectively reduces to two independent solutions
in accordance with two-component formalism.
The field quantization in the two-component formalism is
given in 
\cite{review of electroweak theory} and \cite{my-rnpe1}.
}
We note that there is no distinction between neutrinos and 
anti-neutrinos in the Majorana case, and both the creation and annihilation
operators of the Majorana neutrino appear in $\psi^M$.

\subsection{Effective interaction of neutrinos with electrons}
The neutrino fields in the flavor eigen-basis are related to
those in the mass eigen-basis by a unitary transformation as
\begin{equation}
 \nu_\alpha(x)=\sum_i U_{\alpha i}\,\nu_i(x)
 \,,
\end{equation}
where $\alpha=e,\mu,\tau$ specifies a neutrino flavor, $i=1,2,3$
denotes a mass eigenstate, and $\nu_i(x)=\psi^D(x)$ or 
$\psi^M(x)$ with mass $m_i$. The lepton mixing matrix is written 
as a product of two unitary matrices \cite{nu-matrix}:
\begin{equation}
U=VP\,,
\end{equation}
where 
\begin{equation}
V=\left[\begin{array}{ccc}
          c_{12}c_{13} & s_{12}c_{13} & s_{13}e^{-i\delta}\\
         -s_{12}c_{23}-c_{12}s_{23}s_{13}e^{i\delta} & 
          c_{12}c_{23}-s_{12}s_{23}s_{13}e^{i\delta} &
          s_{23}c_{13}\\
          s_{12}s_{23}-c_{12}c_{23}s_{13}e^{i\delta} & 
          c_{12}c_{23}-s_{12}s_{23}s_{13}e^{i\delta} & 
          c_{23}c_{13}
        \end{array}
  \right]\,,
\end{equation}
with $c_{ij}=\cos\theta_{ij}$ and $s_{ij}=\sin\theta_{ij}$.
The diagonal unitary matrix $P$ may be expressed by
\begin{equation}
P=\mathrm{diag.}(1,e^{i\alpha},e^{i\beta})\,,
\end{equation}
for Majorana neutrinos, and we rotate away the phases $\alpha$
and $\beta$ for Dirac neutrinos.  

\begin{figure*}
\begin{center}
\includegraphics[width=0.6\textwidth]{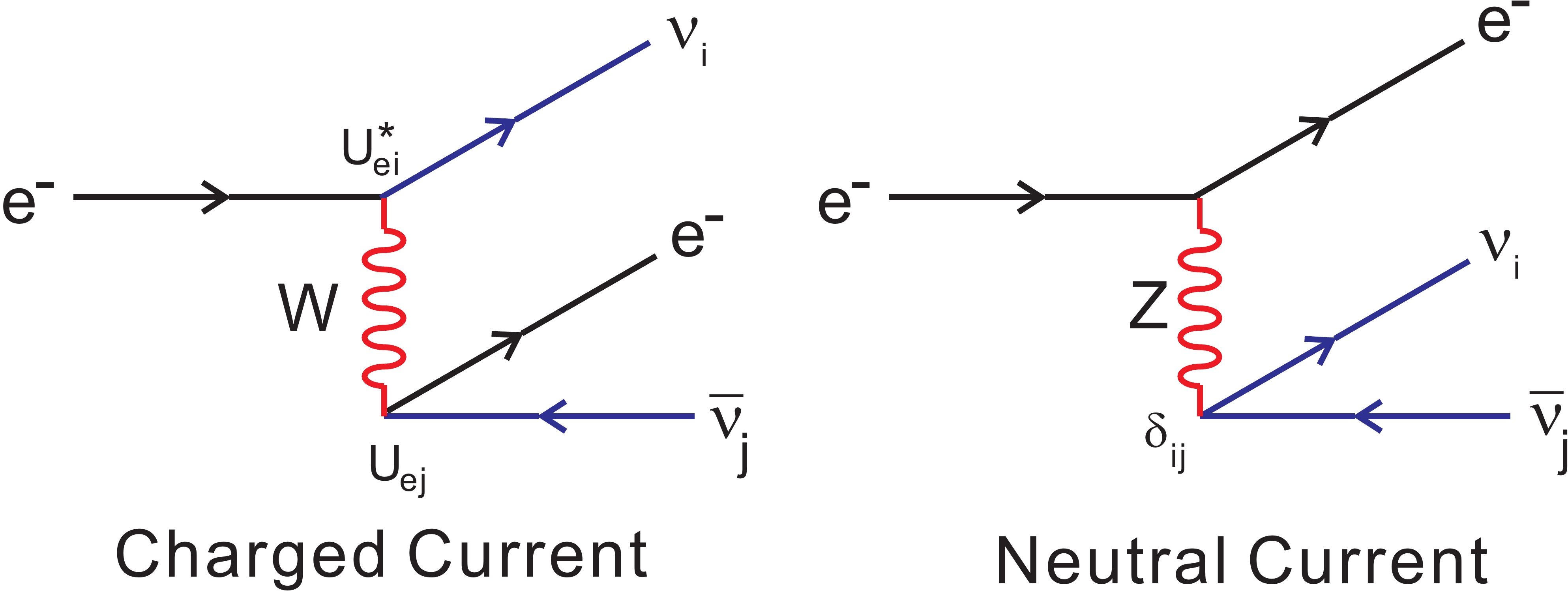}
\end{center} 
\caption{Feynman diagram for $W$ and $Z$ exchange weak interaction
of electron with neutrinos.}
\label{weak interaction diagram} 
\end{figure*}

The effective interaction of neutrinos with the electron field 
arises from $W$ and $Z$ exchange CC and NC interaction
\cite{review of electroweak theory}, as
depicted in Fig.\ref{weak interaction diagram}, and is given by
\begin{equation}
\mathcal{H}_\mathrm{eff}=\frac{G_F}{\sqrt{2}}\,\sum_{i,j}
                       \bar\nu_i\gamma^\mu(1-\gamma_5)\nu_j\,
                       \bar e\gamma_\mu(v_{ij}-a_{ij}\gamma_5)e,
\end{equation}
where
\begin{equation}
v_{ij}=U_{ei}^*U_{ej}-\left(\frac{1}{2}-2\sin^2\theta_W\right)\delta_{ij},\ 
a_{ij}=U_{ei}^*U_{ej}-\frac{1}{2}\delta_{ij}.
\label{eq:v-a-coefficients}
\end{equation}

The interaction above is written using the four component spinor notation.
It has been shown that for non-relativistic atomic electrons
the leading neutrino interaction stems from
the axial-vector part and in the two-component notation
it can be written using the spin operator of the form,
\begin{equation}
\nu^{\dagger} \vec{\sigma} \nu \cdot \phi_e^{\dagger} \vec{\sigma}\phi_e
\,.
\end{equation}
This results in the magnetic type of transition operator
on the electron side.
We may thus use, for the spin matrix element of atomic transitions 
from a bound state $|a\rangle $ to $|b\rangle$, their wave functions 
$\varphi_a(\vec{r}), \varphi_b(\vec{r})$, and compute the overlapping integral of
the kind,
\begin{eqnarray}
&&
\int d^3 r \varphi_a^*(\vec{r})\vec{S}_e \varphi_b(\vec{r})\,.
\end{eqnarray}
In practice, it is better to use experimental data for this
if they are available.


\section{Mathematical structure of Maxwell-Bloch equation}
\label{App:MB-equations}
\subsection{Maxwell-Bloch equation for SR}

We consider equations, Eq. \ref{sr-mb 1} $\sim$ \ref{sr-mb 3}
in the text,
for the single photon SR.
First rescale time, length and field by
\begin{eqnarray}
&&
\tau = \frac{t}{T_R} \,, \hspace{0.5cm}
\xi = \frac{x}{T_R} \,, \hspace{0.5cm}
e = -i dT_R E \,,
\end{eqnarray}
where $T_R$ should have the dimension of time/length.
It is further convenient to introduce the retarded time
$\tau_- = \tau-\xi$ and use $(\tau_-, \xi)$ as independent variables.
In terms of these
the Maxwell-Bloch equation is converted to
\begin{eqnarray}
&&
\partial_{-} R = Z e \,,
\\ &&
\partial_{-}Z = - \frac{1}{2} (e^* R + eR^*) \,,
\\ &&
\partial_{\xi} e = R \,,
\end{eqnarray}
with $T_R = 1/\sqrt{2\pi \omega d^2 n}$
determined from consistency.

Using the conservation law that follows,
\(\:
\partial_{-} (|R|^2 + Z^2 ) = 0
\,,
\:\)
we introduce an angle function $\theta(\tau_-, \xi)$:
\begin{eqnarray}
&&
R = e^{i\varphi}B(\xi)\sin \theta(\tau_-, \xi)
\,, 
Z= B(\xi)\cos \theta(\tau_-, \xi) \,.
\end{eqnarray}
In order to avoid unnecessary complication
we assume real field $e$ and set the phase $\varphi = 0$.
The Maxwell-Bloch equation then becomes
\begin{eqnarray}
&&
\frac{\partial^2}{\partial_- \partial_{\xi}}\theta = B\sin \theta \,,
\\ &&
e = \partial_- \theta \,.
\end{eqnarray}
The first equation is known as the sine-Gordon equation
in 1+1 space-time dimensions.

\begin{figure*}
 \begin{center}
\includegraphics[width=0.4\textwidth]{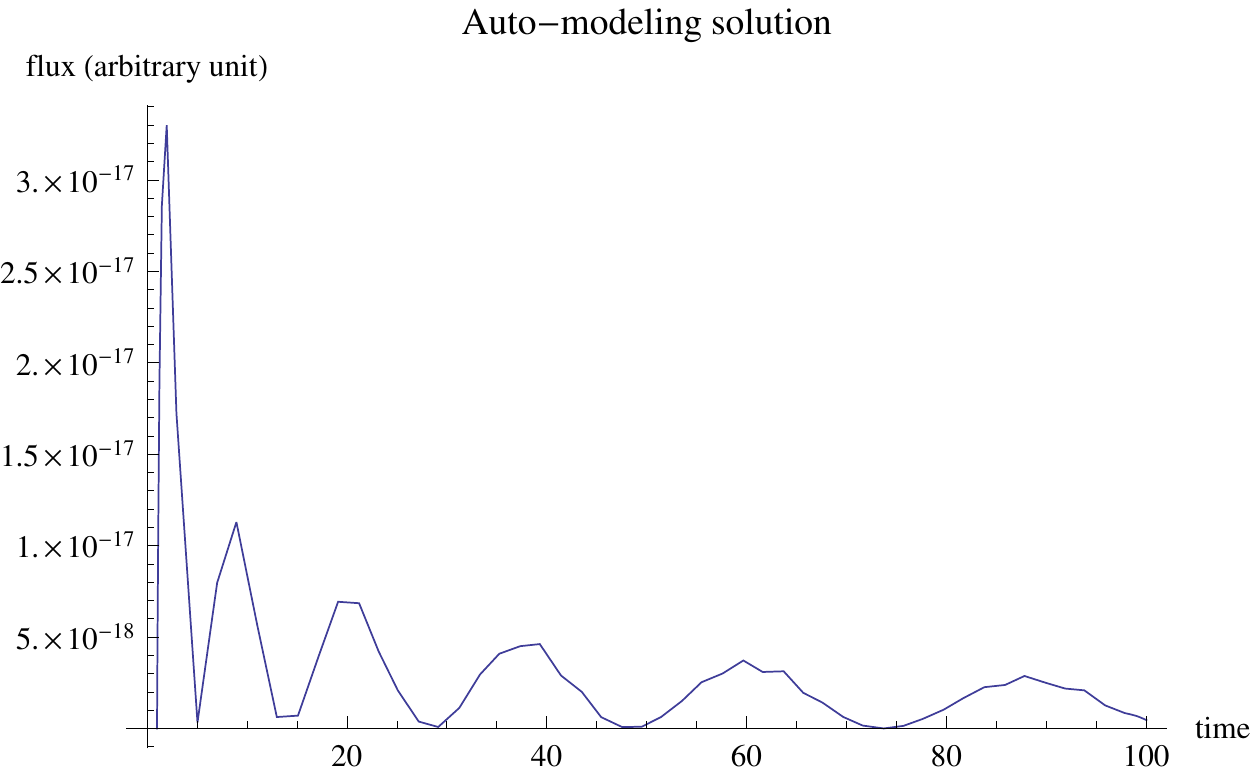}
 \end{center} 
   \caption{Auto-modeling solution numerically computed
   from eq.(\ref{auto-modeling eq}).
   Initial values are $\theta(\tau=2\times 10^{-6})=10^{-8}\,, \theta(\tau=2\times 10^{-6})=0$
and target length = 1. Constant $B \sim 1$ is assumed.
   }
   \label{auto-modeling sol 1}
\end{figure*}

Scaling invariance under
\(\:
\tau_- \rightarrow \beta \tau_-
\,,
\xi \rightarrow \beta^{-1} \xi
\,,
\:\)
exists in this system.
The Burnham-Chiao equation \cite{Burnham-Chiao} is obtained using the variable,
$w = 2\sqrt{\tau_- \xi}=2\sqrt{(\tau -\xi) \xi}$,
\begin{eqnarray}
&&
\frac{d^2 \theta}{dw^2} + \frac{1}{w}\frac{d\theta}{dw}= B \sin \theta
\,.
\label{auto-modeling eq}
\end{eqnarray}
Since this is an ordinary differential equation,
it is much easier to solve than 1+1 dimensional partial differential equation.
This solution is illustrated in Fig.\ref{auto-modeling sol 1}, which
shows interesting ringing structure of emitted pulses.

\subsection{Maxwell-Bloch equation for PSR} 

We recapitulate the master equation for the most
general case of two-color
trigger irradiation of frequencies $\omega_1 + \omega_2 =\epsilon_{eg}$, 
as given in \cite{psr dynamics}.

We introduce the dimensionless unit:
\begin{eqnarray}
&&
\hspace*{-0.5cm}
(\xi\,, \tau) =  (\mu x\,,\mu t)\,,\
\mu(\omega) = \epsilon_{eg}n\alpha_{ge}
                   (\omega, \epsilon_{eg}-\omega)\,,\ 
|e_{L, R}^{(1), (2)}|^2 = \frac{|E_{L, R}^{(1), (2)}|^2}{\epsilon_{eg}n}\,,\
r_i = \frac{R_i}{n}
\,,
\\ &&
\alpha_{ge}(\omega, \epsilon_{eg}-\omega) =
\frac{d_{pe}d_{pg}(\epsilon_{pg}+\epsilon_{pe})}
{(\epsilon_{pe}+ \omega)(\epsilon_{pg}-\omega)}
\,.
\label{off-diagonal identity}
\end{eqnarray}
Assume R-mover field of frequency $\omega_1$ and L-mover of frequency $\omega_2$
(neither R-mover of frequency $\omega_2$ nor L-mover of frequency $\omega_1$).
Note the frequency dependence of coupling parameters
$\alpha_{ab}(\omega_1, \omega_2)$ and the universal relation
$\alpha_{eg}(\omega_1, \omega_2)= \alpha_{ge}(\omega_1, \omega_2)$
for any combination of $\omega_1+ \omega_2= \epsilon_{eg}$.
The master equations for medium polarization and fields are
\begin{eqnarray}
\partial_{\tau} r_1^{(0)} &=& 
 2(\gamma_-^{(1)}|e_R|^2  + \gamma_-^{(2)}|e_L|^2)r_2^{(0)}
+ 4 \Im (e_R e_L)r_3^{(0)}
+ 2\gamma_-^{(12)} e_R e_L^* r_2^{(-)} +  2\gamma_-^{(21)} e_L e_R^* r_2^{(+)}
\nonumber \\ &&
- i (e_L^2- (e_R^*)^2) r_3^{(+)} - i (e_R^2- (e_L^*)^2) r_3^{(-)}
-\frac{r_1^{(0)}}{\tau_2}
\,,
\label{rescaled bloch eq1}
\end{eqnarray}
\begin{equation}
\partial_{\tau} r_1^{(+)} =2\gamma_-^{(12)} e_R e_L^* r_2^{(0)}
- i (e_R^2- (e_L^*)^2) r_3^{(0)}
+2(\gamma_-^{(1)}|e_R|^2  + \gamma_-^{(2)}|e_L|^2)r_2^{(+)} 
+  4 \Im (e_R e_L)r_3^{(+)}
-\frac{r_1^{(+)}}{\tau_2}\,,
\end{equation}
\begin{eqnarray}
&&
\partial_{\tau} r_2^{(0)} = 
-2(\gamma_-^{(1)}|e_R|^2  + \gamma_-^{(2)}|e_L|^2)r_1^{(0)} 
+ 4 \Re (e_R e_L)r_3^{(0)}
- 2\gamma_-^{(12)} e_R e_L^* r_1^{(-)}  - 2\gamma_-^{(21)} e_L e_R^* r_1^{(+)}
\nonumber \\ &&
\hspace*{1cm}
+  (e_L^2 + (e_R^*)^2) r_3^{(+)} +  (e_R^2 + (e_L^*)^2) r_3^{(-)}
-\frac{r_2^{(0)}}{\tau_2}
\,,
\label{rescaled bloch eq2}
\end{eqnarray}
\begin{equation}
\partial_{\tau} r_2^{(+)} =-2\gamma_-^{(12)} e_R e_L^* r_1^{(0)}
+  (e_R^2 + (e_L^*)^2) r_3^{(0)}
-2(\gamma_-^{(1)}|e_R|^2  + \gamma_-^{(2)}|e_L|^2)r_1^{(+)} 
+  4 \Re (e_R e_L)r_3^{(+)}
-\frac{r_2^{(+)}}{\tau_2}\,,
\end{equation}
\begin{eqnarray}
&&
\partial_{\tau} r_3^{(0)} = 
-4 \left( \Re (e_R e_L)r_2^{(0)}  + \Im (e_R e_L )r_1^{(0)}
\right)
+i (e_R^2- (e_L^*)^2) r_1^{(-)} +i (e_L^2- (e_R^*)^2) r_1^{(+)}
\nonumber \\ &&
\hspace*{1cm}
-  (e_L^2 + (e_R^*)^2) r_2^{(+)}  -  (e_R^2 + (e_L^*)^2) r_2^{(-)}
-\frac{r_3^{(0)}+1}{\tau_1}
\,,
\label{rescaled bloch eq3}
\end{eqnarray}
\begin{equation}
\partial_{\tau} r_3^{(+)} =
ir_1^{(0)} (e_R^2- (e_L^*)^2) - r_2^{(0)} (e_R^2 + (e_L^*)^2)
-4 \left( \Re (e_R e_L)r_2^{(+)}  + \Im (e_R e_L )r_1^{(+)}\right)
-\frac{r_3^{(+)}}{\tau_1}\,,
\end{equation}
\begin{equation}
(\partial_{\tau} + \partial_{\xi})e_R = 
 \frac{ia_1}{4}  (\gamma_+^{(1)}  +  \gamma_-^{(1)} r_3^{(0)} ) e_R
  + \frac{i}{4}\gamma_-^{(12)} r_3^{(+)}e_L
+ \frac{ia_{12}}{4}(r_1^{(0)} - ir_2^{(0)})e_L^*
+ \frac{i}{4}(r_1^{(+)} - ir_2^{(+)})e_R^*
\,, 
\label{rescaled quantum field eq1}
\end{equation}
\begin{equation}
(\partial_{\tau} - \partial_{\xi})e_L = 
 \frac{ia_2}{4} (\gamma_+^{(2)}  +  \gamma_-^{(2)} r_3^{(0)} ) e_L
 + \frac{i}{4}\gamma_-^{(21)} r_3^{(-)}e_R
+ \frac{ia_{21}}{4}(r_1^{(0)} - ir_2^{(0)})e_R^*
+ \frac{i}{4}(r_1^{(-)} - ir_2^{(-)})e_L^*
\,.  
\label{rescaled quantum field eq2}
\end{equation}
\begin{equation}
\gamma_{\pm}^{(a)} = \frac{\alpha_{ee}(\omega_a, \omega_a) 
\pm \alpha_{gg}(\omega_a, \omega_a)}{2\alpha_{ge}}
\,, \hspace{0.5cm}
\gamma_{\pm}^{(ab)} = \frac{\alpha_{ee}(\omega_a, \omega_b) 
\pm \alpha_{gg}(\omega_a, \omega_b)}{2\alpha_{ge}}\,,
\end{equation}
\begin{equation}
a_{i} = \frac{2\omega_i}{\epsilon_{eg}}
\,, \hspace{0.5cm}
a_{ij} = \frac{2\omega_j^2}{\omega_i \epsilon_{eg}}
\,,
\end{equation}
with $\alpha_{aa}$ defined by 
\begin{equation}
\alpha_{aa}(\omega_1, \omega_2) = \frac{d_{pa}^2(2\epsilon_{pa}
+ \omega_1 - \omega_2)}{(\epsilon_{pa} + \omega_1)(\epsilon_{pa} - \omega_2)}
\,.
\end{equation}

The single mode equations in the text are readily derived
by taking $\omega_i = \epsilon_{eg}/2, a_i =1, a_{ij}=1$
and all $\gamma_{\pm}^{(ab)} $ $a,b$-independent.

$\;$ \\ 
\section{Molecules for RENP}\label{App:Molecules-RENP}

Molecules are interesting candidates for RENP measurements
from a variety of reasons:
(1) they have a rich structure in energy levels,
in particular energy level spacing of vibrational
bands may give features of the photon energy spectrum
helpful for identification of the RENP process;
(2) closeness of different parity levels may
enhance parity violating effects.

In this Appendix we explain  basic
facts on molecules and present
crude estimates related to molecular RENP, by taking
I$_2$ molecule as an example.
In the fundamental Born-Oppenheimer approximation
\cite{textbook on molecules}
the molecular wave function of diatomic molecules 
consists of three parts; electronic, vibrational, and rotational
parts. Energy scale associated with each of
these is clearly separated reflecting three different
time scales of their motion.

The vibrational part of wave function is usually derived
after electronic wave functions of energy eigenstates are 
calculated taking the  nuclear distance of two atoms fixed.
For each fixed nuclear distance one has different energy
functions corresponding to different electronic states.
These energy functions are called potential curves.
Potential curves have different equilibrium nuclear distances denoted
by $r_e$
(corresponding to different positions of energy minima) and different curvatures
$\omega_e$ at these minima, as illustrated in the conceptual
diagram in Sec.\ref{Sec:RENP-Theory}.
We need to compute matrix elements of electronic 
operators such as electronic spin $\vec{S}$ and 
electric dipole $\vec{d}$.
Even for two different electronic states
these electronic operators have non-trivial overlap
of vibrational wave functions
because of different equilibrium distances and potential curvatures.
This gives rise to the Franck-Condon (FC) factor, as computed
below.

For simplicity we construct as a model of potential curves
the Morse potential from molecular experimental data 
of Table 2 given 
in Sec.\ref{Sec:RENP-Theory} and calculate 
the FC factors we need for RENP spectrum computations.

\subsection{Morse potential and vibrational energy eigenstates}\label{subsection:Morse potential}
\begin{figure}
\begin{minipage}{7.25cm}
 	\begin{center}
	\includegraphics[width=\textwidth]{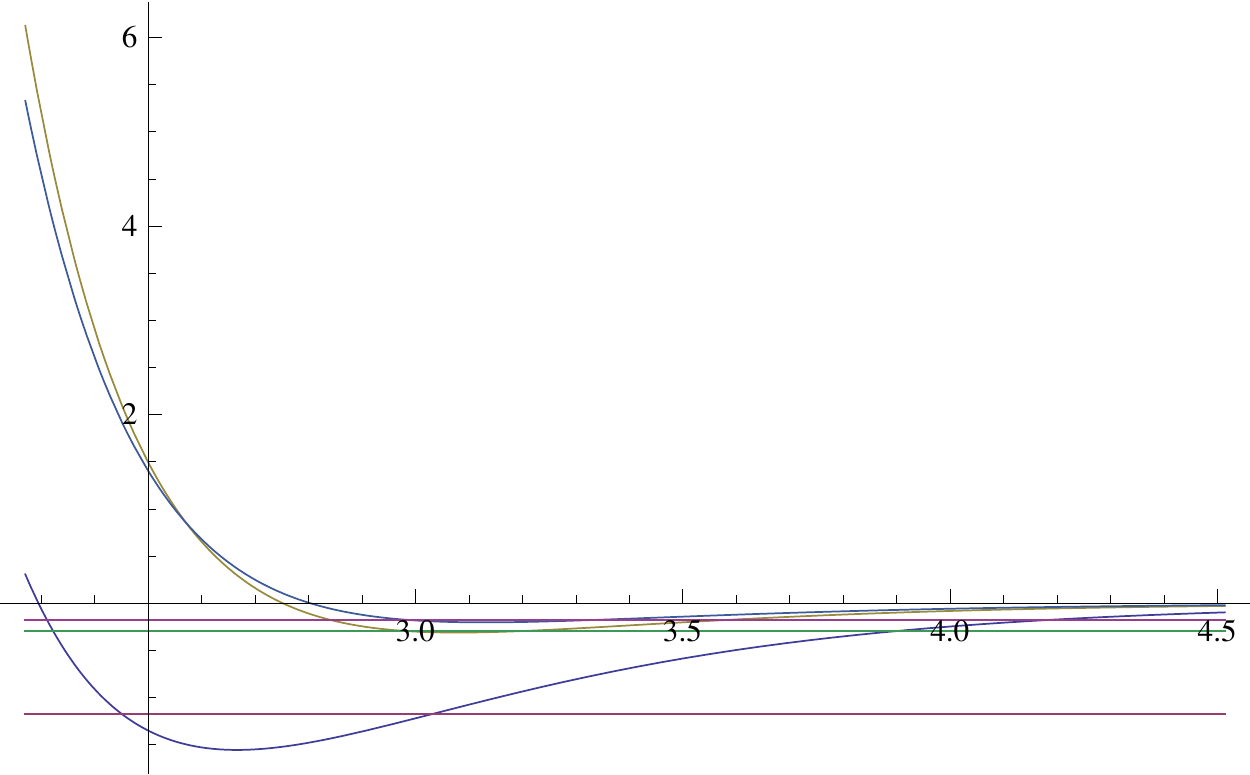}
   	\caption{The Morse potential for states, X(v=15 indicated), A'(v=1) and A(v=2).}
   	\label{morse potential i2}
   	\end{center} 
\end{minipage}
\begin{minipage}{0.5cm}$\;$\end{minipage}
\begin{minipage}{7.25cm}
	\begin{center}
	\includegraphics[width=\textwidth]{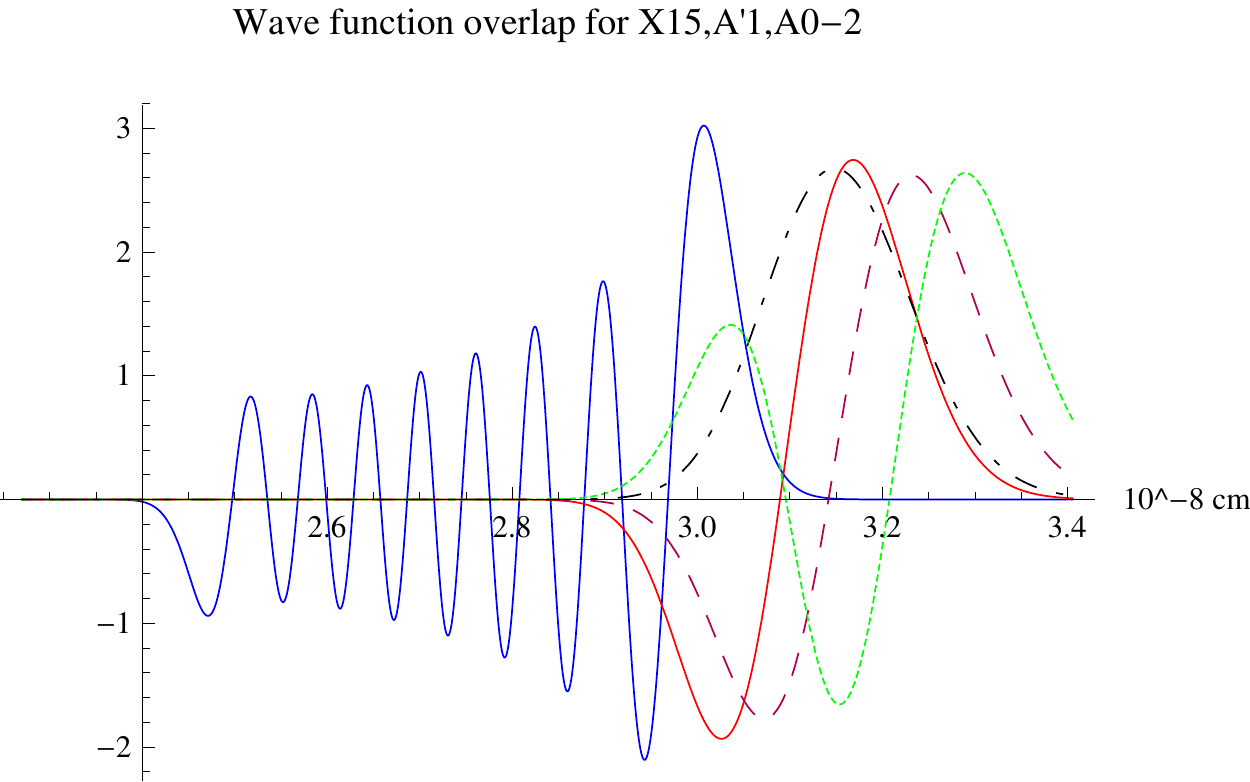} 
   	\caption{Vibrational wave functions for Xv=15, A'v=1 and A$v=0\sim 2$}
        \end{center}
   	\label{vibrational-wfs-ap-ax}
\end{minipage}
\end{figure}

The Morse potential is a three-parameter fit to
the potential curve and is given by
\begin{eqnarray}
&&
V(r) = 
D_e \left( 1 - e^{-a(r-r_e)} \right)^2 - D_e
\,,
\\ &&
a = \sqrt{\frac{m}{2D_e}} \omega_e
\,,
\end{eqnarray}
with $m$ the effective mass of two nuclei,
in this case the half of I atom.
This potential curve is plotted for X, A' and A states,
three lowest electronic states
in Fig.\ref{morse potential i2}, along with three vibrational levels.

This potential has analytic solutions
for the energy eigenvalue problem:
\begin{eqnarray}
&&
E_v = \omega_e (v+\frac{1}{2}) - \frac{\omega_e^2}{4D_e}(v+\frac{1}{2})^2
-D_e
\,, \hspace{0.5cm}
v= 0,1,\cdots [\lambda - \frac{1}{2}]
\,, \hspace{0.5cm}
\lambda = \frac{4 D_e}{\omega_e}
\,.
\end{eqnarray}
The wave functions are given in terms of
the associated Laguerre polynomial $L_v^{\alpha}$. 
Examples of these wave functions are shown in Fig.\ref{vibrational-wfs-ap-ax}.

\subsection{FC factors}

\begin{figure}
 \begin{center}
\includegraphics[width=0.5\textwidth]{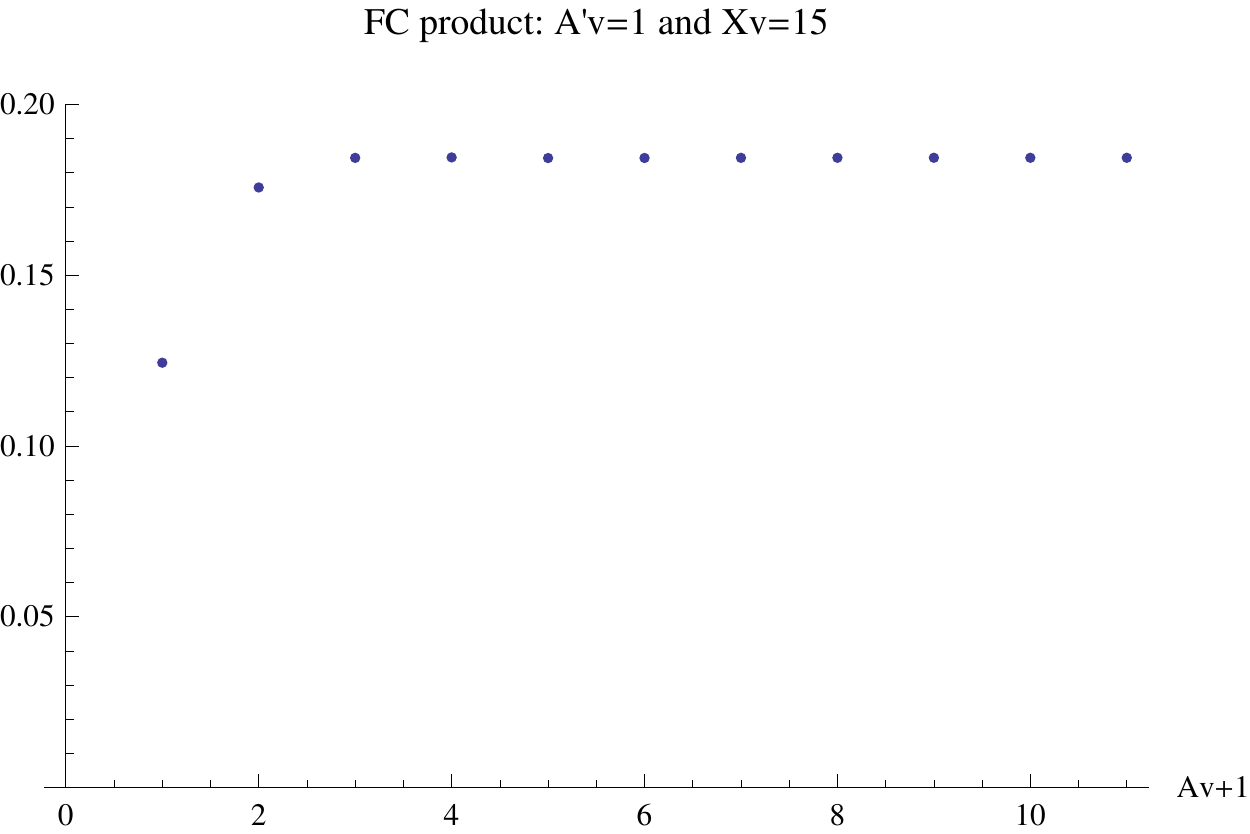}
 \end{center} 
   \caption{FC product between  A' and X:
the intermediate sum is taken up to $Av$
(plotted abscissa is Av+1).
}
   \label{fc product a2x}
\end{figure}

The Franck-Condon (FC) factor is the overlap of vibrational
wave functions, for instance between A'-A, and A-X.
We need a product of FC factors between A'-A and A-X
for RENP amplitude calculation.
Result of FC product between A' and X for summation up to $Av$ 
is shown in Fig.\ref{fc product a2x}, which indicates that 
the approach to the final product value is fast, already seen
$Av \sim 2$.

\subsection{RENP spectrum}

\begin{figure*}
\begin{minipage}{7.25cm}
 	\begin{center}
	\includegraphics[width=\textwidth]{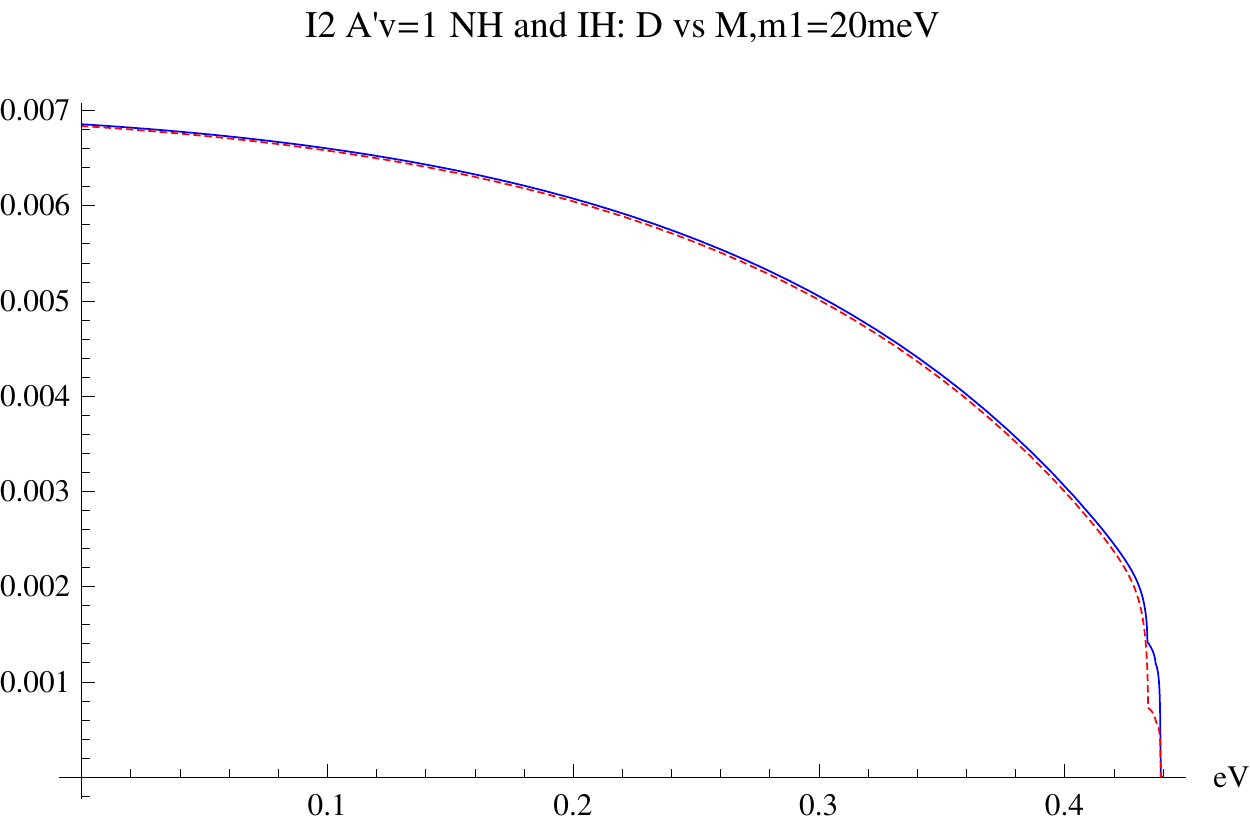}
   	\caption{I$_2$ RENP spectrum between  A' v=1 and Xv=15.
   	IH vs NH is compared, using the smallest neutrino mass 20 meV.
        The vertical scale is in arbitrary units.}
   	\label{i2 renp spectral overall}
        \end{center}  
\end{minipage}
\begin{minipage}{0.5cm}$\;$\end{minipage}
\begin{minipage}{7.25cm}
	\begin{center}
	\includegraphics[width=\textwidth]{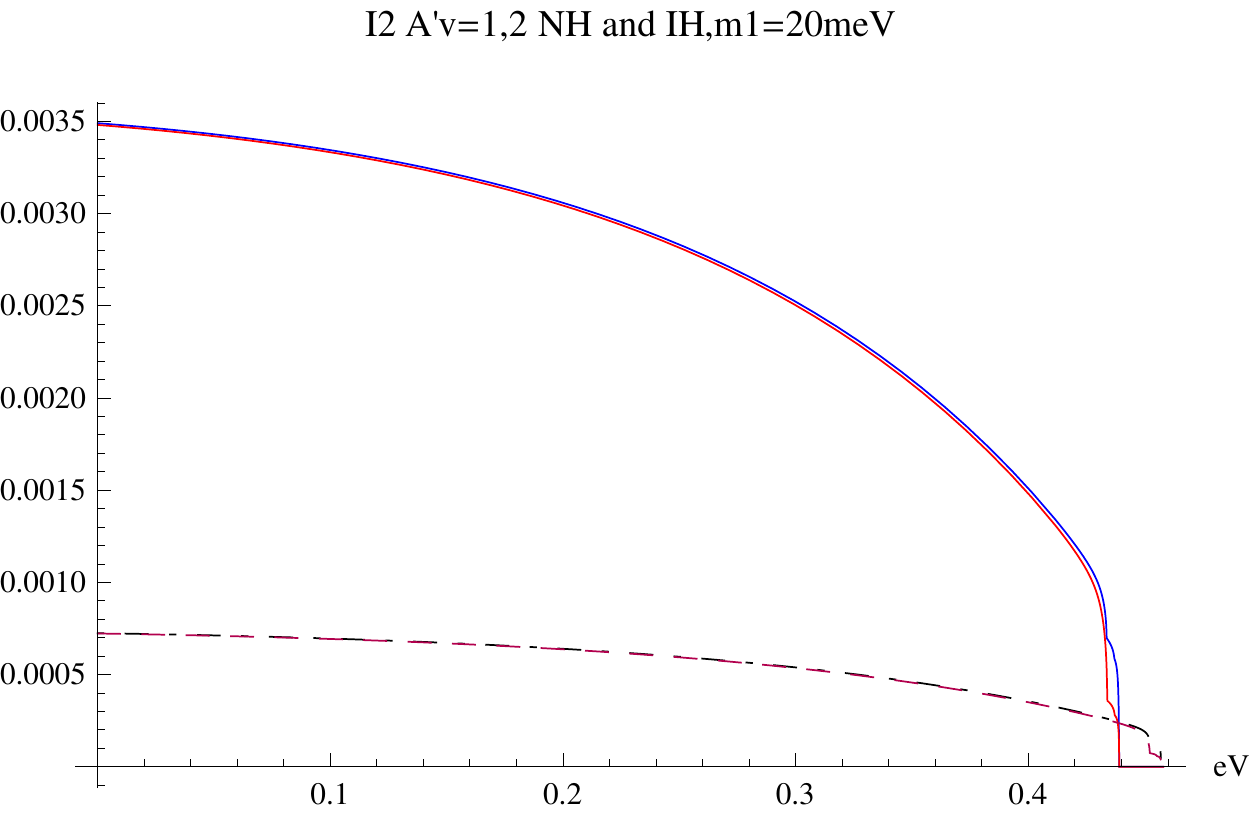}
   	\caption{I$_2$ RENP spectrum between  A'v=1 and Xv=15 in solid,
		and A' v=2 and Xv=14 in dashed.
   		IH vs NH is compared, 
                for two cases of the smallest neutrino mass 2 (in dashed) and 20 (in solid) meV.
                The vertical scale is in arbitrary units. 
   	\label{i2 renp spectral all 115-214}}
        \end{center}
\end{minipage}
\end{figure*}

The RENP spectral shape given in the text is used.
We neglect the spin factor $\vec{S}$ expected to be order unity,
and further ignore the presence of rotational levels.
Examples of the photon spectrum function $I(\omega)$ are shown
in Fig.\ref{i2 renp spectral overall} $\sim$
Fig.\ref{i2 renp spectral all 115-214}.
The molecular spin factor has been taken as a constant
for simplicity.
A more accurate calculation including the correct treatment
of the spin factor is in progress \cite{i2 renp rate}.

\subsection{The I$_{2}$ $A'$ state}
\begin{figure}
\begin{center}
\includegraphics[width=0.5\textwidth]{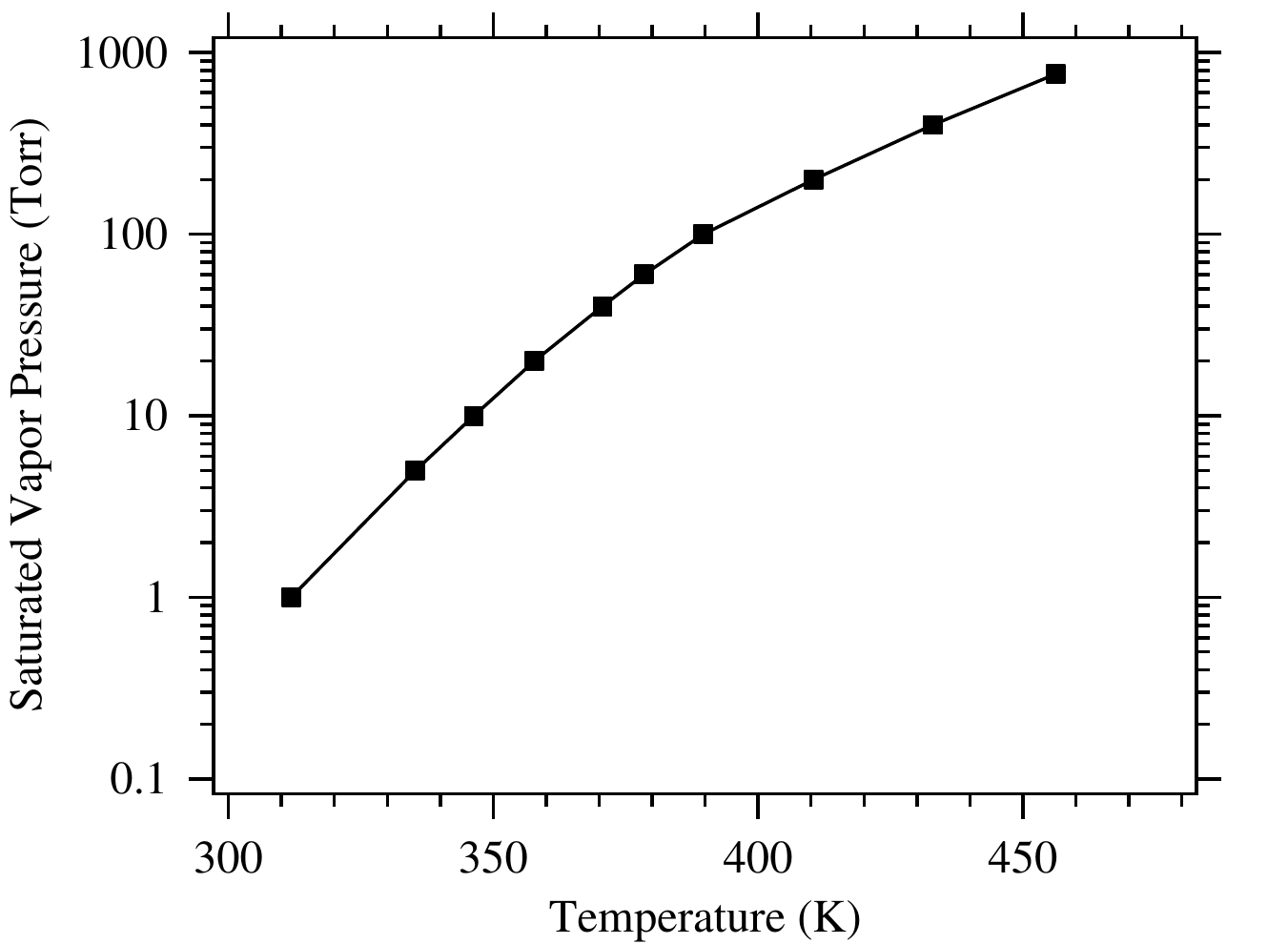}
\end{center}
\caption{Saturated vapor pressure of iodine.\cite{Hydrogen-Iodine}}
\label{hydrogen7}
\end{figure}

In this Subsection, we briefly describe how to produce the metastable $A'$ states.
I$_{2}$ molecules form stable solids at room temperature 
with relatively high vapor pressure ($\sim$ 0.5 Torr). Fig. \ref{hydrogen7} shows saturated vapor pressure of iodine.\cite{Hydrogen-Iodine}
It is possible to prepare 1 atm I$_{2}$ gas by heating up to about 450 K. 
Therefore, I$_{2}$ is an easy-to-handle sample for both gas phase or matrix isolation experiments. 
Actually, many spectroscopic studies of I$_{2}$ has been performed in both phases. 
In gas phase, however, at least three photons are needed to access the $A'$ state from $X$ state. 
For example, $X \rightarrow A \rightarrow a \rightarrow A'$ may be possible by pure optical transition. 
As an alternative pathway, Koffend et al. used collisional 
induced transition to prepare the $A'$ state.\cite{Hydrogen-Iodine2} 
On the other hand, solid environment makes access to the $A'$ state easy. 
It is known that $A'-X$ transition is weakly dipole-allowed by interaction with a matrix. \cite{Hydrogen-Iodine3} 
This may be advantageous for preparing initial population in the $A'$ state while both {\it T$_{1}$} and {\it T$_{2}$} 
in the solid phase are much shorter than in the gas phase.

$\;$ \\ 
\section{Coherence time measurements of para-hydrogen vibrational levels}
\label{App:Experimet-pH2-T2}

\subsection{TRCARS method}
Solid para-hydrogen (pH$_2$) is an attractive target to study the PSR phenomenon.
Its properties of high density ($2.6 \times 10^{22}$ molecules cm$^{-3}$) and long coherence time 
	are well suited to observe explosive type PSR events, as discussed in Sec. 2.
In this Appendix, we focus on the $v=2$ (overtone) vibrational level of pH$_2$, one possible initial state for PSR.
Previous studies have reported that the $v = 1$ (of consisting  H$_2$ molecules) coherence decay exceeds 10 ns \cite{Momose92, Kuroda03, Abram80a, Li98}, that is quite long as a condensed sample.
A direct coherence time measurement is possible by investigating the temporal change in time-domain spectra.
With this method, the coherence decay of the $v=1$ state has been studied already \cite{Li98}.
Below, we describe our time-domain examination of the coherence decay of the overtone $v = 2$ state of solid para-hydrogen.

\begin{figure}
\begin{center}
\includegraphics[width=1\textwidth]{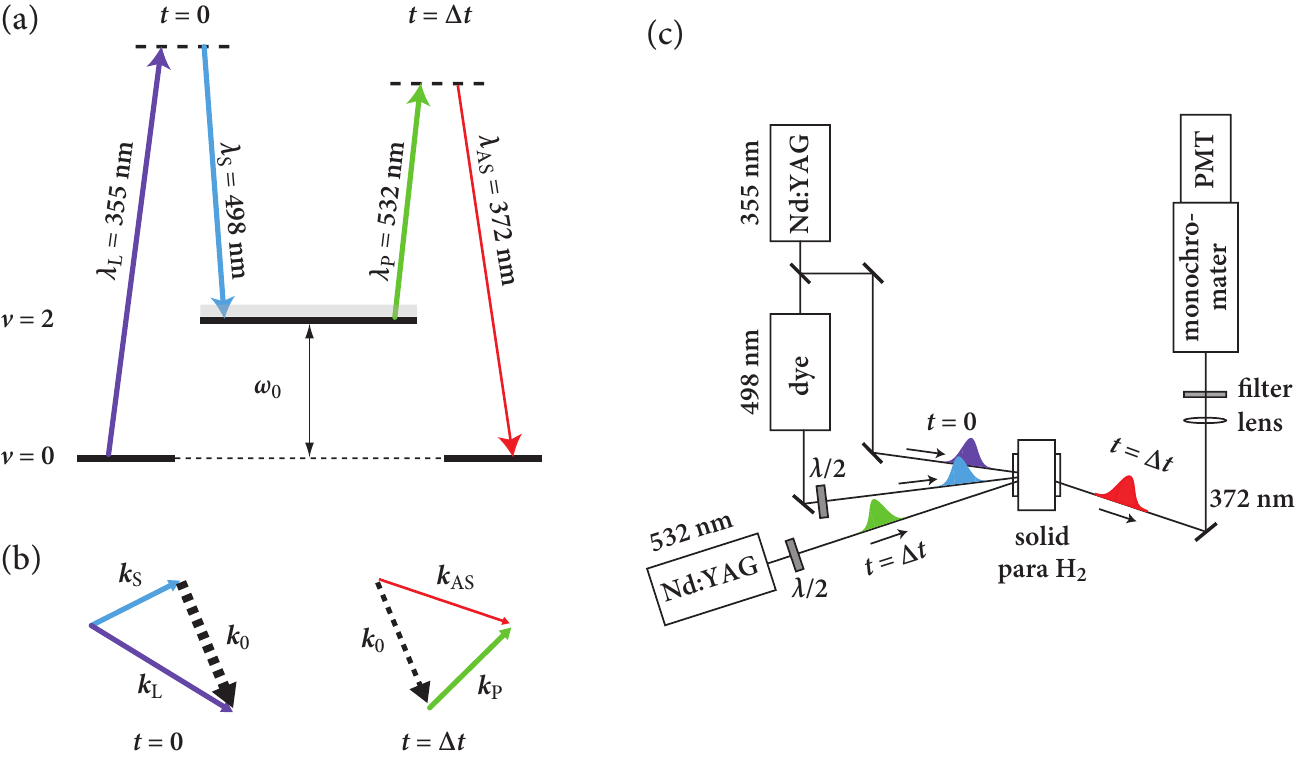}
\end{center}
\caption{
(a) TRCARS scheme for $v = 2$ coherence decay measurement in solid para-hydrogen.
The $v = 2$ coherence produced at $t = 0$ by a pump (denoted by the subscript L) and Stokes (S) pulse pair is monitored by a probe (P) pulse delayed by $t = \Delta t$ generating an anti-Stokes (AS) pulse.
(b) The wavevector relations at $t = 0$ (left) and $t = \Delta t$ (right).
(c) Experimental setup.
Three nano-second pulsed lasers (Nd:YAG (wavelength: 355 nm) as pump, dye (498 nm) as Stokes, and another Nd:YAG (532 nm) as probe were irradiated to a sample of solid para-hydrogen.
The phase-matched anti-Stokes signal (372 nm) was detected by a photomultiplier tube after spectral separation via a UV filter and a monochromator.
}
\label{Kuma_Setup}
\end{figure}

TRCARS (time-resolved coherent anti-Stokes Raman scattering) \cite{Laubereau78} is a technique to trace the coherence decay in a real-time basis for Raman-type transitions.
This is applicable to para-hydrogen vibrational levels since they are Raman-active.
In this technique, the coherence produced by a short excitation pulse pair (pump and Stokes, whose energy difference is equal to the Raman transition frequency $\omega_0$) are monitored by a delayed short probe pulse at $t = \Delta t$ (see Fig.\ \ref{Kuma_Setup} (a)).
At $t = 0$, molecules in the excited sample are in a coherent superposition state between the upper and lower states of the Raman transition.
This superposition state starts to decay after the excitation.
At $t = \Delta t$, the probe pulse and the coherence that remains the sample generate the anti-Stokes pulse along the direction determined by phase-matching condition.
The intensity of this output pulse is proportional to the square of the coherence that remains at the delay  $\Delta t$.
In the case of usual exponential-like decays,
\begin{eqnarray}
I_{AS} (\Delta t) \propto \Biggl| q_0 \cdot \exp \left(  - \frac{\Delta t}{\tau} \right)  \Biggr| ^2, \label{eq:Kuma_decay}
\end{eqnarray}
where $q_0$ is the amplitude of the vibrational coordinate at $t = 0$, and $\tau$ the coherence decay constant.
The energy conservation law (Fig.\ \ref{Kuma_Setup} (a)) and the phase-matching condition (Fig.\ \ref{Kuma_Setup} (b)) are fulfilled in the TRCARS process as follows:
\begin{eqnarray}
\omega_{\mathrm{L}} - \omega_{\mathrm{S}} = \omega_{\mathrm{AS}} - \omega_{\mathrm{P}} = \omega_0, \nonumber\\ 
\mathbf k_{\mathrm{L}} - \mathbf k_{\mathrm{S}} = \mathbf k_{\mathrm{AS}} - \mathbf k_{\mathrm{P}} = \mathbf k_0. \nonumber \label{Kuma_pmc}
\end{eqnarray}
Here $\omega_i$ and $\mathbf k_i$ represent the frequency and the wavevector, related to each other by $|\mathbf k_i | = n(\omega_i) c \omega_i$, where $n(\omega_i)$ is the refractive index of solid para-hydrogen \cite{Perera11} at $\omega_i$.
c the speed of light in vacuum.
The subscripts L, S, AS, and P represent pump, Stokes, anti-Stokes, and probe, respectively.
For $v =2$, $\omega_0$ is 8070.4(1) cm$^{-1}$ in solid hydrogen \cite{Prior72}.
$\mathbf k_0$ is the wavevector generated at $t = 0$ and then probed at $t = \Delta t$. 

The experimental setup used in this study is found elsewhere \cite{Kuma12}.
Briefly, the solid samples were obtained by the method described in Section 4.3.1.
The sample temperature and the ortho hydrogen impurity concentration were controlled for optimization.
For the TRCARS measurement as shown in Fig.\ \ref{Kuma_Setup} (b), three commercially available, nano-second pulsed lasers were used as the pump, Stokes, and probe sources.
After spatial separation from the laser pulses, the anti-stokes signal pulse in the UV region was detected as a function of the delay.

\begin{figure}
\begin{center}
\includegraphics[width=0.5\textwidth]{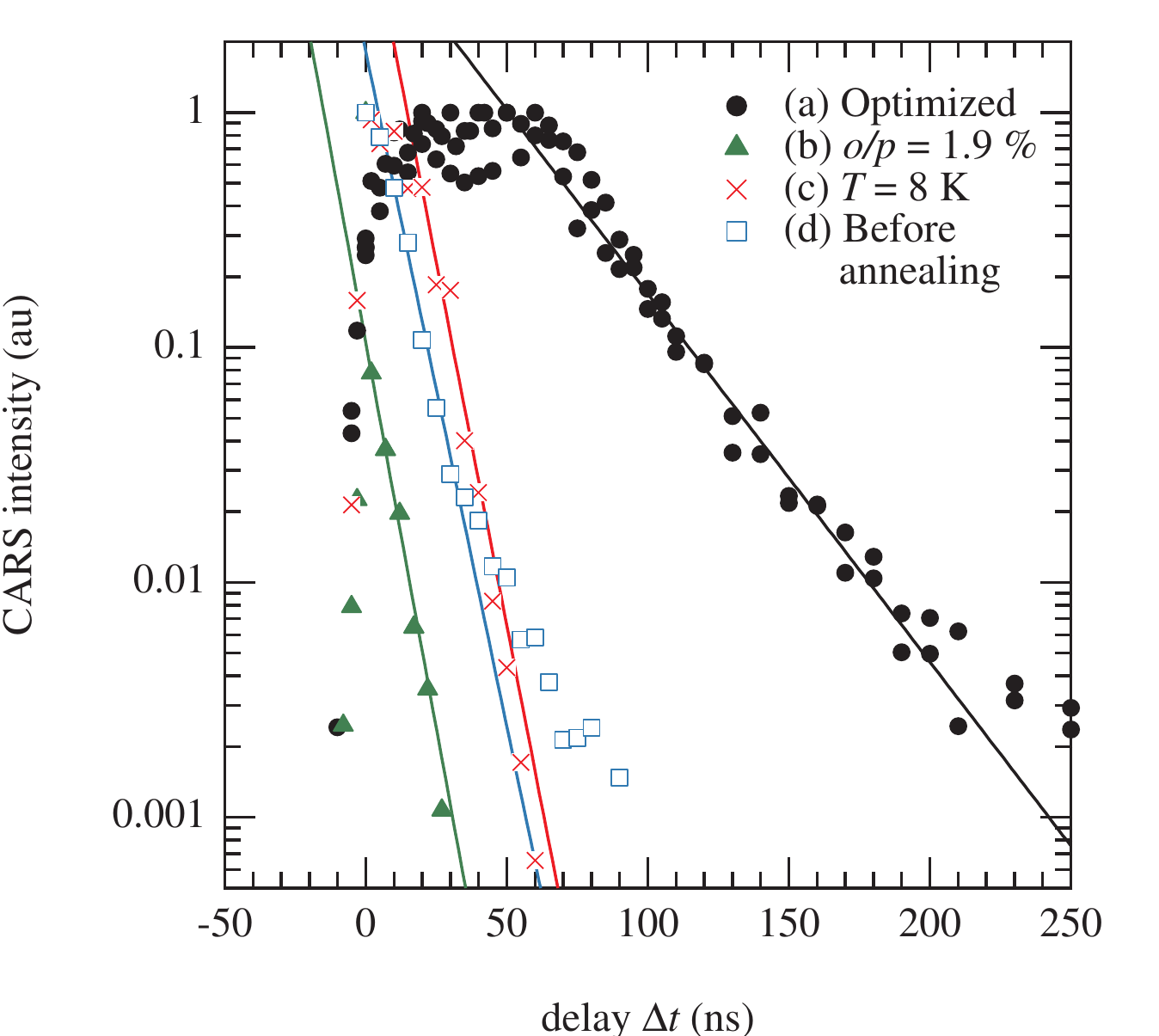}
\end{center}
\caption{
Coherence decay profiles of the vibrational states $v = 2$ of solid para-hydrogen.
(a) The decay measured under the optimized condition (after annealing, at the sample temperature $T = 4$ K, and at the ortho hydrogen concentration $o/p = 0.01 \%$).
The others were obtained under the same condition except for (b) $o/p = 1.9 \%$, (c) $T = 8$ K, and (d) before annealing.
Each profile is normalized by its peak value.
Solid lines are exponential fits to the data, in which the data points corresponding to the normalized intensity below 0.1 were used.
The decay constant $\tau$ of (a) is found to be $\sim 50$ ns.
}
\label{Kuma_v=2}
\end{figure}

\subsection{$v=2$ coherence time of solid para-hydrogen}

Figure \ref{Kuma_v=2} shows the $v = 2$ coherence decay profiles under various conditions.
Before optimization, the decay constants $\tau$ were found to be 10--15 ns (Fig.\ \ref{Kuma_v=2} (b)--(d)).
Ortho hydrogen impurities disturb the translational symmetry of the sample crystal, then an induced inhomogeneous component contributes to the coherence decay \cite{Abram80a} (Fig.\ \ref{Kuma_v=2} (b)).
An additional contribution occurs from elastic scattering of the vibrationally excited molecules by thermal phonons \cite{Kuroda03, Kien03} (Fig.\ \ref{Kuma_v=2} (c)).
This contribution is known to obey a $T^7$ law on the temperature dependence of the decay rates \cite{McCumber63}.
We observed this dependence for the $v=2$ excited state.\cite{Kuma12}
Finally, the intrinsic structural defects induced during sample preparation accelerates the coherence decay (Fig.\ \ref{Kuma_v=2} (d)).

By improving the sample conditions described above, we obtained quite a long decay profile (Fig.\ \ref{Kuma_v=2} (a)).
Here, the ortho concentration was lowered to 0.01 \%, and the sample temperature was cooled down to $T = 4$ K.
To optimize the decay further, an annealing process for half an hour at $T = 9$ K was performed before the measurement.
As seen in Fig.\ \ref{Kuma_v=2} (a), the coherence was kept almost constant up to $\sim 50$ ns just after the excitation, .
This can be accounted for by a transient stimulated Raman process induced on the excitation. \cite{Carman70}
The profile after this plateau has an exponential-like decay.
Fitting this part to Eq.~\ref{eq:Kuma_decay} gives $\tau \sim 50$ ns. 
A deviation from an exponential function at longer delay (also seen in Fig.\ \ref{Kuma_v=2} (d)) is discussed elsewhere.  \cite{Kuma12}
The obtained coherence time for $v = 2$ is comparable to that for $v = 1$.
In the PSR process, a coherent superposition state of the upper and lower states develops in time.
It is required that the disturbance of the coherence is weak during the process.
The long coherence times of $v = 1$ and $v = 2$ show that a nano-second scale measurement can trace PSR signals for solid para-hydrogen when these states are employed.

Generally in solid samples at low temperatures, the dephasing mechanism in the gas phase as Doppler broadening and collisional relaxation is minimized.
Instead, the interatomic/intermolecular distance is close as a few \AA, and the inhomogeneity and the fluctuation of the solid crystal structure causes corresponding variations in the interatomic/intermolecular interaction.
These variations result in fast relaxation usually in a time scale of 10--100 ps.
However, the vibrational states in solid para-hydrogen have such a long coherence time as a few 10 ns, because of a rather weak intermolecular interaction and the high homogeneity due to large zero-point vibration around lattice points.

The coherence time of 50 ns obtained here for the $v=2$ state of solid para-hydrogen satisfies the assumption in a simulative work of PSR. \cite{psr dynamics}
The coherence time can be further improved by a single crystal sample by the pressurized liquid method  \cite{Wallace74}, that was exhibited for $v = 1$.\cite{Hydrogen-press}.
To succeed in PSR observation, the excitation sample density is another important factor.
Additionally, the initial coherence preparation by intense lasers are necessary to observe an explosive PSR.
In the present solid para-hydrogen experiment, the low damage threshold of our sample crystals ($\sim 1 $ MW/cm$^2$) limits the maximum laser intensity.
Crystals with high damage threshold can be obtained by the pressurized liquid method mentioned above.
In order to achieve the initial coherence of $0.1-0.01$ as achieved in the $v= 1$ excitation,\cite{Katsuragawa02} the excitation lasers are required to have a good coherence.
These laser system, as discussed in Sec.\ \ref{Subsubsec:Experimet-pH2-Laser}, is being exploited for PSR in gas para-hydrogen experiment, which is a promising candidate for explosive PSR because the samples are basically damage-free.



$\;$ \\ 
\section{Experimental studies on PSR/RENP targets in condensed phases}
\label{App:Condensed-phases}
%
%
Atomic and molecular systems, examined in our group for PSR/RENP targets other than pH$_{2}$, I$_{2}$, and Xe, are briefly introduced. Condensed matter targets are advantageous in the following three aspects. 1) High density energy storage is possible in a typical number density of 10$^{21}$ cm$^{-3}$ for target atoms and molecules. 2) Coherence control is possible by tuning inter-atomic or inter-molecular interaction to suppress undesired processes for decoherence accompanied by dissipation of energy. 3) Doppler broadening in the spectral linewidth is minimized by keeping the solid sample at low temperatures. Despite a disadvantage for some materials of relatively low damage threshold to laser exposure, solid targets are promising for future experimental research for PSR/RENP. In this Appendix, spectroscopic characterization of condensed-phase atomic and molecular systems are described, i.e., HF molecules in solid pH$_{2}$, atomic N in a carbon cage of C$_{60}$, and Bi atoms in solid Ne. 
%
\subsection{Bismuth in neon matrix}\label{App:Bi-Neon}
\subsubsection{Energy levels of Bi}
\begin{figure}
\begin{center}
\includegraphics[width=0.55\textwidth]{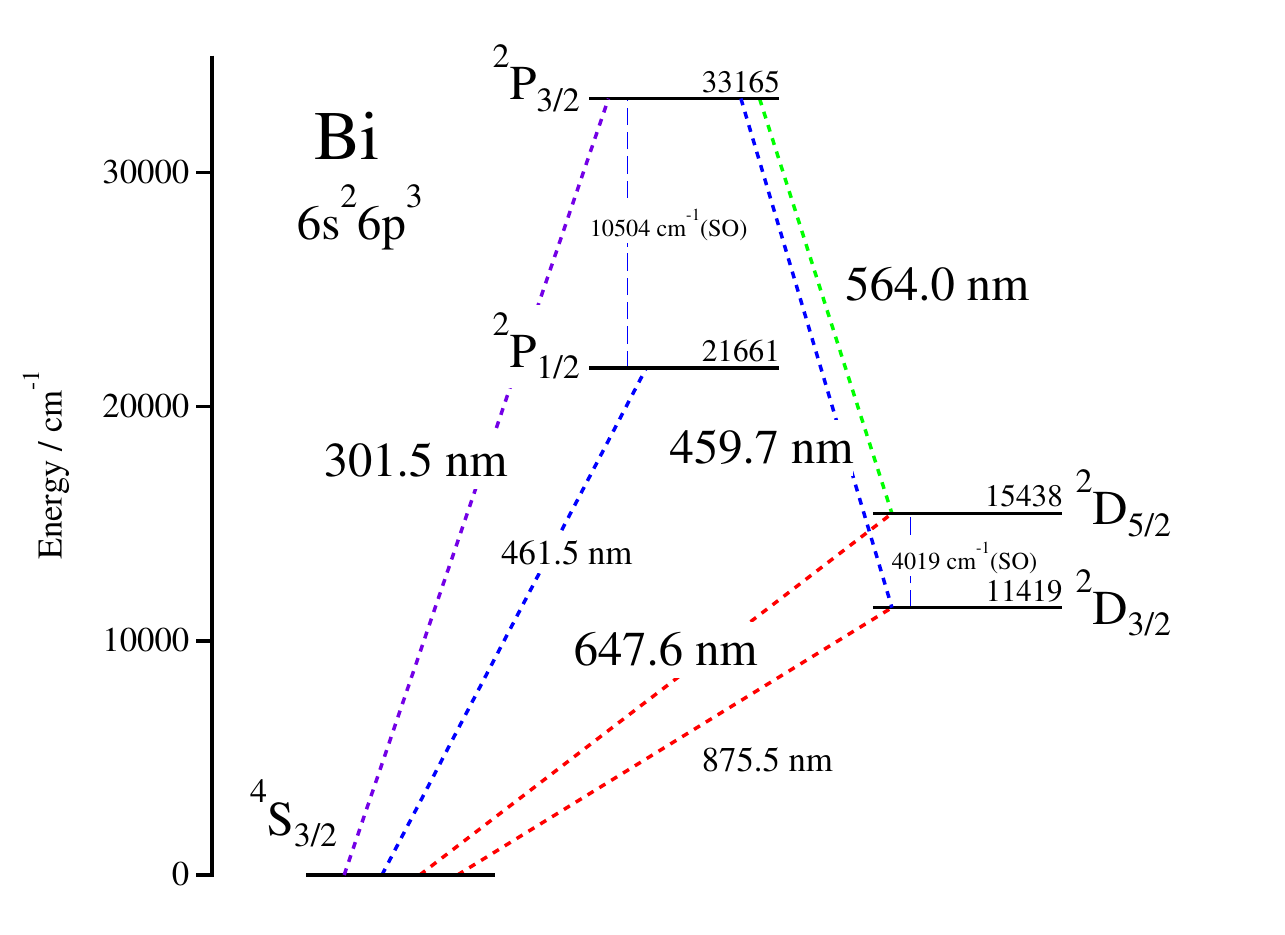}
\end{center}
\caption{Level diagram of bismuth atom (gas phase).}
\label{fBi1}
\end{figure}

As a heavy element with large spin-orbit coupling, bismuth is one of the excitation targets for the observation of radiative emission of neutrino pair (RENP). Figure \ref{fBi1} shows the energy levels for Bi atom. Similarly to the other group-15 elements, the low-lying excited states, $^{2}{\rm P}_{3/2,1/2}$ and $^{2}{\rm P}_{5/2,3/2}$, are stemming from the same electron configuration, ${\rm 6s}^{2}{\rm 6p}^{3}$, as that for the ground state, $^{4}{\rm S}_{3/2}$. Due to the large spin-orbit coupling, however, unlike the lightest group-15 element of nitrogen, different $J$ states for bismuth are largely separated, i.e., 10504 cm$^{-1}$ for $^{2}{\rm P}$ ($J$=3/2 and 1/2) and 4019 cm$^{-1}$ for $^{2}{\rm D}$ ($J$=5/2 and 3/2). As a result of the mixing of different spin states, orbital angular momentum, $L$, and spin angular momentum, $S$, are no longer good quantum numbers. As a result of the mixing of different spin states by the spin-orbit interaction, transitions become possible for a pair of states with the same electron configuration. The radiative lifetimes for these transitions, ${\sim}{\rm 10}^{-2}{-}{\rm 10}^{-1}$ seconds in the gas phase, are long enough to be detected by using an instantaneous excitation with a nanosecond laser pulse. These low-lying excited states were observed in solid Ne matrix \cite{BondybeyCPL1980}. Using higher excited levels accessible by non-resonant two-photon excitation, single-photon super-radiance was observed with Bi atoms in the gas phase \cite{CremerApplPhysB1984}.

\subsubsection{PSR/RENP transition}
In Fig. \ref{fBi1}, two energy levels are noted. One is a metastable $^{2}{\rm D}_{5/2}$ state (${\rm 6p}^{3}$) and the other is the ground $^{4}{\rm S}_{3/2}$ state (${\rm 6p}^{3}$). These two states can be connected via intermediate states at higher energies, e.g., $^{4}{\rm P}_{5/2}$ (${\rm 6p}^{3}{\rm 7s}^{1}$), with a combination of E1 and M1 (practically E1) transitions. The set of these energy levels is an example for a ${\lambda}$-type ladder for paired super-radiance (PSR) using bismuth atoms. If the coherence is developed between the ground state and the metastable excited state, and if an appropriate trigger source is applied to the system, explosive two-photon emission of radiation by PSR or even RENP can be promoted. A PSR rate of ${\sim}$17 kHz was calculated for the number density ${\rm 10}^{16}$ cm$^{-3}$ of excited bismuth atoms.

\subsubsection{Matrix isolation technique}
In order to realize PSR or RENP, to produce a system containing a large number of excited atoms in a small volume is crucial. The use of a solid-state material is demanded for realization of a large atomic density, thus for prolonged observation in the detection of events. Matrix isolation technique has been a spectroscopic tool applicable to reactive species such as atoms, molecules, open-shell radicals, and clusters of metal atoms. The target species are entrapped in small vacancies in rare-gas solids of Ne, Ar, Kr, or Xe. The number density of target atoms can be increased up to ${\rm 10}^{20}$ cm$^{-3}$ under well-isolated conditions. For matrix-isolated species, relaxation pathways to the long-lived metastable state can be enhanced, by which the excited-state population is increased to a substantial fraction or even inverted against the ground state, as recently demonstrated for Yb atoms in solid Ne \cite{XuPRL2011}.
\begin{figure}
\begin{center}
\includegraphics[width=0.9\textwidth]{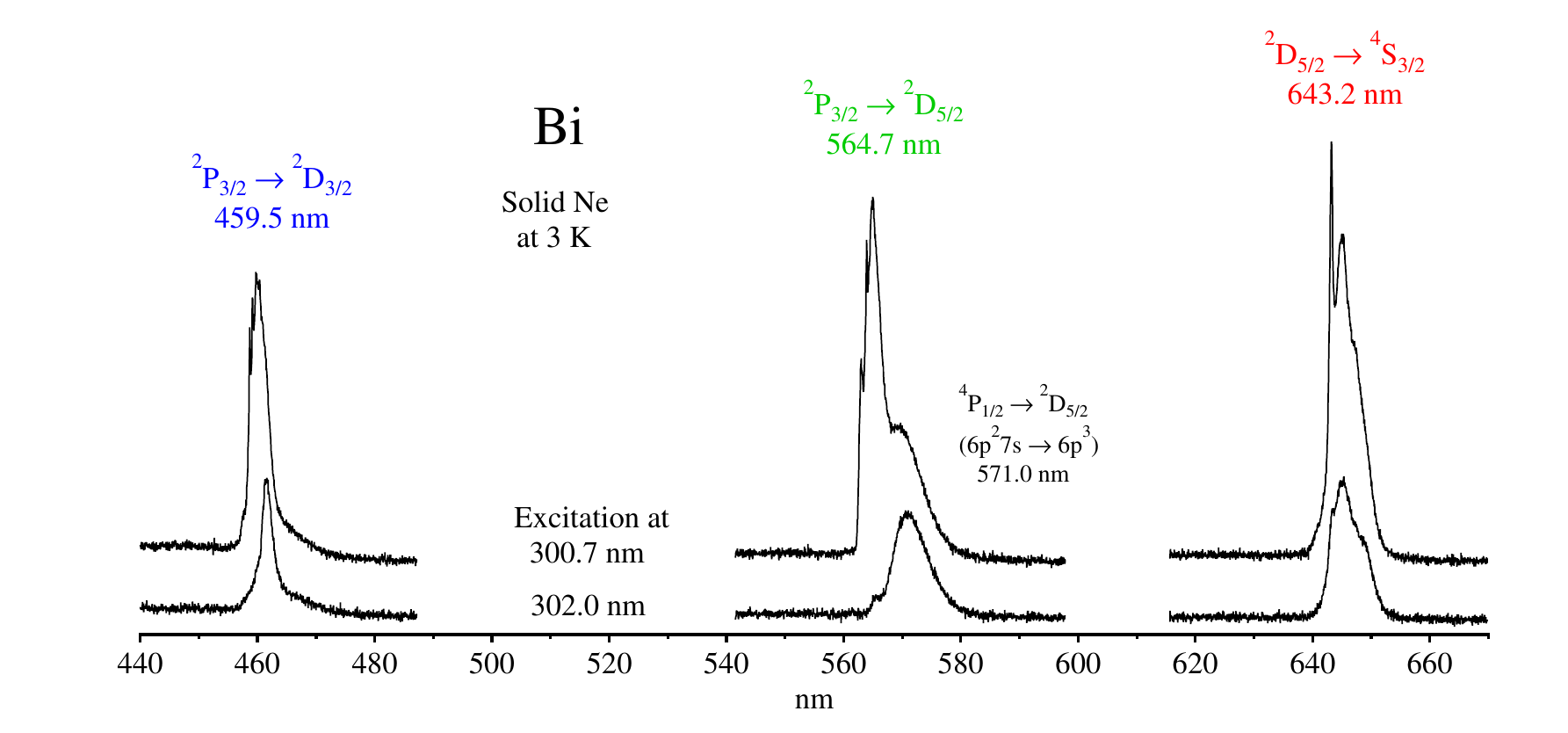}
\end{center}
\caption{Emission spectra of Bi atoms trapped in a solid Ne matrix at 3 K using the excitation with nanosecond laser pulses at 300.7 nm (upper trace) and 302.0 nm (lower trace) for the $^{2}{\rm P}_{3/2}{\leftarrow}^{4}{\rm S}_{3/2}$ transition.}
\label{fBi2}
\end{figure}
%

\subsubsection{Laser induced optical emission}
We performed leaser-induced optical emission spectroscopy to determine energy levels and lifetimes for Bi atoms in solid Ne matrix. Vapor of bismuth emanated from a bismuth-containing molybdenum crucible at ${\sim}$1000 K was co-deposited on a cold surface of sapphire at 3 K together with an excess of neon gas. After the deposition, nanosecond laser pulses tuned at transition wavelengths for Bi atom were irradiated on the solid sample in a grazing angle. The emitted light was dispersed by using a grating spectrometer (Acton SP300i) and detected by using a CCD camera (PI SPEC10) for the observation of emission spectra. A photomultiplier (Hamamatsu R928) was used for the measurement of lifetimes. In order to reduce stray light of the excitation laser, a long-pass optical filter (Schott Glass Filter) was used.

First we excited the $^{2}{\rm P}_{3/2}{\leftarrow}^{4}{\rm S}_{3/2}$ transition. Figure \ref{fBi2} shows typical emission spectra for the excitation at 300.7 nm. Emission bands corresponding to three relaxation pathways were observed. Two of them were the transitions from the upper state of $^{2}{\rm P}_{3/2}$ to the metastable states of $^{2}{\rm D}_{5/2}$ (565 nm) and $^{2}{\rm D}_{3/2}$ (460 nm). The other one was the transition from the metastable state of $^{2}{\rm D}_{5/2}$ (645 nm) to the ground state of $^{4}{\rm S}_{3/2}$. Each band was composed of a few lines, for which the relative intensity varied upon different excitation wavelengths. At the excitation wavelength of 300.7 nm providing maximum emission intensity, relatively sharp emission lines were intensified at the blue edge of the emission band. A broad emission band at 570 nm in the lower trace in Fig. \ref{fBi2} may be associated with another transition of $^{4}{\rm P}_{1/2}{\rightarrow}^{2}{\rm D}_{5/2}$ (${\rm 6p}^{2}{\rm 7s}^{1}{\rightarrow}{\rm 6p}^{3}$), whose upper sate of $^{4}{\rm P}_{1/2}$ locates slightly below the $^{2}{\rm P}_{3/2}$ (${\rm 6p}^{3}$) level. The lifetime for this transition at 570 nm was several times shorter than that for the transition at 565 nm.
\begin{figure}
\begin{center}
\includegraphics[width=0.5\textwidth]{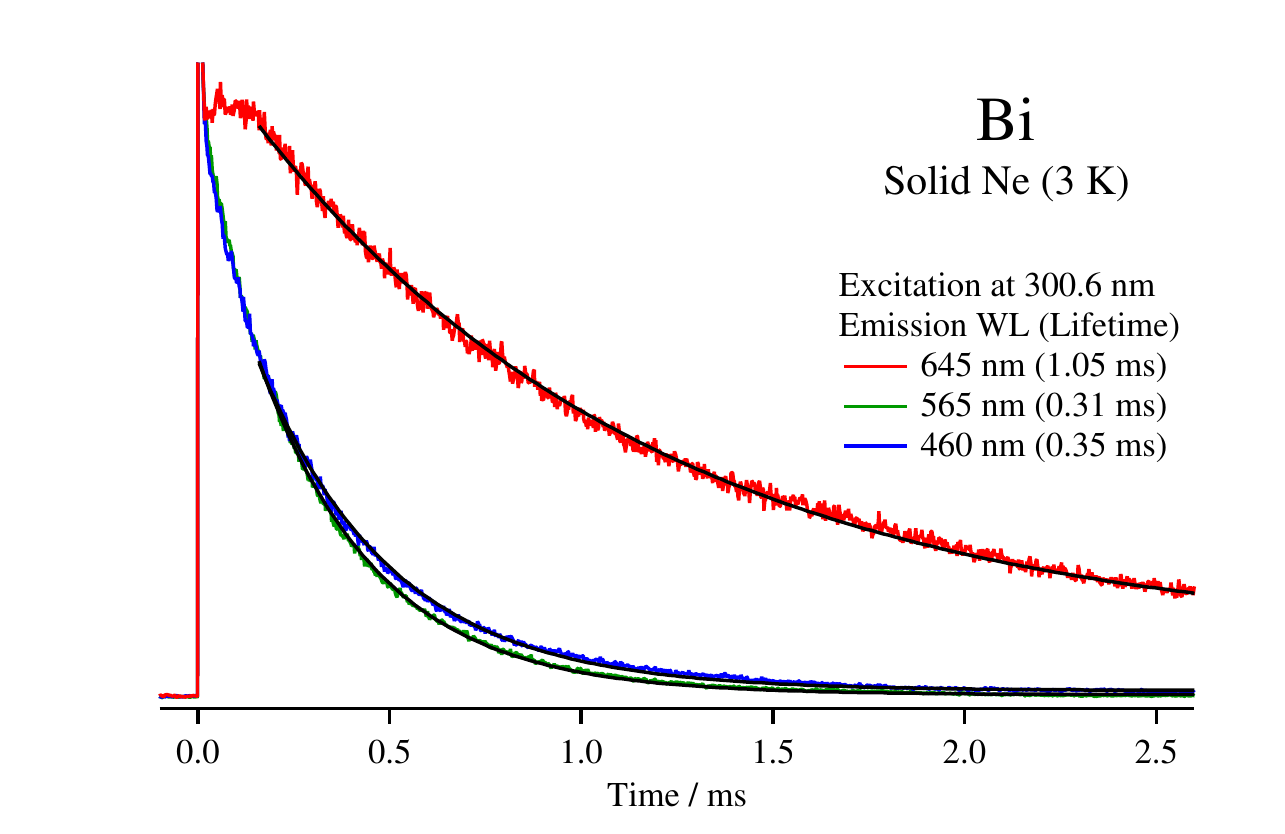}
\end{center}
\caption{Decay of emission bands for Bi in solid Ne with the excitation at 300.6 nm, $^{2}{\rm D}_{5/2}{\rightarrow}^{4}{\rm S}_{3/2}$ (645 nm), $^{2}{\rm P}_{3/2}{\rightarrow}^{2}{\rm D}_{5/2}$ (565 nm), and $^{2}{\rm P}_{3/2}{\rightarrow}^{2}{\rm D}_{3/2}$ (460 nm).}
\label{fBi3}
\end{figure}

Figure \ref{fBi3} depicts decay profiles for the three transitions for Bi atom in solid Ne. The single exponential fits provide lifetimes of 1.05 ms for $^{2}{\rm D}_{5/2}{\rightarrow}^{4}{\rm S}_{3/2}$, 0.31 ms for $^{2}{\rm P}_{3/2}{\rightarrow}^{2}{\rm D}_{5/2}$, and 0.35 ms for $^{2}{\rm P}_{3/2}{\rightarrow}^{2}{\rm D}_{3/2}$. These lifetimes in a Ne matrix are two orders of magnitude shorter than those in the gas phase. The latter two, corresponding to relaxation to the metastable $^{2}{\rm D}_{5/2,3/2}$ states, showed a single exponential decay with a characteristic lifetime of ${\sim}$0.3 ms, while the former one, form the metastable state to the ground state, showed a characteristic feature of rise in ${\sim}$0.2 ms and a decay in a longer lifetime of ${\sim}$1 ms. These profiles are consistent with a picture that the population in the metastable state, $^{2}{\rm D}_{5/2}$, increases via the radiative decay of $^{2}{\rm P}_{3/2}{\rightarrow}^{2}{\rm D}_{5/2}$ in a time scale of ${\sim}$0.3 ms and decreases via that of $^{2}{\rm D}_{5/2}{\rightarrow}^{4}{\rm S}_{3/2}$ in ${\sim}$1 ms.
\begin{figure}
\begin{center}
\includegraphics[width=0.36\textwidth]{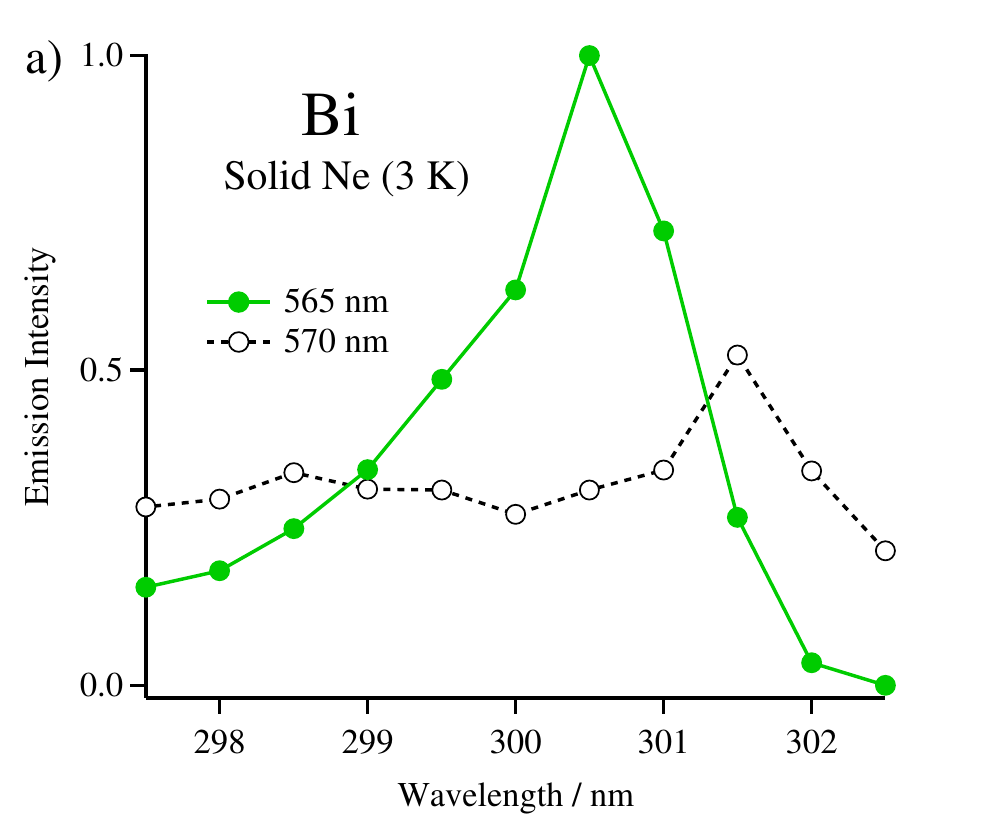}
\includegraphics[width=0.54\textwidth]{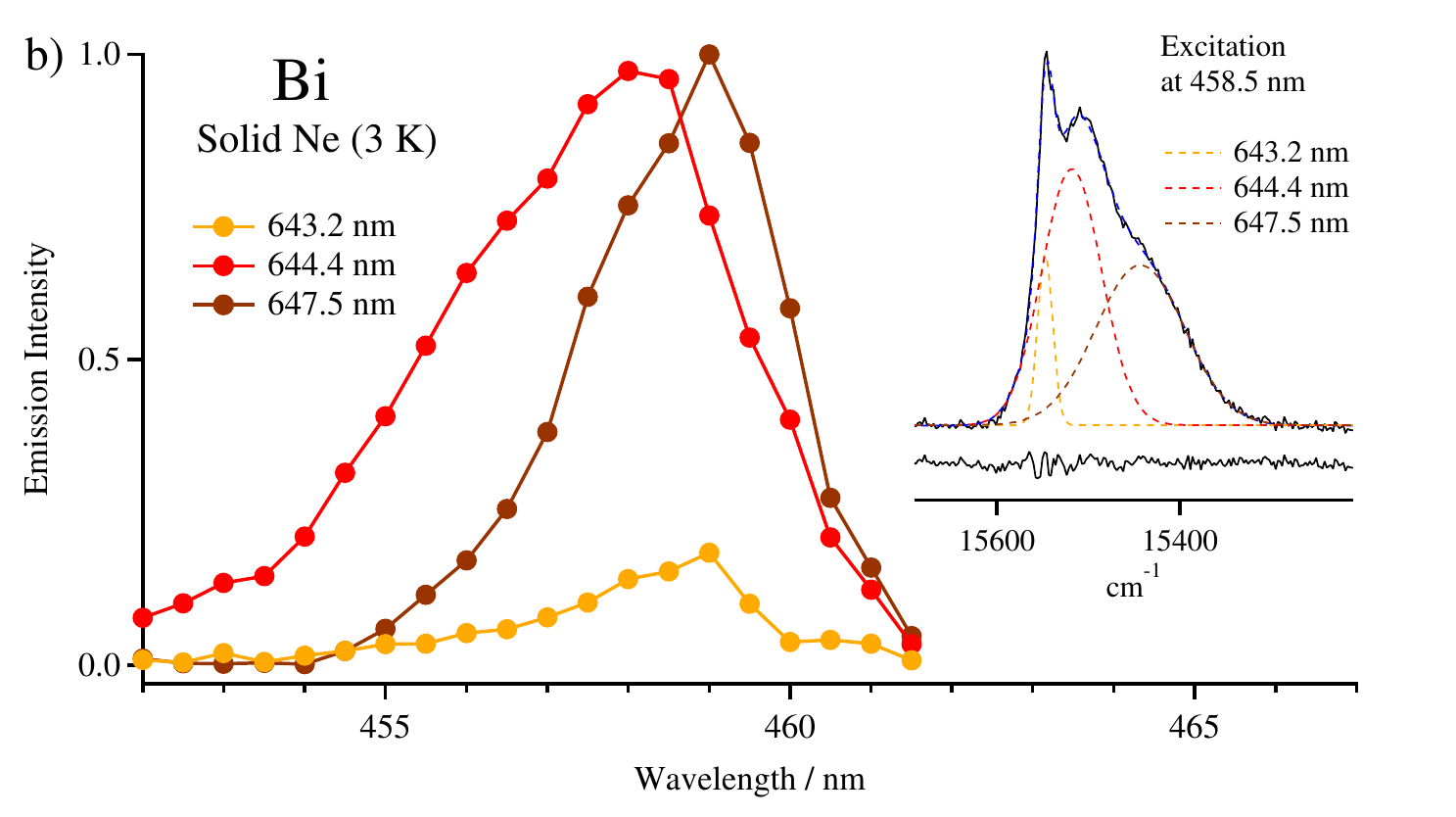}
\end{center}
\caption{Excitation profiles for Bi in solid Ne, a) $^{2}{\rm P}_{3/2}{\leftarrow}^{4}{\rm S}_{3/2}$ (${\sim}$300 nm) and b) $^{2}{\rm P}_{1/2}{\leftarrow}^{4}{\rm S}_{3/2}$ (${\sim}$460 nm). Inset shows a typical emission spectrum with fitting curves by three components.}
\label{fBi4}
\end{figure}

In order to locate the energy levels of $^{2}{\rm P}_{3/2}$ and $^{2}{\rm P}_{1/2}$ states, excitation profiles were examined. By using the photomultiplier detector equipped to the spectrometer, emission intensity at a fixed wavelength was measured and plotted as a function of laser-excitation wavelength. Figure \ref{fBi4} shows the excitation profiles for Bi atoms in solid Ne. 
Corresponding to the absorption for $^{2}{\rm P}_{3/2}{\leftarrow}^{4}{\rm S}_{3/2}$, the maximum was found at 300.5${\pm}$0.25 nm by plotting the emission intensity at 565 nm as shown by closed circles in green in Fig. \ref{fBi4}(a). The bandwidth for this transition was ${\sim}$2 nm (${\sim}$220 cm$^{-1}$). 
The excitation profile for the 570-nm emission band (see the lower trace in Fig. \ref{fBi2}) was much broader as shown by open circles in Fig. \ref{fBi4}(a). 
For $^{2}{\rm P}_{1/2}{\leftarrow}^{4}{\rm S}_{3/2}$, 
the maximum was found at 458.5${\pm}$0.25 nm by plotting the emission intensity at 645 nm as shown by the three profiles in Fig. \ref{fBi4}b, for which spectral decomposition was performed for each of the observed emission spectra at different excitation wavelengths and the excitation profile was plotted for each of the three components in the emission spectrum. The bandwidth for the $^{2}{\rm P}_{1/2}{\leftarrow}^{4}{\rm S}_{3/2}$ transition was ${\sim}$4 nm (${\sim}$190 cm$^{-1}$).

\subsubsection{Linewidth and coherence}
At the blue edge for each of the emission bands in Fig. \ref{fBi2}, a relatively sharp peak is discernible. This peak corresponds to the zero-phonon line for each transition. The linewidth for these relatively sharp lines was ${\sim}$10 cm$^{-1}$, while the total width for the bands was ${\sim}$200 cm$^{-1}$. Even for the narrow lines, the linewidth corresponds to a natural lifetime on the order of ${\sim}{\rm 10}^{-12}$ seconds. 
Since the observed lifetimes in Fig. \ref{fBi3} is much longer, ${\sim}$1 ms, the observed spectral linewidth is ascribed to inhomogeneous broadening due to interaction of the target atoms with the matrix media. With substantial deviations in the transition wavelength, it is considered to be difficult to develop coherence longer than a picosecond in this system.
\begin{table}
\caption{Observed transitions for Bi atoms in solid Ne matrix and comparison with the transition energy in the gas phase. $^{a}$Radiative lifetimes in solid Ne with an excitation at 300.6 nm. $^{b}$Difference of term energies.}
\label{tBi1}
\begin{center}
\begin{tabular}{lllllllll}
\hline
\multicolumn{1}{c}{Transition} && \multicolumn{1}{c}{Wavelength/nm} & \multicolumn{1}{c}{Lifetime/ms$^{a}$} & \multicolumn{1}{c}{Energy/${\rm cm}^{-1}$} & \multicolumn{1}{c}{Energy(gas)/${\rm cm}^{-1}$} \\
\hline
\ $^{2}{\rm P}_{3/2}{\leftarrow}^{4}{\rm S}_{3/2}$ && 300.5 & - & 33280 & 33165 \\
\ $^{2}{\rm P}_{1/2}{\leftarrow}^{4}{\rm S}_{3/2}$ && 458.5 & - & 21810 & 21661 \\
\ $^{2}{\rm P}_{3/2}{\rightarrow}^{2}{\rm D}_{3/2}$ && 459.5 & 1.05 & 21760 & 21746$^{b}$ \\
\ $^{2}{\rm P}_{3/2}{\rightarrow}^{2}{\rm D}_{5/2}$ && 564.7 & 0.31 & 17710 & 17727$^{b}$ \\
\ $^{2}{\rm D}_{5/2}{\rightarrow}^{4}{\rm S}_{3/2}$ && 643.2 & 0.35 & 15550 & 15438 \\
\hline
\end{tabular}
\end{center}
\end{table}

The laser-induced emission spectroscopy revealed energy levels for all the ${\rm 6p}^{3}$ states of Bi atom in solid Ne. The observed transitions in this work are summarized in Table \ref{tBi1}. Term energies for the four low-lying excited states of Bi atom in solid Ne remain intact as those in the gas phase within a bandwidth in the matrix spectra. Lifetimes were revealed to be two orders of magnitude shorter than those in the gas phase. Despite the advantage of the long lifetimes of ${\sim}$1 ms, the apparent inhomogeneous broadening would suppress coherence time shorter than expected. To remove undesired interaction between embedded guest atoms and hosting matrix atoms is crucial for realizing a system for PSR or RENP.

\subsection{HF molecule trapped in solid pH$_{2}$}\label{App:HF-molecule}
\subsubsection{Matrix isolation spectroscopy using solid pH$_{2}$}
Spectroscopy of molecules embedded in condensed phases is called matrix isolation spectroscopy, which has long been used to pilot gas phase spectroscopy. Not only unstable but also stable molecules have been subjected to studies by this method. Rare gas matrices such as solid Ne and solid Ar have been widely used because of their chemical inertness and of relatively weak perturbative interactions. The perturbation in solid rare-gas matrices sometimes makes spectral linewidths broader compared to those in the gas phase spectroscopy, to conceal detailed spectroscopic information. Small molecules in solid pH$_{2}$, on the other hand, exhibit extremely sharp lines to reflect quantized rovibrational states of the entrapped molecules in the condensed phase. The full width at half maximum (FWHM) for the vibrational $\nu_{4}$ band of methane, CH$_{4}$, in solid pH$_{2}$ is as narrow as 0.015 cm$^{-1}$, whose signal is more than one order of magnitude sharper than that observed in conventional solid rare-gas matrices \cite{Hydrogen-CH4}. Such a narrow linewidth of $\sim$0.01 cm$^{-1}$ is a promise of a long relaxation time advantageous for the observation of coherent phenomena.

\subsubsection{Preparation of chemically doped solid pH$_{2}$}
To prepare the doped solid pH$_{2}$, the closed cell method (Sec. 4.2.1) can be used. 
Premixed pH$_{2}$ gas containing small amount of the target molecules is introduced into the cooled closed cell and a transparent doped crystal grows in the same manner as pure pH$_{2}$. 
However, only dopants of small intermolecular interaction like methane are isolated by this method while molecules with strong interaction like HF are difficult to be doped dispersedly. 
In contrast, the rapid deposition method developed by Fajardo et al \cite{Hydrogen-deposition} can isolate a number of molecular species in solid pH$_{2}$. 
They include CH$_{4}$, CH$_{3}$F, CH$_{3}$I, CH$_{2}$ClI, CO, CO$_{2}$, NO, N$_{2}$O, H$_{2}$O, HF, and HCN. 
Several millimeters thick, optically transparent solid pH$_{2}$ can be prepared by this method. 
The pH$_2$ and dopant molecules are co-deposited onto a cold substrate (typically BaF$_2$ at 2$\sim$4 K) placed in a cryostat. 
The substrate is contacted on the copper block with an indium gasket. 
The crystal grows perpendicularly from the substrate and a polycrystalline aggregate of h.c.p. and f.c.c. having their c axes roughly normal to the substrate is produced. 
To remove the metastable f.c.c. crystal, the mixed crystals are heated up to $\sim$ 5 K for 10 to 30 min. 
This annealing aligns c axes normal to the substrate.

\subsubsection{Infrared spectroscopy of HF in solid pH$_{2}$}
Hydrogen fluoride, HF, in solid pH$_{2}$ has been studied in our laboratory as a target for super-radiance in the condensed phase \cite{Hydrogen-HF}. Super-radiance was observed indeed in the gas phase for the rotational transition of HF, where gaseous HF was optically pumped into the rovibrationally excited state ({\it v}=1, {\it J}) for realizing total population inversion against the lower state ({\it v} =1, {\it J-}1), resulting in the super-radiance with far infrared (FIR) emission of photons \cite{first sr}.
\begin{figure}
\begin{center}
\includegraphics[width=0.4\textwidth]{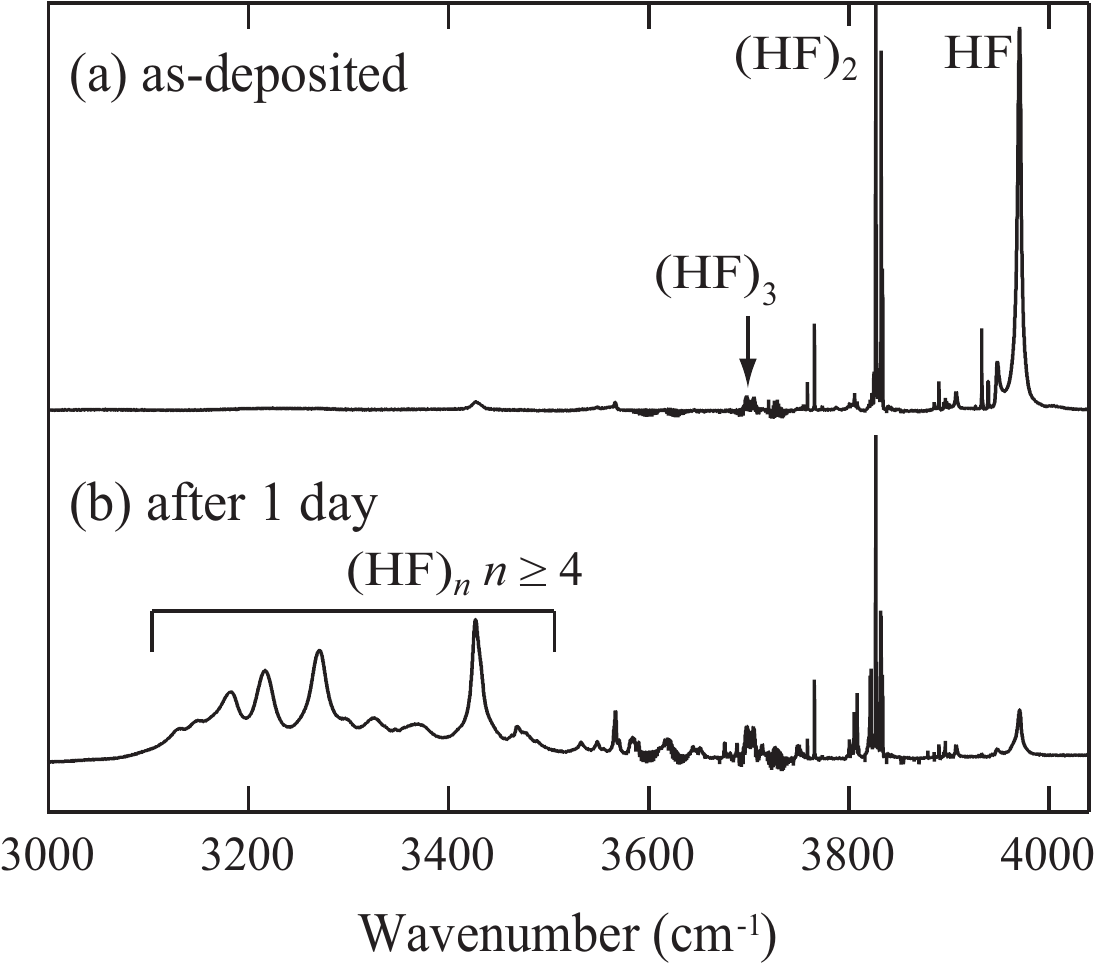}
\end{center}
\caption{IR absorption spectra of HF molecules in solid pH$_{2}$ for the sample a) as prepared and b) 1-day after preparation.}
\label{hydrogen6}
\end{figure}

We measured IR absorption spectra for HF molecules in solid pH$_{2}$ at 3.6 K by using an FTIR spectrometer. 
The observed spectra for HF/pH$_{2}$ are shown in Fig. \ref{hydrogen6}. 
Based on the systematic spectral change caused by the difference in HF/pH$_{2}$ concentration (not shown), we assigned the observed IR absorption lines in the region of 4000-3000 cm$^{-1}$ to isolated molecular HF, dimers (HF)$_{2}$, trimers (HF)$_{3}$, and clusters (HF)$_{n}$ (${n}{\geq}$4) as indicated in Fig. \ref{hydrogen6}.

Time evolution in the IR absorption spectra in Fig. \ref{hydrogen6}(a) and \ref{hydrogen6}(b) indicated that the HF molecules could migrate in the solid pH$_{2}$ at 3.6 K, as was noted by the decrease in the line at 3970 cm$^{-1}$ for isolated HF and the concomitant increase in the bands at lower frequencies for clusters (HF)$_{n}$. The aggregation of HF molecules in solid pH$_{2}$ indicates that the isolated HF molecules cannot be the targets for long-term observation. Moreover, the linewidth for the isolated HF in solid pH$_{2}$ is rather broad, ${\sim}$4 cm$^{-1}$, probably due to matching between rotational energy for HF and phonon energy for solid pH$_{2}$. Note that the linewidth for the signal at 3820 cm$^{-1}$ for the dimer, (HF)$_{2}$, is much narrower than that for the isolated HF, probably because the rotational motion is quenched. Solid pH$_{2}$ is, in general, a good matrix medium for trapping molecules with minimal interaction, whereas, for the small but polar molecule such as HF, another difficulty may arise in migration or rotation-phonon coupling.

\subsection{Nitrogen atom in fullerene C$_{60}$}\label{App:N-C60}
\subsubsection{N@C$_{60}$: discovery and characteristics}
Fullerene C$_{60}$ is a hollow, closed-cage molecule of carbon with a dimension of ${\sim}$1 nm, in which atoms of the other elements can be accommodated. Atoms of lanthanides and some transition metals are encapsulated in larger fullerenes such as C$_{82}$ by carbon arc using a metal-carbon composite rod as an electrode. Encapsulation of atoms in a C$_{60}$ cage of icosahedral symmetry has turned out to require another method, i.e., ion implantation. Li@C$_{60}$ \cite{CampbellAdvMater1999} is a prototype of ion-implanted fullerene molecules (M@C$_{60}$ depicts M inside a C$_{60}$ fullerene cage) \cite{ChaiJPC1991}, while N@C$_{60}$ is an unique member among the endohedral fullerenes \cite{AlmeidaMurphyPRL1996}.

Unlike metallofullerenes having a metal atom locating close to inner walls of the carbon cage, the nitrogen atom in C$_{60}$ is believed to be located at the center of symmetry of the C$_{60}$ molecule. Moreover, the electron spin, $S=3/2$, of atomic nitrogen is retained even in the carbon cage. Furthermore, the phase coherence time in the electronic ground state can be longer than 0.2 ms as revealed by electron paramagnetic resonance (EPR) \cite{MortonJCP2006}. With all these characteristics in mind, one can believe that the interaction between the trapped atom and the hosting cage should be minimized, thus enjoyed by further experimental studies. Once a crystalline form of pure N@C$_{60}$ is obtained, the density of atomic nitrogen amounts to 2.0${\times}{10}^{21}$ cm$^{-3}$, each atom being well isolated as a spin carrier for coherent phenomena.
\begin{figure}
\begin{center}
\includegraphics[width=0.5\textwidth]{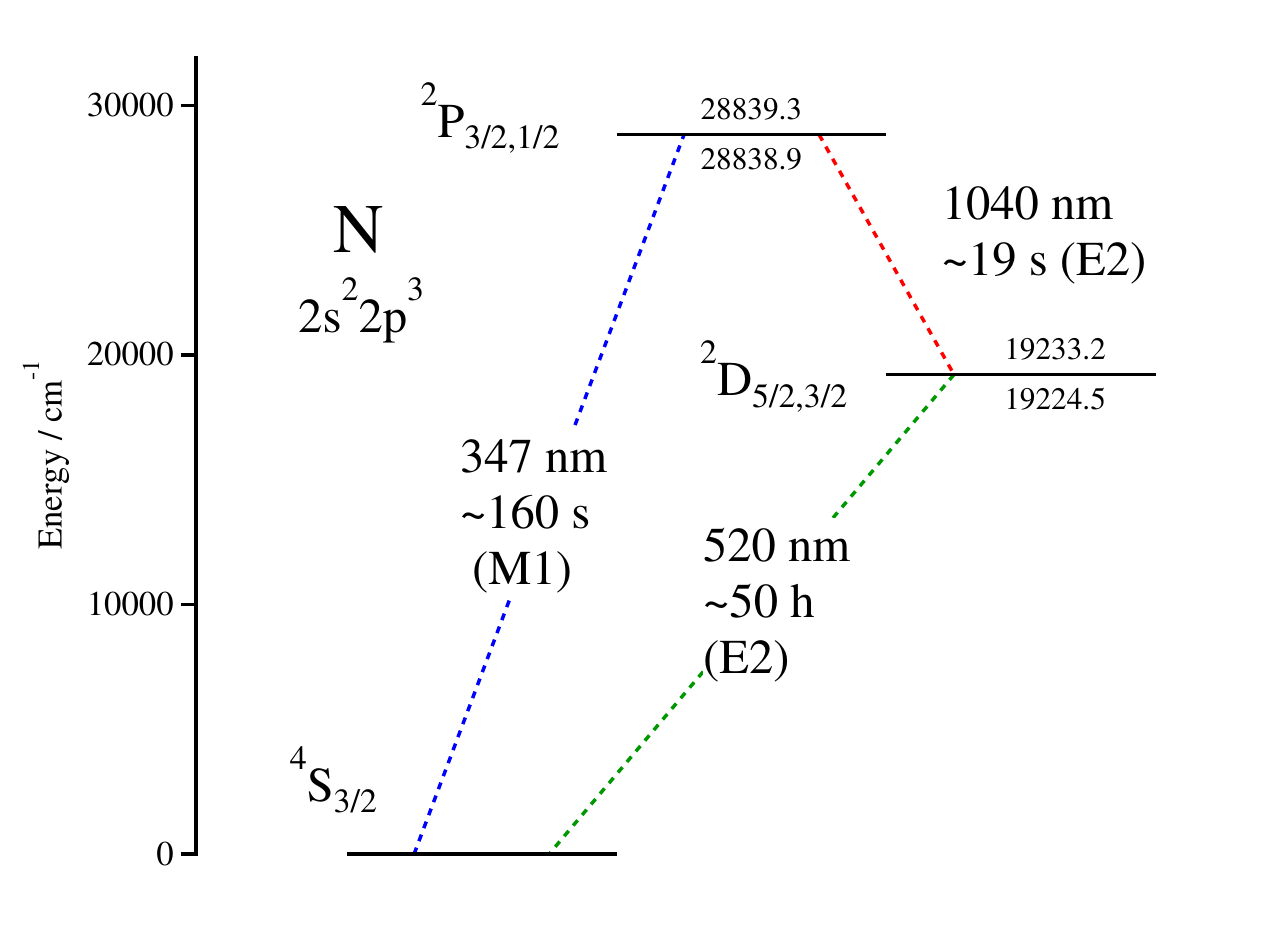}
\end{center}
\caption{Level diagram of nitrogen atom (gas phase).}
\label{fN1}
\end{figure}

\subsubsection{Electronic states of atomic nitrogen}
Figure \ref{fN1} illustrates atomic energy levels for some low-lying electronic states of free nitrogen atom. The excited states, $^{2}{\rm P}_{3/2,1/2}$ and $^{2}{\rm D}_{5/2,3/2}$, as well as the ground state, $^{4}{\rm S}_{3/2}$, are stemming from the same electron configuration of ${\rm 2s}^{2}{\rm 2p}^{3}$. Therefore, transition between any pair of states is forbidden by E1 mechanism. To excite N atom in its $^{2}{\rm P}$ or $^{2}{\rm D}$ states, some higher excited states to which the E1 transition is allowed from the ground state should be involved. Note that transitions between the low-lying electronic states are possible by M1 and/or E2 mechanism, though the rate of these transitions is relatively slow. 

The excitation channel by E1 mechanism is realized by vacuum-UV transitions of 3s${\leftarrow}$2p (120 nm and 113 nm). Pumping to the higher level of $^{4}{\rm P}_{5/2,3/2,1/2}$ is followed by relaxation to the low-lying excited state of $^{2}{\rm P}_{3/2,1/2}$ or $^{2}{\rm D}_{5/2,3/2}$ via transitions induced by spin-orbit interaction. When the incoherent energy loss is minimized, population at the low-lying excited state increases by repetitive pumping at a high rate.

\subsubsection{Preparation of N@C$_{60}$}
Formation of N@C$_{60}$ is conducted under vacuum where C$_{60}$ molecules are bombarded with positively charged ions of nitrogen, which are accelerated to have a kinetic energy of several tens of electron volts. Eventually, some are trapped inside the C$_{60}$ cage and extracted by solvents as a neutral molecule of N@C$_{60}$. The raw material after ion implantation contains at most one thousandth of N@C$_{60}$ in C$_{60}$. Therefore, separation of N@C$_{60}$ from C$_{60}$ is crucial.
\begin{figure}
\begin{minipage}{7.25cm}
	\begin{center}
	\includegraphics[width=\textwidth]{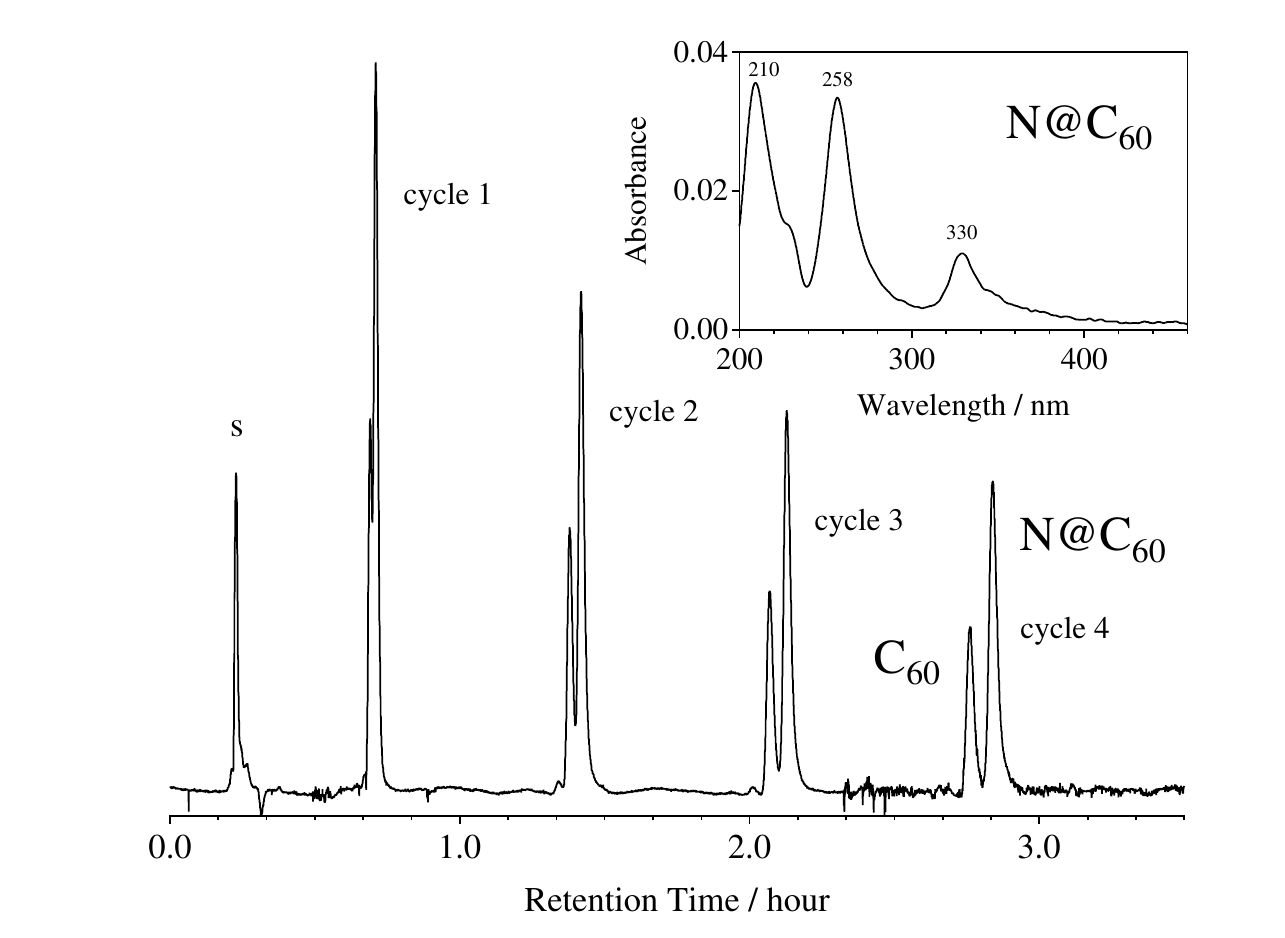}
	\end{center}
	\caption{HPLC chromatogram for separation of N@C$_{60}$ and C$_{60}$, i.e., the 6th recycling step using 5PBB (${\phi}$10${\times}$250 mm${\times}$3 columns), toluene (3.0 mL min$^{-1}$), and a UV detector (333 nm). Inset shows UV-vis absorption spectrum of the fraction at 2h52m in the main panel. Using absorption coefficient for C$_{60}$ (4.8${\times}{\rm 10}^{5}$ L mol$^{-1}$ cm$^{-1}$ at 330 nm) and the volume of the solution (2.5 mL), the amount of isolated N@C$_{60}$ is estimated to be 3${\times}{\rm 10}^{14}$ molecules.}
	\label{fN2}
\end{minipage}
\begin{minipage}{0.5cm}$\;$\end{minipage}
\begin{minipage}{7.25cm}
	\begin{center}
	\includegraphics[width=\textwidth]{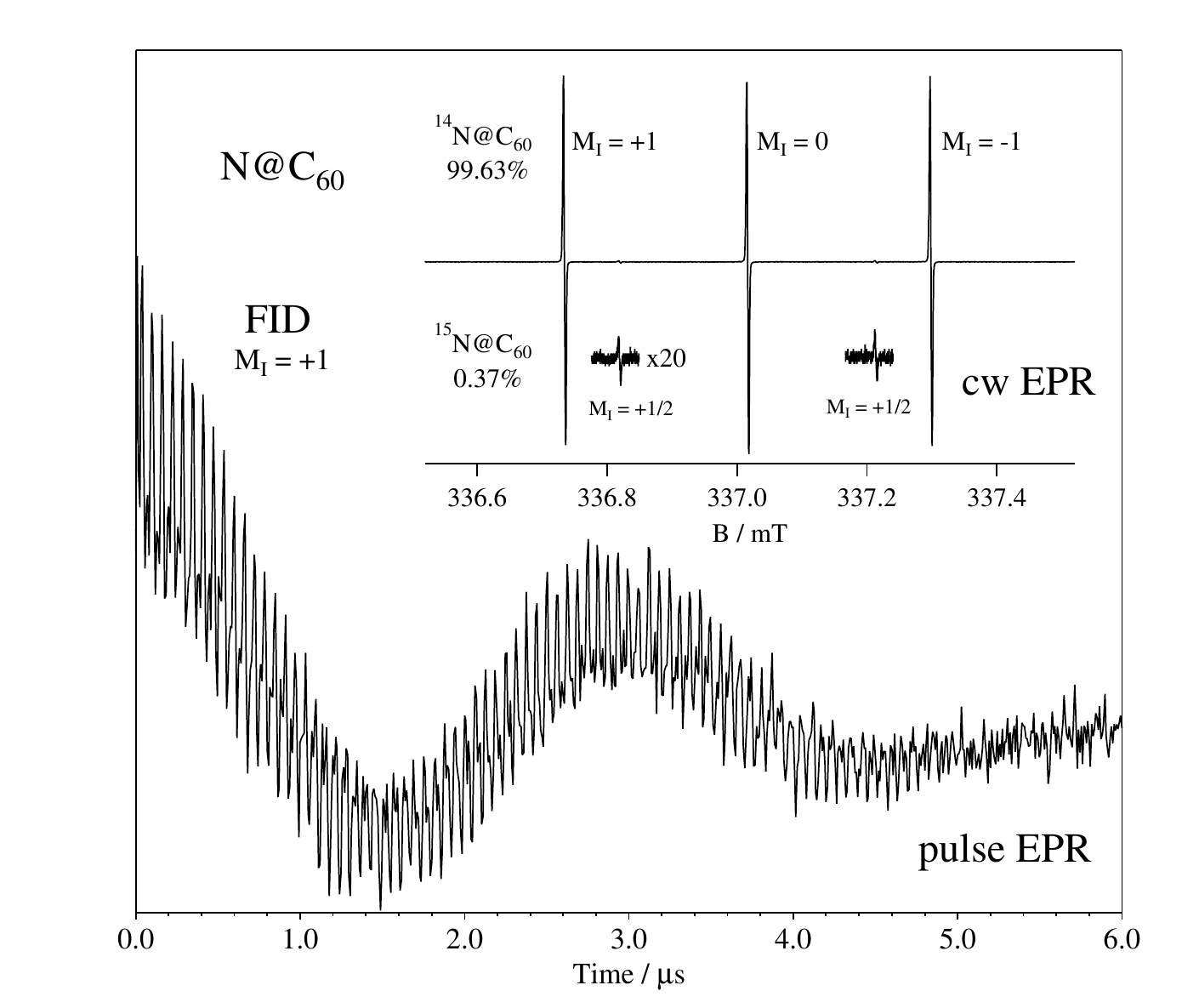}
	\end{center}
	\caption{FID signal for the $M_I=+1$ transition of $^{14}{\rm N@C}_{60}$ ($I=1$) obtained from X-band pulse-EPR measurement at room temperature for the powder sample of N@C$_{60}$/C$_{60}$ mixture (main panel). Inset shows X-band cw-EPR spectrum of purified N@C$_{60}$ in carbon disulfide. The triplet with a hyperfine splitting of 15.8 MHz belongs to $^{14}{\rm N@C}_{60}$ ($I=1$), while the weak doublet with a splitting of 22.1 MHz belongs to $^{15}{\rm N@C}_{60}$ ($I=1/2$).}
	\label{fN3}
\end{minipage}
\end{figure}

We succeeded in isolation of N@C$_{60}$ using a recycling-HPLC system. Figure \ref{fN2} shows a chromatogram for the final step of separation of N@C$_{60}$ and C$_{60}$. Due to the faint difference in interaction between the molecule and the surface of the porous material in the 5PBB column, it takes a few percent longer times for N@C$_{60}$ to pass through the column than those for C$_{60}$. After the repetitive passage of four cycles, the fraction of N@C$_{60}$ was collected for the measurement of optical properties.

The inset in Fig. \ref{fN2} shows UV-vis absorption spectrum of N@C$_{60}$ after purification. The three major bands in the UV are essentially the same as those for empty C$_{60}$, indicating that the interaction between the N atom and the C$_{60}$ molecule is not discernible at all. Since the excited states for N@C$_{60}$, $^{2}{\rm P}$ and $^{2}{\rm D}$, are not accessible directly by the transition of E1 mechanism, corresponding absorption lines are not observed. It is natural to consider that the low-lying excited states of atomic nitrogen in a C$_{60}$ cage remain intact as those in the gas phase. 

\subsubsection{EPR detection}
Here, the EPR properties in the ground state are described briefly to demonstrate the significance of N@C$_{60}$. In the inset in Fig. \ref{fN3}, three narrow lines by cw-EPR measurement correspond to the transitions between different hyperfine states in $^{4}{\rm S}_{3/2}$ of $^{14}{\rm N@C}_{60}$ at room temperature in a solution of carbon disulfide. Among the total 9 lines for the EPR transitions, three lines of constant $M_I=+1,0$, or ${-1}$ are overlapping in each of the three lines in the spectrum. This indicates that there is no noticeable anisotropy for the electron spin in N@C$_{60}$.

The main panel in Fig. \ref{fN3} displays free induction decay (FID) for the pulse-EPR signal for a powder sample of N@C$_{60}$/C$_{60}$ mixture, tuned close to the transition of the $M_I=+1$ line at lower field. The moderate wiggle of a period of ${\sim}$3 ${\mu}$s is due to detuning from the line of $M_I=+1$, while the rapid modulation is due to interference by the signals of $M_I=0$ and $M_I=-1$ which are separated by 15.8 and 31.6 MHz from the $M_I=+1$ line, respectively. Owing to the narrow linewidth even in a solid state, coherence can be maintained at least for several microseconds.

In conclusion, we have developed a system for preparation of purified N@C$_{60}$ in a microgram order, ${\sim}{\rm 10}^{14}$ molecules. UV-vis absorption spectra indicate that the nitrogen atom in a C$_{60}$ cage stays intact as that in the gas phase. The excited-state properties are intriguing in view of the coherent phenomena induced by radiation field.

\vfill\pagebreak

\end{document}